\documentclass[11pt]{article}

\usepackage{booktabs}
\usepackage{fancyhdr}
\usepackage{threeparttable}
\usepackage[table]{xcolor}
\usepackage[colorlinks,bookmarksopen,bookmarksnumbered,citecolor=blue,urlcolor=blue,linkcolor=blue]{hyperref}
\usepackage{float}
\usepackage{amsmath}

\newcommand{\rank}{\mathrm{rank}}

\usepackage{amsthm,amsmath,enumerate,amsbsy,amsfonts,amssymb,mathabx,amscd,graphicx,algorithm, indentfirst}
\usepackage[authoryear]{natbib}

\usepackage{color}
\usepackage{lscape}
\usepackage{rotating}
\usepackage[compact]{titlesec}
\usepackage{lineno}
\usepackage{cancel}
\usepackage{multirow}
\usepackage{setspace}

\definecolor{mygray}{gray}{0.6}

\usepackage{authblk, caption, subcaption}
\usepackage{adjustbox}
\usepackage{enumitem}
\usepackage{mathrsfs}
\usepackage{epsfig,morefloats}
\newtheorem{definition}{Definition}
\newtheorem{theorem}{Theorem}
\newtheorem{theoremprime}{Theorem}[theorem]

\newtheorem{corollary}{Corollary}
\newtheorem{corollaryprime}{Corollary}[corollary]

\newtheorem{lemma}{Lemma}

\newtheorem{proposition}{Proposition}
\newtheorem{remark}{Remark}

\newtheorem{condition}{Condition}
\newtheorem{assumption}{Assumption}
\newtheorem{assumptionprime}{Assumption}[assumption]

\newcommand{\bzero}{\boldsymbol 0}
\newcommand{\bone}{\boldsymbol 1}
\newcommand{\bLambda}{\boldsymbol \Lambda}

\newcommand{\bOmega}{\boldsymbol \Omega}
\newcommand{\bSigma}{\boldsymbol \Sigma}

\newcommand{\bGamma}{\boldsymbol \Gamma}
\newcommand{\bTheta}{\boldsymbol \Theta}

\newcommand{\bXi}{\boldsymbol \Xi}
\newcommand{\bpsi}{\boldsymbol \psi}

\newcommand{\bphi}{\boldsymbol \phi}
\newcommand{\bPhi}{\boldsymbol \Phi}
\newcommand{\bmu}{\boldsymbol \mu}
\newcommand{\bxi}{\boldsymbol \xi}

\newcommand{\bgamma}{\boldsymbol \gamma}

\newcommand{\bbeta}{\boldsymbol \beta}
\newcommand{\bnu}{\boldsymbol \nu}
\newcommand{\bbbeta}{\boldsymbol \eta}

\newcommand{\bvarphi}{\boldsymbol \varphi}

\newcommand{\bvarepsilon}{\boldsymbol \varepsilon}
\newcommand{\bzeta}{\boldsymbol \zeta}

\newcommand{\bchi}{\boldsymbol \chi}

\newcommand{\bkappa}{\boldsymbol \kappa}

\DeclareMathOperator*{\esssup}{ess\,sup}

\newcommand{\be}{{\mathbf e}}
\newcommand{\bbf}{{\mathbf f}}
\newcommand{\bx}{{\mathbf x}}
\newcommand{\by}{{\mathbf y}}
\newcommand{\bu}{{\mathbf u}}
\newcommand{\bv}{{\mathbf v}}

\newcommand{\bw}{{\mathbf w}}
\newcommand{\bg}{{\mathbf g}}

\newcommand{\bz}{{\mathbf z}}
\newcommand{\bbb}{{\mathbf b}}

\newcommand{\bq}{{\mathbf q}}

\newcommand{\bD}{{\bf D}}
\newcommand{\bA}{{\bf A}}
\newcommand{\bB}{{\bf B}}
\newcommand{\bC}{{\bf C}}
\newcommand{\bE}{{\bf E}}
\newcommand{\bI}{{\bf I}}
\newcommand{\bK}{{\bf K}}
\newcommand{\bP}{{\bf P}}
\newcommand{\bS}{{\bf S}}

\newcommand{\bY}{{\bf Y}}
\newcommand{\bZ}{{\bf Z}}
\newcommand{\bR}{{\bf R}}
\newcommand{\bU}{{\bf U}}
\newcommand{\bV}{{\bf V}}
\newcommand{\bQ}{{\bf Q}}

\newcommand{\bM}{{\bf M}}
\newcommand{\bF}{{\bf F}}
\newcommand{\bH}{{\bf H}}
\newcommand{\bL}{{\bf L}}
\newcommand{\bG}{{\bf G}}

\newcommand{\cA}{\scriptscriptstyle {\cal A}}
\newcommand{\cD}{\scriptscriptstyle {\cal D}}
\newcommand{\cF}{\scriptscriptstyle {\cal F}}
\newcommand{\cT}{\scriptscriptstyle {\cal T}}
\newcommand{\sS}{\scriptscriptstyle {\cal S}}
\newcommand{\sL}{\scriptscriptstyle {\cal L}}

\newcommand{\cL}{{\cal L}}
\newcommand{\cM}{{\cal M}}

\newcommand{\cU}{{\cal U}}

\newcommand{\cS}{{\cal S}}

\newcommand{\cR}{{\cal R}}

\newcommand{\cN}{{\cal N}}

\newcommand{\eZ}{\mathbb{Z}}
\newcommand{\eR}{\mathbb{R}}
\newcommand{\eH}{\mathbb{H}}
\newcommand{\eE}{\mathbb{E}}
\newcommand{\eS}{\mathbb{S}}

\newcommand{\eP}{\mathbb{P}}

\newcommand{\cov}{\text{Cov}}
\newcommand{\var}{\text{Var}}

\newcommand{\tr}{\mbox{tr}}
\newcommand{\diag}{\mbox{diag}}

\newcommand{\du}{{\rm d}u}
\newcommand{\dv}{{\rm d}v}
\newcommand{\dz}{{\rm d}z}
\newcommand{\tKer}{{\rm Ker}}
\newcommand{\tIm}{{\rm Im}}

\def\T{{ \mathrm{\scriptscriptstyle T} }}

\def\F{{ \mathrm{\scriptstyle F} }}

\setcitestyle{aysep={,},yysep={;}}

\allowdisplaybreaks
\numberwithin{remark}{section}

\makeatletter
\newcommand*{\rom}[1]{\expandafter\@slowromancap\romannumeral #1@}
\newcommand{\figcaption}{\def\@captype{figure}\caption}
\newcommand{\tabcaption}{\def\@captype{table}\caption}
\makeatother

\makeatletter
\DeclareRobustCommand\widecheck[1]{{\mathpalette\@widecheck{#1}}}
\def\@widecheck#1#2{%
	\setbox\z@\hbox{\m@th$#1#2$}%
	\setbox\tw@\hbox{\m@th$#1%
		\widehat{%
			\vrule\@width\z@\@height\ht\z@
			\vrule\@height\z@\@width\wd\z@}$}%
	\dp\tw@-\ht\z@
	\@tempdima\ht\z@ \advance\@tempdima2\ht\tw@ \divide\@tempdima\thr@@
	\setbox\tw@\hbox{%
		\raise\@tempdima\hbox{\scalebox{1}[-1]{\lower\@tempdima\box
				\tw@}}}%
	{\ooalign{\box\tw@ \cr \box\z@}}}
\makeatother
\RequirePackage[colorlinks,citecolor=blue,urlcolor=blue]{hyperref}

\usepackage{geometry}
\newcommand{\blind}{1}

\begin{document}
	\if1\blind
	{
		\title{\bf 
			Factor-guided estimation of large covariance matrix function with conditional functional sparsity
            \hspace{.2cm}\\
		}
		\author[1]{Dong Li}
		\author[2]{Xinghao Qiao\footnote{The authors’ names are sorted alphabetically, and the corresponding author is Xinghao Qiao (xinghaoq@hku.hk).}}
		\author[1]{Zihan Wang}
		\affil[1]{\it Department of Statistics and Data Science, Tsinghua University, Beijing 100084, China}
          \affil[2]{\it Faculty of Business and Economics, The University of Hong Kong, Hong Kong SAR}
		\setcounter{Maxaffil}{0}
		
		\renewcommand\Affilfont{\itshape\small}
		\date{\vspace{-5ex}}
		\maketitle
	} \fi
	\if0\blind
	{
		\bigskip
		\bigskip
		\bigskip
		\begin{center}
			{\Large\bf Factor-guided estimation of large covariance matrix function with conditional functional sparsity\footnote{Author names are sorted alphabetically.}
		\end{center}
		\medskip
	} \fi

	\setcounter{Maxaffil}{0}
	\renewcommand\Affilfont{\itshape\small}
\bigskip

\begin{abstract}
This paper addresses the fundamental task of estimating covariance matrix functions for high-dimensional functional data/functional time series. We consider two functional factor structures encompassing either functional factors with scalar loadings or scalar factors with functional loadings, and postulate functional sparsity on the covariance of idiosyncratic errors after taking out the common unobserved factors. To facilitate estimation, we rely on the spiked matrix model and its functional generalization, and derive some novel asymptotic identifiability results, based on which we develop DIGIT and FPOET estimators under two functional factor models, respectively. Both estimators involve performing associated eigenanalysis to estimate the covariance of common components, followed by adaptive functional thresholding applied to the residual covariance. We also develop functional information criteria for model selection with theoretical guarantees. The convergence rates of involved estimated quantities are respectively established for DIGIT and FPOET estimators. Numerical studies including extensive simulations and a real data application on functional portfolio allocation are conducted to examine the finite-sample performance of the proposed methodology.
\end{abstract}

\bigskip
\noindent%


\newpage
\onehalfspacing
\section{Introduction}
\label{sec.intro}

With advancements in data collection technology, multivariate functional data/functional time series are emerging in a wide range of scientific and economic applications. 
Examples include different types of brain imaging data in neuroscience, intraday return trajectories for a collection of stocks, age-specific mortality rates across different countries, and daily energy consumption curves from thousands of households, among others. Such data can be represented as
$\by_t(\cdot)=\{y_{t1}(\cdot), \dots, y_{tp}(\cdot)\}^\T$ defined on a compact interval $\cU,$ with the marginal- and cross-covariance operators induced from the associated kernel functions. These operators together form the operator-valued covariance matrix, which is also referred to as the following covariance matrix function for notational simplicity:
$$\bSigma_y=\{\Sigma_{y,jk}(\cdot,\cdot)\}_{p \times p}, ~~~\Sigma_{y,jk}(u,v)=\cov\{y_{tj}(u), y_{tk}(v)\},~~(u,v) \in \cU^2,$$
and we observe stationary $\by_t(\cdot)$ for $t=1, \dots, n.$

The estimation of the covariance matrix function and its inverse is of paramount importance in multivariate functional data/functional time series analysis. An estimator of $\bSigma_y$ is not only of interest in its own right but also essential for subsequent analyses, such as dimension reduction and modeling of $\{\by_t(\cdot)\}.$ Examples include multivariate functional principal components analysis (MFPCA) \cite[]{happ2018}, functional risk management, multivariate functional linear regression \cite[]{chiou2016} and functional linear discriminant analysis \cite[]{xue2024optimal}. 
See Section~\ref{sec.app} for details of some applications. In increasingly available high-dimensional settings where the dimension $p$ diverges with, or is larger than, the number of independent or serially dependent observations $n,$ the sample covariance matrix function $\widehat\bSigma_y^{\sS}$ performs poorly and some regularization is needed. \cite{fang2024adaptive} pioneered this effort by assuming approximate functional sparsity in $\bSigma_y$, where the Hilbert--Schmidt norms of some $\Sigma_{y,jk}$'s are assumed zero or close to zero. Then they applied adaptive functional thresholding to the entries of $\widehat\bSigma_y^{\sS}$ to achieve a consistent estimator of $\bSigma_y.$

Such functional sparsity assumption, however, is restrictive or even unrealistic for many datasets, particularly in finance and economics, where variables often exhibit high correlations.
E.g., in the stock market, the co-movement of intraday return curves \cite[]{horvath2014testing} is typically influenced by a small number of common market factors, leading to highly correlated functional variables. To alleviate the direct imposition of sparsity assumption, we employ the functional factor model (FFM) framework for $\by_t(\cdot),$ which decomposes it into two uncorrelated components, one common $\bchi_t(\cdot)$ driven by low-dimensional latent factors and one idiosyncratic $\bvarepsilon_t(\cdot).$ We consider two types of FFM. The first type, explored in \cite{guo2025factor}, admits the representation with functional factors and scalar loadings:
	\begin{equation}
		\label{eq.model_1}	\by_t(\cdot)=\bchi_t(\cdot)+\bvarepsilon_t(\cdot)=\bB\bbf_t(\cdot)+\bvarepsilon_t(\cdot),~~t=1, \dots, n,
	\end{equation}
	where $\bbf_t(\cdot)$ is a latent stationary $r$-vector of functional factors, $\bB$ is a $p \times r$ matrix of factor loadings and $\bvarepsilon_t(\cdot)$ is a $p$-vector of idiosyncratic errors.
The second type, introduced by \cite{hallin2023factor}, involves scalar factors and functional loadings:
	\begin{equation}
		\label{eq.model_2}\by_t(\cdot)=\bkappa_t(\cdot)+\bvarepsilon_t(\cdot)=\bQ(\cdot)\bgamma_t+\bvarepsilon_t(\cdot), ~~t=1, \dots, n,
	\end{equation}
where $\bgamma_t$ is a latent stationary  $r$-vector of factors and $\bQ(\cdot)$ is a $p \times r$ matrix of functional factor loadings. 
We refer to $\bSigma_f,$ $\bSigma_{\chi}$, $\bSigma_{\kappa}$ and $\bSigma_{\varepsilon}$ as the covariance matrix functions of $\bbf_t,$ $\bchi_t$, $\bkappa_t$ and $\bvarepsilon_t,$ respectively.
One may consider a more generalized FFM with functional factors and operator-valued loadings, see \eqref{eq.model_3} below. However, estimating such a complex structure will introduce elevated errors when estimating the covariance matrix function. Moreover, by employing a basis expansion approach, the estimation of \eqref{eq.model_3} can be reduced to that of \eqref{eq.model_2}, see Remark~\ref{rmk.model3} below. Hence, our paper focuses on FFMs \eqref{eq.model_1} and \eqref{eq.model_2}. 

Within the FFM framework, our goal is to estimate the covariance matrix function $\bSigma_y=\bSigma_{\chi} + \bSigma_\varepsilon$ for model~\eqref{eq.model_1} (or $\bSigma_y=\bSigma_{\kappa} + \bSigma_\varepsilon$ for model~\eqref{eq.model_2}). 
Inspired by \cite{fan2013}, we impose the approximately functional sparsity assumption on $\bSigma_{\varepsilon}$ instead of $\bSigma_y$ directly giving rise to the conditional functional sparsity structure in models~\eqref{eq.model_1} and \eqref{eq.model_2}. To effectively separate $\bchi_t(\cdot)$ (or $\bkappa_t(\cdot)$) from $\bvarepsilon_t(\cdot),$ we rely on the spiked matrix model \cite[]{fan2021recent} and its functional generalization, i.e., a large nonnegative definite matrix or operator-valued matrix $\bLambda=\bL+\bS,$ where $\bL$ is low rank and its nonzero eigenvalues grow fast as $p$ diverges, whereas all the eigenvalues of $\bS$ are bounded or grow much slower. The spikeness pattern ensures that the large signals are concentrated on $\bL,$ which facilitates our estimation procedure. Specifically, for model~\eqref{eq.model_2}, with the decomposition
\begin{equation}
\label{decomp.dcov2}
\underbrace{\bSigma_y(\cdot,\ast)}_{\bLambda} = \underbrace{\bQ(\cdot)\cov(\bgamma_t)\bQ(\ast)^{\T}}_{\bL} + \underbrace{\bSigma_{\varepsilon}(\cdot,\ast)}_{\bS},
\end{equation}
we perform MFPCA based on $\widehat \bSigma_y^{\sS},$ then estimate $\bSigma_{\kappa}$ using the leading $r$ functional principal components and finally propose a novel adaptive functional thresholding procedure to estimate the sparse $\bSigma_{\varepsilon}.$ This results in a Functional Principal Orthogonal complEment Thresholding (FPOET) estimator, extending the POET methodology for large covariance matrix estimation \cite[]{fan2013,wang2021} to the functional domain. 
Alternatively, for model~\eqref{eq.model_1}, considering the violation of nonnegative definiteness in $\bSigma_y(u,v)$ for $u \neq v,$ we utilize the nonnegative definite doubly integrated Gram covariance:
\begin{equation}
\label{decomp.dcov}
{ \underbrace{\int\int\bSigma_y(u,v)\bSigma_y(u,v)^{\T} {\rm d}u {\rm d}v}_{\bLambda}=\underbrace{\bB\Big\{\int\int\bSigma_f(u,v)\bB^{\T}\bB\bSigma_f(u,v)^{\T} {\rm d}u {\rm d}v\Big\}\bB^{\T}}_{\bL} + \underbrace{\text{remaining terms}}_{\bS},}
\end{equation}
which is shown to be identified asymptotically as $p \to \infty.$
We propose to carry out eigenanalysis of the sample version of $\bLambda$ in (\ref{decomp.dcov}) combined with least squares to estimate $\bB,$ $\bbf_t(\cdot)$ and hence $\bSigma_{\chi},$ and then employ the same thresholding method to estimate $\bSigma_{\varepsilon}$. This yields a Doubly Integrated Gram covarIance and Thresholding (DIGIT) estimator.

The new contribution of this paper can be summarized in five key aspects.
First, though our model~\eqref{eq.model_1} shares the same form as the one in \cite{guo2025factor} and aligns with the direction of static factor models in \cite{bai2002determining} and \cite{fan2013}, substantial advances have been made in our methodology and theory: 
(i) We allow weak serial correlations in idiosyncratic components $\bvarepsilon_t(\cdot)$
rather than assuming the white noise. 
(ii) Unlike the autocovariance-based method \cite[]{guo2025factor} for serially dependent data, we leverage the covariance information to propose a more efficient estimation procedure that encompasses independent observations as a special case.
(iii) More importantly, under the pervasiveness assumption, we establish novel asymptotic identifiability in (\ref{decomp.dcov}), where the first $r$ eigenvalues of $\bL$ grow at rate $O(p^2),$ whereas all the eigenvalues of $\bS$ diverge at a rate slower than $O(p^2).$ 

Second, for model~\eqref{eq.model_2}, we extend the standard asymptotically identified covariance decomposition in \cite{bai2002determining} to the functional domain, under the functional counterpart of the pervasiveness assumption. 
Building upon these findings, we provide mathematical insights when the functional factor analysis for models \eqref{eq.model_1} and \eqref{eq.model_2}
and the proposed eigenanalysis of the respective $\bLambda$'s in (\ref{decomp.dcov2}) and (\ref{decomp.dcov}) are approximately the same for high-dimensional functional data. 
Differing from the least squares method in \cite{hallin2023factor}, we develop a novel MFPCA method to estimate model~\eqref{eq.model_2} and also establish the theoretical equivalence of the covariance matrix function estimators based on two methods.

Third, we develop a new adaptive functional thresholding approach to estimate sparse $\bSigma_{\varepsilon}.$ Compared to the competitor in \cite{fang2024adaptive}, our approach requires weaker assumptions while achieving similar finite-sample performance. 
Fourth, with the aid of such thresholding technique combined with our estimation of FFMs~\eqref{eq.model_1} and \eqref{eq.model_2}, we propose two factor-guided covariance matrix function estimators, DIGIT and FPOET, respectively. We derive the associated convergence rates of estimators for $\bSigma_{\varepsilon},$ $\bSigma_y$ and its inverse
under various functional matrix norms. 
Additionally, we introduce fully functional information criteria to select the more suitable model between FFMs~\eqref{eq.model_1} and \eqref{eq.model_2} with theoretical guarantees. 
Last but not least, we establish a new functional risk management framework to account for uncertainties in intraday returns of financial data, where our proposed estimators can be applied.

The rest of the paper is organized as follows. Section \ref{sec.method} presents the corresponding procedures for estimating $\bSigma_y$ under two FFMs and the information criteria used for model selection. 
Section~\ref{sec.theory} provides the asymptotic theory for the estimated quantities. Section~\ref{sec.app} discusses a couple of applications of the proposed estimation. We assess the finite-sample performance of our proposal through extensive simulations in Section~\ref{sec.sim} and a financial data application in Section~\ref{sec.real}. Section~\ref{sec.discussion} discusses some future work.

Throughout the paper, for any matrix $\bM=(M_{ij})_{p \times q},$ we denote its matrices $\ell_1$ norm, $\ell_{\infty}$ norm, operator norm, Frobenius norm and elementwise
$\ell_{\infty}$ norm by $\|\bM\|_1 = \max_j \sum_{i} |M_{ij}|,$ $\|\bM\|_{\infty} = \max_i \sum_{j} |M_{ij}|,$ $\|\bM\|=\lambda_{\max}^{1/2}(\bM^\T\bM),$
$\|\bM\|_{\F}=(\sum_{i,j}M_{ij}^2)^{1/2}$ and $\|\bM\|_{\max}=\max_{i,j} |M_{ij}|,$ respectively.
Let $\eH=L_2(\cU)$ be the Hilbert space of squared integrable functions defined on the compact set $\cU$. We denote its $p$-fold Cartesian product by $\eH^p=\eH\times\cdots\times \eH$ and tensor product by $\mathbb{S}=\eH\otimes \eH.$
For $\bbf=(f_1, \dots, f_p)^{\T},\bg=(g_1, \dots, g_p)^{\T}\in\eH^p,$ we denote the inner product by $\langle\bbf,\bg\rangle=\int_{\cU}\bbf(u)^{\T}\bg(u)\du$ with induced norm $\Vert\cdot\Vert=\langle\cdot,\cdot\rangle^{1/2}.$
For an integral matrix operator $\bK: \eH^p \to \eH^q$ induced from the kernel matrix function $\bK=\{K_{ij}(\cdot,\cdot)\}_{q\times p}$ with each $K_{ij} \in {\mathbb S},$ $\bK(\bbf)(\cdot)=\int_{\cU} \bK(\cdot, u) \bbf(u) {\rm d}u \in {\eH}^q$ for any given $\bbf \in \eH^p.$ For notational economy, we will use $\bK$ to denote both the kernel function and the operator. We define the functional version of matrix $\ell_1$ norm by $\Vert\bK\Vert_{\cS,1}=\max_{j}\sum_i\Vert K_{ij}\Vert_{\cS}$, where, for each $K_{ij}\in \mathbb{S}$, we denote its Hilbert–Schmidt norm by 
$\Vert K_{ij}\Vert_{\cS}=\{\int\int K_{ij}(u,v)^2{\rm d}u{\rm d}v\}^{1/2}$.
Similarly, we define $\Vert\bK\Vert_{\cS,\infty}=\max_{i}\sum_{j}\Vert K_{ij}\Vert_{\cS}$, $\Vert\bK\Vert_{\cS,\F}=\{\sum_{i,j}\Vert K_{ij}\Vert_{\cS}^2\}^{1/2}$ and $\Vert\bK\Vert_{\cS,\max}=\max_{i,j}\Vert K_{ij}\Vert_{\cS}$ as the functional versions of matrix $\ell_{\infty}$, Frobenius and elementwise $\ell_{\infty}$ norms, respectively. We define the operator norm by $\Vert\bK\Vert_{\cal L}=\sup_{\bx\in\mathbb{H}^p,\|\bx\| \leq 1}\Vert\bK (\bx)\Vert$ and the trace norm by $\Vert\bK\Vert_{\cN}=\tr\{\int\bK(u,u)\du\}$ for $p=q.$ For a positive integer $m,$ write $[m]=\{1, \dots, m\}$ and denote by $\bI_m$ the identity matrix of size $m \times m.$ For $x,y \in {\mathbb R},$ we use $x \wedge y = \min(x,y).$
For two positive sequences $\{a_n\}$ and $\{b_n\}$, we write $a_n\lesssim b_n$ or $a_n=O(b_n)$ or $b_n\gtrsim a_n$ if there exists a positive constant $c$ such that $a_n/b_n\le c,$ and $a_n=o(b_n)$ if $a_n/b_n \to 0.$ We write $a_n\asymp b_n$ if and only if $a_n\lesssim b_n$ and $a_n\gtrsim b_n$ hold simultaneously.

\section{Methodology}
\label{sec.method}

\subsection{FFM with functional factors}
\label{subsec.factor_model}
Suppose that $\by_t(\cdot)$ admits FFM representation \eqref{eq.model_1}, where $r$ common functional factors in $\bbf_t(\cdot) = \{f_{t1}(\cdot), \dots, f_{tr}(\cdot)\}^{\T}$ are uncorrelated with the idiosyncratic errors $\bvarepsilon_t(\cdot) = \{\varepsilon_{t1}(\cdot), \dots, \varepsilon_{tp}(\cdot)\}^{\T}$ and $r$ is assumed to be fixed. Then we have 
\begin{equation}
\label{eq.decom}
\bSigma_{y}(u,v) =  \bB \bSigma_{f}(u,v) \bB^{\T} + \bSigma_{\varepsilon}(u,v),~~(u,v) \in \cU^2,
\end{equation}
which is not nonnegative definite for some $u,v.$ To ensure nonnegative definiteness and accumulate covariance information as much as possible, we propose to perform an eigenanalysis of doubly integrated Gram covariance:
\begin{equation}
\label{DIG.decompose}
\bOmega=\int \int \bSigma_{y}(u,v)\bSigma_{y}(u,v)^{\T}{\rm d}u{\rm d}v \equiv\bOmega_{\cL}+\bOmega_{\cR} ,
\end{equation}
where $\bOmega_{\cR}=\int \int \bSigma_{\varepsilon}(u,v)\bSigma_{\varepsilon}(u,v)^{\T} {\rm d}u{\rm d}v+\int \int \bB\bSigma_{f}(u,v) \bB^{\T} \bSigma_{\varepsilon}(u,v)^{\T} {\rm d}u{\rm d}v + \int \int \bSigma_{\varepsilon}(u,v) \bB\bSigma_{f}(u,v)^{\T}\bB^{\T} {\rm d}u{\rm d}v$ and $\bOmega_{\cL}=\bB\{\int \int\bSigma_{f}(u,v)\bB^{\T}\bB\bSigma_{f}(u,v)^{\T} {\rm d}u{\rm d}v\}\bB^{\T}$.
To make the loading matrix $\bB$ asymptotically identifiable in the decomposition (\ref{DIG.decompose}), we impose the following condition.

\begin{assumption}
\label{ass.ind1}
$p^{-1}\bB^{\T}\bB=\bI_{r}$  and 
$\int\int\bSigma_{f}(u,v)\bSigma_{f}(u,v)^{\T}{\rm d}u{\rm d}v=\textup{diag}(\theta_1, \dots, \theta_{r}),$ where there exist some constants $\overline{\theta}>\underline{\theta}>0$ such that $\overline{\theta}>\theta_1>\theta_2>\cdots>\theta_{r}>\underline{\theta}.$ 
\end{assumption}

\begin{remark}
\label{rmk.identif}
Model~\eqref{eq.model_1} exhibits an identifiable issue as it remains unchanged if $\{\bB,\bbf_t(\cdot)\}$ is replaced by $\{\bB\bU,\bU^{-1}\bbf_t(\cdot)\}$ for any invertible matrix $\bU.$ \cite{bai2002determining} assumed two types of normalization for the scalar factor model: one is $p^{-1}\bB^{\T}\bB=\bI_r$ and the other is $\cov(\bbf_t)=\bI_p.$ We adopt the first type for model \eqref{eq.model_1} to simplify the calculation of the low rank matrix $\bOmega_{\cL}$ in (\ref{DIG.decompose}). However, this constraint alone is insufficient to identify $\bB,$ but the space spanned by the columns of $\bB=(\bbb_1, \dots, \bbb_r).$ Hence, we introduce an additional constraint based on the diagonalization of $\int\int\bSigma_{f}(u,v)\bSigma_{f}(u,v)^{\T}\du\dv,$ which is ensured by the fact that any nonnegative-definite matrix can be orthogonally diagonalized. Under Assumption~\ref{ass.ind1}, we can express $\bOmega_{\cL}=\sum_{i=1}^{r}p\theta_i\bbb_i\bbb_i^{\T}$, implying that $\Vert\bOmega_{\cL}\Vert\asymp\Vert\bOmega_{\cL}\Vert_{\min}\asymp p^2$.
\end{remark}

We now elucidate why performing eigenanalysis of $\bOmega$ can be employed for functional factor analysis under model~\eqref{eq.model_1}.
Write $\widetilde{\bB}=p^{-1/2}\bB=(\widetilde{\bbb}_1,\cdots,\widetilde{\bbb}_r),$ which satisfies $\widetilde{\bB}^{\T}\widetilde{\bB}=\bI_r.$ Under Assumption~\ref{ass.ind1}, it holds that
$\bOmega_{\cL}=p^2 \sum_{i=1}^{r} \theta_i \widetilde \bbb_i \widetilde \bbb_i^{\T},$ whose eigenvalue/eigenvector pairs are $\{(p^2\theta_i, \widetilde \bbb_i)\}_{i \in [r]}.$
Let $\lambda_1 \geq \dots \geq \lambda_p$ be the ordered eigenvalues of $\bOmega$ and $\bxi_1, \dots, \bxi_p$ be the corresponding eigenvectors. We then have the following proposition. 
\begin{proposition}
\label{propos.eigenvalues}
Suppose that Assumption~\ref{ass.ind1} and $\|\bOmega_{\cR}\|=o(p^2)$ hold. Then we have\\ (i) $|\lambda_j-p^2\theta_j|\le \|\bOmega_{\cR} \|$ for $j\in[r]$ and $|\lambda_j|\le \|\bOmega_{\cR} \|$ for $j \in [p] \setminus [r];$\\
(ii) $\|\bxi_j-\widetilde{\bbb}_j\|=O(p^{-2}\|\bOmega_{\cR} \|)$ for $j\in[r]$. 
\end{proposition}
		
Proposition~\ref{propos.eigenvalues} indicates that we can distinguish the leading eigenvalues $\{\lambda_j\}_{j \in [r]}$ from the remaining eigenvalues, and ensure the approximate equivalence between eigenvectors $\{\bxi_j\}_{j \in [r]}$ and the normalized factor loading columns $\{\widetilde\bbb_j\}_{j \in [r]},$ provided that $\Vert\bOmega_{\cR}\Vert=o(p^2)$. Towards this, we impose an approximately functional sparsity condition on $\bSigma_{\varepsilon}$ measured through
\begin{equation}
\label{eq.sparsity}		
 s_p=\max_{i\in[p]}\sum_{j=1}^{p}\|\sigma_i\|_{\cN}^{(1-q)/2}\Vert\sigma_j\Vert_{\cN}^{(1-q)/2}\Vert\Sigma_{\varepsilon,ij}\Vert_{\cS}^q, ~~~\text{for some}~q\in[0,1),
\end{equation}
where $\sigma_i(u)=\Sigma_{\varepsilon,ii}(u,u)$ for $u\in\cU$ and $i\in[p].$ 
Specially, when $q=0$ and $\{\Vert\sigma_i\Vert_{\cN}\}$ are bounded, $s_p$ can be simplified to the exact functional sparsity, i.e., $\max_i \sum_j I(\|\Sigma_{\varepsilon,ij}\|_{\cal S}\neq 0).$ 

\begin{remark}
\label{rmk.fs}
(i) Our proposed measure of functional sparsity in \eqref{eq.sparsity} for non-functional data degenerates to the measure of sparsity adopted in \cite{cai2011adaptive}. It is worth mentioning that \cite{fang2024adaptive} introduced a different measure of functional sparsity as
$$\tilde s_p=\max_{i\in[p]}\sum_{j=1}^{p}\Vert\sigma_i\Vert_{\infty}^{(1-q)/2}\Vert\sigma_j\Vert_{\infty}^{(1-q)/2}\Vert\Sigma_{\varepsilon,ij}\Vert_{\cS}^q,$$
where $\Vert\sigma_i\Vert_{\infty}=\sup_{u\in\cU}\sigma_i(u) \geq \|\sigma_i\|_{\cN}.$
As a result, we will use $s_p$ instead of $\tilde s_p.$ \\
(ii) With bounded $\{\Vert\sigma_i\Vert_{\cN}\},$ we can easily obtain $\Vert\bSigma_{\varepsilon}\Vert_{\cS,1}=\Vert\bSigma_{\varepsilon}\Vert_{\cS,\infty}=O(s_p),$ which, along with Lemmas~\ref{lem.ineq2}, \ref{lem.ineq3} in the appendix and Assumption~\ref{ass.ind1}, yields that
\begin{equation*}  \Vert\bOmega_{\cR} \Vert
\le \Vert\bSigma_{\varepsilon}\Vert_{\cS,\infty}\Vert\bSigma_{\varepsilon}\Vert_{\cS,1}+2\left(\Vert\bB\bSigma_{f}\bB^{\T}\Vert_{\cS,\infty}\Vert\bB\bSigma_{f}\bB^{\T}\Vert_{\cS,1}\right)^{1/2}\left(\Vert\bSigma_{\varepsilon}\Vert_{\cS,1}\Vert\bSigma_{\varepsilon}\Vert_{\cS,\infty}\right)^{1/2}=O(s_p^2+ ps_p).
\end{equation*}
Hence, when $s_p=o(p),$ Proposition~\ref{propos.eigenvalues} implies that functional factor analysis under model \eqref{eq.model_1} and eigenanalysis of $\bOmega$ are approximately the same for high-dimensional functional data.
\end{remark}

To estimate model \eqref{eq.model_1}, we assume the number of functional factors (i.e., $r$) is known, and will introduce a data-driven approach to determine it in Section~\ref{subsec.deter_num}.
Without loss of generality, we assume that $\by_t(\cdot)$ has been centered to have mean zero. The sample covariance matrix function of $\bSigma_{y}(\cdot,\cdot)$ is given by $\widehat{\bSigma}_{y}^{\sS}(u,v)=n^{-1}\sum_{t=1}^{n}\by_t(u)\by_t(v)^{\T}$.
Performing eigen-decomposition on the sample version of $\bOmega,$ 
\begin{equation}
\label{Omega.est}
\widehat{\bOmega}=\int\int\widehat{\bSigma}_{y}^{\sS}(u,v)\widehat{\bSigma}_{y}^{\sS}(u,v)^{\T}{\rm d}u{\rm d}v,
\end{equation}
leads to estimated eigenvalues
$\hat \lambda_1, \dots, \hat \lambda_p$ and their associated eigenvectors $\widehat \bxi_1, \dots, \widehat \bxi_p.$
Then the estimated factor loading matrix is
	$\widehat{\bB}=\sqrt{p}(\widehat{\bxi}_1,\dots,\widehat{\bxi}_{r})=(\widehat{\bbb}_1,\dots,\widehat{\bbb}_{r}).$

To estimate functional factors $\{\bbf_t(\cdot)\}_{t \in [n]},$ we minimize the least squares criterion 
\begin{equation}
\label{ls.crit1}
\sum_{t=1}^{n}\Vert\by_t-\widehat\bB\bbf_t\Vert^2
=\sum_{t=1}^{n}\int_{\cU}\{\by_t(u)-\widehat\bB\bbf_t(u)\}^{\T}\{\by_t(u)-\widehat\bB\bbf_t(u)\}{\rm d}u
\end{equation}
with respect to $\bbf_1(\cdot), \dots, \bbf_n(\cdot).$
Setting the functional derivatives to zero, we obtain 
the least squares estimator $\widehat\bbf_t(\cdot)=p^{-1}\widehat\bB^{\T}\by_t(\cdot)$ and estimated idiosyncratic errors in 
$\widehat{\bvarepsilon}_t(\cdot)=(\bI_p-p^{-1}\widehat{\bB}\widehat{\bB}^{\T})\by_t(\cdot)$. Hence, we can obtain sample covariance matrix functions of estimated common factors and estimated idiosyncratic errors as $\widehat{\bSigma}_{f}(u,v)=n^{-1}\sum_{t=1}^n\widehat{\bbf}_t(u)\widehat{\bbf}_t(v)^{\T}$ and $\widehat{\bSigma}_{\varepsilon}(u,v)=\{\widehat{\Sigma}_{\varepsilon,ij}(u,v)\}_{p \times p}=\sum_{t=1}^nn^{-1}\widehat{\bvarepsilon}_t(u)\widehat{\bvarepsilon}_t(v)^{\T},$ respectively. 

Since $\bSigma_{\varepsilon}$ is assumed to be functional sparse, we introduce an adaptive functional thresholding (AFT) estimator of $\bSigma_{\varepsilon}.$ To this end, we define the functional variance factors $\Theta_{ij}(u,v)=\var\{\varepsilon_{ti}(u)\varepsilon_{tj}(v)\}$ for $i,j \in [p],$ whose estimators are
	$$\widehat{\Theta}_{ij}(u,v)=\frac{1}{n}\sum_{t=1}^{n}\big\{\widehat{\varepsilon}_{ti}(u)\widehat{\varepsilon}_{tj}(v)-\widehat{\Sigma}_{\varepsilon,ij}(u,v)\big\}^2,$$
	with $\widehat{\varepsilon}_{ti}(\cdot)=y_{ti}(\cdot)-\widecheck{\bbb}_i^{\T}\widehat{\bbf}_t(\cdot)$ and $\widecheck{\bbb}_i$ being the $i$-th row vector of $\widehat{\bB}.$
We develop an AFT procedure on $\widehat \bSigma_{\varepsilon}$ using entry-dependent functional thresholds that automatically adapt to the variability of $\widehat\Sigma_{\varepsilon,ij}$'s. 
Specifically, the AFT estimator is defined as $\widehat{\bSigma}_{\varepsilon}^{\cA}=\{\widehat{\Sigma}_{\varepsilon,ij}^{\cA}(\cdot,\cdot)\}_{p\times p}$ with 
\begin{equation}
\label{eq.thre}	
\widehat{\Sigma}_{\varepsilon,ij}^{\cA}=\big\Vert\widehat{\Theta}_{ij}^{1/2}\big\Vert_{\cS} \times s_{\lambda}\Big(\widehat{\Sigma}_{\varepsilon,ij}/\big\Vert\widehat{\Theta}_{ij}^{1/2}\big\Vert_{\cS}\Big)\ {\rm with\ }\lambda=\dot{C}\big(\sqrt{\log p /n}+1/\sqrt{p}\big), 
\end{equation}
where $\dot{C}>0$ is a pre-specified constant that can be selected via multifold cross-validation and the order $\sqrt{\log p/n}+1/\sqrt{p}$ is related to the convergence rate of $\widehat{\Sigma}_{\varepsilon,ij}/\big\Vert\widehat{\Theta}_{ij}^{1/2}\big\Vert_{\cS}$ under functional elementwise $\ell_{\infty}$ norm. 
Here $s_{\lambda}$ is a functional thresholding operator with regularization parameter $\lambda\ge0$ \cite[]{fang2024adaptive} and belongs to the class  $s_{\lambda}:{\mathbb S}\to {\mathbb S}$ satisfying: (i) $\Vert s_{\lambda}(Z)\Vert_{\cS}\le c\Vert Y\Vert_{\cS}$ for all $Z,Y\in\mathbb{S}$ that satisfy $\Vert Z-Y\Vert_{\cS}\le\lambda$ and some $c>0$; (ii) $\Vert s_{\lambda}(Z)\Vert_{\cS}=0$ for $\Vert Z\Vert_{\cS}\le\lambda$; (iii) $\Vert s_{\lambda}(Z)-Z\Vert_{\cS}\le\lambda$ for all $Z\in\mathbb{S}$. This class includes functional versions of commonly-adopted thresholding functions, such as soft thresholding, SCAD \cite[]{fan2001}, 
and the adaptive lasso \cite[]{zou2006}.

\begin{remark}
By comparison, \cite{fang2024adaptive} introduced an alternative AFT estimator
\begin{equation}
\label{eq.thre_2}
\widetilde \bSigma_{\varepsilon}^{\cA}=(\widetilde \Sigma_{\varepsilon,ij}^{\cA})_{p \times p}~~\text{with}~~\widetilde{\Sigma}_{\varepsilon, ij}^{\cA}=
\widehat{\Theta}_{ij}^{1/2}\times s_{\lambda}\Big(\widehat{\Sigma}_{\varepsilon,ij}/\widehat{\Theta}_{ij}^{1/2}\Big),
\end{equation}
which uses a single threshold level to functionally threshold standardized entries $\widehat{\Sigma}_{\varepsilon,ij}/\widehat{\Theta}_{ij}^{1/2}$ across all $(i,j),$ resulting in entry-dependent functional thresholds for $\widehat{\Sigma}_{\varepsilon,ij}.$
Since $\widetilde\Sigma_{\varepsilon,jk}^{\cA}$ requires stronger assumptions (see Remark~\ref{rmk.fs} above and the remark for Assumption~\ref{ass.thres} below), we adopt the AFT estimator $\widehat\Sigma_{\varepsilon,jk}^{\cA}$, leading to comparable empirical performance (see Section~\ref{supsec.F} of the supplementary material).
\end{remark}
	
Finally, we obtain a DIGIT estimator of $\bSigma_{y}$ as
\begin{equation}
\label{eq.estimator_1}
\widehat{\bSigma}_{y}^{\cD}(u,v)=\widehat{\bB}\widehat{\bSigma}_{f}(u,v)\widehat{\bB}^{\T}+\widehat{\bSigma}_{\varepsilon}^{\cA}(u,v), ~~(u,v) \in \cU^2.
\end{equation}

\subsection{FFM with functional loadings}
\label{subsec.fun_load}
The structure of FFM is not unique. We could also assume that $\by_t(\cdot)$ satisfies FFM~\eqref{eq.model_2} with scalar factors and functional loadings $\bQ(\cdot)=\{\bq_1(\cdot),\dots,\bq_p(\cdot)\}^{\T}$ with each $\bq_i(\cdot)\in\eH^{r},$ where $r$ common scalar factors $\bgamma_t=(\gamma_{t1},\dots,\gamma_{tr})^{\T}$ are uncorrelated with idiosyncratic errors $\bvarepsilon_t(\cdot)$ and $r$ is assumed to be fixed.

\begin{remark}
\label{rmk.dif}
(i) While both FFMs assume that different components of $\by_t(\cdot)$ are defined in a common domain, FFM~\eqref{eq.model_2} can be generalized to allow different components to reside in different domains that may differ in dimensions (e.g., curves and images). However, this generalization is not feasible for FFM~\eqref{eq.model_1} as it involves integrals beyond the inner product used for FFM~\eqref{eq.model_2}. In almost all real data applications in the existing literature, the corresponding components all lie in the same domain. Therefore, we focus on a conceptually special yet practical common case as in FFMs~\eqref{eq.model_1} and \eqref{eq.model_2}.
\\
(ii) Both \eqref{eq.model_1} and ~\eqref{eq.model_2} yield useful FFMs but are tailored to tackle rather different situations. A crucial aspect of modeling functional time series is to characterize the functional and dynamic structures. FFM~\eqref{eq.model_1} assumes that both structures are inherited from functional time series factors $\{\bbf_t(\cdot)\}$ with scalar loadings for enhanced interpretability. In contrast, FFM~\eqref{eq.model_2} incorporates dynamic information through scalar time series factors $\{\bgamma_t\}$ while preserving infinite-dimensionality in functional loadings of $\bQ(\cdot)$.
\end{remark}

\begin{remark}
\label{rmk.model3}
Consider a more generalized FFM \citep{leng2024covariance} with functional factors and operator-valued loadings:
\begin{equation}	
\label{eq.model_3}
\by_t(\cdot)= \widetilde\bA(\widetilde \bbf_t)(\cdot) + \bvarepsilon_t(\cdot) = \int_\cU\widetilde\bA(\cdot, v)\widetilde \bbf_t(v) {\rm d}v 
+ \bvarepsilon_t(\cdot), ~~t \in [n],
\end{equation}
where $\widetilde \bbf_t(\cdot)$ is a latent stationary $r$-vector of functional factors and $\widetilde\bA(\cdot,\cdot)$ is a $p \times r$ operator-valued matrix. For each $j\in[r],$ we apply the basis expansion approach to the $j$-th component of $\widetilde \bbf_t(\cdot),$ 
$\tilde{f}_{tj}(\cdot)=\bphi_j(\cdot)^{\T}\widetilde\bgamma_{t} + \eta_{tj}(\cdot),$ where $\bphi_j(\cdot)$ is a $d$-vector of basis function, $\widetilde \bgamma_{t}$ is a $d$-vector of basis coefficients, and $\eta_{tj}(\cdot)$ is the truncation error. 
Let $\bPhi(\cdot)=\{\bphi_1(\cdot),\dots,\bphi_{r}(\cdot)\}^{\T},\bbbeta_t(\cdot)=\{\eta_{t1}(\cdot),\dots,\eta_{tr}(\cdot)\}^{\T}$, $\widetilde\bQ(\cdot)=\int_{\cU}\widetilde{\bA}(\cdot,v)\bPhi(v)\dv$ and $\widetilde\bvarepsilon_t(\cdot)=\bvarepsilon_t(\cdot)+\int_{\cU}\widetilde{\bA}(\cdot,v)\bbbeta_t(v)\dv.$ Then, FFM~\eqref{eq.model_3} can be rewritten as $\by_t(\cdot)=\widetilde\bQ(\cdot)\widetilde\bgamma_t+\widetilde\bvarepsilon_t(\cdot),$ which mirrors the form of FFM~\eqref{eq.model_2} with $d$ scalar factors, as the truncation errors are incorporated into the idiosyncratic components. However, the estimation procedure demands strong technical conditions, and estimating the covariance matrix function within a more complex structure can lead to increased accumulated errors.
Hence, our paper focuses on FFM~\eqref{eq.model_2} instead of FFM~\eqref{eq.model_3}. 
\end{remark}
 
Under FFM~\eqref{eq.model_2}, we have the covariance decomposition
	\begin{equation}
		\label{eq.decom_2}
		\bSigma_{y}(u,v) = \bQ(u)\bSigma_{\gamma} \bQ(v)^\T + \bSigma_{\varepsilon}(u,v),~~(u,v) \in \cU^2.
	\end{equation}By the multivariate version of Mercer’s theorem, which serves as the foundation of MFPCA \cite[]{chiou2014,happ2018}, there exists an orthonormal basis consisting of eigenfunctions $\{\bvarphi_i(\cdot)\}_{i=1}^{\infty}$ of $\bSigma_y$ and the associated eigenvalues $\tau_1 \geq \tau_2 \geq \dots \geq0$ such that
\begin{equation}
\label{eq.y_mercer}
    \bSigma_{y}(u,v)=\sum_{i=1}^{\infty}\tau_i\bvarphi_i(u)\bvarphi_i(v)^{\T},~~(u,v) \in \cU^2,
\end{equation}
where the sum converges absolutely and uniformly in both $u$ and $v.$

We now provide mathematical insights into why MFPCA can be applied for functional factor analysis under model~\eqref{eq.model_2}. 
To ensure the identifiability of the decomposition in \eqref{eq.decom_2}, we impose a normalization-type condition similar to Assumption~\ref{ass.ind1}.

\begin{assumptionprime}
\label{ass.ind2}
$\bSigma_{\gamma}=\bI_{r}$ and $p^{-1}\int\bQ(u)^{\T}\bQ(u){\rm d}u=\textup{diag}(\vartheta_1, \dots, \vartheta_{r}),$ where there exist some constants $\overline{\vartheta}>\underline{\vartheta}>0$ such that $\overline{\vartheta}>\vartheta_1>\vartheta_2>\cdots>\vartheta_{r}>\underline{\vartheta}.$  
\end{assumptionprime}

Suppose Assumption~\ref{ass.ind2} holds, and let $\widetilde{\bq}_1(\cdot), \dots, \widetilde{\bq}_{r}(\cdot)$ be the normalized columns of $\bQ(\cdot)$ such that $\Vert\widetilde{\bq}_j\Vert=1$ for $j\in[r]$. By Lemma \ref{lem.eigen_QQ} in the appendix, $\{\widetilde{\bq}_j(\cdot)\}_{j\in[r]}$ are the orthonormal eigenfunctions of the kernel function $\bQ(\cdot)\bQ(\cdot)^{\T}$ with corresponding eigenvalues $\{p\vartheta_j\}_{j=1}^{r}$ and the rest 0. 
We then give the following proposition.
	
\begin{proposition}
\label{propos.eigenvalues_2}
Suppose that Assumption~\ref{ass.ind2} and $\|\bSigma_{\varepsilon}\|_{\cL}=o(p)$ hold. Then we have \\
(i) $\left|\tau_j-p\vartheta_j\right|\le \Vert\bSigma_{\varepsilon}\Vert_{\cL}$ for $j\in[r]$ and $|\tau_j|\le \Vert\bSigma_{\varepsilon}\Vert_{\cL}$ for $j\in[p] \setminus [r]$;\\
(ii) $\Vert\bvarphi_j-\widetilde{\bq}_j\Vert=O(p^{-1}\Vert\bSigma_{\varepsilon}\Vert_{\cL})$ for $j\in[r]$. 
\end{proposition} 

Proposition~\ref{propos.eigenvalues_2} implies that, if we can prove $\Vert\bSigma_{\varepsilon}\Vert_{\cL}=o(p),$  then we can distinguish the principal eigenvalues $\{\tau_j\}_{j \in [r]}$ from the remaining eigenvalues. 
Additionally, the first $r$ eigenfunctions $\{\bvarphi_j(\cdot)\}_{j \in [r]}$ are approximately the same as the normalized columns of $\{\widetilde{\bq}_j(\cdot)\}_{j \in [r]}.$ 
To establish this, we impose the same functional sparsity condition on $\bSigma_{\varepsilon}$ as measured by $s_p$ in \eqref{eq.sparsity}. 
Applying Lemma \ref{lem.op_le_S1}(iii) in the appendix, we have $\Vert\bSigma_{\varepsilon}\Vert_{\cL}\le\Vert\bSigma_{\varepsilon}\Vert_{\cS,1}^{1/2}\Vert\bSigma_{\varepsilon}\Vert_{\cS,\infty}^{1/2}=O(s_p).$ Hence, when $s_p=o(p),$ MFPCA is approximately equivalent to functional factor analysis under model \eqref{eq.model_2} for high-dimensional functional data.  

We now present the estimation procedure assuming that $r$ is known, and we will develop a ratio-based approach to identify $r$ in Section~\ref{subsec.deter_num}. Let $\hat\tau_1 \geq \hat\tau_2 \geq\dots \geq 0$ be the eigenvalues of the sample covariance $\widehat{\bSigma}_{y}^{\sS}$ and $\{\widehat \bvarphi_j(\cdot)\}_{j=1}^{\infty}$ be their corresponding eigenfunctions. Then $\widehat{\bSigma}_{y}^{\sS}$ has the spectral decomposition
$$\widehat{\bSigma}_{y}^{\sS}(u,v)=\sum_{j=1}^{r}\hat{\tau}_j\widehat{\bvarphi}_j(u)\widehat{\bvarphi}_j(v)^{\T}+\widehat{\bR}(u,v),$$
where $\widehat{\bR}(u,v)=\sum_{j=r+1}^{\infty}\hat{\tau}_j\widehat{\bvarphi}_j(u)\widehat{\bvarphi}_j(v)^{\T}$ is the functional principal orthogonal complement. Applying AFT as introduced in Section~\ref{subsec.factor_model} to $\widehat{\bR}$ yields the estimator $\widehat{\bR}^{\cA}.$ We finally obtain a FPOET estimator as
\begin{equation}
\label{eq.estimator_2}
\widehat{\bSigma}_{y}^{\cF}(u,v)=\sum_{j=1}^{r}\hat{\tau}_j\widehat{\bvarphi}_j(u)\widehat{\bvarphi}_j(v)^{\T}+\widehat{\bR}^{\cA}(u,v).
\end{equation}

It is noteworthy that, with $\bSigma_y$ satisfying decompositions~\eqref{eq.decom} and \eqref{eq.decom_2} under FFMs~\eqref{eq.model_1} and \eqref{eq.model_2}, respectively, 
both DIGIT and FPOET methods embrace the fundamental concept of a ``low-rank plus sparse" representation generalized to the functional setting. Consequently, the common estimation steps involve applying PCA or MFPCA to estimate the factor loadings, and applying AFT to estimate sparse $\bSigma_{\varepsilon}$. Essentially, these two methods are closely related, allowing the proposed estimators to exhibit empirical robustness even in cases of model misspecification (See details in Section~\ref{sec.sim}). See also Section~\ref{subsec.relationship} of the supplementary material for a discussion about the relationship between two FFMs. 

We next present an equivalent representation of FPOET estimator \eqref{eq.estimator_2} from a least squares perspective.
We consider solving a constraint least squares minimization problem: 
\begin{equation}
\label{eq.problem}
\{\widehat\bQ(\cdot), \widehat \bGamma\} = \arg\min_{\bQ(\cdot),\bGamma}\int\Vert\bY(u)-\bQ(u)\bGamma^{\T}\Vert_{\F}^2{\rm d}u=\arg\min_{\bQ(\cdot),\bgamma_1, \dots, \bgamma_n}\sum_{t=1}^{n}\Vert\by_t-\bQ\bgamma_t\Vert^2,
\end{equation}
subject to the normalization constraint corresponding to Assumption~\ref{ass.ind2}, i.e.,
  $$\frac{1}{n}\sum_{t=1}^{n}\bgamma_t\bgamma_t^{\T}=\bI_{r} ~~\text{and}~~ \frac{1}{p}\int\bQ(u)^{\T}\bQ(u){\rm d}u\ {\rm is\ diagonal},$$ 
	where $\bY(\cdot)=\{\by_1(\cdot),\dots,\by_n(\cdot)\}$ and $\bGamma^{\T}=(\bgamma_1,\dots,\bgamma_n)$. 
 Given each $\bGamma,$ setting the derivative of the objective in \eqref{eq.problem} w.r.t. $\bQ(\cdot)$ to zero, we obtain the constrained least squares estimator $\widetilde \bQ(\cdot) = n^{-1}\bY(\cdot)\bGamma.$ 
 Plugging it into \eqref{eq.problem}, the objective as a function of $\bGamma$ 
 becomes
 $\int\Vert\bY(u)-n^{-1}\bY(u)\bGamma\bGamma(u)^{\T}\Vert_{\F}^2{\rm d}u=\int{\rm tr}\big\{(\bI_n-n^{-1}\bGamma\bGamma^{\T})\bY(u)^{\T}\bY(u)\big\}{\rm d}u,$
 whose minimizer is equivalent to the maximizer of 
 ${\rm tr}\big[\bGamma^{\T}\big\{\int\bY(u)^{\T}\bY(u){\rm d}u\big\}\bGamma\big].$ This implies that the columns of $n^{-1/2}\widehat{\bGamma}$ are the eigenvectors corresponding to the $r$ largest eigenvalues of  $\int\bY(u)^{\T}\bY(u){\rm d}u \in {\mathbb R}^{n \times n},$ and then $\widehat{\bQ}(\cdot)=n^{-1}\bY(\cdot)\widehat{\bGamma}$. 

Let $\widetilde{\bvarepsilon}_{t}(\cdot)=\by_{t}(\cdot)-\widehat{\bQ}(\cdot)\widehat{\bgamma}_t$ and $\widetilde{\bSigma}_{\varepsilon}(u,v)=n^{-1}\sum_{t=1}^{n}\widetilde{\bvarepsilon}_t(u)\widetilde{\bvarepsilon}_t(v)^{\T}$. Applying our proposed AFT in \eqref{eq.thre} to $\widetilde{\bSigma}_{\varepsilon}$ yields the estimator $\widetilde{\bSigma}_{\varepsilon}^{\cA}$.  Analogous to the decomposition \eqref{eq.decom_2} under Assumption~\ref{ass.ind2}, we propose the following substitution estimator
\begin{equation}
\label{eq.estimator_3}
\widehat{\bSigma}_{y}^{\sL}(u,v)=\widehat{\bQ}(u)\widehat{\bQ}(v)^{\T}+\widetilde{\bSigma}_{\varepsilon}^{\cA}(u,v).
\end{equation}
	
The following proposition reveals the equivalence between the FPOET estimator \eqref{eq.estimator_2} and the constrained least squares estimator \eqref{eq.estimator_3}.
\begin{proposition}
\label{propos.equiv}
Suppose the same regularization parameters are used when applying AFT to $\widehat{\bR}$ and $\widetilde{\bSigma}_{\varepsilon}.$ Then we have $\widehat{\bSigma}_{y}^{\cF}=\widehat{\bSigma}_{y}^{\sL}$ and $\widehat{\bR}^{\cA}=\widetilde{\bSigma}_{\varepsilon}^{\cA}$.
\end{proposition}

\begin{remark}
(i) While our FFM \eqref{eq.model_2} shares the same form as the model studied in \cite{hallin2023factor,tavakoli2023factor}, which focused on the estimation of scalar factors and functional loadings from a least squares viewpoint, the main purpose of this paper lies in the estimation of large covariance matrix function. Consequently, we also propose a least-squares-based estimator of $\bSigma_y,$ which turns out to be equivalent to our FPOET estimator by Proposition \ref{propos.equiv}.\\
(ii) Using a similar procedure, we can also develop an alternative estimator for $\bSigma_y$ under FFM~\eqref{eq.model_1} from a least squares perspective. However, this estimator is distinct from the DIGIT estimator \eqref{eq.estimator_1} and leads to declined estimation efficiency.
See detailed discussion in Section~\ref{digit.ls} of the supplementary material.
\end{remark}

\subsection{Determining the number of factors}
\label{subsec.deter_num}
We have developed the estimation procedures for FFMs \eqref{eq.model_1} and \eqref{eq.model_2}, assuming the known number of functional or scalar factors (i.e., $r$). In this section, we take the frequently-used ratio-based approach \cite[]{lam2012,ahn2013eigenvalue} to determine the value of $r$.

For model \eqref{eq.model_1}, we let $\hat \lambda_1 \geq \dots \geq \hat \lambda_p$ be the ordered eigenvalues of $\widehat{\bOmega}$ in (\ref{Omega.est}), and propose to estimate $r$ by
\begin{equation}
\label{eq.deter_num_1}
\hat{r}^{\cD}=\arg \min_{j\in[r_{1,0}]}
\frac{\hat{\lambda}_{j+1}+\vartheta_{1n}}{\hat{\lambda}_j+\vartheta_{1n}},
\end{equation}
where $\vartheta_{1n}$ provides a lower bound correction to $\hat{\lambda}_j$ for $j>r$ \cite[]{han2022rank} and satisfies Assumption~\ref{ass.mdoel_dete_r_1} below, and $r_{1,0}>r$ is sufficiently large. 
For model \eqref{eq.model_2}, we employ a similar eigenvalue-ratio estimator:
\begin{equation}
\label{eq.deter_num_2}
\hat{r}^{\cF}=\arg \min_{j\in[r_{2,0}]}
\frac{\hat{\tau}_{j+1}+\vartheta_{2n}}{\hat{\tau}_j+\vartheta_{2n}},
\end{equation}
where $\vartheta_{2n}$ provides a lower bound correction to $\hat{\tau}_j$ for $j>r$ and satisfies Assumption~\ref{ass.mdoel_dete_r_2} below, $\{\hat{\tau}_i\}_{i=1}^{\infty}$ represents the ordered eigenvalues of the sample covariance $\widehat{\bSigma}_{y}^{\sS}(\cdot,\cdot),$ and $r_{2,0}>r$ is sufficiently large. We set $\vartheta_{1n}=c_rp^2n^{-4/5}$ and $\vartheta_{2n}=c_rpn^{-4/5}$ for some $c_r>0$ so that Assumptions~\ref{ass.mdoel_dete_r_1} and \ref{ass.mdoel_dete_r_2} are satisfied. In our empirical analysis, we choose $c_r=0.1,$ which consistently yields good finite-sample performance.

\subsection{Model selection criterion}
\label{subsec.selection}
A natural question that arises is which of the two candidate FFMs is more appropriate for modeling $\by_t(\cdot).$ This section develops functional information criteria based on observed data for model selection. 

When $k$ functional factors are estimated under FFM~\eqref{eq.model_1}, motivated from the least squares criterion (\ref{ls.crit1}), we define the mean squared residuals as
$$
V^{\cD}(k)=(pn)^{-1}\sum_{t=1}^{n}\Vert\by_t-p^{-1}\widehat{\bB}_{k}\widehat{\bB}_{k}^{\T}\by_t\Vert^2,
$$
where $\widehat{\bB}_{k}$ is the estimated factor loading matrix by DIGIT. Analogously, when $k$ scalar factors are estimated under FFM~\eqref{eq.model_2}, it follows from the objective function in \eqref{eq.problem} that the corresponding mean squared residuals is
$$
V^{\cF}(k)=(pn)^{-1}\sum_{t=1}^{n}\Vert\by_t-n^{-1}\bY\widehat{\bGamma}_k\widehat{\bgamma}_{t,k}\Vert^2,
$$
where $\widehat{\bGamma}_k^{\T}=(\widehat{\bgamma}_{1,k},\dots,\widehat{\bgamma}_{n,k})$ is formed by estimated factors using FPOET. 

For any given $k,$ we propose the following information  criteria:
\begin{equation}
    \label{eq.criteria}
    {\rm IC}^{\cD}(k)=\log\{V^{\cD}(k)\}+kg^{\cD}(p,n)~~\text{and}~~
    {\rm IC}^{\cF}(k)=\log\{V^{\cF}(k)\}+kg^{\cF}(p,n),
\end{equation}
where ${\rm IC}^{\cD}(k)$ and ${\rm IC}^{\cF}(k)$ are represented as the sum of the log transformation of the average of squared residuals and the penalty term with 
$g^{\cD}(p,n)$ and $g^{\cF}(p,n)$ being the corresponding penalty functions of $(p,n)$ to avoid overparameterization. 
While there is much existing literature \citep[see e.g.,][]{bai2002determining, fan2013} that has adopted this type of criterion for identifying the number of factors in scalar factor models, we propose fully functional criteria for selecting the more appropriate FFM. 
Following \cite{bai2002determining}, we suggest three examples of the same penalty functions $g^{\cD}(p,n)=g^{\cF}(p,n)=g(p,n)$ in \eqref{eq.criteria}, referred to as ${\rm IC}_1$, ${\rm IC}_2$, and ${\rm IC}_3$, respectively, 
        $$
		(i)\ g(p,n)=\frac{p+n}{pn}\log\left(\frac{pn}{p+n}\right),\
		(ii)\ g(p,n)=\frac{p+n}{pn}\log(p\wedge n),\
		(iii)\ g(p,n)=\frac{\log(p\wedge n)}{p\wedge n}.
	$$
For model selection, we define the information criterion difference between two FFMs as
$\Delta {\rm IC}_i={\rm IC}_i^{\cD}(\hat{r}^{\cD})-{\rm IC}_i^{\cF}(\hat{r}^{\cF})$ for $i=1,2,3.$
The negative (or positive) values of $\Delta {\rm IC}_i$'s indicate that FFM~\eqref{eq.model_1} (or FFM~\eqref{eq.model_2}) is more suitable based on the observed data $\{\by_t(\cdot)\}.$ 
See the model selection consistency guarantee in Theorem~\ref{thm.model_sele} under mild requirements on penalty functions, which three examples of $g(p,n)$ satisfy.

In general, instead of using the same penalty functions in ${\rm IC}^{\cD}$ and ${\rm IC}^{\cF},$ we can employ different penalty functions that satisfy the requirements of Theorem~\ref{thm.model_sele} from a model complexity perspective. Specifically, considering the panel nature of the problem with an effective number of observations $pn$ and assuming each function has a complexity of $s_n$, the effective total number of parameters for each model is calculated as the total number of parameters minus the number of constraints in Assumption~\ref{ass.ind1} or \ref{ass.ind2}, which yields $kp+kns_n-k^2$ for FFM~\eqref{eq.model_1}  and $kps_n+kn-k^2$ for FFM~\eqref{eq.model_2}.
Therefore, the feasible options for the penalty functions are $g^{\cD}(p,n)=(p+ns_n-k)/pn$ and $g^{\cF}(p,n)=(ps_n+n-k)/pn.$ However, rigorously determining $s_n$ remains an open question. Theoretically, by imposing specific requirements on $s_n,$ Theorem~\ref{thm.model_sele} can ensure the model selection consistency of ${\rm IC}^{\cD}$ and ${\rm IC}^{\cF}.$  In practice, one can select the leading $s_n$ eigenvalues of the corresponding covariance functions such that the cumulative percentage exceeds a certain threshold or adopt the eigenvalue-ratio-based method.

\section{Theory}
\label{sec.theory}

\subsection{Assumptions}
\label{subsec.ass}
	
The assumptions for models \eqref{eq.model_1} and \eqref{eq.model_2} exhibit a close one-to-one correspondence. For clarity, we will present them separately in a pairwise fashion.

\begin{assumption}
\label{ass.factor_ids}
For model \eqref{eq.model_1}, $\{\bbf_t(\cdot)\}_{t\geq 1}$ and $\{\bvarepsilon_t(\cdot)\}_{t\geq 1}$ are weakly stationary and $\eE\{\varepsilon_{ti}(u)\}=\eE\{\varepsilon_{ti}(u)f_{tj}(v)\}=0$ for all $i \in [p],$ $j \in [r],$ and $(u,v) \in \cU^2.$ 
\end{assumption}
	
\begin{assumptionprime}
\label{ass.factor_ids_2}
For model \eqref{eq.model_2}, $\{\bgamma_t\}_{t\geq 1}$ and $\{\bvarepsilon_t(\cdot)\}_{t\geq 1}$ are weakly stationary and $\eE\{\varepsilon_{ti}(u)\}=\eE\{\varepsilon_{ti}(u)\gamma_{tj}\}=0$ for all $i \in [p],$ $ j \in [r],$ and $u \in \cU.$
\end{assumptionprime}

\begin{assumption}
\label{ass.regu_cond_DIGIT}
For model \eqref{eq.model_1}, there exists some constant $C>0$ such that, for all $j\in[r],$ $t\in[n],$ (i) $\Vert\bbb_j\Vert_{\max}<C;$ (ii) $\eE\Vert p^{-1/2}\bB^{\T}\bvarepsilon_t\Vert^2<C;$ 
(iii) $\Vert\bSigma_{\varepsilon}\Vert_{\cL}<C;$ 
(iv) $\max_{i\in[p]}\|\Sigma_{\varepsilon,ii}\|_{\cN}<C.$ 
\end{assumption}
	
\begin{assumptionprime}
\label{ass.regu_cond_fpoet}
For model \eqref{eq.model_2}, there exists some constant $C'>0$ such that, for all $i\in[p],$ $t,t'\in[n]$: (i) $\Vert\bq_i\Vert<C';$ (ii) $\eE\Vert p^{-1/2}\int\bQ(u)^{\T}\bvarepsilon_t(u)\du\Vert^2<C'$and $\eE|p^{-1/2}\{\langle\bvarepsilon_t,\bvarepsilon_{t'}\rangle-\eE\langle\bvarepsilon_t,\bvarepsilon_{t'}\rangle\}|^4<C'$; (iii) $\Vert\bSigma_{\varepsilon}\Vert_{\cL}<C';$  (iv) $\max_{i\in[p]}\|\Sigma_{\varepsilon,ii}\|_{\cN}<C'.$
\end{assumptionprime}

Assumption~\ref{ass.regu_cond_DIGIT}(i) or \ref{ass.regu_cond_fpoet}(i) requires the functional or scalar factors to be pervasive in the sense they influence a large fraction of the functional outcomes. Such pervasiveness-type assumption is commonly imposed in the literature \cite[]{bai2003inferential,fan2013}. 
Assumptions~\ref{ass.regu_cond_DIGIT}(ii) and \ref{ass.regu_cond_fpoet}(ii) involve weaker moment constraints compared to \cite{fan2013} and are needed to estimate factors and loadings consistently.
Assumption~\ref{ass.regu_cond_DIGIT}(iii) and \ref{ass.regu_cond_fpoet}(iii) generalize the standard conditions for scalar factor models \cite[]{fan2013,wang2021} to the functional domain. Assumptions~\ref{ass.regu_cond_DIGIT}(iv) and \ref{ass.regu_cond_fpoet}(iv) are for technical convenience. However we can relax them by allowing $\max_i\|\Sigma_{\varepsilon,ii}\|_{\cN}$ to grow at some slow rate as $p$ increases. 
	
We use the functional stability measure \cite[]{chang2024autocovariance} to characterize the serial dependence. For $\{\by_t(\cdot)\},$ denote its autocovariance matrix functions by $\bSigma_y^{(h)}(u,v)=\cov\{\by_t(u), \by_{t+h}(v)\}$ for $h \in \mathbb Z$ and $(u,v) \in \cU^2$ and its spectral density matrix function at frequency $\theta \in [-\pi,\pi]$ by 
$\bbf_{y,\theta}(u,v)=(2\pi)^{-1}\sum_{h\in\mathbb{Z}}\bSigma_{y}^{(h)}(u,v)\exp(-ih\theta).$ The functional stability measure of $\{\by_t(\cdot)\}$ is defined as
\begin{equation}
\label{eq.stab_meas}
\cM_{y}=2\pi\cdot\esssup_{\theta\in[-\pi,\pi],\bphi\in\eH_0^p}\frac{\langle\bphi, \bbf_{y,\theta}(\bphi)\rangle}{\langle\bphi,\bSigma_{y}(\bphi)\rangle},
\end{equation}
where $\bSigma_y(\bphi)(\cdot)=\int_{\cU}\bSigma_y(\cdot,v)\bphi(v){\rm d}v$ and $\eH_0^p=\{\bphi\in\eH^p:\langle\bphi,\bSigma_{y}(\bphi)\rangle\in(0,\infty)\}$. When $\by_1(\cdot), \dots, \by_n(\cdot)$ are independent, $\cM_y=1.$
See also \cite{guo2023consistency} for examples satisfying  ${\cal M}_y<\infty,$ such as functional moving average model and functional linear process. Similarly, we can define $\cM_{\varepsilon}$ of $\{\bvarepsilon_t(\cdot)\},$ $\cM_{f}$ of $\{\bbf_t(\cdot)\}$ and $\cM_{\gamma}$ of scalar time series $\{\bgamma_t\}$ \cite[]{basu2015a}. 
To derive exponential-type tails used in convergence analysis, we assume the sub-Gaussianities for functional (or scalar) factors and idiosyncratic components. 
We relegate the definitions of sub-Gaussian (functional) process and multivariate (functional) linear process to Section \ref{supsybsec.subgauss} of the supplementary material. 

\begin{assumption}
\label{ass.concentration}
For  model \eqref{eq.model_1}, (i) $\{\bbf_t(\cdot)\}_{t\in [n]}$ and $\{\bvarepsilon_t(\cdot)\}_{t\in [n]}$ follow sub-Gaussian functional linear processes;
(ii) $\cM_f<\infty, \cM_{\varepsilon}<\infty$ and $\cM_{\varepsilon}^2\log p=o(n)$.
\end{assumption}

\begin{assumptionprime}
\label{ass.concentration_2}
For model~\eqref{eq.model_2}, (i) $\{\bgamma_t\}_{t\in [n]}$ follows sub-Gaussian linear process and $\{\bvarepsilon_t(\cdot)\}_{t\in[n]}$ follows sub-Gaussian functional linear process;
(ii) $\cM_{\gamma}<\infty, \cM_{\varepsilon}<\infty$ and $\cM_{\varepsilon}^2\log p=o(n).$ 
\end{assumptionprime}
	 
\begin{assumption}
\label{ass.thres}
There exists some constant $\tau>0$ such that $\min_{i,j\in[p]}\Vert{\rm Var}(\varepsilon_{ti}\varepsilon_{tj})\Vert_{\cS}\ge\tau.$
\end{assumption}

\begin{assumption}
\label{ass.n.p}
The pair $(n,p)$ satisfies $\cM_{\varepsilon}^2\log p=o(n/\log n)$ and $\log n=o(p)$. 
\end{assumption}

\begin{assumption}
    \label{ass.mdoel_dete_r_1}$\vartheta_{1n}p^{-2}\to0,\vartheta_{1n}\cM_{\varepsilon}^{-2}p^{-2}n\to\infty$ and $\vartheta_{1n}\to\infty,$ where $\vartheta_{1n}$ is specified in \eqref{eq.deter_num_1}.
\end{assumption}

\begin{assumptionprime}
    \label{ass.mdoel_dete_r_2}$\vartheta_{2n}p^{-1}\to0,\vartheta_{2n}\cM_{\varepsilon}^{-2}p^{-1}n\to\infty$ and $\vartheta_{2n}p\to\infty,$ where $\vartheta_{2n}$ is specified in \eqref{eq.deter_num_2}.
\end{assumptionprime}

\begin{assumption}
    \label{ass.mdoel_sele_1}
    For model~\eqref{eq.model_1}, (i) there exists some constant $\underline{\omega}>0$ such that $\omega_1\ge\omega_2\ge\dots\ge\omega_{r_{2,0}+1}\ge\underline{\omega}$, where $\{\omega_i\}_{i=1}^{\infty}$ are the ordered eigenvalues of $\bSigma_f(\cdot,\cdot)$ and $r_{2,0}$ is given in \eqref{eq.deter_num_2}; (ii) $\Vert\bSigma_{\varepsilon}\Vert_{\cN}=o(p).$
\end{assumption}

\begin{assumptionprime}
    \label{ass.mdoel_sele_2}
    For model~\eqref{eq.model_2}, (i) $\rank\big\{\int\bQ(u)\bQ(u)^{\T}\du\big\}\ge r_{1,0}+1,$ where $r_{1,0}$ is given in \eqref{eq.deter_num_1}; (ii) 
    $\Vert\bSigma_{\varepsilon}\Vert_{\cN}=o(p).$
\end{assumptionprime}

Assumption~\ref{ass.thres} is required when implementing AFT, however, it is weaker than the similar assumption $\inf_{(u,v)\in\cU^2}\min_{i,j\in[p]}{\rm Var}[\varepsilon_{ti}(u)\varepsilon_{tj}(v)]\ge\tau$ imposed in \cite{fang2024adaptive}. Assumption~\ref{ass.n.p} allows the high-dimensional case, where $p$ grows exponentially as $n$ increases.
Assumptions~\ref{ass.mdoel_dete_r_1} and \ref{ass.mdoel_dete_r_2} on the lower bound correlations (i.e., $\vartheta_{1n}$ and $\vartheta_{2n}$) are imposed to ensure the consistency of ratio-based estimators for the number of factors. Both assumptions are satisfied by setting $\vartheta_{1n}\asymp p^2n^{-\rho}$ and $\vartheta_{2n}\asymp pn^{-\rho}$ when $\cM_{\varepsilon}=O(1)$ and $n^{\rho}=o(p^2)$ for $\rho \in (0,1).$ 
Assumptions~\ref{ass.mdoel_sele_1} and \ref{ass.mdoel_sele_2} are needed when establishing the model selection consistency for the proposed information criteria.
For model~\eqref{eq.model_1}, Assumption~\ref{ass.mdoel_sele_1}(i) implies that the common covariance $\bSigma_{\chi}(\cdot,\cdot)=\bB\bSigma_f(\cdot,\cdot)\bB^{\T}$ has at least $r_{2,0}+1$ nonzero eigenvalues of order $p$, which cannot be recovered by the FPOET estimator with $\hat{r}^{\cF}\le r_{2,0}$ factors. 
Similarly, for model~\eqref{eq.model_2}, Assumption~\ref{ass.mdoel_sele_2}(i) implies that $\int\int\bSigma_{\kappa}(u,v)\bSigma_{\kappa}(u,v)^{\T}\du\dv=\int\bQ(u)\bQ(u)^{\T}\du$ (which holds under Assumption~\ref{ass.ind2}) has at least $r_{1,0}+1$ nonzero eigenvalues of order $p^2$, which cannot be recovered by the DIGIT estimator with $\hat{r}^{\cD}\le r_{1,0}$ factors. 
To guarantee the asymptotic identifiability of the respective FFMs for model selection, we require Assumptions~\ref{ass.mdoel_sele_1}(ii) and \ref{ass.mdoel_sele_2}(ii) to hold for $\{\bvarepsilon_t(\cdot)\}$.

\subsection{Convergence of estimated loadings and factors}
\label{subsec.con_load}
While the main focus of this paper is to estimate $\bSigma_y$, 
the estimation of factors and loadings remains a crucial aspect, encompassed by DIGIT and FPOET estimators, as well as in many other applications.
We first present various convergence rates of estimated factors and loading matrix when implementing DIGIT. 
For the sake of simplicity, we denote 
$$\varpi_{n,p}=\cM_{\varepsilon}\sqrt{\log p/n}+1/\sqrt{p}.$$ 

\begin{theorem}
\label{thm.load_factor_1}
Suppose that Assumptions \ref{ass.ind1}--\ref{ass.concentration} hold. Then there exists an orthogonal matrix $\bU \in {\mathbb R}^{r\times r}$ such that
(i) 
$\Vert\widehat{\bB}-\bB\bU^{\T}\Vert_{\max}=O_p\left(\varpi_{n,p}\right);$
(ii) $n^{-1}\sum_{t=1}^{n}\Vert\widehat{\bbf}_t-\bU\bbf_t\Vert^2=O_p(\cM_{\varepsilon}^2/n+1/p);$
(iii) $\max_{t\in[n]}\Vert\widehat{\bbf}_t-\bU\bbf_t\Vert=O_p\big(\cM_{\varepsilon}\sqrt{\log n/n}+\sqrt{\log n/p}\big)$.
\end{theorem}
	
The orthogonal matrix $\bU$ above is needed to ensure that $\bbb_j^{\T}\widehat{\bbb}_j\ge0$ for each $j\in[r].$ 
Provided that $\widehat{\bB}\bU\bU^{\T}\widehat{\bbf}_t=\widehat{\bB}\widehat{\bbf}_t,$ the estimation of the common components and $\bSigma_y$ remain unaffected by the choice of $\bU.$ By Theorem \ref{thm.load_factor_1}, we can derive the following corollary, which provides the uniform convergence rate of the estimated common component. Let $\breve{\bbb}_i$ and $\widecheck{\bbb}_i$ denote the $i$-th rows of $\bB$ and $\widehat{\bB},$ respectively.
\begin{corollary}
\label{coro.DIGIT}
Under the assumptions of Theorem \ref{thm.load_factor_1}, we have $\max_{i\in[p],t\in[n]}\Vert\widecheck{\bbb}_i^{\T}\widehat{\bbf}_t-\breve{\bbb}_i^{\T}\bbf_t\Vert=O_p(\varrho),$ where $\varrho=\cM_{\varepsilon}\sqrt{\log n \log p /n}+\sqrt{\log n/p}.$
\end{corollary}
	
In the context of FPOET estimation of factors and loadings, we require an additional asymptotically orthogonal matrix $\bH$ such that $\widehat{\bgamma}_t$ is a valid estimator of $\bH\bgamma_t$. Differing from DIGIT, we follow \cite{bai2003inferential} to construct $\bH$ in a deterministic form. Let $\bV \in {\mathbb R}^{r\times r}$ denote the  diagonal matrix of the first $r$ largest eigenvalues of $\widehat{\bSigma}_{y}^{\sS}$ in a decreasing order. 
Define $\bH=n^{-1}\bV^{-1}\widehat{\bGamma}^{\T}\bGamma\int\bQ(u)^{\T}\bQ(u){\rm d}u.$ 
By Lemma~\ref{lem.HH_bound} of the supplementary material, $\bH$ is asymptotically orthogonal such that $\bI_{r}=\bH^{\T}\bH+o_p(1)=\bH\bH^{\T}+o_p(1)$. 
	
\begin{theoremprime}
\label{thm.load_factor_2}
Suppose that Assumptions \ref{ass.ind2}--\ref{ass.concentration_2} hold.
(i)		$n^{-1}\sum_{t=1}^{n}\Vert\widehat{\bgamma}_t-\bH\bgamma_t\Vert^2=O_p\big(\cM_{\varepsilon}^2/n+1/p\big)$;		
(ii) $\max_{t\in[n]}\Vert\widehat{\bgamma}_t-\bH\bgamma_t\Vert=O_p\big(\cM_{\varepsilon}/\sqrt{n}+\sqrt{\log n/p}\big)$;
(iii) $\max_{i\in[p]}\Vert\widehat{\bq}_i-\bH\bq_i\Vert=O_p(\varpi_{n,p})$.
	\end{theoremprime}
	
\begin{corollaryprime}
\label{coro.fpoet}
Under the assumptions of Theorem~\ref{thm.load_factor_2}, we have $\max_{i\in[p],t\in[n]}\Vert\widehat{\bq}_i^{\T}\widehat{\bgamma}_t-\bq_i^{\T}\bgamma_t\Vert
=O_p(\varrho),$ where $\varrho$ is specified in Corollary~\ref{coro.DIGIT}. 
\end{corollaryprime}

When $\cM_{\varepsilon}=O(1),$ the convergence rates presented in Theorem~\ref{thm.load_factor_1}(i), (ii) for model~\eqref{eq.model_1} are, respectively, consistent to those established in \cite{fan2013} and \cite{bai2002determining}, and the uniform convergence rates presented in Theorem~\ref{thm.load_factor_1}(iii) and Corollary~\ref{coro.DIGIT} are faster than those established in \cite{fan2013}.
Additionally, the rates in Theorem \ref{thm.load_factor_2} and Corollary~\ref{coro.fpoet} for model~\eqref{eq.model_2} align with those in Theorem~\ref{thm.load_factor_1} and Corollary~\ref{coro.DIGIT}.
These uniform convergence rates are essential not only for estimating the FFMs but also for many subsequent high-dimensional learning tasks. 

\begin{theorem}
\label{thm.deter_num}
(i) Under Assumptions~\ref{ass.ind1}--\ref{ass.concentration} and \ref{ass.mdoel_dete_r_1}, $\eP(\hat{r}^{\cD}=r)\to1$ as $p,n \to \infty,$ where $\hat{r}^{\cD}$ is defined in \eqref{eq.deter_num_1}. (ii) Under Assumptions~\ref{ass.ind2}--\ref{ass.concentration_2} and \ref{ass.mdoel_dete_r_2}, $\eP(\hat{r}^{\cF}=r)\to1$ as $p,n \to \infty,$ where $\hat{r}^{\cF}$ is defined in \eqref{eq.deter_num_2}.
\end{theorem}
\begin{theorem}
    \label{thm.model_sele}
    Suppose that $g(p,n)\to0$ and $(\cM_{\varepsilon}^2/n+1/p)^{-1}g(p,n)\to\infty$ as $p,n\to\infty$ for both penalty functions $g^{\cD}(p,n)$ and $g^{\cF}(p,n)$ in \eqref{eq.criteria}. Then, (i) under Assumptions~\ref{ass.ind1}--\ref{ass.concentration}, \ref{ass.mdoel_dete_r_1} and \ref{ass.mdoel_sele_1}, $\eP\big\{{\rm IC}^{\cD}(\hat{r}^{\cD})<{\rm IC}^{\cF}(\hat{r}^{\cF})\big\}\to1$ as $p,n\to\infty$; (ii) under Assumptions~\ref{ass.ind2}--\ref{ass.concentration_2}, \ref{ass.mdoel_dete_r_2} and \ref{ass.mdoel_sele_2}, $\eP\big\{{\rm IC}^{\cD}(\hat{r}^{\cD})>{\rm IC}^{\cF}(\hat{r}^{\cF})\big\}\to1$ as $p,n\to\infty$. 
\end{theorem}
\begin{remark}
With the aid of Theorems~\ref{thm.deter_num} and \ref{thm.model_sele}, our estimators explored in Sections~\ref{subsec.con_load} and \ref{subsec.con_DIGIT} are asymptotically adaptive to the number of factors and the data-generating model. To see this, consider, e.g., model~\eqref{eq.model_2}, and let $\widehat{\kappa}_{ti,\hat r}(\cdot)$ be the estimated common component, and $\widehat\bgamma_{t,\hat r}$ and $\widehat \bq_{i,\hat r}(\cdot)$ be constructed using $\hat{r}^{\cF}$ estimated scalar factors and functional loadings. Then, for any constant $\tilde c>0,{\mathbb P}\big(\varrho^{-1}\max_{i\in[p],t\in[n]}\Vert\widehat{\kappa}_{ti,\hat r}-\kappa_{ti}\Vert >\tilde c\big) \leq {\mathbb P}\big\{\varrho^{-1}\max_{i\in[p],t\in[n]}\Vert\widehat{\bq}_{i}^{\T}\widehat{\bgamma}_{t}-\bq_i^{\T}\bgamma_t\Vert >\tilde c|\hat{r}^{\cF}=r,{\rm IC}^{\cD}(\hat{r}^{\cD})>{\rm IC}^{\cF}(\hat{r}^{\cF})\big\}+{\mathbb P}(\hat{r}^{\cF} \neq r)+\eP\big\{{\rm IC}^{\cD}(\hat{r}^{\cD})\le {\rm IC}^{\cF}(\hat{r}^{\cF})\big\},$ which, combined with Corollary~\ref{coro.fpoet}, implies that 
$\max_{i\in[p],t\in[n]}\Vert\widehat{\kappa}_{ti,\hat r}-\kappa_{ti}\Vert=O_p(\varrho).$
Similar arguments can be applied to other estimated quantities in Sections~\ref{subsec.con_load} and \ref{subsec.con_DIGIT}. Therefore, we assume that the number of factors and data-generating model are known in our asymptotic results.
\end{remark}

\subsection{Convergence of estimated covariance matrix functions}
\label{subsec.con_DIGIT}
Estimating the idiosyncratic covariance matrix function $\bSigma_{\varepsilon}$ is important in factor modeling and subsequent learning tasks. With the help of functional sparsity as specified in \eqref{eq.sparsity}, we can obtain consistent estimators of $\bSigma_{\varepsilon}$ under functional matrix $\ell_1$ norm $\Vert\cdot\Vert_{\cS,1}$ in the high-dimensional scenario. The following rates of convergence based on estimated idiosyncratic components are consistent with the rate based on direct observations of independent functional data \cite[]{fang2024adaptive} when $\cM_{\varepsilon}=O(1)$ and $p \log p \gtrsim n.$
	
\begin{theorem}
\label{thm.idio_DIGIT}
Suppose that Assumptions \ref{ass.ind1}--\ref{ass.n.p} hold. Then, for a sufficiently large constant $\dot{C}$ in \eqref{eq.thre}, 
$
\Vert\widehat{\bSigma}_{\varepsilon}^{\cA}-\bSigma_{\varepsilon}\Vert_{\cS,1}=O_p(\varpi_{n,p}^{1-q}s_p).
$
\end{theorem}
	
\begin{theoremprime}
\label{thm.idio_fpoet}
Suppose that Assumptions \ref{ass.ind2}--\ref{ass.concentration_2}, \ref{ass.thres}, \ref{ass.n.p} hold. Then, for a sufficiently large constant $\dot{C}$ in \eqref{eq.thre}, 
$
\Vert\widehat{\bR}^{\cA}-\bSigma_{\varepsilon}\Vert_{\cS,1}=O_p(\varpi_{n,p}^{1-q}s_p).
$
\end{theoremprime}

When assessing the convergence criteria for our DIGIT and FPOET estimators, it is crucial to note that functional matrix norms such as $\Vert\cdot\Vert_{\cS,1}$ and $\Vert\cdot\Vert_{\cL}$ are not suitable choices. This is because $\widehat{\bSigma}_y$ may not converge to $\bSigma_y$ in these norms for high-dimensional functional data, unless specific structural assumptions are directly imposed on $\bSigma_y$. 
This issue does not arise from the poor performance of estimation methods but rather from the inherent limitation of high-dimensional models. 
To address this, we present convergence rates in functional elementwise $\ell_{\infty}$ norm $\Vert\cdot\Vert_{\cS,\max}.$ 
\begin{theorem}
\label{thm.DIGIT}
Under the assumptions of Theorem \ref{thm.idio_DIGIT}, we have
$\Vert\widehat{\bSigma}_{y}^{\cD}-\bSigma_{y}\Vert_{\cS,\max}=O_p(\varpi_{n,p}).$
\end{theorem}
	
\begin{theoremprime}
\label{thm.fpoet}
Under the assumptions of Theorem \ref{thm.idio_fpoet}, we have
$\Vert\widehat{\bSigma}_{y}^{\cF}-\bSigma_{y}\Vert_{\cS,\max}=O_p(\varpi_{n,p}).$
\end{theoremprime}
	
\begin{remark}
\label{rmk.DIGIT1}

(i) The convergence rates of DIGIT and FPOET estimators (we use $\widehat{\bSigma}_{y}$ to denote both) comprise two terms. The first term $O_p(\cM_{\varepsilon}\sqrt{\log p/n})$ arises from the rate of $\widehat{\bSigma}_y^{\sS},$ while the second term $O_p(p^{-1/2})$ primarily stems from the estimation of unobservable factors. When $\cM_{\varepsilon}=O(1),$ our rate aligns with the result obtained in \cite{fan2013}.\\ 
(ii) Compared to $\widehat{\bSigma}_y^{\sS},$ we observe that using a factor-guided approach results in the same rate in $\|\cdot\|_{\cS, \max}$ as long as $p \log p \gtrsim n.$ Nevertheless, our proposed estimators offer several advantages. 
First, under a functional weighted quadratic norm introduced in Section~\ref{subsec.risk}, which is closely related to functional risk management, 
$\widehat{\bSigma}_y$ converges to $\bSigma_y$ in the high-dimensional case (see Theorem \ref{thm.new_norm}), while $\widehat{\bSigma}_y^{\sS}$ does not achieve this convergence. 
Second, as evidenced by empirical results in Sections \ref{sec.sim} and \ref{sec.real}, $\widehat{\bSigma}_y$ significantly outperforms $\widehat{\bSigma}_y^{\sS}$ in terms of various functional matrix losses. 
\end{remark}

Finally, we explore convergence properties of the inverse covariance matrix function estimation. Based on Section 3.5 of \cite{Bhsing2015}, although the inverse operator $\bSigma_y^{-1}$ may not be well-defined, we instead use the Moore--Penrose inverse. 
Denote the null space of $\bSigma_y$ and its orthogonal complement by $\tKer(\bSigma_y)=\{\bx \in {\mathbb H}^p: \bSigma_y(\bx)={\bf 0}\}$ and ${\tKer(\bSigma_y)}^{\perp}=\{\bx \in {\mathbb H}^p: \langle \bx,\by \rangle =0, \forall \by \in \tKer(\bSigma_y)\},$ respectively. Let $\widetilde{\bSigma}_y$ be the restriction of $\bSigma_y$ to $\tKer(\bSigma_y)^{\perp}$. 
By Definition 3.5.7 of \cite{Bhsing2015}, the Moore--Penrose inverse of $\bSigma_y$ is defined as $\bSigma_y^{\dagger}(\bx)=\widetilde{\bSigma}_y^{-1}(\bx)$ for $\bx\in\tIm(\bSigma_y)$ and $0$ for $\bx\in\tIm(\bSigma_y)^{\perp}.$ 
The similar definition applies to the Moore--Penrose inverses of other covariance matrix operators.
To obtain the inverse DIGIT estimator, we assume  $\tKer(\bSigma_{\varepsilon})=\tKer(\bB\bSigma_f\bB^{\T}),$ which implies that $\tKer(\bSigma_{\varepsilon})=\tKer(\bSigma_{y}),$ and $\widetilde \bSigma_y,$ $\widetilde \bSigma_\varepsilon$ and $\bB\widetilde \bSigma_f\bB^\T$ are all invertible on $\tKer(\bSigma_y)^{\perp}.$
We then rely on \eqref{eq.decom} to apply Sherman--Morrison--Woodbury formula (Theorem 3.5.6 of \cite{Bhsing2015}) to obtain 
its inverse
$\bSigma_y^{\dagger}=\bSigma_{\varepsilon}^{\dagger}-\bSigma_{\varepsilon}^{\dagger}\bB\big(\bSigma_f^{\dagger}+\bB^{\T}\bSigma_{\varepsilon}^{\dagger}\bB\big)^{\dagger}\bB^{\T}\bSigma_{\varepsilon}^{\dagger},$ and the plug-in inverse DIGIT estimator is
$(\widehat{\bSigma}_y^{\cD})^{\dagger}=\widehat{\bSigma}_{\varepsilon}^{\dagger}-\widehat{\bSigma}_{\varepsilon}^{\dagger}\widehat{\bB}\big(\widehat{\bSigma}_f^{\dagger}+\widehat{\bB}^{\T}\widehat{\bSigma}_{\varepsilon}^{\dagger}\widehat{\bB}\big)^{\dagger}\widehat{\bB}^{\T}\widehat{\bSigma}_{\varepsilon}^{\dagger}.$
The plug-in inverse FPOET estimator $(\widehat{\bSigma}_y^{\cF})^{\dagger}$ can be defined similarly by assuming that $\tKer(\bSigma_{\varepsilon})=\tKer(\bQ\bSigma_{\gamma}\bQ^{\T})$.
To make $(\widehat{\bSigma}_y^{\cD})^{\dagger}$ and $(\widehat{\bSigma}_y^{\cF})^{\dagger}$ meaningful inverse covariance matrix estimators, we focus on finite-dimensional functional objects $\{\by_t(\cdot)\}_{t\in[n]},$ i.e., $\Vert\bSigma_y^{\dagger}\Vert_{\cL}$ is bounded. 
Then, both the inverse DIGIT and FPOET estimators are consistent in the operator norm, as presented in the following theorems.

\begin{theorem}
\label{thm.inverse_digit}
Suppose that the assumptions of Theorem \ref{thm.DIGIT} hold, $\varpi_{n,p}^{1-q}s_p=o(1)$, $\tKer(\bSigma_{\varepsilon})=\tKer(\bB\bSigma_f\bB^{\T})$, and both $\Vert\bSigma_{\varepsilon}^{\dagger}\Vert_{\cL}$ and $\Vert\bSigma_{f}^{\dagger}\Vert_{\cL}$ are bounded. 
Then, $\widehat{\bSigma}_y^{\cD}$ has a bounded Moore--Penrose inverse with probability approaching one, and
$\big\Vert(\widehat{\bSigma}_y^{\cD})^{\dagger}-\bSigma_y^{\dagger}\big\Vert_{\cL}=O_p(\varpi_{n,p}^{1-q}s_p).$
\end{theorem}

\begin{theoremprime}
\label{thm.inverse_fpoet}
Suppose that the assumptions of Theorem \ref{thm.fpoet} hold, $\varpi_{n,p}^{1-q}s_p=o(1)$, $\tKer(\bSigma_{\varepsilon})=\tKer(\bQ\bSigma_{\gamma}\bQ^{\T}),$ and $\Vert\bSigma_{\varepsilon}^{\dagger}\Vert_{\cL}$ is bounded.
Then, $\widehat{\bSigma}_y^{\cF}$ has a bounded Moore--Penrose inverse with probability approaching one, and
$\big\Vert(\widehat{\bSigma}_y^{\cF})^{\dagger}-\bSigma_y^{\dagger}\big\Vert_{\cL}=O_p(\varpi_{n,p}^{1-q}s_p).$
\end{theoremprime}

\begin{remark}
\label{rmk.error}
(i) 
The condition that $\Vert\bSigma_{\varepsilon}^{\dagger}\Vert_{\cL}$ and $\Vert\bSigma_{f}^{\dagger}\Vert_{\cL}$ are bounded implies that $\Vert\bSigma_y^{\dagger}\Vert_{\cL}$ is bounded,
which means that $\bSigma_y$ has a finite number of nonzero eigenvalues, denoted as $d<\infty.$ 
Then $\bSigma_y(u,v)=\sum_{i=1}^{d}\tau_i\bvarphi_i(u)\bvarphi_i(v)^{\T}$ with its inverse $\bSigma_y^{\dagger}(u,v)=\sum_{i=1}^{d}\tau_i^{-1}\bvarphi_i(u)\bvarphi_i(v)^{\T}.$
While the inverse of $\widehat{\bSigma}_y^{\sS}$ fails to exhibit convergence even though it operates within finite-dimensional Hilbert space, our factor-guided methods can achieve such convergence. It should be noted that $d$ can be made arbitrarily large relative to $n,$ e.g., $d=2000, n=200.$
Hence, this finite-dimensional assumption does not place a practical constraint on our method.
See applications of inverse covariance matrix function estimation in Sections~\ref{subsec.risk} and ~\ref{subsec.inverse}. \\
(ii) An example that satisfies $\tKer(\bSigma_{\varepsilon})=\tKer(\bB\bSigma_f\bB^{\T})$ is $\bSigma_{\varepsilon}(u,v)=\sum_{j=1}^{d}\lambda_j\bpsi_j(u)\bpsi_j(v)^{\T}$ and $\bSigma_f(u,v)=\sum_{j=1}^{d}\omega_j\bphi_j(u)\bphi_j(v)^{\T},$ where $\lambda_j>0$ and $\omega_j>0$ for $j\in[d]$ and ${\rm span}\{\bpsi_j(\cdot)\}_{j\in[d]}={\rm span}\{\bB\bphi_j(\cdot)\}_{j\in[d]}.$ While the existing literature has not explored the convergence properties of high-dimensional inverse covariance operators, our papers makes a first attempt within the functional factor modeling framework by assuming the finite-dimensional functional objects and the same spaces spanned by the corresponding eigenfunctions. We leave the possible relaxation of these conditions as future research.\\
(iii) Within infinite-dimensional Hilbert space, the inverse operator $\bSigma_y^{\dagger}(u,v)=\sum_{i=1}^{\infty}\tau_i^{-1}\bvarphi_i(u)\bvarphi_i(v)^{\T}$ becomes an unbounded operator, which is discontinuous and cannot be estimated in a meaningful way. However, $\bSigma_y^{\dagger}$ is usually associated with another function/operator, and the composite function/operator in $\tKer(\bSigma_y)^{\perp}$ can reasonably be assumed to be bounded, such as regression function/operator and discriminant direction function in Section \ref{subsec.inverse}. 
Specifically, consider the spectral decomposition \eqref{eq.y_mercer}, which is truncated at $d<\infty,$ i.e.,
$\bSigma_{y,d}(u,v)=\sum_{i=1}^{d} \tau_i \bvarphi_i(u)\bvarphi_i(v)^{\T}.$ Under certain smoothness conditions, such as those on coefficient functions in multivariate functional linear regression \cite[]{chiou2016}, the impact of truncation errors through $\sum_{i=d+1}^{\infty} \tau_i^{-1} \bvarphi_i(u)\bvarphi_i(v)^{\T}$ on associated functions/operators is expected to diminish, ensuring the boundedness of composite functions/operators. Consequently, the primary focus shifts towards estimating the inverse of $\bSigma_{y,d},$ and our results in Theorems~\ref{thm.inverse_digit} and \ref{thm.inverse_fpoet} become applicable.
\end{remark}

Upon observation, a remarkable consistency is evident between DIGIT and FPOET methods developed under different models in terms of imposed regularity assumptions and associated convergence rates, despite the substantially different proof techniques employed.

\section{Applications}
\label{sec.app}

\subsection{Functional risk management}
\label{subsec.risk}
One main task of risk management in the stock market is to estimate the portfolio variance, which can be extended to the functional setting to account for the intraday uncertainties.
Consider a portfolio consisting of $p$ stocks. Let $\bP_t(\cdot)=\{P_{t1}(\cdot),\dots, P_{tp}(\cdot)\}^{\T}, \bZ_t(\cdot)=\{Z_{t1}(\cdot),\dots, Z_{tp}(\cdot)\}^{\T}$ and $\widetilde{\bw}_t(\cdot)=\{\widetilde{w}_{t1}(\cdot),\dots,\widetilde{w}_{tp}(\cdot)\}^{\T}$, where $P_{ti}(u), Z_{ti}(u)$ and $\widetilde{w}_{ti}(u)=Z_{ti}(u)P_{ti}(u)$ respectively denote the price per-share, the quantity held, and the amount of money held for the $i$-th stock at time $u$ on the $t$-th trading day. We assume that $\cU=[0,1]$ and $\widetilde{\bw}_t(0)^{\T}\bone_p=1,$ where $\bone_p=(1, \dots, 1)^\T\in\mathbb R^p,$ meaning that the initial portfolio amount is normalized to facilitate subsequent analysis. Then, the functional portfolio return of the $t$-th day is 
\begin{flalign}
    \label{eq.fun_return}
\big\langle\bZ_t, \nabla\bP_t\big\rangle
&=\int_{0}^{1}\widetilde{\bw}_t(u)^{\T}\nabla\{\log \bP_t(u)\}\du=\widetilde{\bw}_t(u)^{\T}\log\{\bP_t(u)\}~\big|_{0}^{1}-
\big\langle\nabla\widetilde{\bw}_t,\log(\bP_t)\big\rangle
\nonumber\\
&=\widetilde{\bw}_t(1)^{\T}\by_t(1)-
\big\langle\nabla\widetilde{\bw}_t,\by_t\big\rangle,
\end{flalign}
where $\by_t(u)=\log\{\bP_t(u)\}-\log\{\bP_t(0)\},u\in[0,1]$ turns to be the cumulative intraday return (CIDR) trajectory as defined in \cite{horvath2014testing}. 
As discussed in \cite{lou2019tug}, the close-to-close return of the stock can be decomposed into overnight and intraday components, and the profits of popular trading strategies are either earned entirely overnight or entirely intraday. 
Since the CIDR trajectories are only recorded during the trading period, our functional risk management focuses on the intraday strategy, where assets are bought and sold within the same day to profit from short-term price movements, so we close out all positions by the end of the trading day with $\widetilde{\bw}_t(1)=\bzero.$ By \eqref{eq.fun_return}, the functional portfolio return of the $t$-th day is $r_t=\langle\bw_t,\by_t\rangle$ with $\bw_t(\cdot)=-\nabla\widetilde{\bw}_t(\cdot).$ If $w_{ti}(u)=-\nabla\widetilde{w}_{ti}(u)>0$ (or $<0$), it indicates that an amount $w_{ti}(u)$ of the $i$-th stock is sold (or bought) at time $u$. The amount held in each stock at the opening is determined by $\widetilde{\bw}_t(0)=\int_{0}^{1}\bw_t(u)\du.$
The constraint on $\bw_t(\cdot)$ is $\int_{0}^{1}\bw_t(u)^{\T}\bone_p\du=\widetilde{\bw}_t(0)^{\T}\bone_p-\widetilde{\bw}_t(1)^{\T}\bone_p=1,$ and hence $\bw_t(\cdot)$ can be viewed as the functional portfolio allocation vector. 
Despite being formulated within a functional framework, the practical implementation can be achieved through discretization, such as at intervals of every 5 or 10 minutes.
The functional portfolio variance of the $t$-th day is calculated as
$\var(r_t)=\langle \bw_t,\bSigma_y(\bw_t)\rangle.$ 
For a more general $\widetilde{\bw}_t(1),$ the corresponding functional portfolio variance involves the cross-covariance between two terms in \eqref{eq.fun_return}, which largely complicates functional risk management and is left for future research.

For a given $\bw(\cdot),$ the true and perceived variances (i.e., risks) of the functional portfolio are
$\langle\bw,\bSigma_y(\bw)\rangle$ 
and $\langle\bw,\widehat{\bSigma}_y(\bw)\rangle,$ respectively. According to Proposition~\ref{propos.risk1} of the supplementary material, the estimation error of the functional portfolio variance is bounded by
$$
\big|\langle\bw,\widehat{\bSigma}_y(\bw)\rangle-\langle\bw,\bSigma_y(\bw)\rangle\big|\le\Vert\widehat{\bSigma}_y-\bSigma_y\Vert_{\cS,\max}\big(\sum_{i=1}^{p}\Vert w_i\Vert\big)^2,
$$
in which Theorems~\ref{thm.DIGIT} and \ref{thm.fpoet} quantify the maximum approximation error $\Vert\widehat{\bSigma}_y-\bSigma_y\Vert_{\cS,\max}.$ 

In addition to the absolute error between perceived and true risks, we are also interested in quantifying the relative error. To this end, we introduce the functional version of weighted quadratic norm \cite[]{fan2008high}, defined as $\Vert\bK\Vert_{\cS,\Sigma_y}=p^{-1/2}\Vert(\bSigma_y^{\dagger})^{1/2}\bK(\bSigma_y^{\dagger})^{1/2}\Vert_{\cS,\F},$ where $\bK \in \eH^p \otimes \eH^p$ and the normalization factor $p^{-1/2}$ serves the role of $\Vert\bSigma_y\Vert_{\cS,\Sigma_y}=1$. To ensure the validity of this functional norm, we assume that $\bSigma_y$ has a bounded inverse, which does not place a constraint in practice (see Remark~\ref{rmk.error}(i)). 
With such functional norm, the relative error can be measured by
\begin{equation}
\label{eq.norm_Frob}
p^{-1/2}\big\Vert(\bSigma_y^{\dagger})^{1/2}\widehat{\bSigma}_y(\bSigma_y^{\dagger})^{1/2}-\tilde{\bI}_p\big\Vert_{\cS,\F}=\big\Vert\widehat{\bSigma}_y-\bSigma_y\big\Vert_{\cS,\Sigma_y},
\end{equation}
where $\tilde{\bI}_p$ is defined as $\tilde{\bI}_p(\bx)=\bx$ for $\bx\in\tIm(\bSigma_y)$ and 0 for $\bx\in\tIm(\bSigma_y)^{\perp}$.
Provided that $\Vert\widehat{\bSigma}_y^{\sS}-\bSigma_y\Vert_{\cS,\Sigma_y}=O_p(\cM_{\varepsilon}\sqrt{p/n}),$ the sample covariance fails to converge in $\Vert\cdot\Vert_{\cS,\Sigma_y}$ under the high-dimensional setting with $p>n.$ 
On the contrary, the following theorem reveals that our DIGIT estimator $\widehat{\bSigma}_y^{\cD}$ converges to $\bSigma_y$ as long as $\cM_{\varepsilon}^4p=o(n^2)$ and $\varpi_{n,p}^{1-q}s_p=o(1)$. The same result can also be extended to the FPOET estimator.

\begin{theorem}
\label{thm.new_norm}
Under the assumptions of Theorem \ref{thm.inverse_digit}, we have
$
\Vert\widehat{\bSigma}_y^{\cD}-\bSigma_y\Vert_{\cS,\Sigma_y}=O_p\big(\cM_{\varepsilon}^2 p^{1/2} n^{-1}+\varpi_{n,p}^{1-q}s_p\big).
$
\end{theorem}

By Proposition~\ref{propos.risk2} of the supplementary material, the relative error is bounded by
$$
\big|\langle\bw,\widehat{\bSigma}_y(\bw)\rangle/\langle\bw,\bSigma_y(\bw)\rangle-1\big|\le\big\Vert(\bSigma_y^{\dagger})^{1/2}\widehat{\bSigma}_y(\bSigma_y^{\dagger})^{1/2}-\tilde{\bI}_p\big\Vert_{\cL},
$$
which, in conjunction with Theorem \ref{thm.new_norm} and  \eqref{eq.norm_Frob}, controls the maximum relative error.

\subsection{Estimation of regression and discriminant direction functions}
\label{subsec.inverse}

The second application explores multivariate functional linear regression \cite[]{chiou2016}, which involves a scalar response $z_t$ or a functional response
$$z_t(v) = \big\langle \by_t, \bbeta(\cdot, v) \big\rangle + e_t(v),~~ v\in {\cal V},$$ where $\bbeta(\cdot, \cdot)=\{\beta_1(\cdot,\cdot), \dots, \beta_p(\cdot,\cdot)\}^\T$ is an operator-valued coefficient vector to be estimated. 
We can impose certain smoothness condition such that $\bbeta(u,v)=\sum_{i=1}^{\infty} \tilde \tau_i \bvarphi_i(u)\bvarphi_i(v)^{\T}$ is sufficiently smooth relative to $\bSigma_y(u,v)=\sum_{i=1}^{\infty} \tau_i \bvarphi_i(u)\bvarphi_i(v)^{\T},$ ensuring the boundedness of the regression operator $\bbeta(u,v) = \int_{\cU}\bSigma_y^{\dagger}(u, u')\cov\{\by_t(u'), z_t(v)\}{\rm d}u'.$ Replacing relevant terms by their (truncated) sample versions, we obtain 
$\widehat \bbeta(u,v)=n^{-1}\sum_{t=1}^n\int_{\cU}\widehat\bSigma_{y,d}^{\dagger}(u, u')\by_t(u')z_t(v){\rm d}u'.$ This application highlights the need for estimators $\widehat\bSigma_{y,d}^{\dagger},$ as studied in Theorems~\ref{thm.inverse_digit} and \ref{thm.inverse_fpoet}.

The third application delves into linear discriminant analysis for classifying multivariate functional data \cite[]{xue2024optimal} with class labels $w_t=\{1,2\}.$ Specifically, we assume that $\by_t(\cdot)|w_t=1$ and $\by_t(\cdot)|w_t=2$ follow multivariate Gaussian distributions with mean functions $\bmu_1(\cdot)$ and $\bmu_2(\cdot),$ respectively, while sharing a common covariance matrix function $\bSigma_y.$ Our goal is to determine the linear classifier by estimating the discriminant direction function $\int_{\cU}\bSigma_y^{\dagger}(u,v)\{\bmu_1(v) - \bmu_2(v)\} {\rm d}v,$ which takes the same form as the regression function $\bbeta(u)=\int_{\cU}\bSigma_y^{\dagger}(u,v)\cov\{\by_t(v), z_t\}{\rm d}v$ encountered in the second application with a scalar response $z_t.$ By similar arguments as above, both applications call for the use of estimators $\widehat\bSigma_{y,d}^{\dagger}.$

\subsection{Estimation of correlation matrix function}
The fourth application involves estimating the correlation matrix function and its inverse, which are essential in graphical models for truly infinite-dimensional objects, see, e.g., \cite{solea2022copula}. Our proposed covariance estimators can be employed to estimate the corresponding correlation matrix function and its inverse.	
Specifically, let $\bD_{y}(\cdot,\cdot)={\rm diag}\{\Sigma_{y,11}(\cdot,\cdot), \dots, \Sigma_{y,pp}(\cdot,\cdot)\}$ be the $p\times p$ diagonal matrix function. According to \cite{baker1973joint}, there exists a correlation matrix function $\bC_{y}$ with $\Vert\bC_{y}\Vert_{\cL}\le1$ such that $\bSigma_{y}=\bD_{y}^{1/2}\bC_{y}\bD_{y}^{1/2}$. Under certain compactness and smoothness assumptions, $\bC_y$ has a bounded inverse, denoted by $\bTheta_y,$ and its functional sparsity pattern corresponds to the network (i.e., conditional dependence) structure among $p$ components in $\by_t(\cdot);$ see  \cite{solea2022copula}. Although the inverse of the estimator $\widehat\bD_{y}={\rm diag}(\widehat\Sigma_{y,11}, \dots, \widehat\Sigma_{y,pp})$ is not well-defined, we adopt the Tikhonov regularization to estimate $\bC_{y}$ by $\widehat{\bC}_{y}^{(\kappa)}=(\widehat{\bD}_{y}+\kappa\bI_p)^{-1/2}\widehat{\bSigma}_{y}(\widehat{\bD}_{y}+\kappa\bI_p)^{-1/2}$ for some regularization parameter $\kappa>0.$ The estimator of $\bTheta_y$ is then given by
$\widehat{\bTheta}_{y}^{(\kappa)}=\widehat{\bD}_{y}^{1/2}(\widehat{\bSigma}_{y}+\kappa\bI_p)^{-1}\widehat{\bD}_{y}^{1/2}.$ 
Consequently, we can plug into the DIGIT or the FPOET estimator for estimating $\bC_y$ and its inverse $\bTheta_y.$

\section{Simulations}
\label{sec.sim}
For the first data-generating process (denoted as DGP1), 
we generate observed data from model \eqref{eq.model_1}, where the entries of $\bB\in\mathbb{R}^{p\times r}$ are sampled independently from ${\rm Uniform}[-0.75,0.75],$ satisfying Assumption \ref{ass.regu_cond_DIGIT}(i). To mimic the infinite-dimensionality of functional data, each functional factor is generated from $f_{tj}(\cdot)=\sum_{i=1}^{50}i^{-1}\xi_{tji}\phi_i(\cdot)$ for $j \in [r]$ over $\cU=[0,1]$, where $\{\phi_i(\cdot)\}_{i=1}^{50}$ is a $50$-dimensional Fourier basis and basis coefficients $\bxi_{ti}=(\xi_{t1i}, \dots, \xi_{tri})^{\T}$ are generated from a vector autoregressive model, $\bxi_{ti}=\bA\bxi_{t-1,i}+\bu_{ti}$ with $\bA=\{A_{jk}={0.4}^{|j-k|+1}\}_{r\times r}$, and the innovations $\{\bu_{ti}\}_{t \in [n]}$ being sampled independently from ${\cal N}(\mathbf{0}_{r},\bI_{r})$.  
For the second data-generating process (denoted as DGP2), 
we generate observed data from model \eqref{eq.model_2}, where $r$-vector of scalar factors $\bgamma_t$ is generated from a vector autoregressive model, $\bgamma_t=\bA\bgamma_{t-1}+\bu_t$ with $\{\bu_t\}_{t \in [n]}$ being sampled independently from ${\cal N}(\mathbf{0}_{r},\bI_{r}).$ The functional loading matrix $\bQ(\cdot)=\{Q_{jk}(\cdot)\}_{p\times r}$ is generated  by $Q_{jk}(\cdot)=\sum_{i=1}^{50}i^{-1}q_{ijk}\phi_i(\cdot),$
where each $q_{ijk}$ is sampled independently from the ${\cal N}(0,0.3^2),$ satisfying Assumption \ref{ass.regu_cond_fpoet}(i). 

The idiosyncratic components are generated from $\bvarepsilon_t(\cdot)=\sum_{l=1}^{25}2^{-l/2}\bpsi_{tl}\phi_l(\cdot),$ where each $\bpsi_{tl}$ is generated from $\bpsi_{tl}=0.5\bpsi_{t-1,l}+\bzeta_{tl}$ with $\bzeta_{tl}$ being independently sampled from ${\cal N}(\mathbf{0}_{p},\bC_{\zeta})$ with $\bC_{\zeta}=\bD\bC_0\bD$.
Given this autoregressive structure, it can be shown that $\cM_{\varepsilon}=O(1)$.
Here, we set $\bD={\rm diag}(D_1,\dots,D_p),$ where each $D_i$ is generated from $\text{Gamma}(3,1)$. The generation of $\bC_0$ involves the following three steps: (i) we set the diagonal entries of $\breve{\bC}$ to 1, and generate the off-diagonal and symmetrical entries from $\text{Uniform}[0,0.5]$; (ii) we employ hard thresholding \cite[]{cai2011adaptive} on $\breve{\bC}$ to obtain a sparse matrix $\breve{\bC}^{\cT}$, where the threshold level is found as the smallest value such that $\max_{i\in[p]}\sum_{j=1}^{p}I(\breve{C}^{\cT}_{ij} \neq 0)\le p^{1-\alpha}$ for $\alpha \in [0,1]$; (iii) we set $\bC_0=\breve{\bC}^{\cT}+\tilde{\delta}\bI_{p}$ where $\tilde{\delta}=\max\{-\lambda_{\min}(\breve{\bC}),0\} + 0.01$ to guarantee the positive-definiteness of $\bC_0.$ The parameter $\alpha$ controls the sparsity level with larger values yielding sparser structures in $\bC_0$ as well as functional sparser patterns in $\bSigma_{\varepsilon}(\cdot,\cdot).$ This is implied from Proposition~\ref{propos.scenario}(iii) of the supplementary material, whose parts (i) and (ii) respectively specify the true covariance matrix functions of $\by_t(\cdot)$ for DGP1 and DGP2.

\begin{figure}[ht]	
\centering
\includegraphics[width=0.75\textwidth]{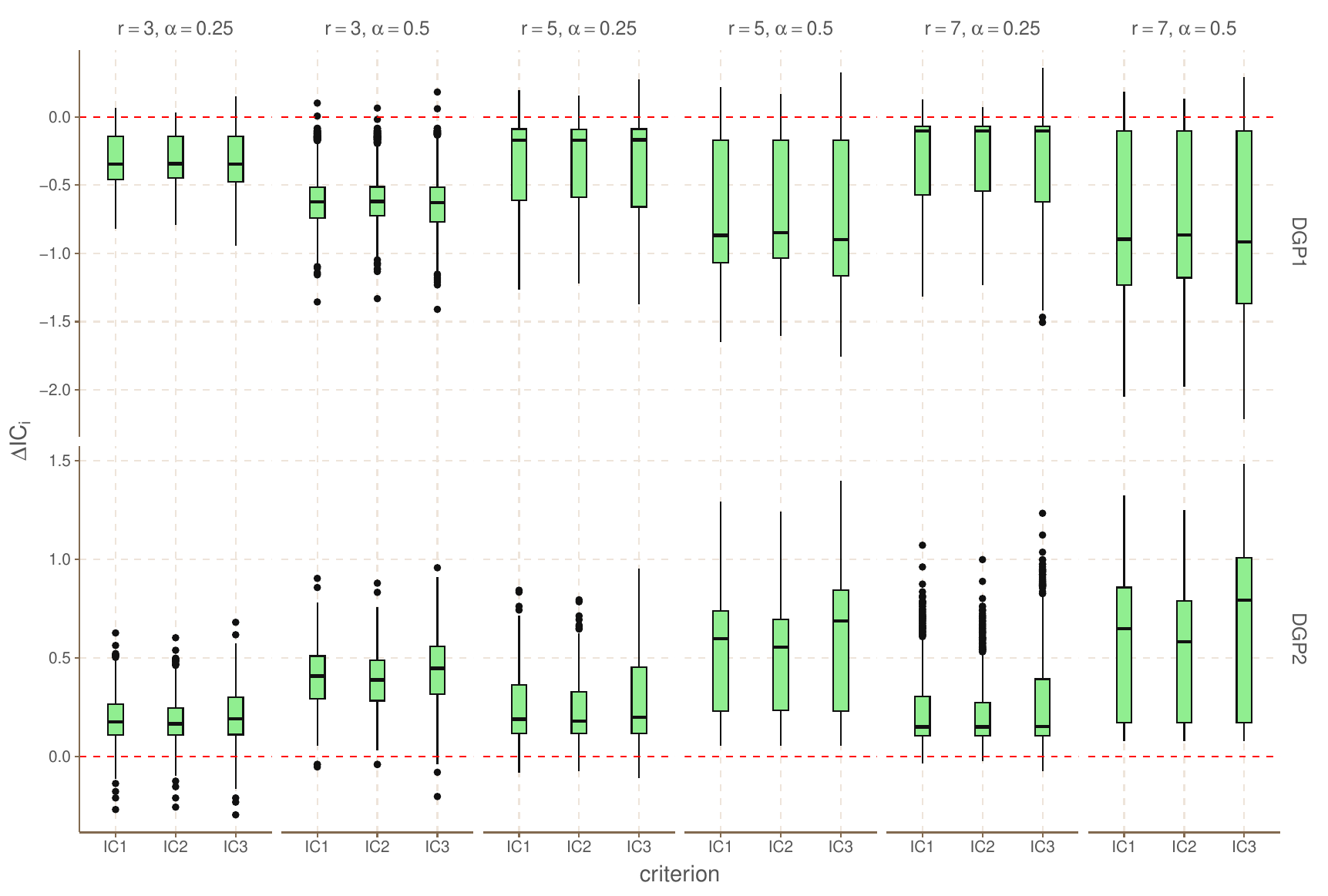}
\caption{\small The boxplots of $\Delta{\rm IC}_i$ ($i \in [3]$) for DGP1 and DGP2 with $p=100,n=50,\alpha=0.25,0.5$ and $r=3,5,7$ over 1000 simulation runs. 
}
\label{fig.0}
\vspace{-0.2cm}
\end{figure}

We firstly assess the finite-sample performance of the proposed information criteria in Section~\ref{subsec.selection} under different combinations of $p, n$ and $\alpha$ for DGP1 and DGP2. Figure~\ref{fig.0} presents boxplots of $\Delta{\rm IC}_i$ ($i=1,2,3$) for two DGPs under the setting $p=100$, $n=50$, $\alpha=0.25,0.5$ and $r=3,5,7.$ See also similar results for $p=100$, $n=50$, $\alpha=0.05,0.1$ in Figure~\ref{fig.model_select_case2} and the corresponding model selection accuracies in Table~\ref{tab.S1} of the supplementary material. We observe a few trends. First, the proposed criteria can lead to high model selection accuracies in all cases.
Second, larger values of $\alpha$ lead to improved model selection accuracy, as they correspond to a higher relative strength of the common components over idiosyncratic components, quantified by $\Vert\bSigma_{\chi}\Vert_{\cS,\F}/\Vert\bSigma_{\varepsilon}\Vert_{\cS,\F} \gtrsim p^{\alpha/2}$ for DGP 1 and $\Vert\bSigma_{\kappa}\Vert_{\cS,\F}/\Vert\bSigma_{\varepsilon}\Vert_{\cS,\F} \gtrsim p^{\alpha/2}$ for DGP 2.
Third, different penalty functions $g(n,p)$ have similar impacts on the information criteria when $p$ and $n$ are relatively large. 

\begin{table}[ht]
\footnotesize
\centering
\caption{\small The average relative frequency estimates for ${\mathbb P}(\hat{r}=r)$ over 1000 simulation runs.}
\label{tab.1}
\vspace{-0.2cm}
\begin{tabular}{lllcccccc}
\toprule
			& & & \multicolumn{2}{c}{$r=3$} & \multicolumn{2}{c}{$r=5$} & \multicolumn{2}{c}{$r=7$}\\
			$\alpha$ & $p$ & $n$ & ${\mathbb P}(\hat{r}^{\cD}=r)$ & ${\mathbb P}(\hat{r}^{\cF}=r)$ & ${\mathbb P}(\hat{r}^{\cD}=r)$ & ${\mathbb P}(\hat{r}^{\cF}=r)$ &
			${\mathbb P}(\hat{r}^{\cD}=r)$ & ${\mathbb P}(\hat{r}^{\cF}=r)$  \\
			\hline
			
			\multirow{4}{*}{$0.25$} & \multirow{2}{*}{$100$} & $100$  & 0.674  & 0.632  & 0.468  & 0.449  & 0.331  & 0.355  \\
			&       & $200$ & 0.757  & 0.726  & 0.622  & 0.571  & 0.466  & 0.483  \\
			& \multirow{2}{*}{$200$} & $100$  & 0.757  & 0.647  & 0.563  & 0.430  & 0.345  & 0.319  \\
			&       & $200$ & 0.846  & 0.797  & 0.717  & 0.668  & 0.560  & 0.521  \\
			\hline
			
			\multirow{4}{*}{$0.50$} & \multirow{2}{*}{$100$} & $100$  & 0.881  & 0.893  & 0.815  & 0.819  & 0.640  & 0.743  \\
			&       & $200$ & 0.936  & 0.921  & 0.891  & 0.908  & 0.805  & 0.828  \\
			& \multirow{2}{*}{$200$} & $100$  & 0.977  & 0.953  & 0.910  & 0.911  & 0.778  & 0.836  \\
			&       & $200$ & 0.970  & 0.962  & 0.954  & 0.958  & 0.898  & 0.928  \\
			\hline
			
			\multirow{4}{*}{$0.75$} & \multirow{2}{*}{$100$} & $100$  & 0.974  & 0.983  & 0.947  & 0.963  & 0.899  & 0.929  \\
			&       & $200$ & 0.979  & 0.985  & 0.975  & 0.978  & 0.951  & 0.971  \\
			& \multirow{2}{*}{$200$} & $100$  & 0.999  & 0.997  & 0.989  & 0.998  & 0.945  & 0.980  \\
			&       & $200$ & 0.997  & 0.999  & 0.998  & 1.000  & 0.992  & 0.997  \\
			\bottomrule
		\end{tabular}
  \vspace{-0.2cm}
\end{table}

Once the more appropriate FFM is selected based on observed data, our next step adopts the ratio-based estimator \eqref{eq.deter_num_1} (or \eqref{eq.deter_num_2}) to determine the number of functional (or scalar) factors. The performance of proposed estimators is then examined in terms of their abilities to correctly identify the number of factors. 
When implementing \eqref{eq.deter_num_1} and \eqref{eq.deter_num_2}, we choose $c_r=0.1$ and $r_{1,0}=r_{2,0}=20$. Additional simulations suggest that the results are not sensitive to the choice of $r_{1,0}$ and $r_{2,0}$, and a small value of $c_r$ leads to good performance.
Table~\ref{tab.1} reports average relative frequencies $\hat r=r$  under different combinations of $r=3,5,7$, $n=100,200,$ $p=100,200$ and $\alpha=0.25, 0.5, 0.75$ for both DGPs.
Several conclusions can be drawn.
First, for fixed $p$ and $n,$ larger values of $\alpha$ enhance the accuracy of identifying $r$.
Second, we observe the phenomenon of ``blessing of dimensionality'' in the sense that the estimation improves as $p$ increases, which is due to the increased information from added components on the factors. 

\begin{figure}[ht]
\centering
\includegraphics[width=0.75\textwidth]{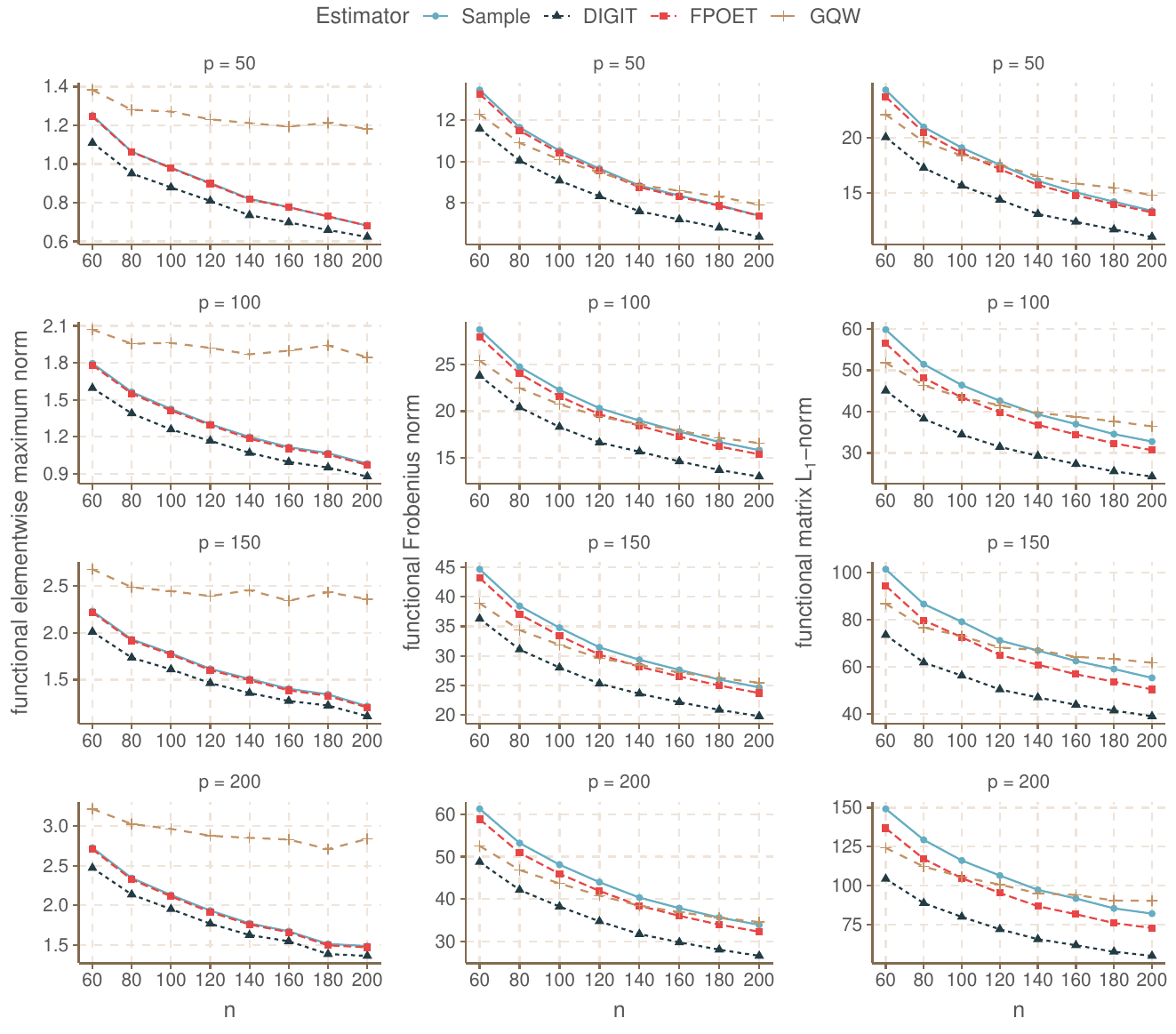}
\caption{\small The average losses of $\widehat \bSigma_y$ in functional elementwise $\ell_{\infty}$ norm (left column), Frobenius norm (middle column) and matrix $\ell_1$ norm (right column) for DGP1 over 1000 simulation runs.}
\label{fig.3}
\vspace{-0.2cm}
\end{figure}

\begin{figure}[ht]	
\centering
\includegraphics[width=0.75\textwidth]{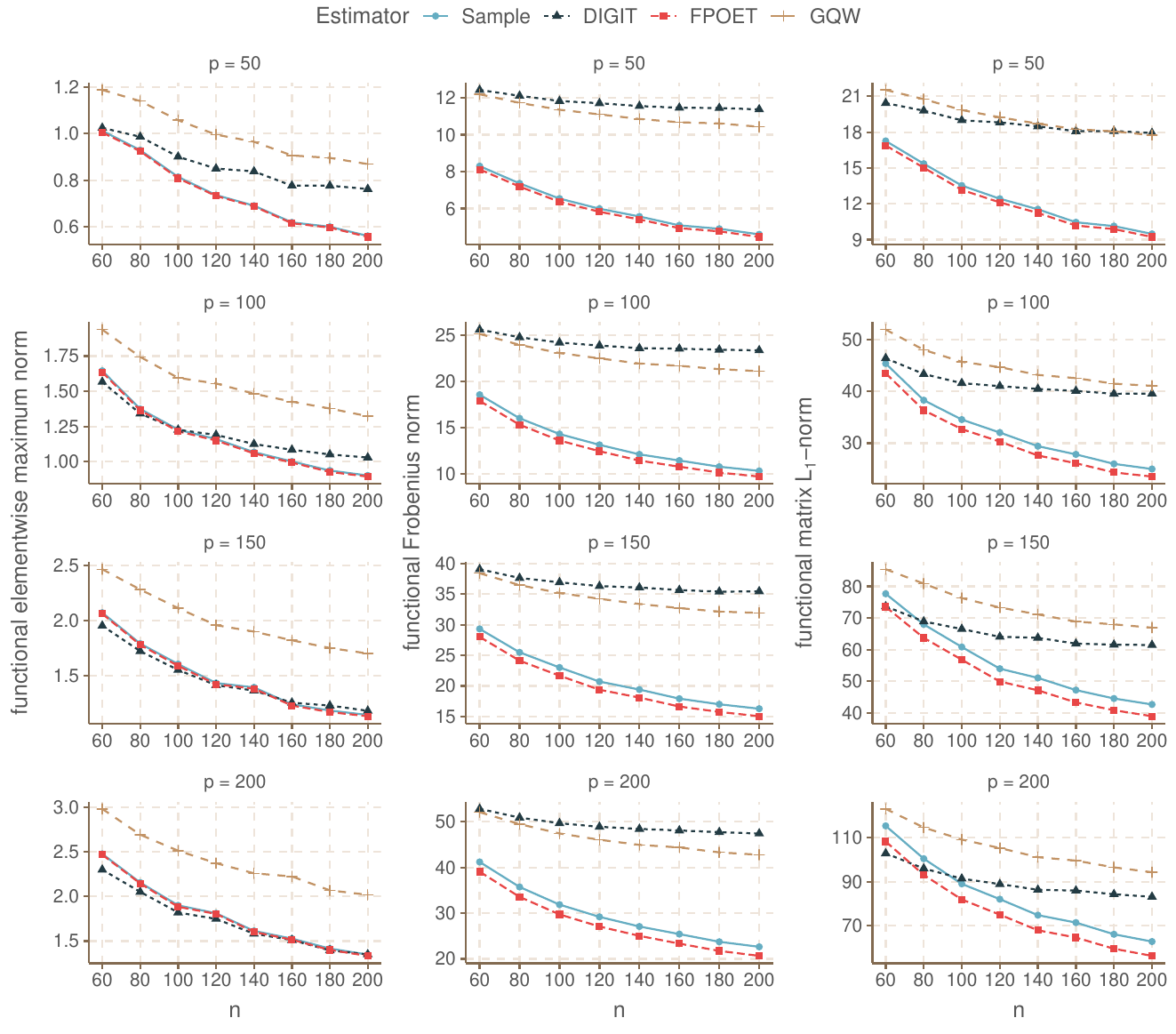}
\caption{\small The average losses of $\widehat \bSigma_y$ in functional elementwise $\ell_{\infty}$ norm (left column), Frobenius norm (middle column) and matrix $\ell_1$ norm (right column) for DGP2 over 1000 simulation runs.}
\label{fig.4}
\vspace{-0.2cm}
\end{figure}

We next compare our proposed AFT estimator in \eqref{eq.thre} with two related methods for estimating the idiosyncratic covariance $\bSigma_{\varepsilon},$ where the details can be found in Section~\ref{supsec.F} of the supplementary material. 
Following \cite{fan2013}, the threshold level under hard thresholding for AFT is selected as $\lambda=\dot{C}(\sqrt{\log p/n}+1/\sqrt{p})$ with $\dot{C}=0.5.$ 
To select the optimal $\dot{C}$, we also implemented a cross-validation method over a candidate set, whose lower bound was determined in a way similar to (4.1) of \cite{fan2013} to ensure the positive definiteness of the AFT estimators.
However, such method incurred heavy computational costs and only gave a very slight improvement. We finally compare our DIGIT and FPOET estimators with two competing methods for estimating the covariance $\bSigma_y.$ The first competitor is the sample covariance estimator $\widehat{\bSigma}_{y}^{\sS}.$ For comparison, we also implement the method of \cite{guo2025factor} in conjunction with our AFT (denoted as GQW). This combined method employs autocovariance-based eigenanalysis to estimate $\bB$ and then follows the similar procedure as DIGIT to estimate $\bbf_t(\cdot)$ and $\bSigma_{\varepsilon}.$
Although DIGIT and GQW estimators (or FPOET estimator) are specifically developed to fit model \eqref{eq.model_1} (or model \eqref{eq.model_2}), we also use them (or it) for estimating $\bSigma_y$ under DGP2 (or DGP1) to evaluate the robustness of each proposal under model misspecifications.
For both DGPs, we set $\alpha=0.5$ and generate $n=60, 80, \dots, 200$ observations of $p=50, 100, 150, 200$ functional variables. 
Figures~\ref{fig.3} and \ref{fig.4} display the numerical summaries of losses measured by functional versions of elementwise $\ell_{\infty}$ norm, Frobenius norm, and matrix $\ell_1$ norm for DGP1 and DGP2, respectively.

A few trends are observable. 
First, for DGP1 (or DGP2) in Figure~\ref{fig.3} (or Figure~\ref{fig.4}), the DIGIT (or FPOET) estimator outperforms the three competitors under almost all functional matrix losses and settings. In high-dimensional large $p$ scenarios, the factor-guided estimators lead to more competitive performance, whereas the results of $\widehat{\bSigma}_{y}^{\sS}$ severely deteriorate especially in terms of functional matrix $\ell_1$ loss.
Second, although both DIGIT and GQW estimators are developed to estimate model~\eqref{eq.model_1}, our proposed DIGIT estimator is prominently superior to the GQW estimator for DGP1 under all scenarios, as seen in Figure~\ref{fig.3}. 
Third, the FPOET estimator exhibits enhanced robustness compared to DIGIT and GQW estimators in the case of model misspecification. In particular, for DGP2, DIGIT and GQW show substantial decline in performance measured by functional Frobenius and matrix $\ell_1$ losses, while, for DGP1, FPOET still achieves reasonably good performance.

\section{Real data analysis}
\label{sec.real}

Our dataset, collected from \url{https://wrds-www.wharton.upenn.edu/}, consists of high-frequency observations of prices for a collection of S\&P100 stocks from $251$ trading days in the year 2017. We removed 2 stocks with missing data so $p=98$ in our analysis. We obtain five-minute resolution prices by using the last transaction price in each five-minute
interval after removing the outliers, and hence convert the trading period (9:30–16:00) to $\cU=[0,1].$ We construct CIDR \cite[]{horvath2014testing} trajectories, in percentage, by $y_{ti}(u_k)=100[\log\{P_{ti}(u_k)\}-\log\{P_{ti}(u_0)\}],$ where $P_{ti}(u_k)$ $(t\in [n], i \in [p], k \in [78])$ denotes the price of the $i$-th stock at the $k$-th five-minute interval ($u_k=k/78$) after the opening time on the $t$-th trading day. We obtain smoothed CIDR curves by expanding the data using a 10-dimensional B-spline basis. The CIDR curves, which always start from zero, not only have nearly the same shape as the original price curves but also enhance the plausibility of the stationarity assumption.
We performed functional KPSS test \cite[]{horvath2014testing} for each stock, and found
no overwhelming evidence (under 1\% significance level) against the stationarity.

For model selection, the information criteria, ${\rm IC}_1^{\cD}=-0.632<{\rm IC}_1^{\cF}=-0.614$, suggests that FFM~\eqref{eq.model_1} is slightly more preferable and implies that the latent factors may exhibit intraday varying patterns. We consider the problem of functional risk management in Section~\ref{subsec.risk}. Our task is to obtain the optimal functional portfolio allocation $\widehat \bw(\cdot)$ by minimizing the perceived risk of the functional portfolio, specifically, 
$$
\widehat \bw = \arg\min_{\bw\in\eH^p}
\big\langle\bw,\widehat{\bSigma}_y(\bw)\big\rangle
~~\text{subject to}~~
\int_{\cU}\bw(u)^{\T}\bone_p\du=1.
$$
Following the derivations in Section~\ref{supsec.opt_port} of the supplementary material, we obtain the solution:
\begin{equation}
    \label{eq.optim_weight}
    \widehat{\bw}(u)=\frac{\int_{\cU}\widehat{\bSigma}_y^{\dagger}(u,v)\bone_p\dv}{\int_{\cU}\int_{\cU}\bone_p^{\T}\widehat{\bSigma}_y^{\dagger}(z,v)\bone_p\dz\dv},~~u\in\cU,
\end{equation}
which allows us to obtain the actual risk. In practical implementation, we treat components of $\by_t(\cdot)$ as finite-dimensional functional objects and hence can obtain bounded inverse $\widehat{\bSigma}_y^{\dagger}$ using the leading eigenpairs of $\widehat{\bSigma}_y$  such that the corresponding cumulative percentage of selected eigenvalues exceeds 95\%.

\begin{table}[ht]
\footnotesize
\centering
\caption{\small Comparisons of the risks of the functional portfolios obtained using DIGIT, FPOET, GQW, POET-based and sample estimators.}
\label{tab.risk}
\vspace{-0.2cm}
\begin{tabular}{llccccccc}
			\toprule
			Estimator & $\hat r$ & July & August & September & October & November & December & Average \\
			\hline
			\multirow{5}{*}{DIGIT} & $1$   & 0.0064  & 0.0127  & 0.0050  & 0.0154  & 0.0081  & 0.0994  & 0.0245  \\
          & 3   & 0.0029  & 0.0082  & 0.0032  & 0.0054  & 0.0094  & 0.0080  & 0.0062  \\
          & 5   & 0.0057  & 0.0136  & 0.0077  & 0.0075  & 0.0137  & 0.0149  & 0.0105  \\
          & 7   & 0.0081  & 0.0105  & 0.0058  & 0.0112  & 0.0096  & 0.0142  & 0.0099  \\
          & 9   & 0.0105  & 0.0124  & 0.0067  & 0.0095  & 0.0070  & 0.0201  & 0.0110  \\
          & 11  & 0.0130  & 0.0129  & 0.0069  & 0.0077  & 0.0085  & 0.0257  & 0.0125  \\
            \hline
            \multirow{5}{*}{FPOET} & 1   & 0.0350  & 0.0136  & 0.0109  & 0.0441  & 0.0378  & 0.0174  & 0.0265  \\
          & 3   & 0.0227  & 0.0207  & 0.0150  & 0.0224  & 0.0280  & 0.0524  & 0.0269  \\
          & 5   & 0.0154  & 0.0139  & 0.0222  & 0.0349  & 0.0275  & 0.0199  & 0.0223  \\
          & 7   & 0.0142  & 0.0162  & 0.0142  & 0.0108  & 0.0283  & 0.0224  & 0.0177  \\
          & 9   & 0.0428  & 0.0180  & 0.0215  & 0.0201  & 0.0306  & 0.0194  & 0.0254  \\
          & 11  & 0.0562  & 0.0224  & 0.0129  & 0.0294  & 0.0342  & 0.0348  & 0.0316  \\
          \hline
           \multirow{5}{*}{GQW} & 1   & 0.0063  & 0.0249  & 0.2104  & 1.6762  & 0.0047  & 0.4441  & 0.3944  \\
          & 3   & 0.0036  & 0.0092  & 0.0045  & 0.0231  & 0.0093  & 0.0105  & 0.0100  \\
          & 5   & 0.0062  & 0.0114  & 0.0061  & 0.0063  & 0.0117  & 0.0074  & 0.0082  \\
          & 7   & 0.0069  & 0.0152  & 0.0064  & 0.0089  & 0.0081  & 0.0155  & 0.0102  \\
          & 9   & 0.0122  & 0.0111  & 0.0063  & 0.0090  & 0.0114  & 0.0120  & 0.0103  \\
          & 11  & 0.0140  & 0.0157  & 0.0087  & 0.0072  & 0.0103  & 0.0251  & 0.0135  \\
          \hline
             \multirow{5}{*}{POET} & 1   & 0.0271  & 0.0272  & 0.0352  & 0.0197  & 0.0242  & 0.0389  & 0.0287  \\
          & 3   & 0.0270  & 0.0342  & 0.0463  & 0.0267  & 0.0257  & 0.0405  & 0.0334  \\
          & 5   & 0.0185  & 0.0269  & 0.0346  & 0.0229  & 0.0273  & 0.0430  & 0.0289  \\
          & 7   & 0.0214  & 0.0275  & 0.0373  & 0.0226  & 0.0243  & 0.0375  & 0.0284  \\
          & 9   & 0.0206  & 0.0250  & 0.0411  & 0.0241  & 0.0209  & 0.0342  & 0.0277  \\
          & 11  & 0.0178  & 0.0271  & 0.0407  & 0.0223  & 0.0229  & 0.0366  & 0.0279  \\
          \hline
          
          Sample &  & 0.0203  & 0.0290  & 0.0372  & 0.0254  & 0.0267  & 0.0310  & 0.0282  \\
			\bottomrule
		\end{tabular}
  \vspace{-0.2cm}
\end{table}

Following the procedure in \cite{fan2013}, on the 1st trading day of each month from July to December, we estimate $\widehat{\bSigma}_y$ using DIGIT, FPOET, GQW and sample estimators based on the historical data comprising CIDR curves of $98$ stocks for the preceding 6 months ($n=126$). We then determine the corresponding optimal portfolio allocation $\widehat{\bw}(u_k)$ for $k \in [78]$. To illustrate the superiority of functional analytic methods, we also introduce a non-functional competing method based on the POET estimator, whose portfolio construction procedure is detailed in Section~\ref{supsec.opt_nonfun} of the supplementary material.
At the end of the month after 21 trading days, we compare actual risks calculated by ${78}^{-2}\sum_{k,k'\in [78]}\widehat{\bw}(u_k)^{\T}\{{21}^{-1}\sum_{t=1}^{21}\by_t(u_k)\by_t(u_{k'})^{\T}\}\widehat{\bw}(u_{k'}).$
Following \cite{fan2013} and \cite{wang2021}, we try $\hat r=1,3,5,7,9$ and $11$ to check the effect of $r$ in out-of-sample performance.
The numerical results are summarized in Table \ref{tab.risk}. Among the functional analytic methods, we observe that the optimal functional portfolio allocation created by DIGIT, FPOET, and GQW result in minimum averaged risks over six months as 0.0062, 0.0177, and 0.0082, respectively, while the sample covariance estimator gives 0.0282. The risk has been significantly reduced by at least 37\% using our factor-guided approach. Additionally, the POET-based method yields a minimum averaged risk of 0.0277, providing empirical evidence for the advantage of functional analytic methods.

\section{Discussions}
\label{sec.discussion}

Our theoretical results are established under a sub-Gaussian condition, which is imposed to facilitate the use of Hanson--Wright-type concentration inequalities  for time series within Hilbert space \cite[]{chang2024autocovariance} in our non-asymptotic analysis. To the best of our knowledge, the existing literature on concentration inequalities for high-dimensional functional time series are all of Hanson--Wright-type. It is thus of interest to relax such sub-Gaussian condition to a weaker finite moment condition beyond functional linear process, and based on which developing more generalized Nagaev-type concentration inequalities for high-dimensional time series \cite[]{zhang2021convergence} within Hilbert space to aid our theoretical analysis. This relaxation will result in allowing the dimension $p$ to grow polynomially rather than exponentially with $n$.
To achieve this, we also need to propose a new functional dependence measure instead of our functional stability measure defined in (\ref{eq.stab_meas}) to characterize how the established concentration results are affected by the complex serial dependence structure.

It is interesting to conduct specification tests such as testing one FFM against another and testing the constancy of the factor or loading functions. Both tests rely on the inferential theory for FFMs, which requires to explore the limiting distributions of estimated quantities. The existing literature on FFMs only studies the estimation of factors, loadings and number of factors, without delving into the corresponding limiting distributions. 
Note that the inferential theory for FFM based on MFPCA can be developed for FFM~\eqref{eq.model_2} given its equivalence to the least squares method. By comparison, the inferential theory for the eigenanalysis of doubly integrated Gram covariance specific to FFM~\eqref{eq.model_1} presents significant challenges. 
Moreover, deriving the limiting distributions under functional domain involves characterizing the magnitude of functional quantities using a suitable functional norm such as $L_2
$ or supremum norm, presenting additional complexities compared to scalar time series.

While our paper focuses on fully observed functional time series, it is also interesting to consider the common practical scenario, where each curve $y_{tj}(\cdot)$ is only partially observed, with errors, at $T_{tj}$ random time points. For densely observed functional time series with $T_{tj}$'s being larger than some order of $n,$ it is customary to apply nonparametric smoothing to the observations from each curve \cite[]{zhang2007}, which results in reconstructed curves $\widehat y_{tj}(\cdot)$'s serving as new inputs for subsequent analysis. When $T_{tj}$'s are bounded, referred to as sparsely observed functional time series, the pre-smoothing step becomes inapplicable. An alternative approach involves local linear surface smoothing by pooling observations together from all curves \cite[]{chen2022}. This method yields smoothed estimates of $\Sigma_{y,jk}(u,v)$'s that can be utilized in our methodological development. 

The aforementioned topics are beyond the scope of the current paper and will be pursued elsewhere.

\section*{Acknowledgements} We are grateful to the editor, the associate editor and two referees for their insightful and valuable comments and suggestions, which have led to significant improvement of our paper. Li's research is partially supported by the NSFC (no.72471127). Qiao’s research is partially supported by the Seed Fund for Basic Research for New Staff at the University of Hong Kong.

\begin{appendix}

\section*{Appendix}

\appendix
\setcounter{equation}{0}
\renewcommand{\theequation}{\Alph{section}.\arabic{equation}}
\renewcommand{\theHequation}{\Alph{section}.\arabic{equation}}
\setcounter{figure}{0}
\renewcommand{\thefigure}{\Alph{section}.\arabic{figure}}
\renewcommand{\theHfigure}{\Alph{section}.\arabic{figure}}
\setcounter{table}{0}
\renewcommand{\thetable}{\Alph{section}.\arabic{table}}
\renewcommand{\theHtable}{\Alph{section}.\arabic{table}}
\setcounter{lemma}{0}
\renewcommand{\thelemma}{\Alph{section}.\arabic{lemma}}
\renewcommand{\theHlemma}{\Alph{section}.\arabic{lemma}}
\setcounter{remark}{0}
\renewcommand{\theremark}{\Alph{section}.\arabic{remark}}
\renewcommand{\theHremark}{\Alph{section}.\arabic{remark}}
\setcounter{definition}{0}
\renewcommand{\thedefinition}{\Alph{section}.\arabic{definition}}
\renewcommand{\theHdefinition}{\Alph{section}.\arabic{definition}}
\setcounter{proposition}{0}
\renewcommand{\theproposition}{\Alph{section}.\arabic{proposition}}
\renewcommand{\theHproposition}{\Alph{section}.\arabic{proposition}}
\setcounter{condition}{0}
\renewcommand{\thecondition}{\Alph{section}.\arabic{condition}}
\renewcommand{\theHcondition}{\Alph{section}.\arabic{condition}}
\setcounter{assumption}{0}
\renewcommand{\theassumption}{\Alph{section}.\arabic{assumption}}
\renewcommand{\theHassumption}{\Alph{section}.\arabic{assumption}}

This appendix contains the technical proofs of the results in Section \ref{sec.method} and the results for FFM~\eqref{eq.model_1} in Section \ref{sec.theory}, i.e., Theorems~\ref{thm.load_factor_1}--\ref{thm.inverse_digit} and Corollary~\ref{coro.DIGIT}. For the technical proofs of the results for FFM~\eqref{eq.model_2} in Section \ref{sec.theory} and the results in Section~\ref{sec.app}, see the supplementary material, which also provides the proofs of some technical lemmas in the appendix, some further derivations, additional simulation results and real data results. Throughout, we denote the multiplications of matrix kernel functions as $\bM=\bK\bG \in \eH^p\otimes\eH^p$ for $\bK,\bG\in\eH^p\otimes\eH^p,$ where $\bM(u,v)=\int_{\cU}\bK(u,w)\bG(w,v){\rm d}w.$

\section{Proofs of theoretical results in Section \ref{sec.method}}
\label{supsec.A}
	
\subsection{Technical lemmas}
We first introduce useful theorems to prove Proposition \ref{propos.eigenvalues}. In the following two lemmas, $\{\lambda_j\}_{j \in [p]}$ are the eigenvalues of $\bSigma \in {\mathbb R}^{p \times p}$ in a descending order and $\{\bxi_j\}_{j \in [p]}$ are the corresponding eigenvectors. Similarly, $\{\widetilde{\lambda}_j\}_{j \in [p]}$ and $\{\widetilde{\bxi}_j\}_{j \in [p]}$ are the corresponding eigenvalues and eigenvectors of $\widetilde{\bSigma} \in {\mathbb R}^{p \times p},$ respectively. 
	
	\begin{lemma}
		\label{lem.weyl}
		(Weyl's theorem; \cite{weyl1912asymptotische}). $|\widetilde{\lambda}_j-\lambda_j|\le\Vert\widetilde{\bSigma}-\bSigma\Vert$ for $j \in [p].$
	\end{lemma}
	
	\begin{lemma}
		\label{lem.sintheta}
		(A useful variant of ${\rm sin}(\theta)$ theorem; \cite{yu2015useful}). If $\widetilde{\bxi}_j^{\T}\bxi_j\ge0$ for $j \in [p]$, then
		$$\Vert\widetilde{\bxi}_j-\bxi_j\Vert\le\frac{\Vert\widetilde{\bSigma}-\bSigma\Vert/\sqrt{2}}{\min\big(|\widetilde{\lambda}_{j-1}-\lambda_j|,|\lambda_j-\widetilde{\lambda}_{j+1}|\big)}.$$
	\end{lemma}
	
The functional version of Weyl's theorem has been studied in Lemmas 4.2 and 4.3 of \cite{Bbosq1}. Let $\{\tau_i\}_{i=1}^{\infty}$ be the eigenvalues of the kernel function $\bSigma(\cdot,\cdot)$ in a descending order and $\{\bvarphi_i(\cdot)\}_{i=1}^{\infty}$ are the corresponding eigenfunctions. Similarly, $\{\widetilde{\tau}_i\}_{i=1}^{\infty}$ and $\{\widetilde{\bvarphi}_i(\cdot)\}_{i=1}^{\infty}$ are the corresponding eigenvalues and eigenfunctions of $\widetilde{\bSigma}(\cdot,\cdot),$ respectively. 
	\begin{lemma}
		\label{lem.weyl_2}
		(Lemma 4.2 in \cite{Bbosq1}). $|\widetilde{\tau}_i-\tau_i|\le\Vert\widetilde{\bSigma}-\bSigma\Vert_{\cL}$ for all $i.$
	\end{lemma}

        \begin{lemma}
		\label{lem.sintheta_2}
		(Lemma 4.3 in \cite{Bbosq1}). If $\langle\widetilde{\bvarphi}_i,\bvarphi_i\rangle\ge0$, then
        $$\Vert\widetilde{\bvarphi}_i-\bvarphi_i\Vert\le\frac{2\sqrt{2}\Vert\widetilde{\bSigma}-\bSigma\Vert_{\cL}}{\min\big(|\widetilde{\tau}_{i-1}-\tau_i|,|\tau_i-\widetilde{\tau}_{i+1}|\big)}.$$
	\end{lemma}

        The following lemmas introduce some useful functional norm inequalities. Their proofs are relegated to the supplementary material.
        
        \begin{lemma}
		\label{lem.ineq1}
		Suppose that $\bK\in\eH^p\otimes\eH^p$ is a Mercer's kernel with the spectral decomposition $\bK(u,v)=\sum_{i=1}^{\infty}\lambda_i\bphi_i(u)\bphi_i(v)^{\T},$ where $\{\lambda_i\}_{i=1}^{\infty}$ are the eigenvalues of $\bK$ in a descending order and $\{\bphi_i(\cdot)\}$ are the corresponding eigenfunctions. Then, we have\\
		(i) ${\rm tr}\left\{\int\int\bK(u,v)\bK(u,v)^{\T}\du\dv\right\}={\rm tr}\{\int\bK\bK^{\T}(u,u)\du\}=\Vert\bK\Vert_{\cS,\F}^2=\sum_{i=1}^{\infty}\lambda_i^2$;\\
		(ii) $\left\Vert\int\int\bK(u,v)\bK(u,v)^{\T}\du\dv\right\Vert=\left\Vert\int\bK\bK^{\T}(u,u)\du\right\Vert=\Vert\bK\Vert_{\cL}^2=\lambda_1^2;$\\
            (iii) $\tr\left\{\int\bK(u,u)\du\right\}=\Vert\bK\Vert_{\cN}=\sum_{i=1}^{\infty}\lambda_i.$
	\end{lemma}
	
	\begin{lemma}
		\label{lem.ineq2}
		Suppose that $\bK,\bG\in\eH^p\otimes\eH^p$ are Mercer's kernels, then we have\\
		(i) $\left\Vert\int\int \bK(u,v)\bG(u,v)^{\T}{\rm d}u{\rm d}v\right\Vert_{1}\le \Vert\bK\Vert_{\cS,1}\Vert\bG\Vert_{\cS,\infty}$;\\
		(ii) $\left\Vert\int\int \bK(u,v)\bG(u,v)^{\T}{\rm d}u{\rm d}v\right\Vert_{\infty}\le\Vert\bK\Vert_{\cS,\infty}\Vert\bG\Vert_{\cS,1}$;\\
		(iii) $\left\Vert\int\int \bK(u,v)\bG(u,v)^{\T}{\rm d}u{\rm d}v\right\Vert\le \left(\Vert\bK\Vert_{\cS,\infty}\Vert\bK\Vert_{\cS,1}\right)^{1/2}\left(\Vert\bG\Vert_{\cS,\infty}\Vert\bG\Vert_{\cS,1}\right)^{1/2}$;\\
		(iv) $\left\Vert\int\int \bK(u,v)\bG(u,v)^{\T}{\rm d}u{\rm d}v\right\Vert\le\left\{\big\Vert\int\int \bK(u,v)\bK(u,v)^{\T}{\rm d}u{\rm d}v\big\Vert\right\}^{1/2}\left\{\big\Vert\int\int \bG(u,v)\bG(u,v)^{\T}{\rm d}u{\rm d}v\big\Vert\right\}^{1/2}$.
	\end{lemma}
	
	\begin{lemma}
        \label{lem.op_le_S1}
		Suppose that $\bSigma=\{\Sigma_{ij}(\cdot,\cdot)\}_{p\times p}$ with $\Sigma_{ij}\in\mathbb{S}$ and $\widetilde{\bSigma}\in\eH^p\otimes\eH^p$ are Mercer's kernels. Then we have (i) $\Vert\bSigma\widetilde{\bSigma}\Vert_{\cL}\le\Vert\bSigma\Vert_{\cL}\cdot\Vert\widetilde{\bSigma}\Vert_{\cL}$, (ii) $\Vert\bSigma\Vert_{\cL}\le\Vert\bSigma\Vert_{\cS,\F}$, and (iii) $\Vert\bSigma\Vert_{\cL}\le\Vert\bSigma\Vert_{\cS,1}^{1/2}\Vert\bSigma\Vert_{\cS,\infty}^{1/2}$. Furthermore, if $\Vert\Sigma_{ij}\Vert_{\cS}=\Vert\Sigma_{ji}\Vert_{\cS}$ for all $i,j\in[p]$, then $\Vert\bSigma\Vert_{\cL}\le\Vert\bSigma\Vert_{\cS,1}$. 
	\end{lemma}

        \begin{lemma}
		\label{lem.trace}
		Suppose that $\bK,\bG\in\eH^p\otimes\eH^p$ are Mercer's kernels, then we have\\
		(i) ${\rm tr}\{\int\bK\bG(u,u)\du\}={\rm tr}\{\int\bG\bK(u,u)\du\}$, i.e., $\Vert\bK\bG\Vert_{\cN}=\Vert\bG\bK\Vert_{\cN}$;\\
		(ii) ${\rm tr}\{\int\bK\bG(u,u)\du\}\le\Vert\bK\Vert_{\cL}{\rm tr}\{\int\bG(u,u)\du\}$, i.e., $\Vert\bK\bG\Vert_{\cN}\le\Vert\bK\Vert_{\cL}\Vert\bG\Vert_{\cN}$;\\
		(iii) $\Vert\bK\bG\Vert_{\cS,\F}\le\Vert\bK\Vert_{\cL}\Vert\bG\Vert_{\cS,\F}$. 
	\end{lemma}

        \begin{lemma}
		\label{lem.ineq3}
		For $\bA\in\mathbb{R}^{p\times q}$ and $\bK=\{K_{ij}(\cdot,\cdot)\}_{q \times q}\in\eH^q\otimes\eH^q$, we have \\
		(i) $\Vert\bA\bK\Vert_{\cS,\max}\le \Vert\bA\Vert_{\infty}\Vert\bK\Vert_{\cS,\max}$, and  $\Vert\bK\bA^{\T}\Vert_{\cS,\max}\le\Vert\bK\Vert_{\cS,\max}\Vert\bA^{\T}\Vert_{1}=\Vert\bA\Vert_{\infty}\Vert\bK\Vert_{\cS,\max}$;\\
		(ii) $\Vert\bA\bK\Vert_{\cS,\F}\le \Vert\bA\Vert_{\F}\Vert\bK\Vert_{\cS,\F}$, and $\Vert\bK\bA^{\T}\Vert_{\cS,\F}\le\Vert\bK\Vert_{\cS,\F}\Vert\bA^{\T}\Vert_{\F}=\Vert\bA\Vert_{\F}\Vert\bK\Vert_{\cS,\F}$;\\
		(iii) $\Vert\bA\bK\Vert_{\cS,\infty}\le \Vert\bA\Vert_{\infty}\Vert\bK\Vert_{\cS,\infty}$, and $\Vert\bK\bA^{\T}\Vert_{\cS,\infty}\le\Vert\bK\Vert_{\cS,\infty}\Vert\bA^{\T}\Vert_{\infty}=\Vert\bA\Vert_{1}\Vert\bK\Vert_{\cS,\infty}$;\\
		(iv) $\Vert\bA\bK\Vert_{\cS,1}\le \Vert\bA\Vert_{1}\Vert\bK\Vert_{\cS,1}$, and $\Vert\bK\bA^{\T}\Vert_{\cS,1}\le\Vert\bK\Vert_{\cS,1}\Vert\bA^{\T}\Vert_{1}=\Vert\bA\Vert_{\infty}\Vert\bK\Vert_{\cS,1}$.
	\end{lemma}
        
        \begin{lemma}
		\label{lem.factor_norm}
		For $\bbf,\bg\in\eH^r$, and $\bA\in\mathbb{R}^{p\times r}$, we have\\ (i) $\Vert\bA\bbf\Vert\le \Vert\bA\Vert\cdot\Vert\bbf\Vert;$\\
		(ii) $\Vert K\Vert_{\cS}\le\Vert\bbf\Vert\cdot\Vert\bg\Vert$ where $K(\cdot,\cdot)\in\mathbb{S}$ is defined as $K(u,v)=\bbf(u)^{\T}\bg(v)$.
	\end{lemma}

	\subsection{Proof of Proposition \ref{propos.eigenvalues}}
	\noindent (i) Note that $\{\lambda_j\}_{j=1}^{p}$ are the non-vanishing eigenvalues of $\bOmega=\int\int\bSigma_{y}(u,v)\bSigma_{y}(u,v)^{\T}{\rm d}u{\rm d}v$, and $\{p^2\theta_j\}_{j=1}^{r}$ are nonzero eigenvalues of $\bOmega_{\cL}$, while the other $p-r$ eigenvalues are zero. Then applying Lemma~\ref{lem.weyl} yields that, for each $j\in[r]$,
	$$|\lambda_j-p^2\theta_j|\le\Vert\bOmega-\bOmega_{\cL}\Vert=\Vert\bOmega_{\cR}\Vert,$$
and for $r+1\le j\le p,$ $|\lambda_j|=|\lambda_j-0|\le\Vert\bOmega_{\cR}\Vert.$ 
	
	\noindent (ii) By Lemma~\ref{lem.sintheta}, for $j\in[r]$ and $\bxi_j^{\T}\widetilde{\bbb}_j\ge0$,
	$$\Vert\bxi_j-\widetilde{\bbb}_j\Vert\le \frac{\Vert\bOmega_{\cR}\Vert/\sqrt{2}}{\min(|\lambda_{j-1}-p^2\theta_j|,|p^2\theta_j-\lambda_{j+1}|)}.$$ 
	Note that there exists a generic constant $c>0$ such that $|\lambda_{j-1}-p^2\theta_j|>p^2|\theta_{j-1}-\theta_j|-|\lambda_{j-1}-p^2\theta_{j-1}|>cp^2$ since $|\lambda_{j-1}-p^2\theta_{j-1}|\le\Vert\bOmega_{\cR}\Vert=o(p^2)$ from part (i). If $j<r$, a similar argument implies that $|p^2\theta_j-\lambda_{j+1}|>cp^2$. If $j=r,|p^2\theta_r-\lambda_{r+1}|>p^2\theta_r-|\lambda_{r+1}|>cp^2$ since $|\lambda_{r+1}|\le\Vert\bOmega_{\cR}\Vert=o(p^2)$ by using part (i) again. Hence, $\min(|\lambda_{j-1}-p^2\theta_j|,|p^2\theta_j-\lambda_{j+1}|)\gtrsim p^2$, and if $\bxi_j^{\T}\widetilde{\bbb}_j\ge0$, we have
	$$\Vert\bxi_j-\widetilde{\bbb}_j\Vert=O(p^{-2}\Vert\bOmega_{\cR}\Vert),\ {\rm for\ }j\in[r].$$
	
	\subsection{Proof of Proposition \ref{propos.eigenvalues_2}}
  To prove Proposition \ref{propos.eigenvalues_2}, we first present a technical lemma with proof in the supplementary material.
        \begin{lemma}
            \label{lem.eigen_QQ}
            Suppose that Assumption~\ref{ass.ind2} holds. Then, $\{p\vartheta_j\}_{j\in[r]}$ are the non-vanishing eigenvalues of $\bQ(\cdot)\bQ(\cdot)^{\T}\in\eH^p\otimes\eH^p$ with the corresponding eigenfunctions $\{\widetilde{\bq}_j(\cdot)\}_{j\in[r]}.$
        \end{lemma}
        
We are now ready to prove Proposition \ref{propos.eigenvalues_2}.\\
	\noindent (i) Note that $\{\tau_i\}_{i=1}^{\infty}$ are the eigenvalues of $\bSigma_{y}(\cdot,\cdot)$, and $\{p\vartheta_j\}_{j=1}^{r}$ are the $r$ non-vanishing eigenvalues of $\bQ\bQ^{\T}(\cdot,\cdot)$ by Lemma~\ref{lem.eigen_QQ}. Applying Lemma \ref{lem.weyl_2}, we have, for $j\in[r]$,
	$$
	\left|\tau_j-p\vartheta_j\right|\le\Vert\bSigma_{y}-\bQ\bQ^{\T}\Vert_{\cL}=\Vert\bSigma_{\varepsilon}\Vert_{\cL},
	$$
	and, for $j>r+1$, $|\tau_i|=|\tau_i-0|\le\Vert\bSigma_{\varepsilon}\Vert_{\cL}$. 

 \noindent (ii) By Lemma~\ref{lem.op_le_S1}(iii), we have $\Vert\bSigma_{\varepsilon}\Vert_{\cL}\le\Vert\bSigma_{\varepsilon}\Vert_{\cS,1}^{1/2}\Vert\bSigma_{\varepsilon}\Vert_{\cS,\infty}^{1/2}=O(s_p)=o(p)$, which yields that $\tau_j\asymp p\vartheta_j\asymp p$ for $j\in[r]$. Under Assumption~\ref{ass.ind2}, $\vartheta_j$ are distinguishable and bounded away from both zero and infinity, then $\min(|p\vartheta_{j-1}-\tau_j|,|\tau_j-p\vartheta_{j+1}|)\asymp p$ for $j\in[r]$. It follows from Lemma \ref{lem.sintheta_2} that $\Vert\bvarphi_j-\widetilde{\bq}_j\Vert=O(p^{-1}\Vert\bSigma_{\varepsilon}\Vert_{\cL})$ for $j\in[r]$.
	
\subsection{Proof of Proposition \ref{propos.equiv}}
        
The sample covariance matrix of estimated idiosyncratic components by using the constrained least squares follows that
$$\widetilde{\bSigma}_{\varepsilon}(u,v)=\frac{1}{n}\{\bY(u)-\widehat{\bQ}(u)\widehat{\bGamma}^{\T}\}\{\bY(v)^{\T}-\widehat{\bGamma}\widehat{\bQ}(v)^{\T}\}=\frac{1}{n}\bY(u)\bY(v)^{\T}-\widehat{\bQ}(u)\widehat{\bQ}(v)^{\T},$$
where we use the normalization condition $n^{-1}\widehat{\bGamma}^{\T}\widehat{\bGamma}=\bI_r$ and $\widehat{\bQ}(\cdot)=n^{-1}\bY(\cdot)\widehat{\bGamma}$. If we can show that $\widehat{\bQ}(u)\widehat{\bQ}(v)^{\T}=\sum_{j=1}^{r}\hat{\tau}_j\widehat{\bvarphi}_j(u)\widehat{\bvarphi}_j(v)^{\T}$, then by the spectral decompositions of the sample covariance estimator $$\widehat{\bSigma}_{y}^{\sS}(u,v)=\frac{1}{n}\bY(u)\bY(v)^{\T}=\sum_{j=1}^{r}\hat{\tau}_j\widehat{\bvarphi}_j(u)\widehat{\bvarphi}_j(v)^{\T}+\widehat{\bR}(u,v)=\widehat{\bQ}(u)\widehat{\bQ}(v)^{\T}+\widetilde{\bSigma}_{\varepsilon}(u,v),$$
we have $\widehat{\bR}(\cdot,\cdot)=\widetilde{\bSigma}_{\varepsilon}(\cdot,\cdot)$. Thus, by applying the adaptive functional thresholding with the same regularization parameters to the same remainder covariance matrix functions, we have $\widehat{\bR}^{\cA}(\cdot,\cdot)=\widetilde{\bSigma}_{\varepsilon}^{\cA}(\cdot,\cdot)$, and then $\widehat{\bSigma}_{y}^{\cF}(\cdot,\cdot)=\widehat{\bSigma}_{y}^{\sL}(\cdot,\cdot)$, which gives the desired result. 

We next show that $\widehat{\bQ}(u)\widehat{\bQ}(v)^{\T}=\sum_{j=1}^{r}\hat{\tau}_j\widehat{\bvarphi}_j(u)\widehat{\bvarphi}_j(v)^{\T}$ holds. To do this, we impose another identifiability condition that can serve as an alternative (see also Remark \ref{rmk.identif}) to Assumption~\ref{ass.ind2}. 

\begin{assumption}
\label{ass.ind3}
$p^{-1}\int\bQ(u)^{\T}\bQ(u){\rm d}u=\bI_r$ and $\bSigma_{\gamma}$ is diagonal with distinct diagonal elements being bounded away from both 0 and $\infty$ as $p\to\infty$.
\end{assumption}

Note that Assumptions~\ref{ass.ind2} and \ref{ass.ind3} can be converted to each other by orthogonal transformation. Thus, for the minimization problem \eqref{eq.problem}, we can use the following two equivalent normalization constraints:
\begin{flalign}
\label{eq.normalization}
\begin{aligned}
&(i)\ n^{-1}\sum_{t=1}^{n}\bgamma_t\bgamma_t^{\T}=\bI_r,\ {\rm and}\ p^{-1}\int\bQ(u)^{\T}\bQ(u){\rm d}u\ {\rm is\ diagonal},\\
& (ii)\ n^{-1}\sum_{t=1}^{n}\bgamma_t\bgamma_t^{\T}\ {\rm is\ diagonal},\ {\rm and\ }p^{-1}\int\bQ(u)^{\T}\bQ(u){\rm d}u=\bI_r.
\end{aligned} &&
\end{flalign}
Note that \eqref{eq.normalization}(i) is used in Section~\ref{subsec.fun_load} to obtain FPOET estimator. Following the similar procedure, we obtain that $\widecheck{\bQ}(\cdot)=\sqrt{p}(\widehat{\bvarphi}_1(\cdot),\dots,\widehat{\bvarphi}_r(\cdot))$ and $\widecheck{\bGamma}=p^{-1}\int\bY(u)^{\T}\widecheck{\bQ}(u){\rm d}u$ is the solution to \eqref{eq.problem} under \eqref{eq.normalization}(ii). One can show that the two solutions under normalization constraints \eqref{eq.normalization}(i) and (ii) are equivalent and can be converted to each other through an orthogonal matrix, i.e., there exists an $r\times r$ orthogonal matrix $\bH$ such that $\widecheck{\bQ}(\cdot)=\widehat{\bQ}(\cdot)\bH$ and $\widecheck{\bGamma}=\widehat{\bGamma}\bH$. Notice that $\widecheck{\bQ}(\cdot)=\sqrt{p}\{\widehat{\bvarphi}_1(\cdot),\dots,\widehat{\bvarphi}_r(\cdot)\}$ and $\widecheck{\bGamma}=p^{-1}\int\bY(u)^{\T}\widecheck{\bQ}(u){\rm d}u$, then we have
	$$
	\begin{aligned}
		n^{-1}\widecheck{\bGamma}^{\T}\widecheck{\bGamma}=&p^{-2}n^{-1}\int\widecheck{\bQ}(u)^{\T}\bY(u){\rm d}u\int\bY(v)^{\T}\widecheck{\bQ}(v){\rm d}v
		=
		p^{-2}\int\int\widecheck{\bQ}(u)^{\T}\{n^{-1}\bY(u)\bY(v)^{\T}\}\widecheck{\bQ}(v){\rm d}u{\rm d}v\\
		=&p^{-1}\int\{\widehat{\bvarphi}_1(u)^{\T},\dots,\widehat{\bvarphi}_r(u)^{\T}\}^{\T}\Big[\int\widehat{\bSigma}_{y}^{\sS}(u,v)\{\widehat{\bvarphi}_1(v),\dots,\widehat{\bvarphi}_r(v)\}{\rm d}v\Big]{\rm d}u\\
		=&p^{-1}\int\{\widehat{\bvarphi}_1(u)^{\T},\dots,\widehat{\bvarphi}_r(u)^{\T}\}^{\T}\{\hat{\tau}_1\widehat{\bvarphi}_1(u),\dots,\hat{\tau}_r\widehat{\bvarphi}_r(u)\}{\rm d}u
		=p^{-1}{\rm diag}(\hat{\tau}_1,\dots,\hat{\tau}_r).
	\end{aligned}
	$$
	Since $\widecheck{\bQ}(\cdot)\widecheck{\bGamma}^{\T}=\widehat{\bQ}(\cdot)\bH\bH^{\T}\widehat{\bGamma}^{\T}=\widehat{\bQ}(\cdot)\widehat{\bGamma}^{\T}$, it follows that
	$$\widehat{\bQ}(u)\widehat{\bQ}(v)^{\T}=n^{-1}\widehat{\bQ}(u)\widehat{\bGamma}^{\T}\widehat{\bGamma}\widehat{\bQ}(v)^{\T}=n^{-1}\widecheck{\bQ}(u)\widecheck{\bGamma}^{\T}\widecheck{\bGamma}\widecheck{\bQ}(v)^{\T}=\sum_{j=1}^{r}\hat{\tau}_j\widehat{\bvarphi}_j(u)\widehat{\bvarphi}_j(u)^{\T}.$$

	\section{Proofs of theoretical results in Section~\ref{sec.theory}}
	
	\subsection{Technical lemmas}
	\begin{lemma}
		\label{lem.rate_sigma_f}
		Under Assumptions \ref{ass.regu_cond_DIGIT}(iv) and \ref{ass.concentration}, we have that,\\
		(i) for any $i,j\in[r],\Vert n^{-1}\sum_{t=1}^{n}f_{ti}f_{tj}-\Sigma_{f,ij}\Vert_{\cS}=O_p(1/\sqrt{n})$, $\Vert n^{-1}\sum_{t=1}^{n}\bbf_t\bbf_t^{\T}-\bSigma_{f}\Vert_{\cS,\max}=O_p(1/\sqrt{n})$;\\
		(ii) for any $i,j\in[p],\Vert n^{-1}\sum_{t=1}^{n}\varepsilon_{ti}\varepsilon_{tj}-\Sigma_{\varepsilon,ij}\Vert_{\cS}=O_p(\cM_{\varepsilon}/\sqrt{n})$, $\Vert n^{-1}\sum_{t=1}^{n}\bvarepsilon_t\bvarepsilon_t^{\T}-\bSigma_{\varepsilon}\Vert_{\cS,\max}=O_p(\cM_{\varepsilon}\sqrt{\log p/n})$;\\
		(iii) for any $i,j\in[p],\Vert n^{-1}\sum_{t=1}^{n}y_{ti}y_{tj}-\Sigma_{y,ij}\Vert_{\cS}=O_p(\cM_{\varepsilon}/\sqrt{n})$, $\Vert n^{-1}\sum_{t=1}^{n}\by_t\by_t^{\T}-\bSigma_{y}\Vert_{\cS,\max}=O_p(\cM_{\varepsilon}\sqrt{\log p/n})$.
	\end{lemma}

We next introduce a lemma to give the perturbation rate in elementwise $\ell_{\infty}$ norm of the eigenvectors if a matrix is perturbed. Suppose that $\bA\in \mathbb{R}^{p\times p}$ is a symmetric matrix. Let the perturbed matrix be $\widetilde{\bA}=\bA+\bE$, where $\bE\in\mathbb{R}^{p\times p}$ is a symmetric perturbation matrix. Suppose the spectral decomposition of $\bA$ is given by $\bA=\sum_{i=1}^{r}\lambda_i\bv_i\bv_i^{\T}+\sum_{i>r}\lambda_i\bv_i\bv_i^{\T}$, where $|\lambda_1|>|\lambda_2|>\dots>|\lambda_p|$. Clearly, $\bA_r=\sum_{i=1}^{r}\lambda_i\bv_i\bv_i^{\T}$ is the best rank-$r$ approximation of $\bA$. Analogously, the spectral decomposition of $\widetilde{\bA}=\sum_{i=1}^{r}\tilde{\lambda}_i\widetilde{\bv}_i\widetilde{\bv}_i^{\T}+\sum_{i>r}\tilde{\lambda}_i\widetilde{\bv}_i\widetilde{\bv}_i^{\T}$. Write $\bV=(\bv_1,\dots,\bv_r)\in\mathbb{R}^{p\times r}$ and $\widetilde{\bV}=(\widetilde{\bv}_1,\dots,\widetilde{\bv}_r)\in\mathbb{R}^{p\times r}$. 
	\begin{lemma}
		\label{lem.sin_max}
		Suppose $\iota$ satisfies $\iota>\Vert\bE\Vert$ and for any $i\in[r]$, the interval $(\lambda_i-\iota,\lambda_i+\iota)$ does not contain any eigenvalues of $\bA$ other than $\lambda_i$. Then, there exists an orthogonal matrix $\bU\in\mathbb{R}^{r\times r}$ such that
		$$\Vert\widetilde{\bV}\bU-\bV\Vert_{\max}=O\left(\frac{r^{5/2}\mu^2\Vert\bE\Vert_{\infty}}{(|\lambda_r|-\Vert\bA-\bA_r\Vert_{\infty})\sqrt{p}}\right),$$
		where $\mu=\mu(\bV)$ is the coherence of $\bV$ defined as $\mu(\bV)=(p/r)\max_{i}\sum_{j=1}^{r}V_{ij}^2$.
	\end{lemma}
	
        The proof of Lemma~\ref{lem.sin_max} can be found in \citet{fan2018eigenvector}. 

	\begin{lemma}
		\label{lem.eigen}
		(Theorem 4.2.5 in \cite{Bhsing2015}). If $\bK(\cdot,\cdot)$ is a compact and nonnegative definite kernel matrix function with associated eigenvalue/eigenfunction pairs $\{\lambda_j,\be_j(\cdot)\}_{j=1}^{\infty}$, then
		$$
		\lambda_k=\max_{\be\in{\rm span}\{\be_1,\dots,\be_{k-1}\}^{\perp}}\frac{\langle\be,\bK(\be)\rangle}{\Vert\be\Vert^2}.
		$$
	\end{lemma}

	\begin{lemma}
		\label{lem.inv_norm}
		Suppose that $\bK,\bG\in\eH^p\otimes\eH^p$ are Mercer's kernels, $\Vert\bG^{\dagger}\Vert_{\cL}<c_n$ and $\Vert\bK-\bG\Vert_{\cL}=o_p(c_n^{-1})$ for a sequence $c_n>0$. Let $\widetilde{\bK}$ be the restriction of $\bK$ to $\tKer(\bG)^{\perp}$. Define $\bK^{\dagger}$ as $\bK^{\dagger}(\bx)=\widetilde{\bK}^{-1}(\bx)$ for $\bx\in\tIm(\bG)$ and $\bK^{\dagger}(\bx)=\bzero$ for $\bx\in\tIm(\bG)^{\perp}.$ 
        Then, we have (i) $\Vert\bK^{\dagger}\Vert_{\cL}<2c_n$ with probability approaching one, and (ii) $\Vert\bK^{\dagger}-\bG^{\dagger}\Vert_{\cL}=O_p(c_n^2)\Vert\bK-\bG\Vert_{\cL}.$
	\end{lemma}
 
        \begin{lemma}
        \label{lem.rate_max_f}
        (i) Under Assumptions~\ref{ass.regu_cond_DIGIT}(ii)(iv) and \ref{ass.concentration}, $\max_{t\in[n]}\Vert\bbf_t\Vert=O_p(\sqrt{\log n})$, $\max_{t\in[n]}\Vert p^{-1/2}\bvarepsilon_t\Vert=O_p(\sqrt{\log n})$ and $\max_{t\in[n]}\Vert p^{-1/2}\bB^{\T}\bvarepsilon_t\Vert=O_p(\sqrt{\log n})$.\\ (ii) Under Assumptions~\ref{ass.regu_cond_fpoet}(ii)(iv) and \ref{ass.concentration_2}, $\max_{t\in[n]}\Vert\bgamma_t\Vert=O_p(\sqrt{\log n})$, $\max_{t\in[n]}\Vert p^{-1/2}\int\bQ(u)^{\T}\bvarepsilon_t(u)\du
        \Vert=O_p(\sqrt{\log n})$ and $\max_{t'\in[n]}|p^{-1/2}\{\langle\bvarepsilon_t,\bvarepsilon_{t'}\rangle-\eE\langle\bvarepsilon_t,\bvarepsilon_{t'}\rangle\}|^2=O_p(\log n)$ for each $t\in[n]$.
        \end{lemma}

	\subsection{Proof of Theorem \ref{thm.load_factor_1}}
	The proof of part (i) of Theorem \ref{thm.load_factor_1} mainly relies on Lemma~\ref{lem.sin_max}. To prove Theorem \ref{thm.load_factor_1}, we first present some technical lemmas. The proofs of Lemmas~\ref{lem.sigma_f_bound}--\ref{lem.rate_R_inf} are provided in the supplementary material.
	
	\begin{lemma}
		\label{lem.sigma_f_bound}
		Suppose that Assumption~\ref{ass.ind1} holds. Then there exist some constants $C_{\max},C_{\infty}>0$ such that (i) $\Vert\bSigma_{f}\Vert_{\cS,\max}\le C_{\max}$, (ii) $\max(\Vert\bSigma_{f}\Vert_{\cS,\infty},\Vert\bSigma_{f}\Vert_{\cS,1},\Vert\bSigma_{f}\Vert_{\cS,\F})\le C_{\infty}$.
	\end{lemma}
	
	\begin{lemma}
		\label{lem.sigma_y_order}
		Suppose that Assumptions~\ref{ass.ind1}--\ref{ass.regu_cond_DIGIT} hold. Then we have (i) $\Vert\bSigma_{y}\Vert_{\cS,\max}\lesssim1$, (ii) $\Vert\bSigma_{y}\Vert_{\cS,\infty}\lesssim p$, and (iii) $\Vert\bSigma_{y}\Vert_{\cS,1}\lesssim p$. 
	\end{lemma}

        \begin{lemma}
        \label{lem.order}
        Supposing that Assumptions~\ref{ass.ind1}--\ref{ass.regu_cond_DIGIT} hold, we have $\Vert\bOmega_{\cL}\Vert\asymp p^2$ and $\Vert\bOmega_{\cR}\Vert\lesssim p$.
        \end{lemma}
	
	\begin{lemma}
		\label{lem.rate_R_inf}
		Under the assumptions of Theorem \ref{thm.load_factor_1}, we have $\Vert\bOmega_{\cR}\Vert_{\infty}\lesssim ps_p=o(p^2)$.
	\end{lemma}
	
	\begin{lemma}
		\label{lem.appro_int}
		Under the assumptions of Theorem \ref{thm.load_factor_1}, we have (i) $\Vert\widehat{\bOmega}-\bOmega\Vert=O_p(\cM_{\varepsilon}p^2\sqrt{1/n})=o_p(p^2),$
		and (ii) $\Vert\widehat{\bOmega}-\bOmega\Vert_{\infty}=O_p(\cM_{\varepsilon}p^2\sqrt{\log p/n})=o_p(p^2).$
	\end{lemma}
	\begin{proof}
		(i) Note that 
		\begin{equation}
			\label{eq.bound_T}
			\begin{aligned}
				\Vert\widehat{\bOmega}-\bOmega\Vert=&\left\Vert\int\int\widehat{\bSigma}_{y}^{\sS}(u,v)\widehat{\bSigma}_{y}^{\sS}(u,v)^{\T}{\rm d}u{\rm d}v-\int\int\bSigma_{y}(u,v)\bSigma_{y}(u,v)^{\T}{\rm d}u{\rm d}v\right\Vert\\
				=&\left\Vert\int\int \left\{\widehat{\bSigma}_{y}^{\sS}(u,v)-\bSigma_{y}(u,v)\right\}\widehat{\bSigma}_{y}^{\sS}(u,v)^{\T} 
				 +  \bSigma_{y}(u,v)\left\{\widehat{\bSigma}_{y}^{\sS}(u,v)^{\T}-\bSigma_{y}(u,v)^{\T}\right\}{\rm d}u{\rm d}v\right\Vert\\
				=&\left\Vert\int\int \left\{\widehat{\bSigma}_{y}^{\sS}(u,v)-\bSigma_{y}(u,v)\right\}\left\{\widehat{\bSigma}_{y}^{\sS}(u,v)^{\T}-\bSigma_{y}(u,v)^{\T}+\bSigma_{y}(u,v)^{\T}\right\} \right.\\
				& + \left. \bSigma_{y}(u,v)\left\{\widehat{\bSigma}_{y}^{\sS}(u,v)^{\T}-\bSigma_{y}(u,v)^{\T}\right\}{\rm d}u{\rm d}v\right\Vert\\
				\le &\left\Vert\int\int\left\{\widehat{\bSigma}_{y}^{\sS}(u,v)-\bSigma_{y}(u,v)\right\}\left\{\widehat{\bSigma}_{y}^{\sS}(u,v)-\bSigma_{y}(u,v)\right\}^{\T}{\rm d}u{\rm d}v\right\Vert \\
				& + 2\left\Vert\int\int\left\{\widehat{\bSigma}_{y}^{\sS}(u,v)-\bSigma_{y}(u,v)\right\}\bSigma_{y}(u,v)^{\T}{\rm d}u{\rm d}v\right\Vert \\
				\le&\Vert\widehat{\bSigma}_{y}^{\sS}-\bSigma_{y}\Vert_{\cL}^2+2\Vert\widehat{\bSigma}_{y}^{\sS}-\bSigma_{y}\Vert_{\cL}\Vert\bSigma_{y}\Vert_{\cL}
				\le\Vert\widehat{\bSigma}_{y}^{\sS}-\bSigma_{y}\Vert_{\cS,\F}^2+2\Vert\widehat{\bSigma}_{y}^{\sS}-\bSigma_{y}\Vert_{\cS,\F}\Vert\bSigma_{y}\Vert_{\cS,\F}\\
				=&\sum_{i=1}^{p}\sum_{j=1}^{p}\Vert n^{-1}\sum_{t=1}^{n}y_{ti}y_{tj}-\Sigma_{y,ij}\Vert_{\cS}^2+2\left(\sum_{i=1}^{p}\sum_{j=1}^{p}\Vert n^{-1}\sum_{t=1}^{n}y_{ti}y_{tj}-\Sigma_{y,ij}\Vert_{\cS}^2\right)^{1/2}\Vert\bSigma_{y}\Vert_{\cS,\F}\\   
				=&O_p\left(\cM_{\varepsilon}p^2\sqrt{1/n}\right)=o_p(p^2),
			\end{aligned}
		\end{equation}
		where the second inequality follows from Lemmas \ref{lem.ineq1}(ii) and \ref{lem.ineq2}(iv), the third inequality follows from Lemma~\ref{lem.op_le_S1}(ii), and the last line follows from Lemma~\ref{lem.rate_sigma_f}(iii), the fact that $\Vert\bK\Vert_{\cS,\F}\le p\Vert\bK\Vert_{\cS,\max}$ and Assumption~\ref{ass.concentration}(ii). \\
		(ii) The argument can be proved in matrix $\ell_{\infty}$ norm following the similar procedure. Specifically,
		$$
		\begin{aligned}
			\Vert\widehat{\bOmega}-\bOmega\Vert_{\infty}\le&\left\Vert\int\int\left\{\widehat{\bSigma}_{y}^{\sS}(u,v)-\bSigma_{y}(u,v)\right\}\left\{\widehat{\bSigma}_{y}^{\sS}(u,v)-\bSigma_{y}(u,v)\right\}^{\T}{\rm d}u{\rm d}v\right\Vert_{\infty} \\
				& + 2\left\Vert\int\int\left\{\widehat{\bSigma}_{y}^{\sS}(u,v)-\bSigma_{y}(u,v)\right\}\bSigma_{y}(u,v)^{\T}{\rm d}u{\rm d}v\right\Vert_{\infty} \\
            \le&\Vert\widehat{\bSigma}_{y}^{\sS}-\bSigma_{y}\Vert_{\cS,\infty}\Vert\widehat{\bSigma}_{y}^{\sS}-\bSigma_{y}\Vert_{\cS,1}+2\Vert\widehat{\bSigma}_{y}^{\sS}-\bSigma_{y}\Vert_{\cS,\infty}\Vert\bSigma_{y}\Vert_{\cS,1}\\
\asymp&p^2\Vert\widehat{\bSigma}_{y}^{\sS}-\bSigma_{y}\Vert_{\cS,\max}
=O_p\big(\cM_{\varepsilon}p^2\sqrt{\log p/n}\big)=o_p(p^2),
		\end{aligned}
		$$
            where the first inequality can be obtained in a way similar to \eqref{eq.bound_T}, the second inequality follows from Lemma~\ref{lem.ineq2}(ii), and the last line follows from Lemma~\ref{lem.rate_sigma_f}(iii) and Assumption~\ref{ass.concentration}(ii). 
	\end{proof}

	\begin{lemma}
		\label{lem.rth_eigen}
		Let $\{\hat{\lambda}_j\}_{j=1}^{p}$ be the eigenvalues of $\widehat{\bOmega}$ in a descending order. Under the assumptions of Theorem \ref{thm.load_factor_1}, it holds that $\hat{\lambda}_r\gtrsim p^2$ with probability approaching one. Furthermore, $\hat{\lambda}_i-\hat{\lambda}_j\gtrsim p^2$ for all $1\le i<j\le r$ with probability approaching one. 
	\end{lemma}
	\begin{proof}
		By Proposition \ref{propos.eigenvalues} and Lemma \ref{lem.order}, the $r$-th largest eigenvalue $\lambda_r$ of $\bOmega$ satisfies $\lambda_r\ge p^2\theta_r-|\lambda_r-p^2\theta_r|\ge p^2\theta_r-\Vert\bOmega_{\cR}\Vert\gtrsim p^2.$
		Applying Lemma~\ref{lem.weyl} yields that
		$
			|\hat{\lambda}_j-\lambda_j|\le \Vert\widehat{\bOmega}-\bOmega\Vert,\ {\rm for\ }j\in[p].
		$
		From Lemma \ref{lem.appro_int}(i), we have $\Vert\widehat{\bOmega}-\bOmega\Vert=o_p(p^2)$ and hence $\hat{\lambda}_r\gtrsim p^2$ with probability approaching one. Furthermore, for all $1\le i<j\le r$, with probability approaching one, $$\hat{\lambda}_i-\hat{\lambda}_j\ge (\lambda_i-\lambda_j)-|\hat{\lambda}_i-\lambda_i|-|\hat{\lambda}_j-\lambda_j|=p^2(\theta_i-\theta_j)-o_p(p^2)\gtrsim p^2.$$
	\end{proof}	

        The following lemma is used to prove Theorems~\ref{thm.model_sele} and \ref{thm.inverse_digit}, and its proof is provided in the supplementary material.
        \begin{lemma}
            \label{lem.rate_B_F}
            Under the assumptions of Theorem~\ref{thm.load_factor_1}, we have $\Vert\widehat{\bB}-\bB\bU^{\T}\Vert_{\F}=O_p(\cM_{\varepsilon}\sqrt{p/n}+1/\sqrt{p}).$
        \end{lemma}
		
We are now ready to prove Theorem~\ref{thm.load_factor_1}.\\
(i) Let $\bE=\widehat{\bOmega}-\bOmega$ be the $p \times p$ perturbation matrix. By Lemma \ref{lem.appro_int}, we have
	$$
	\Vert\bE\Vert_{\infty}\le \Vert \widehat{\bOmega}-\bOmega\Vert_{\infty}=O_p\big(\cM_{\varepsilon}p^2\sqrt{\log p/n}\big)=o_p(p^2).
	$$
	Corresponding to Lemma \ref{lem.sin_max}, here $\bA=\bOmega,\widetilde{\bA}=\widehat{\bOmega}$, and the $r$-th eigenvalue of $\bA$ satisfies $\lambda_r\asymp p^2$ by Proposition \ref{propos.eigenvalues} and Lemma \ref{lem.order}. Then, $\Vert\bA-\bA_r\Vert_{\infty}\le\Vert\bOmega-\bOmega_{\cL}\Vert_{\infty}=\Vert\bOmega_{\cR}\Vert_{\infty}=ps_p=o(p^2)$ from Lemma \ref{lem.rate_R_inf}. Note that $\bV=(\bxi_1,\dots,\bxi_r)\in\mathbb{R}^{p\times r}$, and denote $\bxi_j=(\xi_{1j},\dots,\xi_{pj})^{\T}$. The coherence of $\bV$ is given by
	$$
	\mu=\mu(\bV)=\frac{p}{r}\max_{i\in[p]}\sum_{j=1}^{r}\xi_{ij}^2\le\frac{p}{r}\max_{i\in[p]}\sum_{j=1}^{r}\big(\widetilde{b}_{ij}^2+\Vert\bxi_j-\widetilde{\bbb}_j\Vert^2\big)=O(1),
	$$
	since $\max_{i\in[p]}\widetilde{b}_{ij}=\max_{i\in[p]}p^{-1/2}b_{ij}\le p^{-1/2}\Vert\bB\Vert_{\max}=O(p^{-1/2})$ and $\Vert\bxi_j-\widetilde{\bbb}_j\Vert=O(p^{-1})$ if $\bxi_j^{\T}\widetilde{\bbb}_j\ge0$ by Proposition \ref{propos.eigenvalues}(ii) and Lemma \ref{lem.order}. In addition, supposing that $\iota\approx\Vert\bE\Vert=o_p(p^2)$ but $\iota>\Vert\bE\Vert$, we can show that for any $j\in[r]$, the the interval $(\lambda_j-\iota,\lambda_j+\iota)$ does not contain any eigenvalues of $\bOmega$ other than $\lambda_j$ with probability approaching one. Thus by Lemma \ref{lem.sin_max}, we have for $j\in[r]$, if $\widehat{\bxi}_j^{\T}\bxi_j\ge0$,
	$$
	\Vert\widehat{\bxi}_j-\bxi_j\Vert_{\max}=O_p\left(\frac{r^{5/2}\mu^2\Vert\bE\Vert_{\infty}}{p^2\sqrt{p}}\right)=O_p\left(\cM_{\varepsilon}\sqrt{\frac{\log p}{pn}}\right).
	$$
	For $j\in[r]$, if $\bxi_j^{\T}\widetilde{\bbb}_j\ge0$, we have $\Vert\bxi_j-\widetilde{\bbb}_j\Vert_{\max}\le\Vert\bxi_j-\widetilde{\bbb}_j\Vert=O(p^{-1})$, which implies that $\Vert\widehat{\bxi}_j-\widetilde{\bbb}_j\Vert_{\max}=O_p(\cM_{\varepsilon}\sqrt{\log p/pn}+1/p)$. 
	Since $\widehat{\bB}=\sqrt{p}(\widehat{\bxi}_1,\dots,\widehat{\bxi}_r)$ and $\bB=\sqrt{p}(\widetilde{\bbb}_1,\dots,\widetilde{\bbb}_r)$, one can obtain that there exists an orthogonal matrix $\bU\in\mathbb{R}^{r\times r}$ (the same as that in Lemma \ref{lem.rate_B_F}) such that
	$$
		\Vert\widehat{\bB}-\bB\bU^{\T}\Vert_{\max}=O_p\big(\cM_{\varepsilon}\sqrt{\log p/n}+1/\sqrt{p}\big)=o_p(1),
	$$
        where the matrix $\bU$ is used to adjust the direction so that each $\bbb_j^{\T}\widehat{\bbb}_j\ge0$ for $j\in[r].$\\
(ii) Note that $\widehat{\bbf}_t(\cdot)=p^{-1}\widehat{\bB}^{\T}\by_t(\cdot)=p^{-1}\widehat{\bB}^{\T}\{\bB\bbf_t(\cdot)+\bvarepsilon_t(\cdot)\}$ for $t\in[n]$
and then
	\begin{equation}
		\label{eq.decom_diff_factor}
		\widehat{\bbf}_t(\cdot)-\bU\bbf_t(\cdot)=p^{-1}(\widehat{\bB}^{\T}\bB-\bU\bB^{\T}\bB)\bbf_t(\cdot)+p^{-1}\widehat{\bB}^{\T}\bvarepsilon_{t}(\cdot).
	\end{equation}
	For the first term of \eqref{eq.decom_diff_factor}, applying Lemmas \ref{lem.factor_norm} and \ref{lem.rate_B_F} yields that 
	$$\Vert p^{-1}(\widehat{\bB}^{\T}\bB-\bU\bB^{\T}\bB)\bbf_t\Vert\le p^{-1}\Vert\widehat{\bB}^{\T}-\bU\bB^{\T}\Vert\Vert\bB\Vert\Vert\bbf_t\Vert=O_p\big(\cM_{\varepsilon}\sqrt{1/n}+1/p\big),$$
	since $\Vert\widehat{\bB}^{\T}-\bU\bB^{\T}\Vert\le\Vert\widehat{\bB}-\bB\bU^{\T}\Vert_{\F}=O_p(\cM_{\varepsilon}\sqrt{p/n}+1/\sqrt{p}),\Vert\bB\Vert=\lambda_{\max}^{1/2}(\bB^{\T}\bB)=\sqrt{p}$ and $\Vert\bbf_t\Vert=O_p(1)$. 
For the second term of \eqref{eq.decom_diff_factor}, notice that
	$$
		\Vert p^{-1}\widehat{\bB}^{\T}\bvarepsilon_{t}\Vert=\Vert p^{-1}\bU\bB^{\T}\bvarepsilon_{t}\Vert+\Vert p^{-1}(\widehat{\bB}^{\T}-\bU\bB^{\T})\bvarepsilon_{t}\Vert=O_p(\cM_{\varepsilon}\sqrt{1/n}+1/\sqrt{p}),
	$$
	which follows from $\Vert p^{-1}\bB^{\T}\bvarepsilon_{t}\Vert=O_p(1/\sqrt{p})$ by Assumption~\ref{ass.regu_cond_DIGIT}(ii), $\Vert\widehat{\bB}^{\T}-\bU\bB^{\T}\Vert=O_p(\cM_{\varepsilon}\sqrt{p/n}+1/\sqrt{p})$ by Lemma~\ref{lem.rate_B_F}, and  $\Vert\bvarepsilon_t\Vert=O_p(\sqrt{p})$ since $\eE\Vert\bvarepsilon_t\Vert^2\le p\max_{i\in[p]}\eE\Vert\varepsilon_{ti}\Vert^2=p\max_{i\in[p]}\Vert\Sigma_{\varepsilon,ii}\Vert_{\cN}=O(p)$ by Assumption \ref{ass.regu_cond_DIGIT}(iv). The result follows immediately that for $t\in[n]$,
	$$
	\Vert\widehat{\bbf}_t-\bU\bbf_t\Vert=O_p\left(\cM_{\varepsilon}/\sqrt{n}+1/\sqrt{p}\right),
	$$
	and thus $n^{-1}\sum_{t=1}^{n}\Vert\widehat{\bbf}_t-\bU\bbf_t\Vert^2=O_p(\cM_{\varepsilon}^2/n+1/p)$.\\
(iii) The proof procedure is similar to part (ii). We only need to notice that by Lemma \ref{lem.rate_max_f}(i), we have $\max_{t\in[n]}\Vert\bbf_t\Vert=O_p(\sqrt{\log n}),\max_{t\in[n]}\Vert p^{-1}\bvarepsilon_t\Vert=O_p(\sqrt{\log n/p})$ and $\max_{t\in[n]}\Vert p^{-1}\bB\bvarepsilon_t\Vert=O_p(\sqrt{\log n/p})$.

\subsection{Proof of Corollary \ref{coro.DIGIT}}
By Theorem~\ref{thm.load_factor_1}(i)(iii), Lemma \ref{lem.rate_max_f}(i), and Assumption \ref{ass.regu_cond_DIGIT}(i), we have
	$$
	\begin{aligned}
		\max_{i\in[p],t\in[n]}\Vert\widecheck{\bbb}_i^{\T}\widehat{\bbf}_t-\breve{\bbb}_i^{\T}\bbf_t\Vert\le&\max_{i\in[p]}\Vert\widecheck{\bbb}_i-\bU\breve{\bbb}_i\Vert\cdot\max_{t\in[n]}\Vert\widehat{\bbf}_t\Vert+\max_{i\in[p]}\Vert\breve{\bbb}_i\Vert\cdot\max_{t\in[n]}\Vert\bU^{\T}\widehat{\bbf}_t-\bbf_t\Vert\\
		=&O_p(\varpi_{n,p}\cdot \sqrt{\log n})+O_p( \cM_{\varepsilon}\sqrt{\log n/n}+\sqrt{\log n/p})\\
        =&O_p\big(\cM_{\varepsilon}\sqrt{\log n\log p/n}+\sqrt{\log n/p}\big).
	\end{aligned}
	$$

 \subsection{Proof of Theorem \ref{thm.deter_num}}
    \noindent
    (i) By Proposition \ref{propos.eigenvalues} and Lemma \ref{lem.order}, $|\lambda_j-p^2\theta_j|\le \Vert\bOmega_{\cR}\Vert\lesssim p$ for $j\in[r]$, which implies that $\lambda_j\asymp p^2$ for $j\in[r]$, and $|\lambda_j|\lesssim p$ for $r+1\le j\le r_{1,0}$. By Lemmas \ref{lem.weyl} and \ref{lem.appro_int}, it follows that $|\hat{\lambda}_j-\lambda_j|\le\Vert\widehat{\bOmega}-\bOmega\Vert=O_p(\cM_{\varepsilon}p^2n^{-1/2})$ for $j\in[p].$ Let $\beta_{1n}=\cM_{\varepsilon}p^2n^{-1/2}$ and $\widetilde{\beta}_{1n}=p+\cM_{\varepsilon}p^2n^{-1/2}$. Then, we have $|\hat{\lambda}_j-\lambda_j|=O_p(\beta_{1n})$ for $j\in[r]$ and $|\hat{\lambda}_j|=O_p(\widetilde{\beta}_{1n})$ for $r+1\le j\le r_{1,0}.$ We next verify the following conditions (a), (b) and (c) by Assumptions~\ref{ass.concentration}(ii) and \ref{ass.mdoel_dete_r_1}:
    \begin{enumerate}
        \item[(a)] $(\vartheta_{1n}+\widetilde{\beta}_{1n})/(\lambda_r^2/\lambda_1)\asymp (\vartheta_{1n}+p+\cM_{\varepsilon}p^2n^{-1/2})/p^2\to0$;

        \item[(b)] $\beta_{1n}/\lambda_r\asymp \cM_{\varepsilon}p^2n^{-1/2}/p^2=\cM_{\varepsilon}n^{-1/2}\to0$;

        \item[(c)] $\widetilde{\beta}_{1n}^2/(\vartheta_{1n}\lambda_r)\asymp(p+\cM_{\varepsilon}p^2n^{-1/2})^2/(\vartheta_{1n}p^2)\asymp \vartheta_{1n}^{-1}+\vartheta_{1n}^{-1}\cM_{\varepsilon}^2p^2n^{-1}\to0$.
    \end{enumerate}
    Under (a), (b) and (c), we apply Proposition~1 of \cite{han2022rank} and obtain $\eP(\hat{r}^{\cD}=r) \to 1$ with $\hat{r}^{\cD}$ defined in \eqref{eq.deter_num_1}, which completes the proof of Theorem~\ref{thm.deter_num}(i).

	\noindent
	(ii) By Proposition~\ref{propos.eigenvalues_2}, $|\tau_j-p\vartheta_j|\le\Vert\bSigma_{\varepsilon}\Vert_{\cL}=O(1)$ for $j\in[r],$ which implies that $\tau_j\asymp p$ for $j\in[r]$, and $|\tau_j|=O(1)$ for $r+1\le j\le r_{2,0}$. By Lemma \ref{lem.weyl_2} and the proof of Lemma \ref{lem.rth_eigen_2} of the supplementary material, it follows that $|\hat{\tau}_j-\tau_j|\le\Vert\widehat{\bSigma}_{y}^{\sS}-\bSigma_{y}\Vert_{\cS,\F}=O_p(\cM_{\varepsilon}pn^{-1/2})$ for $j\in[p].$ Let $\beta_{2n}=\cM_{\varepsilon}pn^{-1/2}$ and $\widetilde{\beta}_{2n}=1+\cM_{\varepsilon}pn^{-1/2}$. Then, we have $|\hat{\tau}_j-\tau_j|=O_p(\beta_{2n})$ for $j\in[r]$ and $|\hat{\tau}_j|=O_p(\widetilde{\beta}_{2n})$ for $r+1\le j\le r_{2,0}.$ We next verify the following conditions (d), (e) and (f) by Assumptions~\ref{ass.concentration_2}(ii) and \ref{ass.mdoel_dete_r_2}:
    \begin{enumerate}
        \item[(d)] $(\vartheta_{2n}+\widetilde{\beta}_{2n})/(\tau_r^2/\tau_1)\asymp (\vartheta_{2n}+1+\cM_{\varepsilon}pn^{-1/2})/p\to0$;
        
        \item[(e)] $\beta_{2n}/\tau_r\asymp \cM_{\varepsilon}pn^{-1/2}/p=\cM_{\varepsilon}n^{-1/2}\to0$;

        \item[(f)] $\widetilde{\beta}_{2n}^2/(\vartheta_{2n}\tau_r)\asymp(1+\cM_{\varepsilon}pn^{-1/2})^2/(\vartheta_{2n}p)\asymp \vartheta_{2n}^{-1}p^{-1}+\vartheta_{2n}^{-1}\cM_{\varepsilon}^2pn^{-1}\to0$.
    \end{enumerate}
    Under (d), (e) and (f), we apply Proposition~1 of \cite{han2022rank} and obtain $\eP(\hat{r}^{\cF}=r) \to 1$ with $\hat{r}^{\cF}$ defined in \eqref{eq.deter_num_2}, which completes the proof of Theorem~\ref{thm.deter_num}(ii).

\subsection{Proof of Theorem \ref{thm.model_sele}}
We denote $r_0=r_{1,0}\wedge r_{2,0}$ for simplicity.
\begin{lemma}
    \label{lem.V_D_F}
    (i) Under Assumptions~\ref{ass.ind1}--\ref{ass.concentration} and \ref{ass.mdoel_sele_1}, there exists some constant $c_1>0$ such that $\eP\big\{\log V^{\cD}(r)+c_1\le\min_{k\in[r-1]}\log V^{\cF}(k)\big\}\to1$ and $\eP\big\{\log V^{\cD}(k)+c_1\le\log V^{\cF}(k)\big\}\to1$ for $r\le k\le r_0$.\\
    (ii) Under Assumptions~\ref{ass.ind2}--\ref{ass.concentration_2} and \ref{ass.mdoel_sele_2}, there exists some constant $c_2>0$ such that $\eP\big\{\min_{k\in[r-1]}\log V^{\cD}(k)\ge\log V^{\cF}(r)+c_2\big\}\to1$ and $\eP\big\{\log V^{\cD}(k)\ge\log V^{\cF}(k)+c_2\big\}\to1$ for $r\le k\le r_0$. 
\end{lemma}
\begin{proof}
    Let $\be_{1kt}(\cdot)=\by_t(\cdot)-p^{-1}\widehat{\bB}_{k}\widehat{\bB}_{k}^{\T}\by_t(\cdot)$ and $\be_{2kt}(\cdot)=\by_t(\cdot)-n^{-1}\bY_t(\cdot)\widehat{\bGamma}_k\widehat{\bgamma}_{t,k}$ for $t\in[n]$ and $k\in[r_0],$ where $\widehat{\bB}_{k}$ is the estimated loadings by DIGIT estimator with $k$ functional factors, and $\widehat{\bGamma}_k$ is the estimated factors by FPOET estimator with $k$ scalar factors. By definitions, it can be seen that
    $$
    V^{\cD}(k)=\frac{1}{pn}\sum_{t=1}^{n}\Vert\be_{1kt}\Vert^2=\frac{1}{pn}\sum_{t=1}^{n}\int\be_{1kt}(u)^{\T}\be_{1kt}(u)\du=\frac{1}{p}\tr\left\{\int\frac{1}{n}\sum_{t=1}^{n}\be_{1kt}(u)\be_{1kt}(u)^{\T}\du\right\}=p^{-1}\big\Vert\widehat{\bSigma}_{e,1k}^{\sS}\big\Vert_{\cN},
    $$
    and similarly it follows that $V^{\cF}(k)=p^{-1}\big\Vert\widehat{\bSigma}_{e,2k}^{\sS}\big\Vert_{\cN}$, where $\widehat{\bSigma}_{e,1k}^{\sS}$ and $\widehat{\bSigma}_{e,2k}^{\sS}$ are the sample covariance of $\{\be_{1kt}(\cdot)\}$ and $\{\be_{2kt}(\cdot)\}$, respectively. 
    Notice that $\bchi_t(\cdot)=\bB\bbf_t(\cdot)$ and $\bkappa_t(\cdot)=\bQ(\cdot)\bgamma_t$ are the common components of the two respective FFMs. By Section~\ref{subsec.relationship} of the supplementary material, we have 
    $$
    \bSigma_{\chi}(u,v)=\sum_{i=1}^{\infty}p\omega_i\bpsi_i(u)\bpsi_i(v)^{\T}\quad{\rm and}\quad\bSigma_{\kappa}(u,v)=\sum_{j=1}^{r}p\widecheck{\vartheta}_j\bnu_j(u)\bnu_j(v)^{\T}\quad{\rm for}\quad(u,v)\in\cU^2.
    $$
    (i) For the first argument, note that $\be_{1rt}-\bvarepsilon_t=\bB\bbf_t-\widehat{\bB}\widehat{\bbf}_t$ for $t\in[n]$. On the one hand, by Theorem~\ref{thm.load_factor_1} and Lemma~\ref{lem.rate_B_F}, it can be shown that $n^{-1}\sum_{t=1}^{n}\Vert\bB\bbf_t-\widehat{\bB}\widehat{\bbf}_t\Vert^2=O_p(\cM_{\varepsilon}^2p/n+1)=o_p(p).$ Since $\eE\big(n^{-1}\sum_{t=1}^{n}\Vert\bvarepsilon_t\Vert^2\big)=n^{-1}\sum_{t=1}^{n}\eE\Vert\bvarepsilon_t\Vert^2=o(p)$ by Assumption~\ref{ass.mdoel_sele_1}(ii), we have $n^{-1}\sum_{t=1}^{n}\Vert\bvarepsilon_t\Vert^2=o_p(p).$ By using the inequality $(a+b)^2\le 2(a^2+b^2)$, it follows that $\big\Vert\widehat{\bSigma}_{e,1r}^{\sS}\big\Vert_{\cN}=o_p(p)$ and thus $V^{\cD}(r)=o_p(1).$ On the other hand, by Assumption~\ref{ass.mdoel_sele_1}(i) and Lemma~\ref{lem.ineq1}(iii), it can be shown that for $1\le k<r,n^{-1}\sum_{t=1}^{n}\Vert \be_{2kt}-\bvarepsilon_t\Vert^2\gtrsim p$ with probability approaching one, since $\sum_{i=k+1}^{\infty}p\omega_i\asymp p$ and the leading $k$ eigenfunctions of $\widehat{\bSigma}_y^{\sS}$ cannot recover the space spanned by the eigenfunctions of $\bSigma_{\chi}$ corresponding to the eigenvalues with order $p$. By using the inequality $(a-b)^2\ge a^2/2-b^2$ and $n^{-1}\sum_{t=1}^{n}\Vert\bvarepsilon_t\Vert^2=o_p(p),$ we have $\big\Vert\widehat{\bSigma}_{e,2k}^{\sS}\big\Vert_{\cN}\gtrsim p$ and $V^{\cF}(k)\gtrsim 1$ with probability approaching one for $1\le k<r$. Hence, there exists some small constant $0<c_1<1/2$ such that $\min_{k\in[r-1]}V^{\cF}(k)/V^{\cD}(r)>1+2c_1$ with probability approaching one for all large $p$ and $n,$ which implies that $\log\{\min_{k\in[r-1]}V^{\cF}(k)/V^{\cD}(r)\}\ge c_1$ with probability approaching one by similar argument to the proof of Corollary 1 of \cite{bai2002determining}.

    For the second argument, following the similar procedures to the proof of Lemma 4 in \cite{bai2002determining}, it can be shown that for any fixed $k\ge r,V^{\cD}(r)-V^{\cD}(k)=O_p(\cM_{\varepsilon}^2/n+1/p)=o_p(1),$ which implies that $V^{\cD}(k)=o_p(1)$ for $k\ge r.$ By Assumption~\ref{ass.mdoel_sele_1}(i) and using the similar procedures to the first argument, we have $V^{\cF}(k)\gtrsim 1$ with probability approaching one. The desired results hold accordingly. \\
    (ii) The proofs are similar to part (i). To show $V^{\cD}(k)\gtrsim 1$ with probability approaching one, it relies on Assumption~\ref{ass.mdoel_sele_2}(i) that $\rank\left\{\int\int\bSigma_{\kappa}(u,v)\bSigma_{\kappa}(u,v)^{\T}\du\dv\right\}=\rank\left\{\int\bQ(u)\bQ(u)^{\T}\du\right\}\ge k+1$ for $k\in[r_0].$
\end{proof}

\begin{lemma}
    \label{lem.IC}
    Suppose that $g(p,n)\to0$ and $(\cM_{\varepsilon}^2/n+1/p)^{-1}g(p,n)\to\infty$ as $p,n\to\infty$ for the penalty functions $g^{\cD}(p,n)$ and $g^{\cF}(p,n)$ in \eqref{eq.criteria}. Then, (i) under Assumptions~\ref{ass.ind1}--\ref{ass.concentration}, $\eP\big\{{\rm IC}^{\cD}(r)\le {\rm IC}^{\cD}(k)\big\}\to1$ for $k\in[r_0]$; (ii) under Assumptions~\ref{ass.ind2}--\ref{ass.concentration_2}, $\eP\big\{{\rm IC}^{\cF}(r)\le {\rm IC}^{\cF}(k)\big\}\to1$ for $k\in[r_0].$
\end{lemma}
\begin{proof}
    (ii) Let $\bGamma_k=(\bgamma_{1,k},\dots,\bgamma_{n,k})^{\T}$ be a $n\times k$ matrix. Define the average of squared residuals when using $\{\bgamma_{t,k}\}_{t\in[n]}$ as $k$ known factors for the estimation of FFM~\eqref{eq.model_2}:
    $$
    \widetilde{V}^{\cF}(k,\bGamma_k)=\min_{\bQ(\cdot)}\frac{1}{pn}\sum_{t=1}^{n}\Vert\by_t-\bQ\bgamma_{t,k}\Vert^2=\frac{1}{pn}\sum_{t=1}^{n}\Vert\by_t-n^{-1}\bY\bGamma_k\bgamma_{t,k}\Vert^2.
    $$
    Note that $\widetilde{V}^{\cF}(k,\widehat{\bGamma}_k)=V^{\cF}(k)$. The proof can be organized in the following three steps.
    
    \emph{Step 1}. For any fixed $k\in[r_0],$ define $\bH_k=n^{-1}\bV_k^{-1}\widehat{\bGamma}_k^{\T}\bGamma\int\bQ(u)^{\T}\bQ(u)\du,$ where $\bV_k$ is the diagonal matrix of the first $k$ largest eigenvalues of $\widehat{\bSigma}_y^{\sS}$ in a decreasing order, $\bGamma=(\bgamma_1,\dots,\bgamma_n)^{\T}$ is the true factors and $\widehat{\bGamma}_k=(\widehat{\bgamma}_{1,k},\dots,\widehat{\bgamma}_{n,k})^{\T}$ is formed by $k$ estimated factors using FPOET. Similar to \eqref{eq.decom_H} of the supplementary material, we have
    $$
    \widehat{\bgamma}_{t,k}-\bH_k\bgamma_t=\Big(\frac{\bV_k}{p}\Big)^{-1}\Big\{\frac{1}{n}\sum_{t'=1}^{n}\widehat{\bgamma}_{t',k}\frac{\eE\langle\bvarepsilon_{t'},\bvarepsilon_t\rangle}{p}+\frac{1}{n}\sum_{t'=1}^{n}\widehat{\bgamma}_{t',k}\zeta_{t't}+\frac{1}{n}\sum_{t'=1}^{n}\widehat{\bgamma}_{t',k}\eta_{t't}+\frac{1}{n}\sum_{t'=1}^{n}\widehat{\bgamma}_{t',k}\xi_{t't}\Big\},
    $$
    where $\zeta_{t't},\eta_{t't}$ and $\xi_{t't}$ take the same definitions as \eqref{eq.decom_H}. Analogous to the proofs of Lemma~\ref{lem.conv_decom_H} and Theorem~\ref{thm.load_factor_2}(i), it can be shown that $n^{-1}\sum_{t=1}^{n}\Vert\widehat{\bgamma}_{t,k}-\bgamma_t\Vert^2=O_p(\cM_{\varepsilon}^2/n+1/p)$ for any fixed $k\in[r_0].$ 

    \emph{Step 2}. Following the similar procedures to the proofs of Lemmas 2--4 in \cite{bai2002determining} and combining the result in Step 1, we obtain that: (i) for $k\in[r],\big|\widetilde{V}^{\cF}(k,\widehat{\bGamma}_k)-\widetilde{V}^{\cF}(k,\bH_k\bGamma)\big|=O_p(\cM_{\varepsilon}/\sqrt{n}+1/\sqrt{p})$; (ii) for $k\in[r]$, there exists some constant $c_k'>0$ such that $\widetilde{V}^{\cF}(k,\bH_k\bGamma)-\widetilde{V}^{\cF}(r,\bGamma)\ge c_k'$ with probability approaching one; (ii) for $r+1\le k\le r_0,\big|\widetilde{V}^{\cF}(k,\widehat{\bGamma}_k)-\widetilde{V}^{\cF}(r,\widehat{\bGamma}_r)\big|=O_p(\cM_{\varepsilon}^2/n+1/p).$ 

    \emph{Step 3}. For $k\in[r],$ consider that 
    $$
    \begin{aligned}
        \widetilde{V}^{\cF}(k,\widehat{\bGamma}_k)-\widetilde{V}^{\cF}(r,\widehat{\bGamma}_r)=&\{\widetilde{V}^{\cF}(k,\widehat{\bGamma}_k)-\widetilde{V}^{\cF}(k,\bH_k\bGamma)\}+\{\widetilde{V}^{\cF}(k,\bH_k\bGamma)-\widetilde{V}^{\cF}(r,\bH_r\bGamma)\}\\
        &+\{\widetilde{V}^{\cF}(r,\bH_r\bGamma)-\widetilde{V}^{\cF}(r,\widehat{\bGamma}_r)\}\ge c_k'
    \end{aligned}
    $$
    with probability approaching one, where the first and third terms are both $O_p(\cM_{\varepsilon}/\sqrt{n}+1/\sqrt{p})=o_p(1)$ by the first argument of Step 2 and the second term is large than $c_k'$ with probability approaching one by the second argument of Step 2 since $\widetilde{V}^{\cF}(r,\bH_r\bGamma)=\widetilde{V}^{\cF}(r,\bGamma)$. Thus, there exists some small constant $0<\epsilon_1<1/2$ such that $V^{\cF}(k)/V^{\cF}(r)>1+2\epsilon_1$ with probability approaching one, which implies that $\log\{V^{\cF}(k)/V^{\cF}(r)\}>\epsilon_1$ with probability approaching one. Since $g^{\cF}(p,n)\to0$ as $p,n\to\infty$, we have $\eP\{{\rm IC}^{\cF}(k)>{\rm IC}^{\cF}(r)\}\to1$ for $k\in[r].$ For $r+1\le k\le r_0,$ the third argument of Step 2 shows that $V^{\cF}(k)/V^{\cF}(r)=1+O_p(\cM_{\varepsilon}^2/n+1/p)$, which implies that $\log\{V^{\cF}(k)/V^{\cF}(r)\}=O_p(\cM_{\varepsilon}^2/n+1/p).$ As $(\cM_{\varepsilon}^2/n+1/p)^{-1}g^{\cF}(p,n)\to\infty$, it follows that $\eP\{{\rm IC}^{\cF}(k)>{\rm IC}^{\cF}(r)\}\to1$ for $r+1\le k\le r_0,$ which completes the proof of part (ii).\\
    (i) Let $\bB_k$ be a given $p\times k$ loading matrix of $k$ functional factors. Define 
    $$
    \widetilde{V}^{\cD}(k,\bB_k)=\min_{\bbf_{t,k}(\cdot)\in\eH^k,t\in[n]}\frac{1}{pn}\sum_{t=1}^{n}\Vert\by_t-\bB_k\bbf_{t,k}\Vert^2=\frac{1}{pn}\sum_{t=1}^{n}\Vert\by_t-p^{-1}\bB_k\bB_k^{\T}\by_t\Vert^2.
    $$
    Note that $\widetilde{V}^{\cD}(k,\widehat{\bB}_k)=V^{\cD}(k).$
    The remaining proof procedures are similar to part (ii) and omitted.
\end{proof}

We are now ready to prove Theorem~\ref{thm.model_sele}.\\
(i) Define the following events: $E_1=\{{\rm IC}^{\cD}(\hat{r}^{\cD})={\rm IC}^{\cD}(r)\},E_{2k}=\{{\rm IC}^{\cD}(r)\le {\rm IC}^{\cD}(k)\}$ for $k\in[r_0],E_3=\{{\rm IC}^{\cD}(r)<\min_{k\in[r-1]}{\rm IC}^{\cF}(k)\}$ and $E_{4k}=\{{\rm IC}^{\cD}(k)<{\rm IC}^{\cF}(k)\}$ for $r\le k\le r_0.$ As $p,n\to\infty$, we have $\eP(E_1)\to1$ by Theorem~\ref{thm.deter_num}(i), $\eP(E_{2k})\to1$ by Lemma~\ref{lem.IC}(i), $\eP(E_3)\to1$ by the first argument of Lemma~\ref{lem.V_D_F}(i) and $g^{\cD}(p,n)=o(1),g^{\cF}(p,n)=o(1)$, and $\eP(E_{4k})\to1$ by the second argument of Lemma~\ref{lem.V_D_F}(i) and $g^{\cD}(p,n)=o(1),g^{\cF}(p,n)=o(1)$. Let $E_2=\bigcap_{k=1}^{r_0} E_{2k}$ and $E_4=\bigcap_{k=r}^{r_0} E_{4k}$. Then, under the event $E=E_1\bigcap E_2\bigcap E_3\bigcap E_4$ with $\eP(E)\to1$, it follows that 
$$
{\rm IC}^{\cD}(\hat{r}^{\cD})={\rm IC}^{\cD}(r)= \min_{k\in[r_0]}{\rm IC}^{\cD}(k)<\min_{k\in[r_0]}{\rm IC}^{\cF}(k)\le {\rm IC}^{\cF}(\hat{r}^{\cF}),
$$ 
where the first equality holds under $E_1$, the second equality holds under $E_2$, the first inequality holds under $E_3\bigcap E_4$ since $\min_{k\in[r_0]}{\rm IC}^{\cF}(k)=\min_{k\in[r-1]}{\rm IC}^{\cF}(k)\wedge{\rm IC}^{\cF}(r)\wedge{\rm IC}^{\cF}(r+1)\wedge\dots\wedge{\rm IC}^{\cF}(r_0)>{\rm IC}^{\cD}(r)\wedge{\rm IC}^{\cD}(r+1)\wedge\dots\wedge{\rm IC}^{\cD}(r_0)\ge \min_{k\in[r_0]}{\rm IC}^{\cD}(k)$, and the second inequality holds automatically, which together show the desired result. \\
(ii) The proof is similar to part (i). Define the following events: $\widetilde E_1=\{{\rm IC}^{\cF}(\hat{r}^{\cF})={\rm IC}^{\cF}(r)\},\widetilde E_{2k}=\{{\rm IC}^{\cF}(r)\le {\rm IC}^{\cF}(k)\}$ for $k\in[r_0],\widetilde E_3=\{{\rm IC}^{\cF}(r)<\min_{k\in[r-1]}{\rm IC}^{\cD}(k)\}$ and $\widetilde E_{4k}=\{{\rm IC}^{\cF}(k)<{\rm IC}^{\cD}(k)\}$ for $r\le k\le r_0.$ As $p,n\to\infty$, we have $\eP(\widetilde E_1)\to1$ by Theorem~\ref{thm.deter_num}(ii), $\eP(\widetilde E_{2k})\to1$ by Lemma~\ref{lem.IC}(ii), $\eP(\widetilde E_3)\to1$ by the first argument of Lemma~\ref{lem.V_D_F}(ii) and $g^{\cD}(p,n)=o(1),g^{\cF}(p,n)=o(1)$, and $\eP(\widetilde E_{4k})\to1$ by the second argument of Lemma~\ref{lem.V_D_F}(ii) and $g^{\cD}(p,n)=o(1),g^{\cF}(p,n)=o(1)$. Let $\widetilde E_2=\bigcap_{k=1}^{r_0} \widetilde E_{2k}$ and $\widetilde E_4=\bigcap_{k=r}^{r_0} \widetilde E_{4k}$. Then, under the event $\widetilde E=\widetilde E_1\bigcap \widetilde E_2\bigcap \widetilde E_3\bigcap \widetilde E_4$ with $\eP(\widetilde E)\to1$, it follows that 
$$
{\rm IC}^{\cF}(\hat{r}^{\cF})={\rm IC}^{\cF}(r)= \min_{k\in[r_0]}{\rm IC}^{\cF}(k)<\min_{k\in[r_0]}{\rm IC}^{\cD}(k)\le {\rm IC}^{\cD}(\hat{r}^{\cD}),
$$
where the first equality holds under $\widetilde E_1$, the second equality holds under $\widetilde E_2$, the first inequality holds under $\widetilde E_3\bigcap \widetilde E_4$ since $\min_{k\in[r_0]}{\rm IC}^{\cD}(k)=\min_{k\in[r-1]}{\rm IC}^{\cD}(k)\wedge{\rm IC}^{\cD}(r)\wedge{\rm IC}^{\cD}(r+1)\wedge\dots\wedge{\rm IC}^{\cD}(r_0)>{\rm IC}^{\cF}(r)\wedge{\rm IC}^{\cF}(r+1)\wedge\dots\wedge{\rm IC}^{\cF}(r_0)\ge \min_{k\in[r_0]}{\rm IC}^{\cF}(k)$, and the second inequality holds automatically, which together show the desired result.

\subsection{Proof of Theorem \ref{thm.idio_DIGIT}}
To prove Theorem~\ref{thm.idio_DIGIT}, we first present some technical lemmas with their proofs.

\begin{lemma}
		\label{lem.varep_bound_1}
		Under the  assumptions of Theorem \ref{thm.idio_DIGIT}, it holds that\\
		(i) $\max_{i\in[p]}n^{-1}\sum_{t=1}^{n}\Vert\widehat{\varepsilon}_{ti}-\varepsilon_{ti}\Vert^2=O_p(\varpi_{n,p}^2)$;\\
		(ii) $\max_{i,j\in[p]}\Vert n^{-1}\sum_{t=1}^{n}\widehat{\varepsilon}_{ti}\widehat{\varepsilon}_{tj}-n^{-1}\sum_{t=1}^{n}\varepsilon_{ti}\varepsilon_{tj}\Vert_{\cS}=O_p(\varpi_{n,p})$;\\
		(iii) $\Vert\widehat{\bSigma}_{\varepsilon}-\bSigma_{\varepsilon}\Vert_{\cS,\max}=O_p(\varpi_{n,p})$. 
	\end{lemma}
	\begin{proof}
		(i) 
		Notice that $\widehat{\varepsilon}_{ti}(\cdot)-\varepsilon_{ti}(\cdot)=\{y_{ti}(\cdot)-\breve{\bbb}_i^{\T}\bbf_t(\cdot)\}-\{y_{ti}(\cdot)-\widecheck{\bbb}_i^{\T}\widehat{\bbf}_t(\cdot)\}=(\widecheck{\bbb}_i-\bU\breve{\bbb}_i)^{\T}\widehat{\bbf}_t(\cdot)-\breve{\bbb}_i^{\T}(\bU^{\T}\widehat{\bbf}_t-\bbf_t)(\cdot)$, where $\breve{\bbb}_i$ and $\widecheck{\bbb}_i$ are the $i$-th rows of $\bB$ and $\widehat{\bB}$, respectively. Applying the inequality $(a+b)^2\le 2(a^2+b^2)$ and the Cauchy--Schwarz inequality yields that
		$$
		\begin{aligned}
			\max_{i\in[p]}\frac{1}{n}\sum_{t=1}^{n}\Vert\widehat{\varepsilon}_{ti}-\varepsilon_{ti}\Vert^2\le&2\max_{i\in[p]}\Vert\widecheck{\bbb}_i-\bU\breve{\bbb}_i\Vert^2\frac{1}{n}\sum_{t=1}^{n}\Vert\widehat{\bbf}_t\Vert^2+2\max_{i\in[p]}\Vert\breve{\bbb}_i\Vert^2\frac{1}{n}\sum_{t=1}^{n}\Vert\bU^{\T}\widehat{\bbf}_t-\bbf_t\Vert^2\\
			=&O_p(\varpi_{n,p}^2)+O_p(\cM_{\varepsilon}^2/n+1/p)=O_p(\varpi_{n,p}^2).
		\end{aligned}
		$$
		
		\noindent
		(ii) Notice that $\max_{i\in[p]}\eE\Vert\varepsilon_{ti}\Vert^2=\max_{i\in[p]}\eE\int\varepsilon_{ti}(u)^2\du=\max_{i\in[p]}\int\Sigma_{\varepsilon,ii}(u,u)\du=O(1)$ from Assumption \ref{ass.regu_cond_DIGIT}(iv), thus we have $\max_{i\in[p]}n^{-1}\sum_{t=1}^{n}\Vert\varepsilon_{ti}\Vert^2=O_p(1)$. By the Cauchy--Schwarz inequality,
		$$
		\begin{aligned}
			\max_{i,j\in[p]}\Big\Vert \frac{1}{n}\sum_{t=1}^{n}\widehat{\varepsilon}_{ti}\widehat{\varepsilon}_{tj}-\frac{1}{n}\sum_{t=1}^{n}\varepsilon_{ti}\varepsilon_{tj}\Big\Vert_{\cS}=&\max_{i,j\in[p]}\Big\Vert\frac{1}{n}\sum_{t=1}^{n}(\widehat{\varepsilon}_{ti}-\varepsilon_{ti})\widehat{\varepsilon}_{tj}+\varepsilon_{ti}(\widehat{\varepsilon}_{tj}-\varepsilon_{tj})\Big\Vert_{\cS}\\
			\le&\max_{i\in[p]}\frac{1}{n}\sum_{t=1}^{n}\Vert\widehat{\varepsilon}_{ti}-\varepsilon_{ti}\Vert^2
			+2\Big(\max_{i\in[p]}\frac{1}{n}\sum_{t=1}^{n}\Vert\varepsilon_{ti}\Vert^2\Big)^{1/2}\Big(\max_{j\in[p]}\frac{1}{n}\sum_{t=1}^{n}\Vert\widehat{\varepsilon}_{tj}-\varepsilon_{tj}\Vert^2\Big)^{1/2}\\
			=&O_p(\varpi_{n,p}^2)+O_p(\varpi_{n,p})=O_p(\varpi_{n,p}).
		\end{aligned}
		$$
		
		\noindent
		(iii) The result is immediately implied by part (ii) above and Lemma \ref{lem.rate_sigma_f}(ii).
	\end{proof}
 
	\begin{lemma}
		\label{lem.Theta_bound}
		Under the assumptions of Theorem \ref{thm.idio_DIGIT}, there exist some constants $\Theta_1,\Theta_2>0$ such that with probability approaching one, 
		$$\Theta_1\le\min_{i\in[p],j\in[p]}\Vert\widehat{\Theta}_{ij}^{1/2}\Vert_{\cS}\le\max_{i\in[p],j\in[p]}\Vert\widehat{\Theta}_{ij}^{1/2}\Vert_{\cS}\le\Theta_2.$$
	\end{lemma}
	\begin{proof}
	    See the supplementary material.
	\end{proof}
	
We are now ready to prove Theorem~\ref{thm.idio_DIGIT}. By Lemmas \ref{lem.varep_bound_1}(iii) and \ref{lem.Theta_bound}, we have $\Vert\widehat{\bSigma}_{\varepsilon}-\bSigma_{\varepsilon}\Vert_{\cS,\max}=O_p(\varpi_{n,p})$ and $\max_{ij\in[p]}\Vert\Theta_{ij}\Vert_{\cS}=O_p(1)$. Consequently, for any $\epsilon>0$, there exist some positive constants $N,\Theta_1$ and $\Theta_2$ such that each of events
	$$\Upsilon_1=\left\{\max_{i\in[p],j\in[p]}\left\Vert\widehat{\Sigma}_{\varepsilon,ij}-\Sigma_{\varepsilon,ij}\right\Vert_{\cS}<N\varpi_{n,p}\right\},\ 
	\Upsilon_2=\left\{\Theta_1\le\big\Vert\widehat{\Theta}_{ij}^{1/2}\big\Vert_{\cS}\le\Theta_2,\ {\rm all}\ i,j\in[p]\right\}$$
	hold with probability at least $1-\epsilon$. The thresholding in \eqref{eq.thre} is equivalent to $\widehat{\Sigma}_{\varepsilon,ij}^{\cA}=s_{ij}\big(\widehat{\Sigma}_{\varepsilon,ij}\big)$, where $s_{ij}(\cdot)\equiv s_{\lambda_{ij}}(\cdot)$ with $\lambda_{ij}=\dot{C}\omega_{n,p}\Vert\widehat{\Theta}_{ij}^{1/2}\Vert_{\cS}$ and $\omega_{n,p}=\sqrt{\log p/n}+1/\sqrt{p}$ which is smaller than $\varpi_{n,p}$. For $\dot{C}>2N\Theta_1^{-1}(\varpi_{n,p}/\omega_{n,p})$, under the event $\Upsilon_1\cap\Upsilon_2$, we obtain that
        \vspace{-0.5em}
{\small
	$$
	\begin{aligned}
		&\Vert\widehat{\bSigma}_{\varepsilon}^{\cA}-\bSigma_{\varepsilon}\Vert_{\cS,1}
        =\max_{i\in[p]}\sum_{j=1}^{p}\Vert\widehat{\Sigma}_{\varepsilon,ij}^{\cA}-\Sigma_{\varepsilon,ij}\Vert_{\cS}=\max_{i\in[p]}\sum_{j=1}^{p}\Vert s_{ij}(\widehat{\Sigma}_{\varepsilon,ij})-\Sigma_{\varepsilon,ij}\Vert_{\cS}\\
		\le&\max_{i\in[p]}\sum_{j=1}^{p}\Vert s_{ij}(\widehat{\Sigma}_{\varepsilon,ij})-\widehat{\Sigma}_{\varepsilon,ij}\Vert_{\cS} I\big(\Vert\widehat{\Sigma}_{\varepsilon,ij}\Vert_{\cS}>\dot{C}\omega_{n,p}\Vert\widehat{\Theta}_{ij}^{1/2}\Vert_{\cS}\big)
		+\max_{i\in[p]}\sum_{j=1}^{p}\Vert \widehat{\Sigma}_{\varepsilon,ij}-\Sigma_{\varepsilon,ij}\Vert_{\cS} I\big(\Vert\widehat{\Sigma}_{\varepsilon,ij}\Vert_{\cS}>\dot{C}\omega_{n,p}\Vert\widehat{\Theta}_{ij}^{1/2}\Vert_{\cS}\big)\\
		&+\max_{i\in[p]}\sum_{j=1}^{p}\Vert\Sigma_{\varepsilon,ij}\Vert_{\cS} I\big(\Vert\widehat{\Sigma}_{\varepsilon,ij}\Vert_{\cS}\le\dot{C}\omega_{n,p}\Vert\widehat{\Theta}_{ij}^{1/2}\Vert_{\cS}\big)\\
		\le&\max_{i\in[p]}\sum_{j=1}^{p}\lambda_{ij}I\big(\Vert\widehat{\Sigma}_{\varepsilon,ij}\Vert_{\cS}>\dot{C}\omega_{n,p}\Theta_1\big)+\max_{i\in[p]}\sum_{j=1}^{p}N\varpi_{n,p}I\big(\Vert\widehat{\Sigma}_{\varepsilon,ij}\Vert_{\cS}>\dot{C}\omega_{n,p}\Theta_1\big)
		+\max_{i\in[p]}\sum_{j=1}^{p}\Vert\Sigma_{\varepsilon,ij}\Vert_{\cS}I\big(\Vert\widehat{\Sigma}_{\varepsilon,ij}\Vert_{\cS}\le\dot{C}\omega_{n,p}\Theta_2\big)\\
		\le&(\dot{C}\Theta_2+N)\varpi_{n,p}\max_{i\in[p]}\sum_{j=1}^{p}I\big(\Vert\Sigma_{\varepsilon,ij}\Vert_{\cS}>N\varpi_{n,p}\big)
		+\max_{i\in[p]}\sum_{j=1}^{p}\Vert\Sigma_{\varepsilon,ij}\Vert_{\cS}I\big(\Vert\Sigma_{\varepsilon,ij}\Vert_{\cS}\le(\dot{C}\Theta_2+N)\varpi_{n,p}\big)\\
		\le&(\dot{C}\Theta_2+N)\varpi_{n,p}\max_{i\in[p]}\sum_{j=1}^{p}\frac{\Vert\Sigma_{\varepsilon,ij}\Vert_{\cS}^q}{N^q\varpi_{n,p}^q}I\big(\Vert\Sigma_{\varepsilon,ij}\Vert_{\cS}>N\varpi_{n,p}\big)\\
		&+\max_{i\in[p]}\sum_{j=1}^{p}\Vert\Sigma_{\varepsilon,ij}\Vert_{\cS}\frac{(\dot{C}\Theta_2+N)^{1-q}\varpi_{n,p}^{1-q}}{\Vert\Sigma_{\varepsilon,ij}\Vert_{\cS}^{1-q}}I\big(\Vert\Sigma_{\varepsilon,ij}\Vert_{\cS}\le(\dot{C}\Theta_2+N)\varpi_{n,p}\big)\\
		\le&(\dot{C}\Theta_2+N)\big\{N^{-q}+(\dot{C}\Theta_2+N)^{-q}\big\}\varpi_{n,p}^{1-q}\max_{i\in[p]}\sum_{j=1}^{p}\Vert\Sigma_{\varepsilon,ij}\Vert_{\cS}^q
		\asymp\varpi_{n,p}^{1-q}s_p,
	\end{aligned}
	$$
	}where the third inequality follows from $\dot{C}\Theta_1\omega_{n,p}>2N\varpi_{n,p}$, and the last line follows from the fact that $s_p=\max_{i\in[p]}\sum_{j=1}^{p}\|\sigma_i\|_{\cN}^{(1-q)/2}\Vert\sigma_j\Vert_{\cN}^{(1-q)/2}\Vert\Sigma_{\varepsilon,ij}\Vert_{\cS}^q\asymp\max_{i\in[p]}\sum_{j=1}^{p}\Vert\Sigma_{\varepsilon,ij}\Vert_{\cS}^q$ since $\max_{i\in[p]}\Vert\sigma_i\Vert_{\cN}=\max_{i\in[p]}\int\Sigma_{\varepsilon}(u,u)\du=O(1)$ by Assumption \ref{ass.regu_cond_DIGIT}(iv). Therefore, with probability at least $1-2\epsilon$, $\Vert\widehat{\bSigma}_{\varepsilon}^{\cA}-\bSigma_{\varepsilon}\Vert_{\cS,1}\lesssim \varpi_{n,p}^{1-q}s_p$. Considering that $\epsilon>0$ can be arbitrarily small, we have the desired result
	$$
	\Vert\widehat{\bSigma}_{\varepsilon}^{\cA}-\bSigma_{\varepsilon}\Vert_{\cL}\le\Vert\widehat{\bSigma}_{\varepsilon}^{\cA}-\bSigma_{\varepsilon}\Vert_{\cS,1}=O_p(\varpi_{n,p}^{1-q}s_p). 
	$$

	\subsection{Proof of Theorem~\ref{thm.DIGIT}}
 To prove Theorem~\ref{thm.DIGIT}, we first present a technical lemma with its proof.
	\begin{lemma}
		\label{lem.rate_hatsigma_f_1}
		Suppose that the assumptions of Theorem \ref{thm.DIGIT} hold. For the sample covariance of $\widehat\bbf_t$, i.e., 
  $\widehat{\bSigma}_{f}(u,v)=n^{-1}\sum_{t=1}^{n}\widehat{\bbf}_t(u)\widehat{\bbf}_t(v)^{\T}$, we have
		$$
		\Vert\widehat{\bSigma}_{f}-\bU\bSigma_{f}\bU^{\T}\Vert_{\cS,\max}=O_p(\cM_{\varepsilon}/\sqrt{n}+1/\sqrt{p}).
		$$
	\end{lemma}
	\begin{proof}
	Consider  $\widehat{\bbf}_t(u)\widehat{\bbf}_t(v)^{\T}-\bU\bbf_t(u)\bbf_t(v)^{\T}\bU^{\T}=\big\{\widehat{\bbf}_t(u)-\bU\bbf_t(u)\big\}\widehat{\bbf}_t(v)^{\T}+\bU\bbf_t(u)\big\{\widehat{\bbf}_t(v)-\bU\bbf_t(v)\big\}^{\T}.$ Then  
		$$
		\begin{aligned}
			\Big\Vert \frac{1}{n}\sum_{t=1}^{n}(\widehat{\bbf}_t\widehat{\bbf}_t^{\T}-\bU\bbf_t\bbf_t^{\T}\bU^{\T})\Big\Vert_{\cS,\max}
			\le&\Big\Vert\frac{1}{n}\sum_{t=1}^{n}(\widehat{\bbf}_t-\bU\bbf_t)\widehat{\bbf}_t^{\T}\Big\Vert_{\cS,\max}+\Big\Vert\frac{1}{n}\sum_{t=1}^{n}\bU\bbf_t(\widehat{\bbf}_t-\bU\bbf_t)^{\T}\Big\Vert_{\cS,\max}\\
			\le&\Big(\frac{1}{n}\sum_{t=1}^{n}\Vert\widehat{\bbf}_t-\bU\bbf_t\Vert^2\Big)^{1/2}\left(\frac{1}{n}\sum_{t=1}^{n}\Vert\widehat{\bbf}_t\Vert^2\right)^{1/2} \\
			&+ \Big(\frac{1}{n}\sum_{t=1}^{n} \Vert\widehat{\bbf}_t-\bU\bbf_t\Vert^2\Big)^{1/2}\left(\frac{1}{n}\sum_{t=1}^{n}\Vert\bU\bbf_t\Vert^2\right)^{1/2} =
			O_p(\varpi_{n,p}),
		\end{aligned}
		$$
		where the second inequality follows from the Cauchy--Schwarz inequality, and the last line follows from $n^{-1}\sum_{t=1}^{n}\Vert\widehat{\bbf}_t-\bU\bbf_t\Vert^2=O_p(\cM_{\varepsilon}^2/n+1/p)$ by Theorem~\ref{thm.load_factor_1}(ii), and $n^{-1}\sum_{t=1}^{n}\Vert\bU\bbf_t\Vert^2=O_p(1)$ since $\Vert\bU\Vert=1$ and $\eE\Vert\bbf_t\Vert^2=O(1)$. Together with Lemma \ref{lem.rate_sigma_f}(i), the desired result follows immediately.
	\end{proof}

We are now ready to prove Theorem \ref{thm.DIGIT}. Consider that 
	$$
        \begin{aligned}
            &\bB\bSigma_{f}\bB^{\T}-\widehat{\bB}\widehat{\bSigma}_{f}\widehat{\bB}^{\T}=\bB\bU^{\T}\bU\bSigma_{f}\bU^{\T}\bU\bB^{\T}-\widehat{\bB}\widehat{\bSigma}_{f}\widehat{\bB}^{\T}\\
            =&\bB\bU^{\T}(\bU\bSigma_{f}\bU^{\T}-\widehat{\bSigma}_{f})\bU\bB^{\T}+(\bB\bU^{\T}-\widehat{\bB})\widehat{\bSigma}_{f}\bU\bB^{\T}+\widehat{\bB}\widehat{\bSigma}_{f}(\bU\bB^{\T}-\widehat{\bB}^{\T}).
        \end{aligned}
	$$
	Then we have
	\begin{equation}
		\label{eq.DIGIT_com_con}
		\begin{aligned}
			&\Vert\bB\bSigma_{f}\bB^{\T}-\widehat{\bB}\widehat{\bSigma}_{f}\widehat{\bB}^{\T}\Vert_{\cS,\max}\\
            \le&\Vert\bB\bU^{\T}\Vert_{\infty}\Vert\bU\bSigma_{f}\bU^{\T}-\widehat{\bSigma}_{f}\Vert_{\cS,\max}\Vert\bU\bB^{\T}\Vert_{1}
			+2\Vert\bB\bU^{\T}-\widehat{\bB}\Vert_{\infty}(\Vert\bU\bSigma_{f}\bU^{\T}\Vert_{\cS,\max}+\Vert\bU\bSigma_{f}\bU^{\T}-\widehat{\bSigma}_{f}\Vert_{\cS,\max})\Vert\bB\bU^{\T}\Vert_{\infty}\\
			\le&r^3C^2\Vert\bU\bSigma_{f}\bU^{\T}-\widehat{\bSigma}_{f}\Vert_{\cS,\max}+2r^{5/2}C(rC_{\max}
			+\Vert\bU\bSigma_{f}\bU^{\T}-\widehat{\bSigma}_{f}\Vert_{\cS,\max})\Vert\bB\bU^{\T}-\widehat{\bB}\Vert_{\max}\\
			=&O_p(\cM_{\varepsilon}/\sqrt{n}+1/\sqrt{p})+O_p(\varpi_{n,p})=O_p(\varpi_{n,p}),
		\end{aligned}
	\end{equation}	
	where the first inequality follows from Lemma \ref{lem.ineq3}(i), the second inequality follows from $\Vert\bB\bU^{\T}\Vert_{\infty}\le r\Vert\bB\bU^{\T}\Vert_{\max}\le r\Vert\bB\Vert_{\max}\Vert\bU\Vert_{\infty}\le r^{3/2}C$ provided that $\Vert\bU\Vert_{\infty}\le\sqrt{r}\Vert\bU\Vert=\sqrt{r}$, $\Vert \bB\bU^{\T}-\widehat{\bB}\Vert_{\infty}\le r\Vert\bB\bU^{\T}-\widehat{\bB}\Vert_{\max}$ and $\Vert\bU\bSigma_{f}\bU^{\T}\Vert_{\cS,\max}\le C_{\max}\Vert\bU\Vert_{\infty}^2\le rC_{\max}$ in Lemma \ref{lem.sigma_f_bound}(i), and the last line follows from Lemma \ref{lem.rate_hatsigma_f_1} and $\Vert\bB\bU^{\T}-\widehat{\bB}\Vert_{\max}=O_p(\varpi_{n,p})$ in Theorem~\ref{thm.load_factor_1}(i). Then note that
	$$
	\begin{aligned}
		\Vert\widehat{\bSigma}_{\varepsilon}^{\cA}-\bSigma_{\varepsilon}\Vert_{\cS,\max}\le&\Vert\widehat{\bSigma}_{\varepsilon}^{\cA}-\widehat{\bSigma}_{\varepsilon}\Vert_{\cS,\max}+\Vert\widehat{\bSigma}_{\varepsilon}-\bSigma_{\varepsilon}\Vert_{\cS,\max}
		\le\max_{i,j\in[p]}(\Vert\widehat{\Theta}_{ij}^{1/2}\Vert_{\cS}\lambda)+O_p(\varpi_{n,p})
		=O_p(\varpi_{n,p}),
	\end{aligned}	
	$$
	where the last line follows from Lemma \ref{lem.varep_bound_1}(iii), the choice of $\lambda=\dot{C}(\sqrt{\log p/n}+\sqrt{1/p})\lesssim\varpi_{n,p}$, and the fact $\max_{i,j\in[p]}\Vert\widehat{\Theta}_{ij}^{1/2}\Vert_{\cS}=O_p(1)$ by Lemma \ref{lem.Theta_bound}. By combining \eqref{eq.decom}, \eqref{eq.estimator_1}, and \eqref{eq.DIGIT_com_con}, we obtain the desired result.

 \subsection{Proof of Theorem~\ref{thm.inverse_digit}}
For the sake of brevity, in this section, we suppose that the orthogonal matrix $\bU$ in Theorem~\ref{thm.load_factor_1} and Lemmas~\ref{lem.rate_B_F}--\ref{lem.rate_hatsigma_f_1} is an identity matrix, which means, when we perform eigen-decomposition on $\widehat{\bOmega}$, we can always select the correct direction of $\widehat{\bxi}_j$ to ensure $\widehat{\bxi}_j^{\T}\widetilde{\bbb}_j\ge0$. The proofs of Theorems \ref{thm.idio_DIGIT} and \ref{thm.DIGIT} verify that the choice of $\bU$ does not affect the theoretical results. In this section, $(\widehat{\bSigma}_{\varepsilon}^{\cA})^{\dagger},\widehat{\bSigma}_f^{\dagger}$ and $\{\widehat{\bSigma}_f^{\dagger}+\widehat{\bB}^{\T}(\widehat{\bSigma}_{\varepsilon}^{\cA})^{\dagger}\widehat{\bB}\}^{\dagger}$ are defined in a similar way to the inverse in Lemma~\ref{lem.inv_norm} associated with $\bSigma_{\varepsilon},\bSigma_f$ and $\bSigma_f^{\dagger}+\bB^{\T}\bSigma_{\varepsilon}^{\dagger}\bB$, respectively. To prove Theorem~\ref{thm.inverse_digit}, we first present some technical lemmas with their proofs.

\begin{lemma}
    \label{lem.idio_inver}
    Under the assumptions of Theorem \ref{thm.inverse_digit}, then, $\widehat{\bSigma}_{\varepsilon}^{\cA}$ has a bounded Moore--Penrose inverse with probability approaching one, and $\big\Vert(\widehat{\bSigma}_{\varepsilon}^{\cA})^{\dagger}-\bSigma_{\varepsilon}^{\dagger}\big\Vert_{\cL}=O_p(\varpi_{n,p}^{1-q}s_p).$
\end{lemma}
\begin{proof}
    Provided that $\varpi_{n,p}^{1-q}s_p=o(1)$ and $\Vert\bSigma_{\varepsilon}^{\dagger}\Vert_{\cL}<c_3$ for some constant $c_3>0$, we combine Lemma \ref{lem.inv_norm} and Theorem \ref{thm.idio_DIGIT} to yield that $\Vert(\widehat{\bSigma}_{\varepsilon}^{\cA})^{\dagger}\Vert_{\cL}<2c_3$ with probability approaching one, and thus $\widehat{\bSigma}_{\varepsilon}^{\cA}$ has a bounded Moore--Penrose inverse with probability approaching one together with the desired result $\big\Vert(\widehat{\bSigma}_{\varepsilon}^{\cA})^{\dagger}-\bSigma_{\varepsilon}^{\dagger}\big\Vert_{\cL}=O_p(\varpi_{n,p}^{1-q}s_p).$
\end{proof}
 
	\begin{lemma}
		\label{lem.inverse1}
		Under the assumptions of Theorem \ref{thm.inverse_digit}, $$\Vert\widehat{\bB}^{\T}(\widehat{\bSigma}_{\varepsilon}^{\cA})^{\dagger}\widehat{\bB}-\bB^{\T}\bSigma_{\varepsilon}^{\dagger}\bB\Vert_{\cL}=O_p(p\varpi_{n,p}^{1-q}s_p)=o_p(p).$$
	\end{lemma}
	\begin{proof}
		Consider
		$$
		\begin{aligned}
			\Vert\widehat{\bB}^{\T}(\widehat{\bSigma}_{\varepsilon}^{\cA})^{\dagger}\widehat{\bB}-\bB^{\T}\bSigma_{\varepsilon}^{\dagger}\bB\Vert_{\cL}\le&2\Vert(\widehat{\bB}-\bB)^{\T}(\widehat{\bSigma}_{\varepsilon}^{\cA})^{\dagger}\widehat{\bB}\Vert_{\cL}+\Vert\bB^{\T}\{(\widehat{\bSigma}_{\varepsilon}^{\cA})^{\dagger}-\bSigma_{\varepsilon}^{\dagger}\}\bB\Vert_{\cL}\\
			\le&2\Vert\widehat{\bB}-\bB\Vert\Vert(\widehat{\bSigma}_{\varepsilon}^{\cA})^{\dagger}\Vert_{\cL}\Vert\bB\Vert+\Vert\bB\Vert^2\Vert(\widehat{\bSigma}_{\varepsilon}^{\cA})^{\dagger}-\bSigma_{\varepsilon}^{\dagger}\Vert_{\cL}\\
			=&O_p(p\varpi_{n,p})+O_p(p\varpi_{n,p}^{1-q}s_p)=O_p(p\varpi_{n,p}^{1-q}s_p)=o_p(p),
		\end{aligned}
		$$
        where the last line follows from Lemmas~\ref{lem.rate_B_F} and \ref{lem.idio_inver}. 
	\end{proof}

	\begin{lemma}
		\label{lem.inverse2}
		Under the assumptions of Theorem \ref{thm.inverse_digit},\\
		(i) $\big\Vert(\bSigma_f^{\dagger}+\bB^{\T}\bSigma_{\varepsilon}^{\dagger}\bB)^{\dagger}\big\Vert_{\cL}=O(p^{-1})$;\\
		(ii) $\big\Vert\{\widehat{\bSigma}_f^{\dagger}+\widehat{\bB}^{\T}(\widehat{\bSigma}_{\varepsilon}^{\cA})^{\dagger}\widehat{\bB}\}^{\dagger}\big\Vert_{\cL}=O_p(p^{-1})$.
	\end{lemma}
	\begin{proof}
		(i) Note that
        $\Vert(\bSigma_f^{\dagger}+\bB^{\T}\bSigma_{\varepsilon}^{\dagger}\bB)^{\dagger}\Vert_{\cL}\le\Vert(\bB^{\T}\bSigma_{\varepsilon}^{\dagger}\bB)^{\dagger}\Vert_{\cL}\le\{\lambda_{\min}(\bB^{\T}\bB)\}^{-1}\Vert\bSigma_{\varepsilon}\Vert_{\cL}=O(p^{-1}),
		$
		where the first inequality follows from the fact $\bSigma_f$ is a Mercer's kernel.\\
		(ii) Since $\Vert\bSigma_f^{\dagger}\Vert_{\cL}<c_4$ and $\Vert\widehat{\bSigma}_f-\bSigma_f\Vert_{\cL}=O_p(\varpi_{n,p})=o_p(1)$, by Lemma \ref{lem.inv_norm}, we have $\Vert\widehat{\bSigma}_f^{\dagger}-\bSigma_f^{\dagger}\Vert_{\cL}=O_p(\varpi_{n,p})$. Thus, by Lemma \ref{lem.inverse1},$
		\big\Vert\big\{\widehat{\bSigma}_f^{\dagger}+\widehat{\bB}^{\T}(\widehat{\bSigma}_{\varepsilon}^{\cA})^{\dagger}\widehat{\bB}\big\}-\big\{\bSigma_f^{\dagger}+\bB^{\T}\bSigma_{\varepsilon}^{\dagger}\bB\big\}\big\Vert_{\cL}=o_p(p). 
		$
		Combining Lemma \ref{lem.inv_norm} and part (i), we obtain that $\big\Vert\{\widehat{\bSigma}_f^{\dagger}+\widehat{\bB}^{\T}(\widehat{\bSigma}_{\varepsilon}^{\cA})^{\dagger}\widehat{\bB}\}^{\dagger}\big\Vert_{\cL}=O_p(p^{-1})$. 
	\end{proof}
	
	We are now ready to prove Theorem~\ref{thm.inverse_digit}. Using the functional version of the Sherman--Morrison--Woodbury formula, we have $\Vert(\widehat{\bSigma}_y^{\cD})^{\dagger}-\bSigma_y^{\dagger}\Vert_{\cL}\le \sum_{k=1}^{4}L_k$, where 
	$$
	\begin{aligned}
		L_1=&\big\Vert(\widehat{\bSigma}_{\varepsilon}^{\cA})^{\dagger}-\bSigma_{\varepsilon}^{\dagger}\big\Vert_{\cL},\\
		L_2=&\big\Vert\big\{(\widehat{\bSigma}_{\varepsilon}^{\cA})^{\dagger}\widehat{\bB}-\bSigma_{\varepsilon}^{\dagger}\bB\big\}\big\{\widehat{\bSigma}_f^{\dagger}+\widehat{\bB}^{\T}(\widehat{\bSigma}_{\varepsilon}^{\cA})^{\dagger}\widehat{\bB}\big\}^{\dagger}\widehat{\bB}^{\T}(\widehat{\bSigma}_{\varepsilon}^{\cA})^{\dagger}\big\Vert_{\cL},\\
		L_3=&\big\Vert\bSigma_{\varepsilon}^{\dagger}\bB\big\{\widehat{\bSigma}_f^{\dagger}+\widehat{\bB}^{\T}(\widehat{\bSigma}_{\varepsilon}^{\cA})^{\dagger}\widehat{\bB}\big\}^{\dagger}\big\{\widehat{\bB}^{\T}(\widehat{\bSigma}_{\varepsilon}^{\cA})^{\dagger}-\bB^{\T}\bSigma_{\varepsilon}^{\dagger}\big\}\big\Vert_{\cL},\\
		L_4=&\big\Vert\bSigma_{\varepsilon}^{\dagger}\bB\big[\big\{\widehat{\bSigma}_f^{\dagger}+\widehat{\bB}^{\T}(\widehat{\bSigma}_{\varepsilon}^{\cA})^{\dagger}\widehat{\bB}\big\}^{\dagger}-\big\{\bSigma_f^{\dagger}+\bB^{\T}\bSigma_{\varepsilon}^{\dagger}\bB\big\}^{\dagger}\big]\bB^{\T}\bSigma_{\varepsilon}^{\dagger}\big\Vert_{\cL}.
	\end{aligned}
	$$
    Clearly, $L_1=O_p(\varpi_{n,p}^{1-q}s_p)$ by Lemma \ref{lem.idio_inver}. Then, note that $\Vert(\widehat{\bSigma}_{\varepsilon}^{\cA})^{\dagger}\widehat{\bB}-\bSigma_{\varepsilon}^{\dagger}\bB\Vert_{\cL}\le\Vert(\widehat{\bSigma}_{\varepsilon}^{\cA})^{\dagger}-\bSigma_{\varepsilon}^{\dagger}\Vert_{\cL}\Vert\widehat{\bB}\Vert+\Vert\bSigma_{\varepsilon}^{\dagger}\Vert_{\cL}\Vert\widehat{\bB}-\bB\Vert=O_p(\sqrt{p}\varpi_{n,p}^{1-q}s_p)$. From Lemma \ref{lem.inverse2}, we obtain that $L_2\asymp L_3=O_p(\varpi_{n,p}^{1-q}s_p)$. Lastly, since $\Vert(\bSigma_f^{\dagger}+\bB^{\T}\bSigma_{\varepsilon}^{\dagger}\bB)^{\dagger}\Vert_{\cL}=O(p^{-1})$ and $\Vert\{\widehat{\bSigma}_f^{\dagger}+\widehat{\bB}^{\T}(\widehat{\bSigma}_{\varepsilon}^{\cA})^{\dagger}\widehat{\bB}\}-\{\bSigma_f^{\dagger}+\bB^{\T}\bSigma_{\varepsilon}^{\dagger}\bB\}\Vert_{\cL}=O_p(p\varpi_{n,p}^{1-q}s_p)=o_p(p)$, we apply Lemma \ref{lem.inv_norm} to obtain that
	$$
	\left\Vert\{\widehat{\bSigma}_f^{\dagger}+\widehat{\bB}^{\T}(\widehat{\bSigma}_{\varepsilon}^{\cA})^{\dagger}\widehat{\bB}\}^{\dagger}-\{\bSigma_f^{\dagger}+\bB^{\T}\bSigma_{\varepsilon}^{\dagger}\bB\}^{\dagger}\right\Vert_{\cL}=O_p(p^{-2})O_p(p\varpi_{n,p}^{1-q}s_p)=O_p(p^{-1}\varpi_{n,p}^{1-q}s_p),
	$$
	which implies that $L_4=O_p(\varpi_{n,p}^{1-q}s_p)$. Combining the above results, $\widehat{\bSigma}_y^{\cD}$ has a bounded Moore--Penrose inverse with probability approaching one, and
	$
	\big\Vert(\widehat{\bSigma}_y^{\cD})^{\dagger}-\bSigma_y^{\dagger}\big\Vert_{\cL}=O_p(\varpi_{n,p}^{1-q}s_p).
	$ 
\end{appendix}

\onehalfspacing
\bibliographystyle{dcu}
\bibliography{main}

 \newpage
 \onehalfspacing
	\begin{center}
		{\noindent \bf \large Supplementary Material to ``Factor-guided estimation of large covariance matrix function with conditional functional sparsity"}\\
	\end{center}
	\begin{center}
		{\noindent Dong Li, Xinghao Qiao and Zihan Wang}
	\end{center}
	\bigskip
	
	\setcounter{page}{1}
    \setcounter{theorem}{0}
	\renewcommand{\thetheorem}{S.\arabic{theorem}}
    \renewcommand{\theHtheorem}{S.\arabic{theorem}}
	\setcounter{section}{0}
	\renewcommand{\thesection}{S.\arabic{section}}
    \renewcommand{\theHsection}{S.\arabic{section}}
	\setcounter{lemma}{0}
	\renewcommand{\thelemma}{S.\arabic{lemma}}
    \renewcommand{\theHlemma}{S.\arabic{lemma}}
	\setcounter{equation}{0}
	\renewcommand{\theequation}{S.\arabic{equation}}
    \renewcommand{\theHequation}{S.\arabic{equation}}
	\setcounter{remark}{0}
	\renewcommand{\theremark}{S.\arabic{remark}}
    \renewcommand{\theHremark}{S.\arabic{remark}}
	\setcounter{definition}{0}
	\renewcommand{\thedefinition}{S.\arabic{definition}}
    \renewcommand{\theHdefinition}{S.\arabic{definition}}
	\setcounter{proposition}{0}
	\renewcommand{\theproposition}{S.\arabic{proposition}}
    \renewcommand{\theHproposition}{S.\arabic{proposition}}
	\setcounter{figure}{0}
	\renewcommand{\thefigure}{S.\arabic{figure}}
    \renewcommand{\theHfigure}{S.\arabic{figure}}
	\setcounter{table}{0}
	\renewcommand{\thetable}{S.\arabic{table}}
    \renewcommand{\theHtable}{S.\arabic{table}}
    \setcounter{condition}{0}
	\renewcommand{\thecondition}{S.\arabic{condition}}
    \renewcommand{\theHcondition}{S.\arabic{condition}}
    \setcounter{assumption}{0}
	\renewcommand{\theassumption}{S.\arabic{assumption}}
    \renewcommand{\theHassumption}{S.\arabic{assumption}}

This supplementary material contains the proofs of the technical lemmas in the appendix in Section~\ref{supsec.L}, and the remaining technical proofs in Sections~\ref{supsec.B}--\ref{supsec.D}, further derivations in Section~\ref{supsec.E}, additional simulation results in Section~\ref{supsec.F} and additional real data result in Section~\ref{supsec.G}.

\section{Proofs of theoretical lemmas in the appendix}
\label{supsec.L}

\subsection{Proof of Lemma~\ref{lem.ineq1}}
\noindent
		(i) Note that $\int\int\bK(u,v)\bK(u,v)^{\T}\du\dv=\int\sum_{i=1}^{\infty}\lambda_i^2\bphi_i(u)\bphi_i(u)^{\T}\du$, and thus
		$$
		{\rm tr}\Big\{\int\int\bK(u,v)\bK(u,v)^{\T}\du\dv\Big\}={\rm tr}\Big\{\sum_{i=1}^{\infty}\lambda_i^2\int\bphi_i(u)^{\T}\bphi_i(u)\du\Big\}=\sum_{i=1}^{\infty}\lambda_i^2.
		$$
		The equality ${\rm tr}\left\{\int\int\bK(u,v)\bK(u,v)^{\T}\du\dv\right\}=\Vert\bK\Vert_{\cS,\F}^2$ can be verified by simple calculation. The first equality can be obtained by $\bK(u,v)^{\T}=\bK^{\T}(v,u)$ and the multiplication of kernel functions.\\
		(ii) Similarly, 
		$$
		\begin{aligned}
			\Big\Vert\int\int\bK(u,v)\bK(u,v)^{\T}\du\dv\Big\Vert=&\Big\Vert\int\sum_{i=1}^{\infty}\lambda_i^2\bphi_i(u)\bphi_i(u)^{\T}\du\Big\Vert=\lambda_{\max}\Big\{\int\sum_{i=1}^{\infty}\lambda_i^2\bphi_i(u)\bphi_i(u)^{\T}\du\Big\}\\
			=&\lambda_{\max}\Big\{\bLambda^2\int\bPhi(u)\bPhi(u)^{\T}\du\Big\}=\lambda_{\max}\Big\{\bLambda^2\int\bPhi(u)^{\T}\bPhi(u)\du\Big\}\\
			=&\lambda_{\max}(\bLambda^2)=\lambda_1^2=\Vert\bK\Vert_{\cL}^2,
		\end{aligned}
		$$
		where $\bLambda={\rm diag}(\lambda_1,\lambda_2,\dots),\bPhi(\cdot)=\{\bphi_1(\cdot),\bphi_2(\cdot),\dots\},$ and the fact that the $\infty\times\infty$ matrix $\int\bPhi(u)^{\T}\bPhi(u)\du$ shares the same nonzero eigenvalues with the $p\times p$ matrix $\int\bPhi(u)\bPhi(u)^{\T}\du,$ which can be obtained following the proof of Proposition 2 in \textcolor{blue}{Bathia et al.} (\textcolor{blue}{2010}).\\
        (iii) The equality holds by the definition of trace norm.

\subsection{Proof of Lemma~\ref{lem.ineq2}}
\noindent
		(i) Note that
		\begin{equation}
			\label{eq.lem.ineq21}
			\begin{aligned}
				\Big\Vert\int\int \bK(u,v)\bG(u,v)^{\T}{\rm d}u{\rm d}v\Big\Vert_1=&\max_{j\in[p]}\sum_{i=1}^{p}\left|\int\int\sum_{k=1}^{p} K_{ik}(u,v)G_{jk}(u,v)\du\dv\right|\\
				\le&\max_{j\in[p]}\sum_{i=1}^{p}\sum_{k=1}^{p}\Vert K_{ik}\Vert_{\cS}\Vert G_{jk}\Vert_{\cS}\\
				\le&\Big(\max_{k\in[p]}\sum_{i=1}^{p}\Vert K_{ik}\Vert_{\cS}\Big)\Big(\max_{j\in[p]}\sum_{k=1}^{p}\Vert G_{jk}\Vert_{\cS}\Big)\\
				=&\Vert\bK\Vert_{\cS,1}\Vert\bG\Vert_{\cS,\infty}.
			\end{aligned}
		\end{equation}\\
		(ii) By similar arguments, we obtain that
		\begin{equation}
			\label{eq.lem.ineq22}
			\Big\Vert\int\int \bK(u,v)\bG(u,v)^{\T}{\rm d}u{\rm d}v\Big\Vert_{\infty}\le\Vert\bK\Vert_{\cS,\infty}\Vert\bG\Vert_{\cS,1}.
		\end{equation}\\
		(iii) The inequality follows immediately from \eqref{eq.lem.ineq21}, \eqref{eq.lem.ineq22}, the matrix norm inequality $\Vert\bA\Vert^2\le\Vert\bA\Vert_{\infty}\Vert\bA\Vert_1$ for any $p\times p$ matrix $\bA$ and the choice of $\bA=\int\int\bK(u,v)\bG(u,v)^{\T}\du\dv$.\\
		(iv) An application of H$\ddot{{\rm o}}$lder's inequality yields the result.

\subsection{Proof of Lemma~\ref{lem.op_le_S1}}
\noindent
            (i) By Lemma~\ref{lem.eigen} or Theorem 4.2.5 in \textcolor{blue}{Hsing and Eubank} (\textcolor{blue}{2015}), we have
            $$
            \begin{aligned}
                \Vert\bSigma\widetilde{\bSigma}\Vert_{\cL}=&\max_{\bx\in\eH^p}\frac{\big\langle\bx,\bSigma\widetilde{\bSigma}(\bx)\big\rangle}{\Vert\bx\Vert^2}=\max_{\bx\in\eH^p}\frac{\Big\langle\widetilde{\bSigma}^{1/2}(\bx),\bSigma\big\{\widetilde{\bSigma}^{1/2}(\bx)\big\}\Big\rangle}{\Big\Vert\widetilde{\bSigma}^{1/2}(\bx)\Big\Vert^2}\cdot\frac{\big\langle\bx,\widetilde{\bSigma}(\bx)\big\rangle}{\Vert\bx\Vert^2}\\
                \le&\max_{\by\in\eH^p}\frac{\big\langle\by,\bSigma(\by)\big\rangle}{\Vert\by\Vert^2}\max_{\bx\in\eH^p}\frac{\big\langle\bx,\widetilde{\bSigma}(\bx)\big\rangle}{\Vert\bx\Vert^2}=\Vert\bSigma\Vert_{\cL}\cdot\Vert\widetilde{\bSigma}\Vert_{\cL}.
            \end{aligned}
            $$
            \\
            (ii) By Lemma~\ref{lem.ineq1}, $\Vert\bSigma\Vert_{\cL}=\lambda_1$ and $\Vert\bSigma\Vert_{\F}=\sqrt{\sum_{i\ge1}\lambda_i^2}$, where $\{\lambda_i\}_{i=1}^{\infty}$ are the eigenvalues of $\bSigma$ in a descending order. Apparently, $\Vert\bSigma\Vert_{\cL}\le\Vert\bSigma\Vert_{\cS,\F}$ holds. \\\
            (iii) By Lemmas~\ref{lem.ineq1}(ii) and \ref{lem.ineq2}(iii), $\Vert\bSigma\Vert_{\cL}^2=\big\Vert\int\int\bSigma(u,v)\bSigma(u,v)^{\T}\du\dv\big\Vert\le\Vert\bSigma\Vert_{\cS,1}\Vert\bSigma\Vert_{\cS,\infty}$. Furthermore, if $\Vert\Sigma_{ij}\Vert_{\cS}=\Vert\Sigma_{ji}\Vert_{\cS}$ for all $i,j\in[p]$, we have $\Vert\bSigma\Vert_{\cS,1}=\Vert\bSigma\Vert_{\cS,\infty}$, and thus $\Vert\bSigma\Vert_{\cL}\le\Vert\bSigma\Vert_{\cS,1}$ holds.

\subsection{Proof of Lemma~\ref{lem.trace}}
\noindent
		(i) Note that 
		$$
		\begin{aligned}
			{\rm tr}\Big\{\int\bK\bG(u,u)\du\Big\}=&{\rm tr}\Big\{\int\int\bK(u,v)\bG(v,u)\du\dv\Big\}=\int\int{\rm tr}\{\bK(u,v)\bG(v,u)\}\du\dv\\
			=&{\rm tr}\Big\{\int\int\bG(v,u)\bK(v,u)\dv\du\Big\}={\rm tr}\Big\{\int\bG\bK(v,v)\dv\Big\}.
		\end{aligned}
		$$\\
		(ii) Suppose that $\bK(u,v)=\sum_{i=1}^{\infty}\lambda_i\bphi_i(u)\bphi_i(v)^{\T}$ and $\bG(u,v)=\sum_{j=1}^{\infty}\omega_j\bpsi_j(u)\bpsi_j(v)^{\T}$ where $\{\bphi_i(\cdot)\}$ and $\{\bpsi_j(\cdot)\}$ are both orthonormal basis functions. Then, we have
		$$
		\begin{aligned}
			{\rm tr}\Big\{\int\bK\bG(u,u)\du\Big\}=&{\rm tr}\Big\{\int\int\bK(u,v)\bG(v,u)\du\dv\Big\}\\
			=&\sum_{i=1}^{\infty}\sum_{j=1}^{\infty}\lambda_i\omega_j\int\bphi_i(v)^{\T}\bpsi_j(v)\dv\int\bpsi_j(u)^{\T}\bphi_i(u)\du\\
			=&\sum_{i=1}^{\infty}\sum_{j=1}^{\infty}\lambda_i\omega_j\left|\langle\bphi_i,\bpsi_j\rangle\right|^2\le\sum_{i=1}^{\infty}\lambda_i\omega_i\\
			\le&\left(\max_{i}\lambda_i\right)\sum_{j=1}^{\infty}\omega_j=\Vert\bK\Vert_{\cL}\Vert\bG\Vert_{\cN}
			=\Vert\bK\Vert_{\cL}{\rm tr}\Big\{\int\bG(u,u)\du\Big\},
		\end{aligned}
		$$
		where the first inequality follows by using similar arguments to prove von Neumann's trace inequality (see \textcolor{blue}{Carlsson}, \textcolor{blue}{2021}). \\
		(iii) From Lemma \ref{lem.ineq1}(i) and the part (ii) above, we have
		$$
		\begin{aligned}
			\Vert\bK\bG\Vert_{\cS,\F}^2=&{\rm tr}\Big\{\int\bK\bG\bG^{\T}\bK^{\T}(u,u)\du\Big\}={\rm tr}\Big\{\int\bK^{\T}\bK\bG\bG^{\T}(u,u)\du\Big\}\\
			\le&\Vert\bK\bK^{\T}\Vert_{\cL}{\rm tr}\Big\{\int\bG\bG^{\T}(u,u)\du\Big\}=\Vert\bK\Vert_{\cL}^2\Vert\bG\Vert_{\cS,\F}^2,
		\end{aligned}
		$$
		which implies the desired result.

\subsection{Proof of Lemma~\ref{lem.ineq3}}
\noindent
		(i) It follows that
		$$
		\begin{aligned}
			\Vert\bA\bK\Vert_{\cS,\max}=&\max_{i\in[p],j\in[q]}\Big\Vert \sum_{k=1}^{q}A_{ik}K_{kj}\Big\Vert_{\cS}\le
			\max_{i\in[p],j\in[q]}\sum_{k=1}^{q}|A_{ik}|\Vert K_{kj}\Vert_{\cS}\\
			\le&\Big(\max_{i\in[p]}\sum_{k=1}^{q}|A_{ik}|\Big)\cdot\Vert \bK\Vert_{\cS,\max}=\Vert\bA\Vert_{\infty}\Vert\bK\Vert_{\cS,\max}.
		\end{aligned}
		$$
		Further,
		$$
		\begin{aligned}
			\Vert\bK\bA^{\T}\Vert_{\cS,\max}=&\Vert(\bK\bA^{\T})^{\T}\Vert_{\cS,\max}=\Vert\bA\bK^{\T}\Vert_{\cS,\max}\le\Vert\bA\Vert_{\infty}\Vert\bK^{\T}\Vert_{\cS,\max}\\
			=&\Vert\bA\Vert_{\infty}\Vert\bK\Vert_{\cS,\max}=\Vert\bA^{\T}\Vert_{1}\Vert\bK\Vert_{\cS,\max}.
		\end{aligned}
		$$\\
		(ii) It follows that
		$$
		\begin{aligned}
			\Vert\bA\bK\Vert_{\cS,\F}^2=&\sum_{i=1}^{p}\sum_{j=1}^{q}\Big\Vert\sum_{k=1}^{q}A_{ik}K_{kj}\Big\Vert_{\cS}^2\le \sum_{i=1}^{p}\sum_{j=1}^{q}\Big(\sum_{k=1}^{q}A_{ik}^2\sum_{k=1}^{q}\Vert K_{kj}\Vert_{\cS}^2\Big)\\
			=&\sum_{i=1}^{p}\sum_{j=1}^{q}\Big(\sum_{k=1}^{q}\sum_{l=1}^{q}A_{ik}^2\Vert K_{lj}\Vert_{\cS}^2\Big)=\left(\sum_{i=1}^{p}\sum_{k=1}^{q}A_{ik}^2\right)\left(\sum_{l=1}^{q}\sum_{j=1}^{q}\Vert K_{lj}\Vert_{\cS}^2\right)\\
			=&\Vert\bA\Vert_{\F}\Vert\bK\Vert_{\cS,\F},
		\end{aligned}
		$$
		where the inequality follows from the Cauchy--Schwarz inequality. Furthermore,
		$$
		\begin{aligned}
			\Vert\bK\bA^{\T}\Vert_{\cS,\F}=\Vert(\bK\bA^{\T})^{\T}\Vert_{\cS,\F}=\Vert\bA\bK^{\T}\Vert_{\cS,\F}\le\Vert\bA\Vert_{\F}\Vert\bK\Vert_{\cS,\F}=\Vert\bK\Vert_{\cS,\F}\Vert\bA^{\T}\Vert_{\F}.
		\end{aligned}
		$$\\
		(iii)\&(iv) It follows that
		$$
		\begin{aligned}
			\Vert\bA\bK\Vert_{\cS,\infty}=&\max_{i\in[p]}\sum_{k=1}^{q}\sum_{j=1}^{q}\Vert A_{ik}K_{kj}\Vert_{\cS}=\max_{i\in[p]}\sum_{k=1}^{q}\sum_{j=1}^{q}|A_{ik}|\Vert K_{kj}\Vert_{\cS}\\
			\le&\max_{i\in[p]}\sum_{k=1}^{q}\sum_{j=1}^{q}|A_{ik}|\max_{k'\in[q]}\Vert K_{k'j}\Vert_{\cS}\\
			=&\Big(\max_{i\in[p]}\sum_{k=1}^{q}|A_{ik}|\Big)\Big(\max_{k'\in[q]}\sum_{j=1}^{q}\Vert K_{k'j}\Vert_{\cS}\Big)
			=\Vert\bA\Vert_{\infty}\Vert\bK\Vert_{\cS,\infty}.
		\end{aligned}
		$$
		Furthermore,
		$$\Vert\bK\bA^{\T}\Vert_{\cS,1}= \Vert\bA\bK^{\T}\Vert_{\cS,\infty}\le\Vert\bA\Vert_{\infty}\Vert\bK^{\T}\Vert_{\cS,\infty}=\Vert\bA\Vert_{\infty}\Vert\bK\Vert_{\cS,1}.$$
		The other two arguments can be proved similarly.

\subsection{Proof of Lemma~\ref{lem.factor_norm}}
\noindent
		(i) By the definition, it follows that $$\Vert\bA\bbf\Vert=\Big\{\int\bbf_t(u)^{\T}\bA^{\T}\bA\bbf_t(u){\rm d}u\Big\}^{1/2}\le\Big\{\int\lambda_{\max}(\bA^{\T}\bA)\bbf_t(u)^{\T}\bbf_t(u){\rm d}u\Big\}^{1/2}=\Vert\bA\Vert\cdot\Vert\bbf\Vert.$$
		
		\noindent
		(ii) By the Cauchy--Schwarz inequality,
		$$
		\begin{aligned}
			\Vert K\Vert_{\cS}=&\Big[\int\int\{\bbf(u)^{\T}\bg(v)\}^2\du\dv\Big]^{1/2}=\Big[\int\int\big\{\sum_{j=1}^{r}f_{j}(u)g_{j}(v)\big\}^2\du\dv\Big]^{1/2}\\
			\le&\Big\{\int\int\sum_{j=1}^{r}f_j(u)^2\sum_{j=1}^{r}g_j(v)^2\du\dv\Big\}^{1/2}=\Big\{\int\bbf(u)^{\T}\bbf(u)\du\int\bg(v)^{\T}\bg(v)\dv\Big\}^{1/2}\\
			=&\Vert\bbf\Vert\cdot\Vert\bg\Vert.
		\end{aligned}
		$$

\subsection{Proof of Lemma~\ref{lem.eigen_QQ}}
\noindent
            Let $\widecheck{\bq}_j(\cdot)$ for $j\in[r]$ be the columns of $\bQ(\cdot)$. By Assumption~\ref{ass.ind2}, we know that $\langle\widecheck{\bq}_i,\widecheck{\bq}_j\rangle=p\vartheta_i I(i=j)$, which implies that $\widetilde{\bq}_i(\cdot)=(p\vartheta_j)^{-1/2}\widecheck{\bq}_j(\cdot)$. Then, for $j\in[r],$
            $$
            \int\{\bQ(u)\bQ(v)^{\T}\}\widetilde{\bq}_j(v)\dv=\sum_{i=1}^{r}\widecheck{\bq}_i(u)\langle\widecheck{\bq}_i,\widetilde{\bq}_j\rangle=p\vartheta_j\widetilde{\bq}_j(u),
            $$
            which indicates that $p\vartheta_j$ is the eigenvalue of $\bQ(\cdot)\bQ(\cdot)^{\T}$ with the corresponding eigenfunction $\widetilde{\bq}_j(\cdot).$ Since $\langle\widetilde{\bq}_i,\widetilde{\bq}_j\rangle=I(i=j)$ for $i,j\in[r]$, we can expand $\{\widetilde{\bq}_j(\cdot)\}_{j\in[r]}$ into a set of orthonormal basis functions in $\eH^p$, denoted as $\{\widetilde{\bq}_j(\cdot)\}_{j=1}^{\infty}$. Considering that $\langle\widetilde{\bq}_j,\widetilde{\bq}_k\rangle=0$ for any $j\le r$ and $k\ge r+1$, we obtain that $\int\{\bQ(u)\bQ(v)^{\T}\}\widetilde{\bq}_k(v)\dv=0$ for $k\ge r+1$. Thus, the rest eigenvalues of $\bQ(\cdot)\bQ(\cdot)^{\T}$ are zero.

\subsection{Proof of Lemma~\ref{lem.rate_sigma_f}}
\noindent
		(i)\&(ii) See Theorem 2 and equations (2.15) in \textcolor{blue}{Fang et al.} (\textcolor{blue}{2022}) for the corresponding proofs.\\
		(iii) The autocovariance matrix functions of $\{\by_t(\cdot)\}_{t\in\mathbb{Z}}$ at lag $h$ satisfy $\bSigma_{y}^{(h)}(\cdot,\cdot)=\bB\bSigma_{f}^{(h)}(\cdot,\cdot)\bB^{\T}+\bSigma_{\varepsilon}^{(h)}(\cdot,\cdot)$, and the corresponding spectral density matrix function at frequency $\theta \in [-\pi,\pi]$ is given by
		$$
		\begin{aligned}
			\bbf_{y,\theta}=&\frac{1}{2\pi}\sum_{h\in\mathbb{Z}}\bSigma_{y}^{(h)}\exp(-ih\theta)=\frac{1}{2\pi}\sum_{h\in\mathbb{Z}}\bB\bSigma_{f}^{(h)}\bB^{\T}\exp(-ih\theta)+\frac{1}{2\pi}\sum_{h\in\mathbb{Z}}\bSigma_{\varepsilon}^{(h)}\exp(-ih\theta)
			=\bB \bbf_{f,\theta}\bB^{\T}+\bbf_{\varepsilon,\theta}.
		\end{aligned}
		$$
		By definition, the functional stability measure of $\{\by_t(\cdot)\}_{t\in\mathbb{Z}}$ is
		$$
		\begin{aligned}
			\cM_{y}=&2\pi\cdot \esssup_{\theta\in[-\pi,\pi],\bphi\in\eH_0^p}\frac{\langle\bphi,\bbf_{y,\theta}(\bphi)\rangle}{\langle\bphi,\bSigma_{y}(\bphi)\rangle}\\
			=&2\pi\cdot\esssup_{\theta\in[-\pi,\pi],\bphi\in\eH_0^p}\frac{\int\int\bphi(u)^{\T}\bbf_{y,\theta}(u,v)\bphi(v){\rm d}u{\rm d}v}{\int\int\bphi(u)^{\T}\bSigma_{y}(u,v)\bphi(v){\rm d}u{\rm d}v}\\
			\le&2\pi\cdot\esssup_{\theta\in[-\pi,\pi],\bphi\in\eH_0^p}\frac{\int\int\bphi(u)^{\T}\bB \bbf_{f,\theta}(u,v)\bB^{\T}\bphi(v){\rm d}u{\rm d}v}{\int\int\bphi(u)^{\T}\bB \bSigma_{f}(u,v)\bB^{\T}\bphi(v){\rm d}u{\rm d}v}
			+		    2\pi\cdot\esssup_{\theta\in[-\pi,\pi],\bphi\in\eH_0^p}\frac{\int\int\bphi(u)^{\T}\bbf_{\varepsilon,\theta}(u,v)\bphi(v){\rm d}u{\rm d}v}{\int\int\bphi(u)^{\T}\bSigma_{\varepsilon}(u,v)\bphi(v){\rm d}u{\rm d}v}\\
			\le&2\pi\cdot\esssup_{\theta\in[-\pi,\pi],\bxi\in\eH_0^r}\frac{\langle\bxi,\bbf_{f,\theta}(\bxi)\rangle}{\langle\bxi,\bSigma_{f}(\bxi)\rangle}+2\pi\cdot\esssup_{\theta\in[-\pi,\pi],\bphi\in\eH_0^p}\frac{\langle\bphi,\bbf_{\varepsilon,\theta}(\bphi)\rangle}{\langle\bphi,\bSigma_{\varepsilon}(\bphi)\rangle}\\
			\asymp&\cM_{f}+\cM_{\varepsilon}\asymp\cM_{\varepsilon}.
		\end{aligned}
		$$
		The other conditions imposed by \textcolor{blue}{Fang et al.} (\textcolor{blue}{2022}) for $\{\by_t(\cdot)\}_{t\in\mathbb{Z}}$ can be easily verified. Then the desired results can be obtained immediately by combining the above results.

\subsection{Proof of Lemma~\ref{lem.inv_norm}}
\noindent
(i) Notice that $\langle\bx,\bK(\bx)\rangle=\int\int\bx(u)^{\T}\bK(u,v)\bx(v)\du\dv$. Then for any $\bx\in\eH^p$ with $\Vert\bx\Vert=1$, we have
		$$
		\left|\int\int\bx(u)^{\T}\{\bK(u,v)-\bG(u,v)\}\bx(v)\du\dv\right|\le\Vert\bx\Vert^2\cdot\Vert\bK-\bG\Vert_{\cL}=o_p(c_n^{-1})
		$$
		Thus, for a sufficiently large $n$, $\langle\bx,\bK(\bx)\rangle\ge\langle\bx,\bG(\bx)\rangle-c_n^{-1}/2>c_n^{-1}/2$ for any $\bx\in\tKer(\bG)^{\perp}$ and $\Vert\bx\Vert=1$ with probability approaching one, since $\langle\bx,\bG(\bx)\rangle\ge\big(\Vert\bG^{\dagger}\Vert_{\cL}\big)^{-1}>c_n^{-1}$ by using Lemma \ref{lem.eigen}. Hence, by the definition of $\bK^{\dagger},$ it follows that $\Vert\bK^{\dagger}\Vert_{\cL}<2c_n$ with probability approaching one. \\
(ii) Notice that
		$$
		\begin{aligned}
			\Vert\bK^{\dagger}-\bG^{\dagger}\Vert_{\cL}=&\max_{\bx\in\tIm(\bG),\Vert\bx\Vert=1}\langle\bx,\bK^{\dagger}(\bx)-\bG^{\dagger}(\bx)\rangle=\max_{\bx\in\tIm(\bG),\Vert\bx\Vert=1}\langle\bx,\bK^{\dagger}(\bG-\bK)\bG^{\dagger}(\bx)\rangle\\
            \le&\Vert\bG^{\dagger}\Vert_{\cL}\cdot\max_{\by\in\tKer(\bG)^{\perp},\Vert\by\Vert=1}\langle\by,\bK^{\dagger}(\bG-\bK)(\by)\rangle
            \le\Vert\bK-\bG\Vert_{\cL}\Vert\bG^{\dagger}\Vert_{\cL}\cdot\max_{\bz\in\tIm(\bG),\Vert\bz\Vert=1}\langle\bz,\bK^{\dagger}(\bz)\rangle\\
            \le&\Vert\bK^{\dagger}\Vert_{\cL}\Vert\bK-\bG\Vert_{\cL}\Vert\bG^{\dagger}\Vert_{\cL}
			=O_p(c_n^2)\Vert\bK-\bG\Vert_{\cL},
		\end{aligned}
		$$
	where the last equality follows from part (i).

\subsection{Proof of Lemma~\ref{lem.rate_max_f}}
\noindent
        The proofs of this lemma utilize the tail probabilities of sub-Gaussian process as presented in Section~\ref{supsybsec.subgauss}. \\
        (i) 
        Notice that $\{\bbf_t(\cdot)\}$ and $\{\bvarepsilon_t(\cdot)\}$ follow the sub-Gaussian functional linear process (see Section~\ref{supsybsec.subgauss}) by Assumption~\ref{ass.concentration}, with $\eE\Vert\bbf_t\Vert^2=O(1)$ and $\eE\Vert p^{-1/2}\bB^{\T}\bvarepsilon_t\Vert^2=O(1)$ by Assumption~\ref{ass.regu_cond_DIGIT}(ii). By Bonferroni's method, it yields that, for each $j\in[r]$ and any given $\eta>0$,
        \begin{equation}
            \label{eq.lemma_B16}
            \eP\left\{\max_{t\in[n]}\big(\Vert f_{tj}\Vert^2-\eE\Vert f_{tj}\Vert^2\big)\ge \eta\right\}\le\sum_{t=1}^{n}\eP\left(\Vert f_{tj}\Vert^2-\eE\Vert f_{tj}\Vert^2\ge \eta\right)
        \le2n\exp\{-c\min(\eta^2,\eta)\},
        \end{equation}
        where $c>0$ is some constant and the second inequality follows from Lemma 5 of \textcolor{blue}{Fang et al.} (\textcolor{blue}{2022}). Letting $\eta=\log n$, \eqref{eq.lemma_B16} shows that $\max_{t\in[n]}\Vert\bbf_t\Vert^2=O_p(\log n)+\eE\Vert\bbf_t\Vert^2$, which implies that $\max_{t\in[n]}\Vert\bbf_t\Vert=O_p(\sqrt{\log n})$. For the second argument, to satisfy Condition 10 in \textcolor{blue}{Fang et al.} (\textcolor{blue}{2022}) and apply their Lemma 5, alternatively we consider that, for each $j\in[r]$ and any given $\eta>0,$
        $$
        \eP\left\{\max_{t\in[n]}\big(p^{-1}\Vert \bbb_j^{\T}\bvarepsilon_t\Vert^2-p^{-1}\eE\Vert\bbb_j^{\T}\bvarepsilon_t\Vert^2\big)\ge \eta\right\}\le2n\exp\{-c\min(\eta^2,\eta)\},
        $$
        which implies that $\max_{t\in[n]}\Vert p^{-1/2}\bB^{\T}\bvarepsilon_t\Vert=O_p(\sqrt{\log n}).$ Given that $\eE\Vert p^{-1/2}\bvarepsilon_t\Vert^2=O(1)$ by Assumption~\ref{ass.regu_cond_DIGIT}(iv), we can also show that $\max_{t\in[n]}\Vert p^{-1/2}\bvarepsilon_t\Vert=O_p(\sqrt{\log n})$ similarly.\\
        (ii) The three arguments $\max_{t\in[n]}\Vert\bgamma_t\Vert=O_p(\sqrt{\log n}),\max_{t\in[n]}\Vert p^{-1/2}\int\bQ(u)^{\T}\bvarepsilon_t(u)\du\Vert=O_p(\sqrt{\log n})$ and $\max_{t'\in[n]}|p^{-1/2}\{\langle\bvarepsilon_t,\bvarepsilon_{t'}\rangle-\eE\langle\bvarepsilon_t,\bvarepsilon_{t'}\rangle\}|^2=O_p(\log n)$ for $t\in[n]$ can be proved similarly to part (i), since $\{\bgamma_t\}$ and $\{\varepsilon_t(\cdot)\}$ are sub-Gaussian (functional) linear processes by Assumption~\ref{ass.concentration_2}, with moment conditions $\eE\Vert\bgamma_t\Vert^2=O(1),\eE\Vert p^{-1/2}\int\bQ(u)^{\T}\bvarepsilon_t(u)\du\Vert^2=O(1)$ and $\eE|p^{-1/2}\{\langle\bvarepsilon_t,\bvarepsilon_{t'}\rangle-\eE\langle\bvarepsilon_t,\bvarepsilon_{t'}\rangle\}|^4=O(1)$ by Assumption~\ref{ass.regu_cond_fpoet}(ii).

\subsection{Proof of Lemma~\ref{lem.sigma_f_bound}}
\noindent
		(i) In Assumption~\ref{ass.ind1}, we assume that
		$$\int\int\bSigma_{f}(u,v)\bSigma_{f}(u,v)^{\T}{\rm d}u{\rm d}v={\rm diag}(\theta_1,\dots,\theta_r),$$
		i.e.,
		$$\int\int\sum_{j=1}^{r}\Sigma_{f,ij}(u,v)^2{\rm d}u{\rm d}v=\theta_i,\ {\rm for\ }i\in[r].$$
		Then we have
		$$
		\begin{aligned}
			\Vert\bSigma_{f}\Vert_{\cS,\max}^2=\max_{i\in[r],j\in[r]}\Vert\Sigma_{f,ij}\Vert_{\cS}^2=\max_{i\in[r],j\in[r]}\int\int\Sigma_{f,ij}(u,v)^2{\rm d}u{\rm d}v
			\le\max_{i\in[r]}\int\int\sum_{j=1}^{r}\Sigma_{f,ij}(u,v)^2{\rm d}u{\rm d}v=\max_{i\in[r]}\theta_i=\theta_1,
		\end{aligned}
		$$
		which implies that $\Vert\bSigma_{f}\Vert_{\cS,\max}\le\theta_1^{1/2}\equiv C_{\max}$. \\
		(ii) Note that $\bSigma_{f}(u,v)\in\mathbb{R}^{r\times r}$, we have $\max(\Vert\bSigma_{f}\Vert_{\cS,\infty},\Vert\bSigma_{f}\Vert_{\cS,1},\Vert\bSigma_{f}\Vert_{\cS,\F})\le r\Vert\bSigma_{f}\Vert_{\cS,\max}\le r\theta_1^{1/2}\equiv C_{\infty}$.

\subsection{Proof of Lemma~\ref{lem.sigma_y_order}}
\noindent
		(i) By Lemma \ref{lem.ineq3}(i) and the fact $\Vert\bSigma_{\varepsilon}\Vert_{\cS,\max}\le \Vert\bSigma_{\varepsilon}\Vert_{\cL}$, we have
		$$
		\begin{aligned}
			\Vert\bSigma_{y}\Vert_{\cS,\max}=&\Vert\bB\bSigma_{f}\bB^{\T}+\bSigma_{\varepsilon}\Vert_{\cS,\max}\le \Vert\bB\bSigma_{f}\bB^{\T}\Vert_{\cS,\max} + \Vert\bSigma_{\varepsilon}\Vert_{\cS,\max}\\
			\le&\Vert\bB\Vert_{\infty}\Vert\bSigma_{f}\Vert_{\cS,\max}\Vert\bB^{\T}\Vert_{1}+\Vert\bSigma_{\varepsilon}\Vert_{\cS,\max}\lesssim r^2C^2C_{\max}+O(1)\asymp1. 
		\end{aligned}
		$$
		(ii) By Lemma \ref{lem.ineq3}(iii), we have
		$$
		\begin{aligned}
			\Vert\bSigma_{y}\Vert_{\cS,\infty}=&\Vert\bB\bSigma_{f}\bB^{\T}+\bSigma_{\varepsilon}\Vert_{\cS,\infty}\le \Vert\bB\bSigma_{f}\bB^{\T}\Vert_{\cS,\infty} + \Vert\bSigma_{\varepsilon}\Vert_{\cS,\infty}\\
			\le& \Vert\bB\Vert_{\infty}\Vert\bB^{\T}\Vert_{\infty}\Vert\bSigma_{f}\Vert_{\cS,\infty}+\Vert\bSigma_{\varepsilon}\Vert_{\cS,\infty}\le rpC^2C_{\infty}+s_p\lesssim p.
		\end{aligned}
		$$
		Part (iii) can be proved similarly.

\subsection{Proof of Lemma~\ref{lem.order}}
\noindent
            For the first part of the lemma, notice that $\Vert\bOmega_{\cL}\Vert\le\Vert\bOmega_{\cL}\Vert_{\F}\le\sqrt{r}\Vert\bOmega_{\cL}\Vert$ where $r$ is the rank of $\bOmega_{\cL}$, so $\Vert\bOmega_{\cL}\Vert\asymp\Vert\bOmega_{\cL}\Vert_{\F}$, and 
	$$\begin{aligned}
		\Vert\bOmega_{\cL}\Vert_{\F}^2=\left\Vert p\bB\int\int\bSigma_{f}(u,v)\bSigma_{f}(u,v)^{\T}{\rm d}u{\rm d}v\bB^{\T}\right\Vert_{\F}^2
		=p^4{\rm tr}\left(\diag\{\theta_1,\dots,\theta_r\}\diag\{\theta_1,\dots,\theta_r\}^{\T}\right)=p^4\sum_{j=1}^{r}\theta_i^2\asymp p^4,
	\end{aligned}$$
	where the second equality follows from Assumption~\ref{ass.ind1} that $\int\int\bSigma_{f}(u,v)\bSigma_{f}(u,v)^{\T}{\rm d}u{\rm d}v=\textup{diag}\{\theta_1, \dots, \theta_r\}$ and $\bB^{\T}\bB=p\bI_{r}$. Thus we have $\Vert\bOmega_{\cL}\Vert\asymp p^2$. 
	For the second part, we have
	$$
	\begin{aligned}
		\Vert\bOmega_{\cR}\Vert\le&\left\Vert\int\int\bSigma_{\varepsilon}(u,v)\bSigma_{\varepsilon}(u,v)^{\T}{\rm d}u{\rm d}v\right\Vert\\
		&+\left\Vert\int\int\bB\bSigma_{f}(u,v)\bB^{\T}\bSigma_{\varepsilon}(u,v)^{\T}{\rm d}u{\rm d}v\right\Vert + 
		\left\Vert\int\int\bSigma_{\varepsilon}(u,v)\bB\bSigma_{f}(u,v)^{\T}\bB^{\T}{\rm d}u{\rm d}v\right\Vert\\
		\le&\Vert\bSigma_{\varepsilon}\Vert_{\cL}^2+2\Vert\bB\bSigma_{f}\bB^{\T}\Vert_{\cL}\Vert\bSigma_{\varepsilon}\Vert_{\cL}
		=O(p),
	\end{aligned}
	$$
	where the second inequality follows from Lemmas \ref{lem.ineq1}(ii) and \ref{lem.ineq2}(iv), and the last line follows from Assumption~\ref{ass.regu_cond_DIGIT}(iii) and Lemmas \ref{lem.op_le_S1}(i)(ii) and \ref{lem.sigma_f_bound}(ii).

\subsection{Proof of Lemma~\ref{lem.rate_R_inf}}
\noindent
		Notice that
		$$
		\begin{aligned}
			\Vert\bOmega_{\cR}\Vert_{\infty}\le&\Big\Vert\int\int\bSigma_{\varepsilon}(u,v)\bSigma_{\varepsilon}(u,v)^{\T}{\rm d}u{\rm d}v\Big\Vert_{\infty}\\
			&+\Big\Vert\int\int\bB\bSigma_{f}(u,v)\bB^{\T}\bSigma_{\varepsilon}(u,v)^{\T}{\rm d}u{\rm d}v\Big\Vert_{\infty} + 
			\Big\Vert\int\int\bSigma_{\varepsilon}(u,v)\bB\bSigma_{f}(u,v)^{\T}\bB^{\T}{\rm d}u{\rm d}v\Big\Vert_{\infty}\\
			\le&\Vert\bSigma_{\varepsilon}\Vert_{\cS,\infty}\Vert\bSigma_{\varepsilon}\Vert_{\cS,1}+2\Vert\bB\bSigma_{f}\bB^{\T}\Vert_{\cS,\infty}\Vert\bSigma_{\varepsilon}\Vert_{\cS,1}\\
			\le&\Vert\bSigma_{\varepsilon}\Vert_{\cS,\infty}\Vert\bSigma_{\varepsilon}\Vert_{\cS,1}+2\Vert\bSigma_{\varepsilon}\Vert_{\cS,1}\Vert\bB\Vert_{\infty}\Vert\bB^{\T}\Vert_{\infty}\Vert\bSigma_{f}\Vert_{\cS,\infty}\\
			\le&s_p^2+2rC^2C_{\infty}s_pp\lesssim s_p^2+ps_p\lesssim ps_p=o(p^2),
		\end{aligned}
		$$
		where the second inequality follows from Lemma \ref{lem.ineq2}(ii), the third inequality follows from Lemma \ref{lem.ineq3}(iii), and the fourth inequality follows from Lemma \ref{lem.sigma_f_bound}.

\subsection{Proof of Lemma~\ref{lem.rate_B_F}}
\noindent
            By Proposition \ref{propos.eigenvalues}(ii) and Lemma \ref{lem.order}, if $\bxi_j^{\T}\widetilde{\bbb}_j\ge0$, then
	\begin{equation}
		\label{eq.conv_b}
		\Vert\bxi_j-\widetilde{\bbb}_j\Vert=O_p(p^{-2}\Vert\bOmega_{\cR}\Vert)=O_p(p^{-1}),\ {\rm for\ }j\in[r].
	\end{equation}
	Applying Lemma~\ref{lem.sintheta} yields that, if $\widehat{\bxi}_j^{\T}\bxi_j\ge0$, we have
	\begin{equation}
		\label{eq.sin_xi}
		\Vert\widehat{\bxi}_j-\bxi_j\Vert\le\frac{\Vert\widehat{\bOmega}-\bOmega\Vert/\sqrt{2}}{\min\big(\big|\hat{\lambda}_{j-1}-\lambda_j\big|,\big|\lambda_j-\hat{\lambda}_{j+1}\big|\big)},
	\end{equation}
	where $\{\hat{\lambda}_j\}_{j=1}^{p}$ are the eigenvalues of $\widehat{\bOmega}$ in a descending order, and $\{\widehat{\bxi}_j\}_{j=1}^{p}$ are their corresponding eigenvectors. Then, for $j\in[r]$, we have
	$|\hat{\lambda}_{j-1}-\lambda_j|\ge|\hat{\lambda}_{j-1}-\hat{\lambda}_j|-|\lambda_j-\hat{\lambda}_j|$, where the first term $|\hat{\lambda}_{j-1}-\hat{\lambda}_j|\gtrsim p^2$ with probability approaching one by Lemma \ref{lem.rth_eigen}, and the second term $|\lambda_j-\hat{\lambda}_j|=o_p(p^2)$ by Lemmas~\ref{lem.weyl} and \ref{lem.appro_int}(i). Hence, $|\hat{\lambda}_{j-1}-\lambda_j|\gtrsim p^2$ with probability approaching one for all $j\in[r]$. We can also show the similar result for $|\lambda_j-\hat{\lambda}_{j+1}|$ if $j\in[r-1]$. If $j=r,|\lambda_r-\hat{\lambda}_{r+1}|>\lambda_r-\hat{\lambda}_{r+1}=p^2\theta_r-\hat{\lambda}_{r+1}\gtrsim p^2$ since $\hat{\lambda}_{r+1}=o_p(p^2)$, which can be implied by Proposition~\ref{propos.eigenvalues} and Lemma~\ref{lem.order} that $\lambda_{r+1}=o(p^2)$, and Lemmas~\ref{lem.weyl} and \ref{lem.appro_int}(i) that $|\hat{\lambda}_{r+1}-\lambda_{r+1}|=o_p(p^2)$. Thus, $$\min\big(|\hat{\lambda}_{j-1}-\lambda_j|,|\lambda_j-\hat{\lambda}_{j+1}|\big)\gtrsim p^2.$$ 
	Applying \eqref{eq.sin_xi}, Lemma \ref{lem.appro_int}(i) and the above argument, we have, if $\widehat{\bxi}_j^{\T}\bxi_j\ge0$, then
	$$
	\Vert\widehat{\bxi}_j-\bxi_j\Vert=O_p\big(\cM_{\varepsilon}\sqrt{1/n}\big),\ {\rm for\ }j\in[r].
	$$
	Combining with \eqref{eq.conv_b} we have, if $\widehat{\bxi}_j^{\T}\widetilde{\bbb}_j\ge0$, then
	$$
	\Vert\widehat{\bxi}_j-\widetilde{\bbb}_j\Vert=O_p\big(\cM_{\varepsilon}\sqrt{1/n}+1/p\big),\ {\rm for\ }j\in[r].
	$$
	Since $\widehat{\bbb}_j=\sqrt{p}\widehat{\bxi}_j$ and $\bbb_j=\sqrt{p}\widetilde{\bbb}_j$, one can obtain that there exists an orthogonal matrix $\bU\in\mathbb{R}^{r\times r}$ such that
	$$
		\Vert\widehat{\bB}-\bB\bU^{\T}\Vert_{\F}=O_p\big(\cM_{\varepsilon}\sqrt{p/n}+1/\sqrt{p}\big),
	$$
        where the matrix $\bU$ is used to adjust the direction so that each $\bbb_j^{\T}\widehat{\bbb}_j\ge0$ for $j\in[r].$

\subsection{Proof of Lemma~\ref{lem.Theta_bound}}
\noindent
		We first prove the upper bound. By the definition of $\widehat{\Theta}_{ij}$, we have
		$$
		\begin{aligned}
			\widehat{\Theta}_{ij}(u,v)=&\frac{1}{n}\sum_{t=1}^{n}\Big\{\widehat{\varepsilon}_{ti}(u)\widehat{\varepsilon}_{tj}(v)-\frac{1}{n}\sum_{s=1}^{n}\widehat{\varepsilon}_{si}(u)\widehat{\varepsilon}_{sj}(v)\Big\}^2\\
			\le&\frac{2}{n}\sum_{t=1}^{n}\Big\{\widehat{\varepsilon}_{ti}(u)\widehat{\varepsilon}_{tj}(v)-\Sigma_{\varepsilon,ij}(u,v)\Big\}^2+2\max_{i\in[p],j\in[p]}\Big\{\Sigma_{\varepsilon,ij}(u,v)-\frac{1}{n}\sum_{s=1}^{n}\widehat{\varepsilon}_{si}(u)\widehat{\varepsilon}_{sj}(v)\Big\}^2,
		\end{aligned}
		$$
		which implies that 
		$$
		\begin{aligned}
			\Vert\widehat{\Theta}_{ij}^{1/2}\Vert_{\cS}^2=&\int\int\widehat{\Theta}_{ij}(u,v)\du\dv\le\frac{2}{n}\int\int\sum_{t=1}^{n}\left\{\widehat{\varepsilon}_{ti}(u)\widehat{\varepsilon}_{tj}(v)-\Sigma_{\varepsilon,ij}(u,v)\right\}^2\du\dv+2\Vert\widehat{\bSigma}_{\varepsilon}-\bSigma_{\varepsilon}\Vert_{\cS,\max}^2\\
			=&\frac{2}{n}\int\int\sum_{t=1}^{n}\Big\{\widehat{\varepsilon}_{ti}(u)\widehat{\varepsilon}_{tj}(v)-\Sigma_{\varepsilon,ij}(u,v)\Big\}^2\du\dv+o_p(1),
		\end{aligned}
		$$
		where the last line follows from Lemma \ref{lem.varep_bound_1}. 
	Moreover $$
		\begin{aligned}
			&\sum_{t=1}^{n}\Big\{\widehat{\varepsilon}_{ti}(u)\widehat{\varepsilon}_{tj}(v)-\Sigma_{\varepsilon,ij}(u,v)\Big\}^2\\
   =&\sum_{t=1}^{n}\Big[\big\{\widehat{\varepsilon}_{ti}(u)-\varepsilon_{ti}(u)\big\}\widehat{\varepsilon}_{tj}(v)+\varepsilon_{ti}(u)\big\{\widehat{\varepsilon}_{tj}(v)-\varepsilon_{tj}(v)\big\}
			+\varepsilon_{ti}(u)\varepsilon_{tj}(v)-\Sigma_{\varepsilon,ij}(u,v)\Big]^2\\
			\le&4\sum_{t=1}^{n}\big\{\widehat{\varepsilon}_{ti}(u)-\varepsilon_{ti}(u)\big\}^2\widehat{\varepsilon}_{tj}(v)^2+4\sum_{t=1}^{n}\varepsilon_{ti}(u)^2\big\{\widehat{\varepsilon}_{tj}(v)-\varepsilon_{tj}(v)\big\}^2
			+2\sum_{t=1}^{n}\big\{\varepsilon_{ti}(u)\varepsilon_{tj}(v)-\Sigma_{\varepsilon,ij}(u,v)\big\}^2\\
			\le&4\max_{i\in[p],t\in[n]}\big\{\widehat{\varepsilon}_{ti}(u)-\varepsilon_{ti}(u)\big\}^2\max_{j\in[p]}\Big[\sum_{t=1}^{n}2\big\{\widehat{\varepsilon}_{tj}(v)-\varepsilon_{tj}(v)\big\}^2+3\varepsilon_{tj}(v)^2\Big]
			+2\sum_{t=1}^{n}\big\{\varepsilon_{ti}(u)\varepsilon_{tj}(v)-\Sigma_{\varepsilon,ij}(u,v)\big\}^2.
		\end{aligned}
		$$
		Here, we bound each term above as follows: (a) by Corollary \ref{coro.DIGIT}, we have $\max_{i\in[p],t\in[n]}\Vert\widehat{\varepsilon}_{ti}-\varepsilon_{ti}\Vert^2=O_p(\varrho^2)=o_p(1)$ under Assumption \ref{ass.n.p}; (b) by Lemma \ref{lem.varep_bound_1}(i), we have $\max_{j\in[p]}n^{-1}\sum_{t=1}^{n}\Vert\widehat{\varepsilon}_{tj}-\varepsilon_{tj}\Vert^2=O_p(\varpi_{n,p}^2)=o_p(1)$; (c) by Lemma \ref{lem.rate_sigma_f}(ii) and Assumption \ref{ass.regu_cond_DIGIT}(iv), we have $\max_{j\in[p]}n^{-1}\sum_{t=1}^{n}\Vert\varepsilon_{tj}\Vert^2\le o_p(1)+\max_{j\in[p]}\int\Sigma_{\varepsilon,jj}(u,u)\du=O_p(1)$. Combining these results yields that$$\Vert\widehat{\Theta}_{ij}^{1/2}\Vert_{\cS}^2\le\frac{2}{n}\int\int\sum_{t=1}^{n}\big\{\varepsilon_{ti}(u)\varepsilon_{tj}(v)-\Sigma_{\varepsilon,ij}(u,v)\big\}^2\du\dv+o_p(1).$$
		Similar arguments as those in the proof of Lemma 2 of \textcolor{blue}{Cai and Liu} (\textcolor{blue}{2011}) results in
		$$
		\max_{i\in[p],j\in[p]}\Big\Vert\frac{1}{n}\sum_{t=1}^{n}(\varepsilon_{ti}\varepsilon_{tj}-\Sigma_{\varepsilon,ij})^2-{\rm Var}(\varepsilon_{ti}\varepsilon_{tj})\Big\Vert_{\cS}=o_p(1).
		$$
		Combining with Assumption \ref{ass.thres} implies that $\max_{i\in[p],j\in[p]}\Vert n^{-1}\sum_{t=1}^{n}(\varepsilon_{ti}\varepsilon_{tj}-\Sigma_{\varepsilon,ij})^2\Vert_{\cS}$ is bounded away from both zero and infinity with probability approaching one. Therefore, $\max_{i,j\in[p]}\Vert\widehat{\Theta}_{ij}^{1/2}\Vert_{\cS}$ is bounded away from infinity with probability approaching one.
		
	We next prove the lower bound. Notice that 
		$$
		\begin{aligned}
			\frac{1}{n}\sum_{t=1}^{n}\Big\{\varepsilon_{ti}(u)\varepsilon_{tj}(v)-\Sigma_{\varepsilon,ij}(u,v)\Big\}^2
            \le&4\sum_{t=1}^{n}\Big\{\varepsilon_{ti}(u)\varepsilon_{tj}(v)-\widehat{\varepsilon}_{ti}(u)\widehat{\varepsilon}_{tj}(v)\Big\}^2
			+4\sum_{t=1}^{n}\Big\{\widehat{\varepsilon}_{ti}(u)\widehat{\varepsilon}_{tj}(v)-\frac{1}{n}\sum_{s=1}^{n}\widehat{\varepsilon}_{si}(u)\widehat{\varepsilon}_{sj}(v)\Big\}^2\\
			&+2\sum_{t=1}^{n}\Big\{\frac{1}{n}\sum_{s=1}^{n}\widehat{\varepsilon}_{si}(u)\widehat{\varepsilon}_{sj}(v)-\Sigma_{\varepsilon,ij}(u,v)\Big\}^2,
		\end{aligned}
		$$
		which implies that 
		$$
		\begin{aligned}
			\frac{1}{n}\int\int\sum_{t=1}^{n}\Big\{\varepsilon_{ti}(u)\varepsilon_{tj}(v)-\Sigma_{\varepsilon,ij}(u,v)\Big\}^2\du\dv
            \le\frac{4}{n}\int\int\sum_{t=1}^{n}[\varepsilon_{ti}(u)\varepsilon_{tj}(v)-\widehat{\varepsilon}_{ti}(u)\widehat{\varepsilon}_{tj}(v)]^2\du\dv
			+4\Vert\widehat{\Theta}_{ij}^{1/2}\Vert_{\cS}^2+o_p(1),
		\end{aligned}
		$$
		where the LHS is bounded away from both zero and infinity uniformly in $i,j$. Then, 
		$$
		\begin{aligned}
			\sum_{t=1}^{n}\Big\{\varepsilon_{ti}(u)\varepsilon_{tj}(v)-\widehat{\varepsilon}_{ti}(u)\widehat{\varepsilon}_{tj}(v)\Big\}^2
   \le&2\sum_{t=1}^{n}\varepsilon_{ti}(u)^2\big\{\varepsilon_{tj}(v)-\widehat{\varepsilon}_{tj}(v)\big\}^2+2\sum_{t=1}^{n}\widehat{\varepsilon}_{tj}(v)^2\big\{\varepsilon_{ti}(v)-\widehat{\varepsilon}_{ti}(u)\big\}^2\\
			\le&4\max_{i\in[p],t\in[n]}\big\{\widehat{\varepsilon}_{ti}(u)-\varepsilon_{ti}(u)\big\}^2\max_{j\in[p]}\sum_{t=1}^{n}\Big[\big\{[\widehat{\varepsilon}_{tj}(v)-\varepsilon_{tj}(v)\big\}^2+\varepsilon_{tj}(v)^2\Big].
		\end{aligned}
		$$
		As demonstrated in the proof of the upper bound above, we have
		$$
		\frac{1}{n}\int\int\sum_{t=1}^{n}\big\{\varepsilon_{ti}(u)\varepsilon_{tj}(v)-\widehat{\varepsilon}_{ti}(u)\widehat{\varepsilon}_{tj}(v)\big\}^2\du\dv=o_p(1).
		$$
		Hence, $\min_{i\in[p],j\in[p]}\Vert\widehat{\Theta}_{ij}^{1/2}\Vert_{\cS}$ is bounded away from zero with probability approaching one.

\section{Proofs of theoretical results in Section~\ref{sec.theory}}
\label{supsec.B}

\subsection{Technical lemmas and their proofs}
	\begin{lemma}
		\label{lem.rate_sigma_f_2}
		Under Assumptions \ref{ass.regu_cond_fpoet}(iv) and \ref{ass.concentration_2}, we have that,\\
		(i) for any $i,j\in[r],|n^{-1}\sum_{t=1}^{n}\gamma_{ti}\gamma_{tj}-\Sigma_{\gamma,ij}|=O_p(1/\sqrt{n})$, $\Vert n^{-1}\sum_{t=1}^{n}\bgamma_t\bgamma_t^{\T}-\bSigma_{\gamma}\Vert_{\max}=O_p(1/\sqrt{n})$;\\
		(ii) for any $i,j\in[p],\Vert n^{-1}\sum_{t=1}^{n}\varepsilon_{ti}\varepsilon_{tj}-\Sigma_{\varepsilon,ij}\Vert_{\cS}=O_p(\cM_{\varepsilon}/\sqrt{n})$, $\Vert n^{-1}\sum_{t=1}^{n}\bvarepsilon_t\bvarepsilon_t^{\T}-\bSigma_{\varepsilon}\Vert_{\cS,\max}=O_p(\cM_{\varepsilon}\sqrt{\log p/n})$;\\
		(iii) for any $i\in[p],j\in[r],\Vert n^{-1}\sum_{t=1}^{n}\varepsilon_{ti}\gamma_{tj}\Vert=O_p(\cM_{\varepsilon}/\sqrt{n})$,  $\max_{i\in[p],j\in[r]}\Vert n^{-1}\sum_{t=1}^{n}\varepsilon_{ti}\gamma_{tj}\Vert=O_p(\cM_{\varepsilon}\sqrt{\log p/n})$.
	\end{lemma}
	\begin{proof}
	For parts (i) and (ii), see Theorem 2 and equations (2.15) in \textcolor{blue}{Fang et al.} (\textcolor{blue}{2022}) for the corresponding proofs. For part (iii), see Remark~3 and equation (2.16) in \textcolor{blue}{Fang et al.} (\textcolor{blue}{2022}) for a proof. 
	\end{proof}

	\subsection{Proof of Theorem \ref{thm.load_factor_2}}
 To prove Theorem \ref{thm.load_factor_2}, we first present some technical lemmas with their proofs.
	\begin{lemma}
		\label{lem.ids_weak_serial}
		Under Assumption \ref{ass.concentration_2}, it holds that		$$\max_{t\in[n]}\sum_{t'=1}^{n}\frac{|\eE\langle\bvarepsilon_{t'},\bvarepsilon_t\rangle|}{p}=O(\cM_{\varepsilon}),\ {\rm\ and\ }\max_{t',t\in[n]}\frac{|\eE\langle\bvarepsilon_{t'},\bvarepsilon_t\rangle|}{p}=O(\cM_{\varepsilon}).$$ 
	\end{lemma}
	\begin{proof}
		From Assumption \ref{ass.concentration_2}, the functional stability measure of $\{\bvarepsilon_{t}(\cdot)\}_{t\in\mathbb{Z}}$ is bounded ($\cM_{\varepsilon}<\infty$), and we would like to associate it with the equation of interest in this lemma. Since $\{\bvarepsilon_{t}(\cdot)\}_{t\in\mathbb{Z}}$ is stationary, we have, uniformly in $n$,  
		$$
		\begin{aligned}
			\max_{t\in[n]}\sum_{t'=1}^{n}\frac{|\eE\langle\bvarepsilon_{t'},\bvarepsilon_t\rangle|}{p}\le&\max_{t\in[n]}\frac{1}{p}\sum_{i=1}^{p}\sum_{t'=1}^{n}|\eE\langle\bvarepsilon_{t'i},\bvarepsilon_{ti}\rangle|\le\max_{t\in[n]}\max_{i\in[p]}\sum_{t'=1}^{n}|\eE\langle\bvarepsilon_{t'i},\bvarepsilon_{ti}\rangle|\\
			\le&\max_{i\in[p]}\sum_{t'=-\infty}^{\infty}
			\left|\eE\int\varepsilon_{1i}(u)\varepsilon_{t'i}(u)\du\right|\\
			\le&\max_{i\in[p]}\sum_{t'=-\infty}^{\infty}\left\{\eE\int\varepsilon_{1i}(u)\varepsilon_{t'i}(u)\du\cdot\int\varepsilon_{1i}(v)\varepsilon_{t'i}(v)\dv\right\}^{1/2}\\
			\le&\max_{i\in[p]}\sum_{t'=-\infty}^{\infty}\eE\int\int\varepsilon_{1i}(u)\varepsilon_{t'i}(v)\du\dv\\
			=&\max_{i\in[p]}\sum_{h\in\mathbb{Z}}\int\int\bphi_i(u)^{\T}\bSigma_{\varepsilon}^{(h)}(u,v)\bphi_i(v)\du\dv\\
			=&2\pi\cdot\max_{i\in[p]}\langle\bphi_i,\bbf_{\varepsilon,\theta=0}(\bphi_i)\rangle\le2\pi\omega_0^{\varepsilon}\cdot\max_{i\in[p]}\frac{\langle\bphi_i,\bbf_{\varepsilon,\theta=0}(\bphi_i)\rangle}{\langle\bphi_i,\bSigma_{\varepsilon}(\bphi_i)\rangle}\\
			\le&2\pi\omega_0^{\varepsilon}\cdot\esssup_{\theta\in[-\pi,\pi],\bphi\in\eH_{0,\varepsilon}^p}\frac{\langle\bphi,\bbf_{\varepsilon,\theta}(\bphi)\rangle}{\langle\bphi,\bSigma_{\varepsilon}(\bphi)\rangle}=\omega_0^{\varepsilon}\cM_{\varepsilon}=O(\cM_{\varepsilon}),
		\end{aligned}
		$$ 
		where $\bphi_i(\cdot)=(0,\dots,1,\dots)^{\T}$ with its $i$-th element being 1 and the rest being 0, $\eH_{0,\varepsilon}^p=\{\bphi\in\eH^p:\langle\bphi,\bSigma_{\varepsilon}(\bphi)\rangle\in(0,\infty)\}$, $\bbf_{\varepsilon,\theta}$ is the spectral density matrix function of $\{\bvarepsilon_{t}(\cdot)\}_{t\in\mathbb{Z}}$ defined in Section \ref{subsec.ass}, and $\omega_0^{\varepsilon}=\max_{j\in[p]}\int\Sigma_{\varepsilon,jj}(u,u){\rm d}u$. Furthermore, we also obtain that
		$$\max_{t',t\in[n]}\frac{|\eE\langle\bvarepsilon_{t'},\bvarepsilon_t\rangle|}{p}\le\max_{t\in[n]}\sum_{t'=1}^{n}\frac{|\eE\langle\bvarepsilon_{t'},\bvarepsilon_t\rangle|}{p}=O(\cM_{\varepsilon}).
		$$ 
	\end{proof}

	Recall the definition of the asymptotically orthogonal matrix $\bH$ introduced in Section \ref{subsec.con_load}. Applying the equation (C.2) in \textcolor{blue}{Fan et al.} (\textcolor{blue}{2013}) or (A.1) in \textcolor{blue}{Bai} (\textcolor{blue}{2003}), we have
	\begin{equation}
		\label{eq.decom_H}
		\widehat{\bgamma}_t-\bH\bgamma_t=\Big(\frac{\bV}{p}\Big)^{-1}\Big\{\frac{1}{n}\sum_{t'=1}^{n}\widehat{\bgamma}_{t'}\frac{\eE\langle\bvarepsilon_{t'},\bvarepsilon_t\rangle}{p}+\frac{1}{n}\sum_{t'=1}^{n}\widehat{\bgamma}_{t'}\zeta_{t't}+\frac{1}{n}\sum_{t'=1}^{n}\widehat{\bgamma}_{t'}\eta_{t't}+\frac{1}{n}\sum_{t'=1}^{n}\widehat{\bgamma}_{t'}\xi_{t't}\Big\},
	\end{equation}
	where 
	$$
	\begin{aligned}
		\zeta_{t't}=\frac{1}{p}\langle\bvarepsilon_{t'},\bvarepsilon_t\rangle-\frac{1}{p}\eE\langle\bvarepsilon_{t'},\bvarepsilon_t\rangle,\quad
		\eta_{t't}=\frac{1}{p}\bgamma_{t'}^{\T}\sum_{i=1}^{p}\int\bq_i(u)\varepsilon_{ti}(u)\du,\quad\xi_{t't}=\frac{1}{p}\bgamma_t^{\T}\sum_{i=1}^{p}\int\bq_i(u)\varepsilon_{t'i}(u)\du.
	\end{aligned}
	$$ 
	
	\begin{lemma}
		\label{lem.conv_decom_H_max}
		Under the assumptions of Theorem \ref{thm.load_factor_2}, it holds that\\
		(i) $\max_{t\in[n]}\Vert (np)^{-1}\sum_{t'=1}^{n}\widehat{\bgamma}_{t'}\eE\langle\bvarepsilon_{t'},\bvarepsilon_t\rangle\Vert=O_p(\cM_{\varepsilon}/\sqrt{n})$;\\
		(ii) $\max_{t\in[n]}\Vert n^{-1}\sum_{t'=1}^{n}\widehat{\bgamma}_{t'}\zeta_{t't}\Vert=O_p(\sqrt{\log n/p})$;\\
		(iii) $\max_{t\in[n]}\Vert n^{-1}\sum_{t'=1}^{n}\widehat{\bgamma}_{t'}\eta_{t't}\Vert=O_p(\sqrt{\log n/p})$; \\
		(iv) $\max_{t\in[n]}\Vert n^{-1}\sum_{t'=1}^{n}\widehat{\bgamma}_{t'}\xi_{t't}\Vert=O_p(\sqrt{\log n/p})$.
	\end{lemma}
	\begin{proof}
		(i) By the Cauchy--Schwarz inequality and the fact that $n^{-1}\sum_{t=1}^{n}\Vert\widehat{\bgamma}_t\Vert^2=O_p(1)$,
		$$
		\begin{aligned}
			\max_{t\in[n]}\Big\Vert\frac{1}{np}\sum_{t'=1}^{n}\widehat{\bgamma}_{t'}\eE\langle\bvarepsilon_{t'},\bvarepsilon_t\rangle\Big\Vert\le&\max_{t\in[n]}\left[\frac{1}{n}\sum_{t'=1}^{n}\Vert\widehat{\bgamma}_{t'}\Vert^2\frac{1}{n}\sum_{t'=1}^{n}\Big\{\frac{\eE\langle\bvarepsilon_{t'},\bvarepsilon_t\rangle}{p}\Big\}^2\right]^{1/2}\\
			\le&O_p(1)\max_{t\in[n]}\left[\frac{1}{n}\sum_{t'=1}^{n}\Big\{\frac{\eE\langle\bvarepsilon_{t'},\bvarepsilon_t\rangle}{p}\Big\}^2\right]^{1/2}\\
			\le&O_p(1)\max_{t',t\in[n]}\left|\frac{\eE\langle\bvarepsilon_{t'},\bvarepsilon_t\rangle}{p}\right|^{1/2}\max_{t\in[n]}\left\{\frac{1}{n}\sum_{t'=1}^{n}\left|\frac{\eE\langle\bvarepsilon_{t'},\bvarepsilon_t\rangle}{p}\right|\right\}^{1/2}\\
			=&O_p(\cM_{\varepsilon}/\sqrt{n}),
		\end{aligned}
		$$
		where the last equality follows from Lemma \ref{lem.ids_weak_serial}.\\
		(ii) By the Cauchy--Schwarz inequality and the fact that $n^{-1}\sum_{t=1}^{n}\Vert\widehat{\bgamma}_t\Vert^2=O_p(1)$,
		$$
		\begin{aligned}
			\max_{t\in[n]}\Big\Vert\frac{1}{n}\sum_{t'=1}^{n}\widehat{\bgamma}_{t'}\zeta_{t't}\Big\Vert\le&\max_{t\in[n]}\frac{1}{n}\Big(\sum_{t'=1}^{n}\Vert\widehat{\bgamma}_{t'}\Vert^2\sum_{t'=1}^{n}\zeta_{t't}^2\Big)^{1/2}=O_p(1)\Big(\max_{t\in[n]}\frac{1}{n}\sum_{t'=1}^{n}\zeta_{t't}^2\Big)^{1/2}\\
			=&O_p(1)\left\{\max_{t\in[n]}\frac{1}{n}\sum_{t'=1}^{n}\Big(\frac{1}{p}\langle\bvarepsilon_{t'},\bvarepsilon_t\rangle-\frac{1}{p}\eE\langle\bvarepsilon_{t'},\bvarepsilon_t\rangle\Big)^2\right\}^{1/2}=O_p(\sqrt{\log n/p}),
		\end{aligned}
		$$
		where the last equality follows from Lemma~\ref{lem.rate_max_f}(ii). \\
		(iii) By the Cauchy--Schwarz inequality and the fact that $\Vert n^{-1}\sum_{t'=1}^{n}\widehat{\bgamma}_{t'}\bgamma_{t'}^{\T}\Vert=O_p(1)$,
		$$\max_{t\in[n]}\left\Vert\frac{1}{n}\sum_{t'=1}^{n}\widehat{\bgamma}_{t'}\eta_{t't}\right\Vert\le\left\Vert\frac{1}{n}\sum_{t'=1}^{n}\widehat{\bgamma}_{t'}\bgamma_{t'}^{\T}\right\Vert\max_{t\in[n]}\left\Vert\frac{1}{p}\sum_{i=1}^{p}\int\bq_i(u)\varepsilon_{ti}(u)\du\right\Vert=O_p(\sqrt{\log n/p}),$$
	 where the last equality follows from Lemma~\ref{lem.rate_max_f}(ii).\\
		(iv) By Assumption~\ref{ass.regu_cond_fpoet}(ii), we can show that $\Vert (np)^{-1}\sum_{t'=1}^{n}\sum_{i=1}^{p}\int\bq_i(u)\varepsilon_{t'i}(u)\du\widehat{\bgamma}_{t'}\Vert=O_p(1/\sqrt{p})$. Additionally, $\max_{t\in[n]}\Vert\bgamma_t\Vert=O_p(\sqrt{\log n})$ by Lemma~\ref{lem.rate_max_f}(ii).  
        The desired result follows immediately that
		$$\max_{t\in[n]}\Big\Vert\frac{1}{n}\sum_{t'=1}^{n}\widehat{\bgamma}_{t'}\xi_{t't}\Big\Vert\le\max_{t\in[n]}\Vert\bgamma_t\Vert\Big\Vert\frac{1}{np}\sum_{t'=1}^{n}\sum_{i=1}^{p}\int\bq_i(u)\varepsilon_{t'i}(u)\du\widehat{\bgamma}_{t'}\Big\Vert=O_p(\sqrt{\log n/p}).$$
	\end{proof}
	
	\begin{lemma}
		\label{lem.conv_decom_H}
		Denote $\widehat{\bgamma}_t=(\widehat{\gamma}_{t1},\dots,\widehat{\gamma}_{tr})^{\T}$. Under the assumptions of Theorem \ref{thm.load_factor_2}, it holds that, for $i\in[r]$,\\
		(i) $n^{-1}\sum_{t=1}^{n}[(np)^{-1}\sum_{t'=1}^{n}\widehat{\gamma}_{t'i}\eE\langle\bvarepsilon_{t'},\bvarepsilon_t\rangle]^2=O_p(\cM_{\varepsilon}^2/n)$;\\
		(ii) $n^{-1}\sum_{t=1}^{n}(n^{-1}\sum_{t'=1}^{n}\widehat{\gamma}_{t'i}\zeta_{t't})^2=O_p(1/p)$;\\
		(iii) $n^{-1}\sum_{t=1}^{n}(n^{-1}\sum_{t'=1}^{n}\widehat{\gamma}_{t'i}\eta_{t't})^2=O_p(1/p)$;\\
		(iv) $n^{-1}\sum_{t=1}^{n}(n^{-1}\sum_{t'=1}^{n}\widehat{\gamma}_{t'i}\xi_{t't})^2=O_p(1/p)$.
	\end{lemma}
	\begin{proof}
		(i) By the Cauchy--Schwarz inequality and the fact that $\sum_{t'=1}^{n}\widehat{\gamma}_{t'i}^2=n$,
		$$
		\begin{aligned}
			\frac{1}{n}\sum_{t=1}^{n}\left(\frac{1}{n}\sum_{t'=1}^{n}\widehat{\gamma}_{t'i}\frac{\eE\langle\bvarepsilon_{t'},\bvarepsilon_t\rangle}{p}\right)^2\le&\frac{1}{n}\sum_{t=1}^{n}\frac{1}{n}\Big(\sum_{t'=1}^{n}\widehat{\gamma}_{t'i}^2\Big)\frac{1}{n}\sum_{t'=1}^{n}\left(\frac{\eE\langle\bvarepsilon_{t'},\bvarepsilon_t\rangle}{p}\right)^2\\
			=&\frac{1}{n}\sum_{t=1}^{n}\frac{1}{n}\sum_{t'=1}^{n}\left(\frac{\eE\langle\bvarepsilon_{t'},\bvarepsilon_t\rangle}{p}\right)^2\le\max_{t\in[n]}\frac{1}{n}\sum_{t'=1}^{n}\left(\frac{\eE\langle\bvarepsilon_{t'},\bvarepsilon_t\rangle}{p}\right)^2\\
			\le&\max_{t',t\in[n]}\left|\frac{\eE\langle\bvarepsilon_{t'},\bvarepsilon_t\rangle}{p}\right|\max_{t\in[n]}\frac{1}{n}\sum_{t'=1}^{n}\left|\frac{\eE\langle\bvarepsilon_{t'},\bvarepsilon_t\rangle}{p}\right|=O(\cM_{\varepsilon}^2/n),
		\end{aligned}
		$$
		where the last equality follows from Lemma \ref{lem.ids_weak_serial}.
		
		\noindent
		(ii) By the Cauchy--Schwarz inequality and the fact that $\sum_{t'=1}^{n}\widehat{\gamma}_{t'i}^2=n$,
		$$
		\begin{aligned}
			\frac{1}{n}\sum_{t=1}^{n}\Big(\frac{1}{n}\sum_{t'=1}^{n}\widehat{\gamma}_{t'i}\zeta_{t't}\Big)^2=&\frac{1}{n^3}\sum_{t',l\in[n]}\Big\{\widehat{\gamma}_{t'i}\widehat{\gamma}_{li}\Big(\sum_{t=1}^{n}\zeta_{t't}\zeta_{lt}\Big)\Big\}\\
			\le&\frac{1}{n^3}\Big\{\sum_{t',l\in[n]}\widehat{\gamma}_{t'i}^2\widehat{\gamma}_{li}^2\sum_{t',l\in[n]}\Big(\sum_{t=1}^{n}\zeta_{t't}\zeta_{lt}\Big)^2\Big\}^{1/2}\\
			\le&\frac{1}{n^3}\sum_{t'=1}^{n}\widehat{\gamma}_{t'i}^2\Big\{\sum_{t',l\in[n]}\Big(\sum_{t=1}^{n}\zeta_{t't}\zeta_{lt}\Big)^2\Big\}^{1/2}=\frac{1}{n^2}\Big\{\sum_{t',l\in[n]}\Big(\sum_{t=1}^{n}\zeta_{t't}\zeta_{lt}\Big)^2\Big\}^{1/2}.
		\end{aligned}
		$$
		Notice that $\eE\big\{\sum_{t',l\in[n]}(\sum_{t=1}^{n}\zeta_{t't}\zeta_{lt})^2\big\}=n^2\eE(\sum_{t=1}^{n}\zeta_{t't}\zeta_{lt})^2\le n^4\max_{t',t}\eE|\zeta_{t't}|^4$, and by Assumption \ref{ass.regu_cond_fpoet}(ii) we have $\max_{t',t}\eE|\zeta_{t't}|^4=O(1/p^2)$, which yields the desired result by using Chebyshev's inequality. 
		
		\noindent
		(iii) By the Cauchy--Schwarz inequality, and the facts that $\sum_{t'=1}^{n}\widehat{\gamma}_{t'i}^2=n$ and $n^{-1}\sum_{t'=1}^{n}\Vert\bgamma_{t'}\Vert^2=O_p(1)$,
		$$
		\begin{aligned}
			\frac{1}{n}\sum_{t=1}^{n}\Big(\frac{1}{n}\sum_{t'=1}^{n}\widehat{\gamma}_{t'i}\eta_{t't}\Big)^2\le&\Big\Vert\frac{1}{n}\sum_{t'=1}^{n}\widehat{\gamma}_{t'i}\bgamma_{t'}^{\T}\Big\Vert^2\frac{1}{n}\sum_{t=1}^{n}\Big\Vert\frac{1}{p}\sum_{j=1}^{p}\int\bq_j(u)\varepsilon_{tj}(u)\du\Big\Vert^2\\
			\le&\Big(\frac{1}{n}\sum_{t'=1}^{n}\widehat{\gamma}_{t'i}^2\frac{1}{n}\sum_{t'=1}^{n}\Vert\bgamma_{t'}\Vert^2\Big)\frac{1}{np^2}\sum_{t=1}^{n}\Big\Vert\sum_{j=1}^{p}\int\bq_j(u)\varepsilon_{tj}(u)\du\Big\Vert^2\\
			=&O_p(1)\cdot\frac{1}{np^2}\sum_{t=1}^{n}\Big\Vert\sum_{j=1}^{p}\int\bq_j(u)\varepsilon_{tj}(u)\du\Big\Vert^2.
		\end{aligned}
		$$
		Notice $\eE\Vert\sum_{j=1}^{p}\int\bq_j(u)\varepsilon_{tj}(u)\du\Vert^2=O(p)$ by Assumption~\ref{ass.regu_cond_fpoet}(ii), which implies $n^{-1}\eE[\sum_{t=1}^{n}\Vert\sum_{j=1}^{p}\int\bq_j(u)\varepsilon_{tj}(u)\du\Vert^2]=O(p)$ and yields the result. 
		
		\noindent
		(iv) By the Cauchy--Schwarz inequality, and the facts that $\sum_{t'=1}^{n}\widehat{\gamma}_{t'i}^2=n$ and $n^{-1}\sum_{t'=1}^{n}\Vert\bgamma_{t'}\Vert^2=O_p(1)$,
		$$
		\begin{aligned}
			\frac{1}{n}\sum_{t=1}^{n}\Big(\frac{1}{n}\sum_{t'=1}^{n}\widehat{\gamma}_{t'i}\xi_{t't}\Big)^2\le&\Big(\frac{1}{n}\sum_{t=1}^{n}\Vert\bgamma_t\Vert^2\Big)\Big\Vert\frac{1}{np}\sum_{t'=1}^{n}\widehat{\gamma}_{t'i}\sum_{j=1}^{p}\int\bq_j(u)\varepsilon_{t'j}(u)\du\Big\Vert^2\\
			\le&O_p(1)\cdot\Big(\frac{1}{n}\sum_{t'=1}^{n}\widehat{\gamma}_{t'i}^2\Big)\frac{1}{np^2}\sum_{t'=1}^{n}\Big\Vert\sum_{j=1}^{p}\int\bq_j(u)\varepsilon_{t'j}(u)\du\Big\Vert^2=O_p(1/p),
		\end{aligned}
		$$
		where the result in the last equality has been obtained in part (iii). 
	\end{proof}
	
	\begin{lemma}
		\label{lem.rth_eigen_2}
		Let $\{\hat{\tau}_j\}_{j=1}^{r}$ be the first $r$ largest eigenvalues of $\widehat{\bSigma}_{y}^{\sS}(\cdot,\cdot)$ in a descending order. Under the assumptions of Theorem \ref{thm.load_factor_2}, it holds that $\hat{\tau}_r\gtrsim p$ with probability approaching one.
	\end{lemma}
	\begin{proof}
		By Proposition \ref{propos.eigenvalues_2}, we obtain that
		$$
		\tau_r\ge\Vert p\vartheta_r\Vert^2-\left|\tau_r-p\vartheta_r\right|\gtrsim p-s_p\asymp p.
		$$
		To show $\hat{\tau}_r\gtrsim p$ with probability approaching one, it suffices to show that $|\hat{\tau}_r-\tau_r|=o_p(p)$. By applying Lemma \ref{lem.weyl_2} again, we only need to show $\Vert\widehat{\bSigma}_{y}^{\sS}-\bSigma_{y}\Vert_{\cS,\F}=o_p(p)$. Note that
		$$
		\begin{aligned}
			\Vert\widehat{\bSigma}_{y}^{\sS}-\bSigma_{y}\Vert_{\cS,\F}=&\Big\Vert\frac{1}{n}\sum_{t=1}^{n}(\bQ\bgamma_t+\bvarepsilon_{t})(\bQ\bgamma_t+\bvarepsilon_{t})^{\T}-\bQ\bQ^{\T}-\bSigma_{\varepsilon}\Big\Vert_{\cS,\F}\\
			\le&\Big\Vert\bQ\big(\frac{1}{n}\sum_{t=1}^{n}\bgamma_t\bgamma_t^{\T}-\bI_r\big)\bQ^{\T}\Big\Vert_{\cS,\F}+\Big\Vert\frac{1}{n}\sum_{t=1}^{n}\bvarepsilon_{t}\bvarepsilon_{t}^{\T}-\bSigma_{\varepsilon}\Big\Vert_{\cS,\F}\\
			&+\Big\Vert\bQ\big(n^{-1}\sum_{t=1}^{n}\bgamma_t\bvarepsilon_{t}^{\T}\big)\Big\Vert_{\cS,\F}+\Big\Vert\big(\frac{1}{n}\sum_{t=1}^{n}\bvarepsilon_{t}\bgamma_t^{\T}\big)\bQ^{\T}\Big\Vert_{\cS,\F}\\
			\le&\Big\Vert n^{-1}\sum_{t=1}^{n}\bgamma_t\bgamma_t^{\T}-\bI_r\Big\Vert_{\F}\cdot\big\Vert\bQ\bQ^{\T}\big\Vert_{\cS,\F}+\Big(\sum_{i=1}^{p}\sum_{j=1}^{p}\big\Vert n^{-1}\sum_{t=1}^{n}\bvarepsilon_{ti}\bvarepsilon_{tj}-\Sigma_{\varepsilon,ij}\big\Vert_{\cS}^2\Big)^{1/2}\\
			&+2\Big(\sum_{i=1}^{p}\sum_{j=1}^{r}\big\Vert n^{-1}\sum_{t=1}^{n}\varepsilon_{ti}\gamma_{tj}\big\Vert^2\Big)^{1/2}\cdot\sqrt{p}\max_{i\in[p]}\Vert\bq_i\Vert\\
			=&O_p(p/\sqrt{n})+O_p(p\cM_{\varepsilon}\sqrt{1/n})+O_p(p\cM_{\varepsilon}\sqrt{1/n})=o_p(p),
		\end{aligned}
		$$
		where the second inequality follows from Lemma \ref{lem.ineq3}(ii), the fact $\Vert\bK\Vert_{\cS,\F}\le p\Vert\bK\Vert_{\cS,\max}$ any $\bK(\cdot,\cdot) \in {\eH}^p \otimes {\eH}^p$, and the Cauchy--Schwarz inequality. The last line of the above equation follows from Lemma \ref{lem.rate_sigma_f_2}, $\cM_{\varepsilon}^2=o(n)$, and the fact 
		$$
		\Vert\bQ\bQ^{\T}\Vert_{\cS,\F}=\Big[\sum_{i=1}^{p}\sum_{j=1}^{p}\int\big\{\bq_i(u)^{\T}\bq_j(v)\big\}^2\du\dv\Big]^{1/2}\le
		p\max_{i\in[p]}\Vert\bq_i\Vert^2\asymp p.
		$$
		Therefore, we have obtained that $\hat{\tau}_r\gtrsim p$ with probability approaching one.
	\end{proof}
	
	\begin{lemma}
		\label{lem.HH_bound}
		Under the assumptions of Theorem \ref{thm.load_factor_2}, it holds that\\
		(i) $\Vert\bH\Vert=O_p(1)$;\\
		(ii) $\bH\bH^{\T}=\bI_{r}+O_p(\cM_{\varepsilon}/\sqrt{n}+1/\sqrt{p})$;\\
		(iii) $\bH^{\T}\bH=\bI_{r}+O_p(\cM_{\varepsilon}/\sqrt{n}+1/\sqrt{p})$.
	\end{lemma}
	\begin{proof}
		(i)
		By Lemma \ref{lem.rth_eigen_2}, $\Vert \bV^{-1}\Vert=\hat{\tau}_r^{-1}=O_p(p^{-1})$. Also, $\Vert\widehat{\bGamma}\Vert=\lambda_{\max}^{1/2}(\widehat{\bGamma}^{\T}\widehat{\bGamma})=\lambda_{\max}^{1/2}(n\bI_r)=\sqrt{n}$ from the normalization \eqref{eq.normalization}, and $\Vert\bGamma\Vert=\lambda_{\max}^{1/2}(\bGamma^{\T}\bGamma)=\lambda_{\max}^{1/2}(\sum_{t=1}^{n}\bgamma_t\bgamma_t^{\T})=O_p(\sqrt{n})$ by Lemma \ref{lem.rate_sigma_f_2}(i). In addition, $\Vert\int\bQ(u)^{\T}\bQ(u)\du\Vert=O(p)$. By the definition of $\bH$, i.e., $\bH=n^{-1}\bV^{-1}\widehat{\bGamma}^{\T}\bGamma\int\bQ(u)^{\T}\bQ(u){\rm d}u$, we have $\Vert\bH\Vert=O_p(1)$, which is also satisfied for $\Vert\bH\Vert_{\F}$ since $\bH\in\mathbb{R}^{r\times r}$. 
		
		\noindent
		(ii) Notice that
		\begin{equation}
			\label{eq.HH_bound}
			\Vert\bH\bH^{\T}-\bI_r\Vert_{\F}\le\Big\Vert\bH\bH^{\T}-\frac{1}{n}\sum_{t=1}^{n}\bH\bgamma_t\bgamma_t^{\T}\bH^{\T}\Big\Vert_{\F}+\Big\Vert\frac{1}{n}\sum_{t=1}^{n}\bH\bgamma_t\bgamma_t^{\T}\bH^{\T}-\bI_r\Big\Vert_{\F}.
		\end{equation}
		In \eqref{eq.HH_bound}, the first term can be bound by $\Vert\bH\bH^{\T}-n^{-1}\sum_{t=1}^{n}\bH\bgamma_t\bgamma_t^{\T}\bH^{\T}\Vert_{\F}\le\Vert\bH\Vert_{\F}^2\Vert\bI_r-n^{-1}\sum_{t=1}^{n}\bgamma_t\bgamma_t^{\T}\Vert_{\F}=O_p(1/\sqrt{n})$ by Lemma \ref{lem.rate_sigma_f_2}(i). The second term can be bounded by
		$$
		\begin{aligned}
			\Big\Vert\frac{1}{n}\sum_{t=1}^{n}\bH\bgamma_t\bgamma_t^{\T}\bH^{\T}-\bI_r\Big\Vert_{\F}=&\Big\Vert\frac{1}{n}\sum_{t=1}^{n}\bH\bgamma_t\bgamma_t^{\T}\bH^{\T}-\frac{1}{n}\sum_{t=1}^{n}\widehat{\bgamma}_t\widehat{\bgamma}_t^{\T}\Big\Vert_{\F}\\
			\le&\Big\Vert\frac{1}{n}\sum_{t=1}^{n}(\bH\bgamma_t-\widehat{\bgamma}_t)\bgamma_t^{\T}\bH^{\T}\Big\Vert_{\F}+\Big\Vert\frac{1}{n}\sum_{t=1}^{n}\widehat{\bgamma}_t(\widehat{\bgamma}_t^{\T}-\bgamma_t^{\T}\bH^{\T})\Big\Vert_{\F}\\
			\le&\Big(\frac{1}{n}\sum_{t=1}^{n}\Vert\bH\bgamma_t-\widehat{\bgamma}_t\Vert^2\frac{1}{n}\sum_{t=1}^{n}\Vert\bH\bgamma_t\Vert^2\Big)^{1/2}
   +\Big(\frac{1}{n}\sum_{t=1}^{n}\Vert\bH\bgamma_t-\widehat{\bgamma}_t\Vert^2\frac{1}{n}\sum_{t=1}^{n}\Vert\widehat{\bgamma}_t\Vert^2\Big)^{1/2}\\
			=&O_p(\cM_{\varepsilon}/\sqrt{n}+1/\sqrt{p}),
		\end{aligned}
		$$
		where the third line follows from Cauchy--Schwarz inequality, and the last line follows from Theorem \ref{thm.load_factor_2}(i) and the fact $n^{-1}\sum_{t=1}^{n}\Vert\widehat{\bgamma}_t\Vert^2=O_p(1)$. 
		
		\noindent
		(iii) From part (ii), we have $\bH\bH^{\T}=\bI_r+O_p(\cM_{\varepsilon}/\sqrt{n}+1/\sqrt{p})$ and $\Vert\bH\Vert=O_p(1).$ Therefore $$\bH\bH^{\T}\bH=\bH+O_p(\cM_{\varepsilon}/\sqrt{n}+1/\sqrt{p}).$$ Also, $\Vert\bH^{-1}\Vert\le\Vert\bH\Vert+o_p(1)\Vert\bH^{-1}\Vert$, which implies that $\Vert\bH^{-1}\Vert=O_p(1)$. Multiplying the LHS of the above by $\bH^{-1}$ yields that $\bH^{\T}\bH=\bI_r+O_p(\cM_{\varepsilon}/\sqrt{n}+1/\sqrt{p})$. 
	\end{proof}
	
We are now ready to prove Theorem \ref{thm.load_factor_2}.\\
(i) By Lemma \ref{lem.rth_eigen_2}, the diagonal elements of $\bV/p={\rm diag}(\hat{\tau}_1/p,\dots,\hat{\tau}_r/p)$ are bounded away from 0. By the inequality $(a+b+c+d)^2\le 4(a^2+b^2+c^2+d^2)$, equation \eqref{eq.decom_H} and Lemma \ref{lem.conv_decom_H}, we have
	$$
	\begin{aligned}
		\max_{i\in[r]} \frac{1}{n}\sum_{t=1}^{n}(\widehat{\bgamma}_t-\bH\bgamma_t)_i^2\lesssim&\max_{i\in[r]}\frac{1}{n}\sum_{t=1}^{n}\Big(\frac{1}{n}\sum_{t'=1}^{n}\widehat{\gamma}_{t'i}\frac{\eE\langle\bvarepsilon_{t'},\bvarepsilon_t\rangle}{p}\Big)^2+\max_{i\in[r]}\frac{1}{n}\sum_{t=1}^{n}\Big(\frac{1}{n}\sum_{t'=1}^{n}\widehat{\gamma}_{t'i}\zeta_{t't}\Big)^2\\
		&+\max_{i\in[r]}\frac{1}{n}\sum_{t=1}^{n}\Big(\frac{1}{n}\sum_{t'=1}^{n}\widehat{\gamma}_{t'i}\eta_{t't}\Big)^2+\max_{i\in[r]}\frac{1}{n}\sum_{t=1}^{n}\Big(\frac{1}{n}\sum_{t'=1}^{n}\widehat{\gamma}_{t'i}\xi_{t't}\Big)^2\\
		=&O_p\left(\cM_{\varepsilon}^2/n+1/p\right).
	\end{aligned}
	$$
	The desired result immediately follows that
	$$\frac{1}{n}\sum_{t=1}^{n}\Vert\widehat{\bgamma}_t-\bH\bgamma_t\Vert^2\le r\max_{i\in[r]}\frac{1}{n}\sum_{t=1}^{n}(\widehat{\bgamma}_t-\bH\bgamma_t)_i^2=O_p\left(\cM_{\varepsilon}^2/n+1/p\right).$$
(ii) Note $\Vert(\bV/p)^{-1}\Vert=O(1)$. Applying the inequality $(a+b+c+d)^2\le 4(a^2+b^2+c^2+d^2)$, equation \eqref{eq.decom_H} and Lemma \ref{lem.conv_decom_H_max}, we have
	$$
	\begin{aligned}
		\max_{t\in[n]}\Vert\widehat{\bgamma}_t-\bH\bgamma_t\Vert\lesssim&
		\max_{t\in[n]}\Big\Vert \frac{1}{np}\sum_{t'=1}^{n}\widehat{\bgamma}_{t'}\eE\langle\bvarepsilon_{t'},\bvarepsilon_t\rangle\Big\Vert+\max_{t\in[n]}\Big\Vert \frac{1}{n}\sum_{t'=1}^{n}\widehat{\bgamma}_{t'}\zeta_{t't}\Big\Vert\\
		&+\max_{t\in[n]}\Big\Vert\frac{1}{n}\sum_{t'=1}^{n}\widehat{\bgamma}_{t'}\eta_{t't}\Big\Vert+\max_{t\in[n]}\Big\Vert\frac{1}{n}\sum_{t'=1}^{n}\widehat{\bgamma}_{t'}\xi_{t't}\Big\Vert\\
		=&O_p(\cM_{\varepsilon}/\sqrt{n}+\sqrt{\log p/n}).
	\end{aligned}
	$$
(iii) Using the facts that $\widehat{\bq}_i(\cdot)=n^{-1}\sum_{t=1}^{n}y_{ti}(\cdot)\widehat{\bgamma}_t$ and $y_{ti}(\cdot)=\bq_i(\cdot)^{\T}\bgamma_t+\varepsilon_{ti}(\cdot)$, we have, for $i\in[p]$  
	\begin{equation}
		\label{eq.bHb_bound}
		\begin{aligned}
			\widehat{\bq}_i(\cdot)-\bH\bq_i(\cdot)=&\frac{1}{n}\sum_{t=1}^{n}y_{ti}(\cdot)\widehat{\bgamma}_t-\frac{1}{n}\sum_{t=1}^{n}\bH\bgamma_t\big\{y_{ti}(\cdot)-\bq_i(\cdot)^{\T}\bgamma_t+\varepsilon_{ti}(\cdot)\big\}-\bH\bq_i(\cdot)\\
			=&\frac{1}{n}\sum_{t=1}^{n}\bH\bgamma_t\varepsilon_{ti}(u)+\frac{1}{n}\sum_{t=1}^{n}y_{ti}(\cdot)(\widehat{\bgamma}_t-\bH\bgamma_t)+\bH\Big(\frac{1}{n}\sum_{t=1}^{n}\bgamma_t\bgamma_t^{\T}-\bI_r\Big)\bq_i(\cdot).
		\end{aligned}
	\end{equation}
	The first term in \eqref{eq.bHb_bound} can be bounded by
	$$
	\max_{i\in[p]}\Big\Vert\frac{1}{n}\sum_{t=1}^{n}\bH\bgamma_t\varepsilon_{ti}\Big\Vert\le\Vert\bH\Vert\max_{i\in[p]}\Big\{\sum_{j=1}^{r}\Big\Vert\frac{1}{n}\sum_{t=1}^{n}\gamma_{tj}\varepsilon_{ti}\Big\Vert^2\Big\}^{1/2}=O_p\left(\cM_{\varepsilon}\sqrt{\log p/n}\right),
	$$
	where the inequality follows from Lemma \ref{lem.factor_norm}(i), and the last equality follows from Lemmas \ref{lem.rate_sigma_f_2}(iii) and \ref{lem.HH_bound}(i). For the second term, since $\Sigma_{y,ii}=\bq_i^{\T}\bSigma_{\gamma}\bq_i+\Sigma_{\varepsilon,ii}$ with $\max_{i\in[p]}\Vert\bq_i\Vert=O(1)$ and $\Vert\bSigma_{\varepsilon}\Vert_{\cS,\max}\le \Vert\bSigma_{\varepsilon}\Vert_{\cL}=O(1)$ by Assumption~\ref{ass.regu_cond_fpoet}(i)(iii), we have $\max_{i\in[p]}\Vert\Sigma_{y,ii}\Vert_{\cS}=\max_{i\in[p]}\eE\Vert y_{ti}\Vert^2=O(1)$, and thus $\max_{i\in[p]}n^{-1}\sum_{t=1}^{n}\Vert y_{ti}\Vert^2=O_p(1)$ by Chebyshev's inequality. Using the Cauchy--Schwarz inequality in the second term of \eqref{eq.bHb_bound}, we obtain that
	$$
	\max_{i\in[p]}\Big\Vert\frac{1}{n}\sum_{t=1}^{n}y_{ti}(\widehat{\bgamma}_t-\bH\bgamma_t)\Big\Vert\le\max_{i\in[p]}\Big(\frac{1}{n}\sum_{t=1}^{n}\Vert y_{ti}\Vert^2\Big)^{1/2}\Big(\frac{1}{n}\sum_{t=1}^{n}\Vert\widehat{\bgamma}_t-\bH\bgamma_t\Vert^2\Big)^{1/2}=O_p\big(\cM_{\varepsilon}/\sqrt{n}+1/\sqrt{p}\big).
	$$
	In addition, $\Vert\bH\Vert=O_p(1)$ from Lemma \ref{lem.HH_bound}(i), $\Vert n^{-1}\sum_{t=1}^{n}\bgamma_t\bgamma_t^{\T}-\bI_r\Vert=O_p(1/\sqrt{n})$ from Lemma \ref{lem.rate_sigma_f_2}(i) and $\max_{i\in[p]}\Vert\bq_i\Vert=O(1)$ from Assumption \ref{ass.regu_cond_fpoet}(i) yield that the third term is of order $O_p(1/\sqrt{n})$. Combining the above results, we obtain that $$\max_{i\in[p]}\Vert\widehat{\bq}_i-\bH\bq_i\Vert=O_p\big(\cM_{\varepsilon}\sqrt{\log p/n}+1/\sqrt{p}\big)=O_p(\varpi_{n,p}).$$

	\subsection{Proof of Corollary \ref{coro.fpoet}}
	By Theorem \ref{thm.load_factor_2}(ii)(iii), Lemmas \ref{lem.HH_bound}(ii) and \ref{lem.rate_max_f}(ii), we have
	$$
	\begin{aligned}
		\max_{i\in[p],t\in[n]}\Vert\widehat{\bq}_i^{\T}\widehat{\bgamma}_t-\bq_i^{\T}\bgamma_t\Vert\le&\max_{i\in[p]}\Vert\widehat{\bq}_i-\bH\bq_i\Vert\cdot\max_{t\in[n]}\Vert\widehat{\bgamma}_t-\bH\bgamma_t\Vert
		+\max_{i\in[p]}\Vert\bH\bq_i\Vert\cdot\max_{t\in[n]}\Vert\widehat{\bgamma}_t-\bH\bgamma_t\Vert\\
		&+\max_{i\in[p]}\Vert\widehat{\bq}_i-\bH\bq_i\Vert\cdot\max_{t\in[n]}\Vert\bH\bgamma_t\Vert
+\max_{i\in[p]}\Vert\bq_i\Vert\cdot\max_{t\in[n]}\Vert\bgamma_t\Vert\cdot\Vert\bH^{\T}\bH-\bI_{r}\Vert\\
		=&O_p\Big\{\varpi_{n,p}\cdot(\cM_{\varepsilon}/\sqrt{n}+\sqrt{\log n/p})\Big\}+O_p\big(\cM_{\varepsilon}/\sqrt{n}+\sqrt{\log n/p}\big)\\
		&+O_p\big(\varpi_{n,p}\cdot \sqrt{\log n}\big)+O_p\Big\{\sqrt{\log n}\cdot(\cM_{\varepsilon}/\sqrt{n}+1/\sqrt{p})\Big\}\\
		=&O_p\big\{\cM_{\varepsilon}\sqrt{\log n\log p/n}+\sqrt{\log n/p}\big\}.
	\end{aligned}
	$$

	\subsection{Proof of Theorem \ref{thm.idio_fpoet}}
  To prove Theorem~\ref{thm.idio_fpoet}, we first present a technical lemma with its proof.
	\begin{lemma}
		\label{lem.varep_bound_2}
		Under the assumptions of Theorem \ref{thm.idio_fpoet}, it holds that\\
		(i) $\max_{i\in[p]}n^{-1}\sum_{t=1}^{n}\Vert\widehat{\varepsilon}_{ti}-\varepsilon_{ti}\Vert^2=O_p(\varpi_{n,p}^2)$;\\
		(ii) $\max_{i,j\in[p]}\Vert n^{-1}\sum_{t=1}^{n}\widehat{\varepsilon}_{ti}\widehat{\varepsilon}_{tj}-n^{-1}\sum_{t=1}^{n}\varepsilon_{ti}\varepsilon_{tj}\Vert_{\cS}=O_p(\varpi_{n,p})$;\\
		(iii) $\Vert\widetilde{\bSigma}_{\varepsilon}-\bSigma_{\varepsilon}\Vert_{\cS,\max}=O_p(\varpi_{n,p})$. 
	\end{lemma}
	\begin{proof}
		(i) Note that $\varepsilon_{ti}(\cdot)-\widehat{\varepsilon}_{ti}(\cdot)=\big\{y_{ti}(\cdot)-\bq_i(\cdot)^{\T}\bgamma_t\big\}-\big\{y_{ti}(\cdot)-\widehat{\bq}_i(\cdot)^{\T}\widehat{\bgamma}_t\big\}=\widehat{\bq}_i(\cdot)^{\T}\widehat{\bgamma}_t-\bq_i(\cdot)^{\T}\bgamma_t$, which can be decomposed as $\widehat{\bq}_i(\cdot)^{\T}\widehat{\bgamma}_t-\bq_i(\cdot)^{\T}\bgamma_t=\big\{\widehat{\bq}_i(\cdot)^{\T}-\bq_i(\cdot)^{\T}\bH\big\}\widehat{\bgamma}_t+\bq_i(\cdot)^{\T}\bH^{\T}(\widehat{\bgamma}_t-\bH\bgamma_t)+\bq_i(\cdot)^{\T}(\bH^{\T}\bH-\bI_r)\bgamma_t$. Applying the inequality $(a+b+c)^2\le 3a^2+3b^2+3c^2$ and the Cauchy--Schwarz inequality yields that
		$$
		\begin{aligned}
			\max_{i\in[p]}\frac{1}{n}\sum_{t=1}^{n}\Vert\widehat{\varepsilon}_{ti}-\varepsilon_{ti}\Vert^2\le&3\max_{i\in[p]}\Vert\widehat{\bq}_i-\bH\bq_i\Vert^2\frac{1}{n}\sum_{t=1}^{n}\Vert\widehat{\bgamma}_t\Vert^2
			+3\max_{i\in[p]}\Vert\bq_i\Vert^2\Vert\bH\Vert^2\frac{1}{n}\sum_{t=1}^{n}\Vert\widehat{\bgamma}_t-\bH\bgamma_t\Vert^2\\
			&+3\max_{i\in[p]}\Vert\bq_i\Vert^2\Vert\bH^{\T}\bH-\bI_r\Vert^2\sum_{t=1}^{n}\Vert\bgamma_t\Vert^2\\
			=&O_p(\varpi_{n,p}^2)+O_p(\cM_{\varepsilon}^2/n+1/p)=O_p(\varpi_{n,p}^2).
		\end{aligned}
		$$
		
		\noindent
		(ii) Notice that $\max_{i\in[p]}\eE\Vert\varepsilon_{ti}\Vert^2=\max_{i\in[p]}\eE\int\varepsilon_{ti}(u)^2\du=\max_{i\in[p]}\int\Sigma_{\varepsilon,ii}(u,u)\du=O(1)$ from Assumption \ref{ass.regu_cond_fpoet}(iv), thus we have $\max_{i\in[p]}n^{-1}\sum_{t=1}^{n}\Vert\varepsilon_{ti}\Vert^2=O_p(1)$. By the Cauchy--Schwarz inequality, we have
		$$
		\begin{aligned}
			\max_{i,j\in[p]}\Big\Vert \frac{1}{n}\sum_{t=1}^{n}\widehat{\varepsilon}_{ti}\widehat{\varepsilon}_{tj}-\frac{1}{n}\sum_{t=1}^{n}\varepsilon_{ti}\varepsilon_{tj}\Big\Vert_{\cS}=&\max_{i,j\in[p]}\Big\Vert\frac{1}{n}\sum_{t=1}^{n}(\widehat{\varepsilon}_{ti}-\varepsilon_{ti})\widehat{\varepsilon}_{tj}+\varepsilon_{ti}(\widehat{\varepsilon}_{tj}-\varepsilon_{tj})\Big\Vert_{\cS}\\
			\le&\max_{i\in[p]}\frac{1}{n}\sum_{t=1}^{n}\Vert\widehat{\varepsilon}_{ti}-\varepsilon_{ti}\Vert^2
			+2\Big(\max_{i\in[p]}\frac{1}{n}\sum_{t=1}^{n}\Vert\varepsilon_{ti}\Vert^2\Big)^{1/2}\Big(\max_{j\in[p]}\frac{1}{n}\sum_{t=1}^{n}\Vert\widehat{\varepsilon}_{tj}-\varepsilon_{tj}\Vert^2\Big)^{1/2}\\
			=&O_p(\varpi_{n,p}^2)+O_p(\varpi_{n,p})=O_p(\varpi_{n,p}).
		\end{aligned}
		$$
		
		\noindent
		(iii) By part (ii) and Lemma \ref{lem.rate_sigma_f_2}(ii), the result follows immediately.
	\end{proof}

         We are now ready to prove Theorem \ref{thm.idio_fpoet}. By Corollary \ref{coro.fpoet} and Lemma \ref{lem.varep_bound_2}(i), we can follow nearly the same procedure as in the proof of~Theorem \ref{thm.idio_DIGIT} to show the similar argument that there exist some constants $C_1,C_2>0$ such that with probability approaching one,
	$$C_1\le\min_{i\in[p],j\in[p]}\Vert\widetilde{\Theta}_{ij}^{1/2}\Vert_{\cS}\le\max_{i\in[p],j\in[p]}\Vert\widetilde{\Theta}_{ij}^{1/2}\Vert_{\cS}\le C_2.$$ Together with Lemmas \ref{lem.varep_bound_2}(iii), we can show that for any $\epsilon>0$, there exist some positive constant $N$ such that each of events
	$$\widetilde{\Upsilon}_1=\left\{\max_{i\in[p],j\in[p]}\left\Vert\widetilde{\Sigma}_{\varepsilon,ij}-\Sigma_{\varepsilon,ij}\right\Vert_{\cS}<N\varpi_{n,p}\right\},\ 
	\widetilde{\Upsilon}_2=\left\{C_1\le\big\Vert\widetilde{\Theta}_{ij}^{1/2}\big\Vert_{\cS}\le C_2,\ {\rm all}\ i,j\in[p]\right\}$$
	hold with probability at least $1-\epsilon$. Then for $\dot{C}>2NC_1^{-1}(\varpi_{n,p}/\omega_{n,p})$ and under the event $\widetilde{\Upsilon}_1\cap\widetilde{\Upsilon}_2$, we obtain that $\Vert\widetilde{\bSigma}_{\varepsilon}^{\cA}-\bSigma_{\varepsilon}\Vert_{\cS,1}\lesssim\varpi_{n,p}^{1-q}s_p$ by using the same way as the proof of Theorem \ref{thm.idio_DIGIT}. By Proposition~\ref{propos.equiv}, we know that $\widehat{\bR}^{\cA}=\widetilde{\bSigma}_{\varepsilon}^{\cA}$. Therefore, with probability at least $1-2\epsilon$, $\Vert\widehat{\bR}^{\cA}-\bSigma_{\varepsilon}\Vert_{\cS,1}\lesssim \varpi_{n,p}^{1-q}s_p$. Considering that $\epsilon>0$ can be arbitrarily small, we have 
	$\Vert\widehat{\bR}^{\cA}-\bSigma_{\varepsilon}\Vert_{\cL}\le\Vert\widehat{\bR}^{\cA}-\bSigma_{\varepsilon}\Vert_{\cS,1}=O_p(\varpi_{n,p}^{1-q}s_p).$

	\subsection{Proof of Theorem \ref{thm.fpoet}}
Under Assumption~\ref{ass.ind2}, we have $\bSigma_{y}(u,v)=\bQ(u)\bQ(v)^{\T}+\bSigma_{\varepsilon}(u,v)$. By the Cauchy--Schwarz inequality and Lemma \ref{lem.factor_norm}, we have
	$$
	\begin{aligned}
		\Vert\widehat{\bQ}\widehat{\bQ}^{\T}-\bQ\bQ^{\T}\Vert_{\cS,\max}=&\max_{i,j\in[p]}\Vert\widehat{\bq}_i^{\T}\widehat{\bq}_j-\bq_i^{\T}\bq_j\Vert_{\cS}\\
		\le&\max_{i,j\in[p]}\left\{\Vert(\widehat{\bq}_i-\bH\bq_i)^{\T}\widehat{\bq}_j\Vert_{\cS}+\Vert\bq_i^{\T}\bH^{\T}(\widehat{\bq}_j-\bH\bq_j)\Vert_{\cS}+\Vert\bq_i^{\T}(\bH^{\T}\bH-\bI_r)\bq_j\Vert_{\cS}\right\}\\
		\le&\max_{i\in[p]}\Vert\widehat{\bq}_i-\bH\bq_i\Vert^2+2\Vert\bH\Vert\max_{i\in[p]}\Vert\bq_j\Vert\Vert\widehat{\bq}_i-\bH\bq_i\Vert
		+\Vert\bH^{\T}\bH-\bI_r\Vert\max_{i\in[p]}\Vert\bq_i\Vert^2\\=&O_p(\varpi_{n,p}^2)+O_p(\varpi_{n,p})+O_p(\cM_{\varepsilon}/\sqrt{n}+1/\sqrt{p})=O_p(\varpi_{n,p}),
	\end{aligned}
	$$
	where the last line follows from Theorem \ref{thm.load_factor_2}(iii) and Lemma \ref{lem.HH_bound}. Then by Lemma \ref{lem.varep_bound_2}, we have $\Vert\widetilde{\bSigma}_{\varepsilon}-\bSigma_{\varepsilon}\Vert_{\cS,\max}=\max_{i,j\in[p]}\Vert\widetilde{\Sigma}_{\varepsilon,ij}-\Sigma_{\varepsilon,ij}\Vert_{\cS}=O_p(\varpi_{n,p})$, and hence
	$$
		\Vert\widetilde{\bSigma}_{\varepsilon}^{\cA}-\bSigma_{\varepsilon}\Vert_{\cS,\max}\le\Vert\widetilde{\bSigma}_{\varepsilon}^{\cA}-\widetilde{\bSigma}_{\varepsilon}\Vert_{\cS,\max}+\Vert\widetilde{\bSigma}_{\varepsilon}-\bSigma_{\varepsilon}\Vert_{\cS,\max}
		\le\max_{i,j\in[p]}(\Vert\widetilde{\Theta}_{ij}^{1/2}\Vert_{\cS}\lambda)+O_p(\varpi_{n,p})
		=O_p(\varpi_{n,p}),
	$$
	where $\widetilde{\Theta}_{ij}(u,v)\equiv n^{-1}\sum_{t=1}^{n}\big\{\widehat{\varepsilon}_{ti}(u)\widehat{\varepsilon}_{tj}(v)-\widetilde{\Sigma}_{\varepsilon,ij}(u,v)\big\}^2,$ the last line follows from Lemma \ref{lem.varep_bound_2}(iii), the choice of $\lambda=\dot{C}(\sqrt{\log p/n}+\sqrt{1/p})\lesssim\varpi_{n,p}$, and the fact $\max_{i,j\in[p]}\Vert\widetilde{\Theta}_{ij}^{1/2}\Vert_{\cS}=O_p(1)$ that can be proved following a similar argument compared to the proof of Lemma \ref{lem.Theta_bound}. The desired result follows immediately.

\subsection{Proof of Theorem~\ref{thm.inverse_fpoet}}

Given $\tKer(\bSigma_{\varepsilon})=\tKer(\bQ\bSigma_{\gamma}\bQ^{\T})$, with the covariance decomposition for FPOET estimator $\bSigma_y(u,v)=\bQ(u)\bSigma_{\gamma}\bQ(v)^{\T}+\bSigma_{\varepsilon}(u,v),$ we apply Sherman--Morrison--Woodbury formula (Theorem 3.5.6 of \textcolor{blue}{Hsing and Eubank} (\textcolor{blue}{2015})) to obtain its inverse $\bSigma_y^{\dagger}=\bSigma_{\varepsilon}^{\dagger}-\bSigma_{\varepsilon}^{\dagger}\bQ(\bI_r+\bQ^{\T}\bSigma_{\varepsilon}^{\dagger}\bQ)^{\dagger}\bQ^{\T}\bSigma_{\varepsilon}^{\dagger}$ (note that under Assumption~\ref{ass.ind2}, $\bSigma_{\gamma}=\bI_r$ and $\widehat{\bSigma}_{\gamma}=\bI_r$). Then, the plug-in inverse FPOET estimator is defined as $(\widehat{\bSigma}_y^{\cF})^{\dagger}=(\widehat{\bR}^{\cA})^{\dagger}-(\widehat{\bR}^{\cA})^{\dagger}\widehat{\bQ}\{\bI_r+\widehat{\bQ}^{\T}(\widehat{\bR}^{\cA})^{\dagger}\widehat{\bQ}\}^{\dagger}\widehat{\bQ}^{\T}(\widehat{\bR}^{\cA})^{\dagger}.$ In this section, $(\widehat{\bR}^{\cA})^{\dagger}$ and $\{\bI_r+\widehat{\bQ}^{\T}(\widehat{\bR}^{\cA})^{\dagger}\widehat{\bQ}\}^{\dagger}$ are defined in a similar way to the inverse in Lemma~\ref{lem.inv_norm} associated with $\bSigma_{\varepsilon}$ and $\bI_r+\bQ^{\T}\bSigma_{\varepsilon}^{\dagger}\bQ$, respectively.
To prove Theorem~\ref{thm.inverse_fpoet}, we first present some technical lemmas with their proofs. 

\begin{lemma}
    \label{lem.idio_inver_fpoet}
    Under the assumptions of Theorem \ref{thm.inverse_fpoet}, then, $\widehat{\bR}^{\cA}$ has a bounded Moore--Penrose inverse with probability approaching one, and $\big\Vert(\widehat{\bR}^{\cA})^{\dagger}-\bSigma_{\varepsilon}^{\dagger}\big\Vert_{\cL}=O_p(\varpi_{n,p}^{1-q}s_p).$
\end{lemma}
\begin{proof}
    Provided that $\varpi_{n,p}^{1-q}s_p=o(1)$ and $\Vert\bSigma_{\varepsilon}^{\dagger}\Vert_{\cL}>c_5$ for some constant $c_5>0$, we combine Lemma \ref{lem.inv_norm} and Theorem \ref{thm.idio_fpoet} to yield that $\Vert(\widehat{\bR}^{\cA})^{\dagger}\Vert_{\cL}<2c_5$ with probability approaching one, 
 and thus $\widehat{\bR}^{\cA}$ has a bounded Moore--Penrose inverse with probability approaching one together with the desired result $\big\Vert(\widehat{\bR}^{\cA})^{\dagger}-\bSigma_{\varepsilon}^{\dagger}\big\Vert_{\cL}=O_p(\varpi_{n,p}^{1-q}s_p).$
\end{proof}

\begin{lemma}
    \label{lem.inverse1_fpoet}
    Under the assumptions of Theorem~\ref{thm.inverse_fpoet}, 
    $$
    \Vert\widehat{\bQ}^{\T}(\widehat{\bR}^{\cA})^{\dagger}\widehat{\bQ}-\bH\bQ^{\T}\bSigma_{\varepsilon}^{\dagger}\bQ\bH^{\T}\Vert_{\cL}=O_p(p\varpi_{n,p}^{1-q}s_p)=o_p(p).
    $$
\end{lemma}
\begin{proof}
    In model~\eqref{eq.model_2}, $\bQ(\cdot)$ can be viewed as a bounded linear operator from $\eR^r$ to $\eH^p,$ and thus we can also regard it as a kernel matrix function satisfying $\bQ(u,v)\equiv\bQ(u),\forall u,v\in\cU.$ From this perspective, $\Vert\bQ\Vert_{\cS,\F}^2=\sum_{i=1}^{p}\Vert\bq_i\Vert^2=\int\bQ(u)^{\T}\bQ(u)\du=p(\vartheta_1+\dots+\vartheta_r)\asymp p$ under Assumption~\ref{ass.ind2}. By Theorem~\ref{thm.load_factor_2}(iii), $\Vert\widehat{\bQ}-\bQ\bH^{\T}\Vert_{\cS,\F}=\big\{\sum_{i=1}^p\Vert\widehat{\bq}_i-\bH\bq_i\Vert^2\big\}^{1/2}=O_p(\sqrt{p}\varpi_{n,p}).$ Hence,
    $$
		\begin{aligned}
			\Vert\widehat{\bQ}^{\T}(\widehat{\bR}^{\cA})^{\dagger}\widehat{\bQ}-\bH\bQ^{\T}\bSigma_{\varepsilon}^{\dagger}\bQ\bH^{\T}\Vert_{\cL}\le&2\Vert(\widehat{\bQ}-\bQ\bH^{\T})^{\T}(\widehat{\bR}^{\cA})^{\dagger}\widehat{\bQ}\Vert_{\cL}+\Vert\bH\bQ^{\T}\{(\widehat{\bR}^{\cA})^{\dagger}-\bSigma_{\varepsilon}^{\dagger}\}\bQ\bH^{\T}\Vert_{\cL}\\
			\le&2\Vert\widehat{\bQ}-\bQ\bH^{\T}\Vert_{\cS,\F}\Vert(\widehat{\bR}^{\cA})^{\dagger}\Vert_{\cL}\Vert\bQ\Vert_{\cS,\F}+\Vert\bQ\Vert_{\cS,\F}^2\Vert\bH\Vert^2\Vert(\widehat{\bR}^{\cA})^{\dagger}-\bSigma_{\varepsilon}^{\dagger}\Vert_{\cL}\\
			=&O_p(p\varpi_{n,p})+O_p(p\varpi_{n,p}^{1-q}s_p)=O_p(p\varpi_{n,p}^{1-q}s_p)=o_p(p),
		\end{aligned}
		$$
		where the second inequality follows from Lemma~\ref{lem.op_le_S1}(i)(ii), and the last line follows from Lemmas~\ref{lem.HH_bound} and \ref{lem.idio_inver_fpoet}.
\end{proof}

\begin{lemma}
    \label{lem.inverse2_fpoet}
    Under the assumptions of Theorem~\ref{thm.inverse_fpoet},\\
    (i) $\Vert(\bI_r+\bH\bQ^{\T}\bSigma_{\varepsilon}^{\dagger}\bQ\bH^{\T})^{\dagger}\Vert_{\cL}=O(p^{-1});$\\
    (ii) $\Vert\{\bI_r+\widehat{\bQ}^{\T}(\widehat{\bR}^{\cA})^{\dagger}\widehat{\bQ}\}^{\dagger}\Vert_{\cL}=O_p(p^{-1});$\\
    (iii) $\Vert(\bI_r+\bQ^{\T}\bSigma_{\varepsilon}^{\dagger}\bQ)^{\dagger}\Vert_{\cL}=O(p^{-1});$\\
    (iv) $\Vert\{(\bH\bH^{\T})^{\dagger}+\bQ^{\T}\bSigma_{\varepsilon}^{\dagger}\bQ\}^{\dagger}\Vert_{\cL}=O(p^{-1}).$
\end{lemma}
\begin{proof}
    (i) By Lemma~\ref{lem.HH_bound}, with probability approaching one, $\lambda_{\min}(\bH\bH^{\T})$ is bounded away from 0. Hence,
    $$
        \Vert(\bI_r+\bH\bQ^{\T}\bSigma_{\varepsilon}^{\dagger}\bQ\bH^{\T})^{\dagger}\Vert_{\cL}\le\Vert(\bH\bQ^{\T}\bSigma_{\varepsilon}^{\dagger}\bQ\bH^{\T})^{\dagger}\Vert_{\cL}
        \le\{\lambda_{\min}(\bH^{\T}\bH)\}^{-1}\Vert\bSigma_{\varepsilon}\Vert_{\cL}\Vert(\bQ\bQ^{\T})^{\dagger}\Vert_{\cL}=O(p^{-1}),
    $$
    where $\Vert(\bQ\bQ^{\T})^{\dagger}\Vert_{\cL}=(p\vartheta_r)^{-1}$ by Assumption~\ref{ass.ind2} and $\Vert\bSigma_{\varepsilon}\Vert_{\cL}=O(1)$ by Assumption~\ref{ass.regu_cond_fpoet}(iii).\\
    (ii) The result follows from part (i) and Lemmas~\ref{lem.inv_norm} and \ref{lem.inverse1_fpoet}.\\
    (iii) The result follows from a similar argument to that for part (i).\\
    (iv) The result follows from part (iii) and Lemmas~\ref{lem.inv_norm} and \ref{lem.HH_bound}.
\end{proof}

We are now ready to prove Theorem \ref{thm.inverse_fpoet}. Using the functional version of Sherman--Morrison--Woodbury formula, we have $\Vert(\widehat{\bSigma}_y^{\cF})^{\dagger}-\bSigma_y^{\dagger}\Vert_{\cL}\le \sum_{k=1}^{6}L_k$, where
$$
	\begin{aligned}
		L_1=&\big\Vert(\widehat{\bR}^{\cA})^{\dagger}-\bSigma_{\varepsilon}^{\dagger}\big\Vert_{\cL},\\
		L_2=&\big\Vert\{(\widehat{\bR}^{\cA})^{\dagger}-\bSigma_{\varepsilon}^{\dagger}\}\widehat{\bQ}\{\bI_r+\widehat{\bQ}^{\T}(\widehat{\bR}^{\cA})^{\dagger}\widehat{\bQ}\}^{\dagger}\widehat{\bQ}^{\T}(\widehat{\bR}^{\cA})^{\dagger}\big\Vert_{\cL},\\
		L_3=&\big\Vert\{(\widehat{\bR}^{\cA})^{\dagger}-\bSigma_{\varepsilon}^{\dagger}\}\widehat{\bQ}\{\bI_r+\widehat{\bQ}^{\T}(\widehat{\bR}^{\cA})^{\dagger}\widehat{\bQ}\}^{\dagger}\widehat{\bQ}^{\T}\bSigma_{\varepsilon}^{\dagger}\big\Vert_{\cL},\\
		L_4=&\big\Vert\bSigma_{\varepsilon}^{\dagger}(\widehat{\bQ}-\bQ\bH^{\T})\{\bI_r+\widehat{\bQ}^{\T}(\widehat{\bR}^{\cA})^{\dagger}\widehat{\bQ}\}^{\dagger}\widehat{\bQ}^{\T}\bSigma_{\varepsilon}^{\dagger}\big\Vert_{\cL},\\
            L_5=&\big\Vert\bSigma_{\varepsilon}^{\dagger}(\widehat{\bQ}-\bQ\bH^{\T})\{\bI_r+\widehat{\bQ}^{\T}(\widehat{\bR}^{\cA})^{\dagger}\widehat{\bQ}\}^{\dagger}\bH\bQ^{\T}\bSigma_{\varepsilon}^{\dagger}\big\Vert_{\cL},\\
            L_6=&\big\Vert\bSigma_{\varepsilon}^{\dagger}\bQ\bH^{\T}\big[\{\bI_r+\widehat{\bQ}^{\T}(\widehat{\bR}^{\cA})^{\dagger}\widehat{\bQ}\}^{\dagger}-(\bI_r+\bH\bQ^{\T}\bSigma_{\varepsilon}^{\dagger}\bQ\bH^{\T})^{\dagger}\big]\bH\bQ^{\T}\bSigma_{\varepsilon}^{\dagger}\big\Vert_{\cL}.
	\end{aligned}
	$$
        Combining with Lemmas~\ref{lem.idio_inver_fpoet} and \ref{lem.inverse2_fpoet}, the desired result follows from a similar argument to the proof of Theorem~\ref{thm.inverse_digit}.

\section{Proofs of theoretical results in Section~\ref{sec.app}}
\label{supsec.D}

For the sake of brevity and readability, in this section, we suppose that the orthogonal matrix $\bU$ in Theorem~\ref{thm.load_factor_1} and Lemmas~\ref{lem.rate_B_F}--\ref{lem.rate_hatsigma_f_1} is an identity matrix, which means that, when we perform eigen-decomposition on $\widehat{\bOmega}$, we can always select the correct direction of $\widehat{\bxi}_j$ to ensure $\widehat{\bxi}_j^{\T}\widetilde{\bbb}_j\ge0$. The proofs in Section~\ref{sec.theory} verify that the choice of $\bU$ does not affect the theoretical results. 

\subsection{Proposition \ref{propos.risk1}--\ref{propos.risk2} and their proofs}
The following two propositions are used in Section~\ref{subsec.risk} to quantify the maximum absolute and 
relative approximation errors of the functional portfolio variance.

        \begin{proposition}
        \label{propos.risk1}
        Let $\bSigma=\{\Sigma_{ij}(\cdot,\cdot)\}_{p \times p},$ and $\widehat\bSigma=\{\widehat\Sigma_{ij}(\cdot,\cdot)\}_{p \times p}$ with each $\Sigma_{ij}, \widehat\Sigma_{ij} \in \eS.$ For any fixed $\bw(\cdot)\in\eH^p,$ we have
        $$
        \Big|\langle\bw,\widehat{\bSigma}(\bw)\rangle-\langle\bw,\bSigma(\bw)\rangle\Big|\le\Vert\widehat{\bSigma}-\bSigma\Vert_{\cS,\max}\Big(\sum_{i \in [p]}\Vert w_i\Vert\Big)^2.
        $$
        \end{proposition}

        \begin{proof}
	Consider that
	$$
	\begin{aligned}
		\langle\bw,\bSigma(\bw)\rangle=&\int\int\sum_{i\in[p]}\sum_{j\in[p]}w_i(u)w_j(v)\Sigma_{ij}(u,v)\du\dv\\
		\le&\sum_{i\in[p]}\sum_{j\in[p]}\left\{\int\int\Sigma_{ij}(u,v)^2\du\dv\right\}^{1/2}\left\{\int\int w_i(u)^2w_j(v)^2\du\dv\right\}^{1/2}\\
		\le&\max_{i\in[p],j\in[p]}\Vert\Sigma_{ij}\Vert_{\cS}\cdot\sum_{i\in[p]}\sum_{j\in[p]}\left\{\int w_i(u)^2\du\right\}^{1/2}\left\{\int w_j(v)^2\dv\right\}^{1/2}\\
		=&\Vert\bSigma\Vert_{\cS,\max}\cdot\sum_{i\in[p]}\sum_{j\in[p]}\Vert w_i\Vert\Vert w_j\Vert=\Vert\bSigma\Vert_{\cS,\max}\Big(\sum_{i \in [p]}\Vert w_i\Vert\Big)^2.
	\end{aligned}
	$$
	Thus, $\Big|\langle\bw,\widehat{\bSigma}(\bw)\rangle-\langle\bw,\bSigma(\bw)\rangle\Big|=\Big|\langle\bw,(\widehat{\bSigma}-\bSigma)\bw)\rangle\Big|\le\Vert\widehat{\bSigma}-\bSigma\Vert_{\cS,\max}(\sum_{i=1}^{p}\Vert w_i\Vert)^2$. 
        \end{proof}

        \begin{proposition}
	\label{propos.risk2}
	Suppose $\bSigma$ and $\widehat{\bSigma}$ are Mercer's kernels and $\bSigma$ has a bounded inverse. For any fixed $\bw(\cdot)\in\tKer(\bSigma)^{\perp},$ we have
	$$
	\left|\frac{\langle\bw,\widehat{\bSigma}(\bw)\rangle}{\langle\bw,\bSigma(\bw)\rangle}-1\right|\le\big\Vert(\bSigma^{\dagger})^{1/2}\widehat{\bSigma}(\bSigma^{\dagger})^{1/2}-\tilde\bI_p\big\Vert_{\cL}.
	$$
	\end{proposition}

\begin{proof}
	For any given $\bw\in\tKer(\bSigma)^{\perp}$, we denote $\bx=\bSigma^{1/2}\bw\in\tIm(\bSigma)$ and $\bw=(\bSigma^{\dagger})^{1/2}\bx$, provided that $\bSigma$ has a bounded inverse. Consider that
    $$
    \begin{aligned}
        \langle\bw,\bSigma(\bw)\rangle=&\int\int\bw(u)^{\T}\bSigma(u,v)\bw(v)\du\dv=\int\int\bw(u)^{\T}\left\{\int\bSigma^{1/2}(u,w)\bSigma^{1/2}(w,v){\rm d}w\right\}\bw(v)\du\dv\\
        =&\int\left\{\int\bw(u)^{\T}\bSigma^{1/2}(u,w)\du\right\}\left\{\int\bSigma^{1/2}(w,v)\bw(v)\dv\right\}{\rm d}w=\int\bx(w)^{\T}\bx(w){\rm d}w=\Vert\bx\Vert^2
    \end{aligned}
    $$
    The relative error can be bounded by
	$$
	\begin{aligned}
		\left|\frac{\langle\bw,\widehat{\bSigma}(\bw)\rangle}{\langle\bw,\bSigma(\bw)\rangle}-1\right|=&\left|\frac{\langle\bw,\widehat{\bSigma}(\bw)\rangle-\langle\bw,\bSigma(\bw)\rangle}{\langle\bw,\bSigma(\bw)\rangle}\right|\\
		=&\frac{\left|\langle\bx,(\bSigma^{\dagger})^{1/2}(\widehat{\bSigma}-\bSigma)(\bSigma^{\dagger})^{1/2}(\bx)\rangle\right|}{\Vert\bx\Vert^2}\\
		\le&\Vert(\bSigma^{\dagger})^{1/2}\widehat{\bSigma}(\bSigma^{\dagger})^{1/2}-\tilde\bI_p\Vert_{\cL},
	\end{aligned}
	$$
	where the last line follows from Lemma \ref{lem.eigen}.
\end{proof}

\subsection{Proof of Theorem~\ref{thm.new_norm}}

Since there exist constants $c_3$ and $c_4>0$ such that $\Vert\bSigma_{\varepsilon}^{\dagger}\Vert_{\cL}<c_3,\Vert\bSigma_f^{\dagger}\Vert_{\cL}<c_4$, we can obtain that $\Vert\bSigma_y^{\dagger}\Vert_{\cL}=O(1)$, and thus for any $\bK\in\eH^p\otimes\eH^p,\Vert\bK\Vert_{\cS,\Sigma_y}^2=O(p^{-1})\Vert\bK\Vert_{\cS,\F}^2$ by Lemma \ref{lem.trace}(iii). 

To prove Theorem \ref{thm.new_norm}, we first present some technical lemmas with their proofs.

	\begin{lemma}
		\label{lem.BSB_bound}
		Under the assumptions of Theorem \ref{thm.new_norm}, we have $\Vert\bB^{\T}\bSigma_y^{\dagger}\bB\Vert_{\cL}=O(1)$.
	\end{lemma}
	\begin{proof}
		By Theorem 3.5.6 of \textcolor{blue}{Hsing and Eubank} (\textcolor{blue}{2015}), we obtain that
		$$
		\bSigma_y^{\dagger}=\bSigma_{\varepsilon}^{\dagger}-\bSigma_{\varepsilon}^{\dagger}\bB(\bSigma_f^{\dagger}+\bB^{\T}\bSigma_{\varepsilon}^{\dagger}\bB)^{\dagger}\bB^{\T}\bSigma_{\varepsilon}^{\dagger}.
		$$
		Then it follows that
            \begin{equation}
                \label{eq.ByB}
                \begin{aligned}
    			\bB^{\T}\bSigma_y^{\dagger}\bB=&\bB^{\T}\bSigma_{\varepsilon}^{\dagger}\bB-\bB^{\T}\bSigma_{\varepsilon}^{\dagger}\bB(\bSigma_f^{\dagger}+\bB^{\T}\bSigma_{\varepsilon}^{\dagger}\bB)^{\dagger}\bB^{\T}\bSigma_{\varepsilon}^{\dagger}\bB\\
    			=&\bB^{\T}\bSigma_{\varepsilon}^{\dagger}\bB(\bSigma_f^{\dagger}+\bB^{\T}\bSigma_{\varepsilon}^{\dagger}\bB)^{\dagger}\bSigma_f^{\dagger}\\
    			=&\bSigma_f^{\dagger}-\bSigma_f^{\dagger}(\bSigma_f^{\dagger}+\bB^{\T}\bSigma_{\varepsilon}^{\dagger}\bB)^{\dagger}\bSigma_f^{\dagger},
    		\end{aligned}
            \end{equation}
		where the last two equalities follow from the assumption $\tKer(\bSigma_{\bvarepsilon})=\tKer(\bB\bSigma_f\bB^{\T})$. \eqref{eq.ByB} also implies that $\bSigma_f^{\dagger}\succeq\bSigma_f^{\dagger}(\bSigma_f^{\dagger}+\bB^{\T}\bSigma_{\varepsilon}^{\dagger}\bB)^{\dagger}\bSigma_f^{\dagger}$ since $\bB^{\T}\bSigma_y^{\dagger}\bB\succeq0.$ Here, for two Mercer's kernels $\bK,\bG\in\eH^r\otimes\eH^r$, we denote $\bK\succeq\bG$ as the eigenvalues of $\bK-\bG$ are nonnegative, i.e., $\bK-\bG$ is still a Mercer's kernel. Similar to the monotonicity of matrix spectral norm, it can be shown that the operator norm is monotone, i.e., $\bK\succeq\bG$ implies $\Vert\bK\Vert_{\cL}\ge\Vert\bG\Vert_{\cL}$. Thus, from \eqref{eq.ByB} we have   
		$$
		\Vert\bB^{\T}\bSigma_y^{\dagger}\bB\Vert_{\cL}\le\Vert\bSigma_f^{\dagger}\Vert_{\cL}+\Vert\bSigma_f^{\dagger}(\bSigma_f^{\dagger}+\bB^{\T}\bSigma_{\varepsilon}^{\dagger}\bB)^{\dagger}\bSigma_f^{\dagger}\Vert_{\cL}\le2\Vert\bSigma_f^{\dagger}\Vert_{\cL}\le 2c_4=O(1),
		$$
	where the second inequality follows from the monotonicity of the operator norm.
	\end{proof}
	
	\begin{lemma}
		\label{lem.new_norm_bound}
		Under the assumptions of Theorem \ref{thm.new_norm}, it follows that\\
		(i) $\Vert\widehat{\bSigma}_{\varepsilon}^{\cA}-\bSigma_{\varepsilon}\Vert_{\cS,\Sigma_y}^2=O_p(\varpi_{n,p}^{2-2q}s_p^2)$;\\
		(ii) $\Vert(\widehat{\bB}-\bB)\widehat{\bSigma}_f(\widehat{\bB}-\bB)^{\T}\Vert_{\cS,\Sigma_y}^2=O_p(\cM_{\varepsilon}^4p/n^2+1/p^3)$;\\
		(iii) $\Vert\bB\widehat{\bSigma}_f(\widehat{\bB}-\bB)^{\T}\Vert_{\cS,\Sigma_y}^2=O_p(\cM_{\varepsilon}^2/n+1/p^2)$;\\
		(iv) $\Vert\bB(\widehat{\bSigma}_f-\bSigma_f)\bB^{\T}\Vert_{\cS,\Sigma_y}^2=O_p\{\cM_{\varepsilon}/(\sqrt{n}p)+1/p^{3/2}\}$.
	\end{lemma}
	\begin{proof}
		(i) Since all the eigenvalues of $\bSigma_y$ are bounded away from zero, and by Theorem \ref{thm.idio_DIGIT}, $$\Vert\widehat{\bSigma}_{\varepsilon}^{\cA}-\bSigma_{\varepsilon}\Vert_{\cS,\Sigma_y}^2\asymp p^{-1}\Vert\widehat{\bSigma}_{\varepsilon}^{\cA}-\bSigma_{\varepsilon}\Vert_{\cS,\F}^2\asymp\Vert\widehat{\bSigma}_{\varepsilon}^{\cA}-\bSigma_{\varepsilon}\Vert_{\cL}^2=O_p(\varpi_{n,p}^{2-2q}s_p^2).$$
		(ii) By applying Lemmas~\ref{lem.trace}(iii) and \ref{lem.rate_B_F}, we have
		$$\Vert(\widehat{\bB}-\bB)\widehat{\bSigma}_f(\widehat{\bB}-\bB)^{\T}\Vert_{\cS,\Sigma_y}^2\le p^{-1}\Vert\widehat{\bB}-\bB\Vert_{\F}^4\Vert\widehat{\bSigma}_f\Vert_{\cL}^2\Vert\bSigma_y^{\dagger}\Vert_{\cL}^2=O_p(\cM_{\varepsilon}^4p/n^2+1/p^3).$$
		(iii) Consider
		$$
		\begin{aligned}
			\Vert\bB\widehat{\bSigma}_f(\widehat{\bB}-\bB)^{\T}\Vert_{\cS,\Sigma_y}^2=&p^{-1}{\rm tr}\left\{\int\left[\widehat{\bSigma}_f(\widehat{\bB}-\bB)^{\T}\bSigma_y^{\dagger}(\widehat{\bB}-\bB)\widehat{\bSigma}_f\bB^{\T}\bSigma_y^{\dagger}\bB\right](u,u)\du\right\}\\
			\le&p^{-1}\Vert\bB^{\T}\bSigma_y^{\dagger}\bB\Vert_{\cL}\Vert\bSigma_y^{\dagger}\Vert_{\cL}\Vert\widehat{\bB}-\bB\Vert_{\F}^2\Vert\widehat{\bSigma}_f\Vert_{\cN}^2\\
            =&O_p(\cM_{\varepsilon}^2/n+1/p^2),
		\end{aligned}
		$$
		where the first inequality follows from Lemma \ref{lem.trace}(ii) and the last line follows from Lemmas \ref{lem.BSB_bound} and \ref{lem.rate_B_F}. \\
		(iv) A similar argument shows that
		$$
		\begin{aligned}
			\Vert\bB(\widehat{\bSigma}_f-\bSigma_f)\bB^{\T}\Vert_{\cS,\Sigma_y}^2=&p^{-1}{\rm tr}\left\{\int\left[(\widehat{\bSigma}_f-\bSigma_f)\bB^{\T}\bSigma_y^{\dagger}\bB(\widehat{\bSigma}_f-\bSigma_f)\bB^{\T}\bSigma_y^{\dagger}\bB\right](u,u)\du\right\}\\
			\le&p^{-1}\Vert\bB^{\T}\bSigma_y^{\dagger}\bB\Vert_{\cL}^2\Vert\widehat{\bSigma}_f-\bSigma_f\Vert_{\cL}\Vert\widehat{\bSigma}_f-\bSigma_f\Vert_{\cN}\\
            =&O_p\{\cM_{\varepsilon}/(\sqrt{n}p)+1/p^{3/2}\},
		\end{aligned}
		$$
		where the first inequality follows from Lemma \ref{lem.trace}(ii) and the last line follows from Lemmas \ref{lem.BSB_bound} and \ref{lem.rate_hatsigma_f_1}. 
	\end{proof}
	
	We are now ready to prove Theorem \ref{thm.new_norm}. By Lemma \ref{lem.new_norm_bound}, 
	$$
	\begin{aligned}
		\Vert\widehat{\bSigma}_y^{\cD}-\bSigma_y\Vert_{\cS,\Sigma_y}^2\le&2\Vert\widehat{\bSigma}_{\varepsilon}^{\cA}-\bSigma_{\varepsilon}\Vert_{\cS,\Sigma_y}^2+2\Vert(\widehat{\bB}-\bB)\widehat{\bSigma}_f(\widehat{\bB}-\bB)^{\T}\Vert_{\cS,\Sigma_y}^2\\
		&+4\Vert\bB\widehat{\bSigma}_f(\widehat{\bB}-\bB)^{\T}\Vert_{\cS,\Sigma_y}^2+2\Vert\bB(\widehat{\bSigma}_f-\bSigma_f)\bB^{\T}\Vert_{\cS,\Sigma_y}^2\\
		=&O_p\left(\frac{\cM_{\varepsilon}^4p}{n^2}+\varpi_{n,p}^{2-2q}s_p^2\right),
	\end{aligned}
	$$
	which then implies that
	$$\Vert\widehat{\bSigma}_y^{\cD}-\bSigma_y\Vert_{\cS,\Sigma_y}=O_p\left(\frac{\cM_{\varepsilon}^2\sqrt{p}}{n}+\varpi_{n,p}^{1-q}s_p\right).$$

\section{Proposition~\ref{propos.scenario} and its proof}
The following proposition supporting Section~\ref{sec.sim} gives the true covariance matrix functions for two DGPs and the functional sparsity condition.

\begin{proposition}
\label{propos.scenario}
(i) For $\by_t(\cdot)$ generated from model \eqref{eq.model_1}, 
$$
\bSigma_{y}(u,v)=\bB\big\{\sum_{i=1}^{50}i^{-2}\phi_i(u)\phi_i(v)(\bI_r-\bA^2)^{-1}\big\}\bB^{\T}+\sum_{l=1}^{25}3^{-1}2^{2-l}\phi_l(u)\phi_l(v)\bC_{\zeta}.
$$
(ii) For $\by_t(\cdot)$ generated from model \eqref{eq.model_2}, 
$$
\bSigma_{y}(u,v)=\bQ(u)(\bI_r-\bA^2)^{-1}\bQ(v)^{\T}+\sum_{l=1}^{25}3^{-1}2^{2-l}\phi_l(u)\phi_l(v)\bC_{\zeta}.
$$
(iii) The functional sparsity condition on $\bSigma_{\varepsilon}$ as specified in \eqref{eq.sparsity} satisfies $s_p\lesssim p^{1-\alpha}$ for $\alpha \in [0,1]$ and $q=0$. 
\end{proposition}

\begin{proof}       
\noindent (i) Let $\bXi_t=(\bxi_{t1},\dots,\bxi_{t,50})\in\eR^{r\times 50}$ and $\bphi(\cdot)=\{1^{-1}\phi_1(\cdot),2^{-1}\phi_2(\cdot),\dots,50^{-1}\phi_{50}(\cdot)\}^{\T}$. Note that $\eE\{\bbf_t(\cdot)\}=\mathbf{0}$ and 
$$
\bSigma_{f}(u,v)={\rm Cov}\{\bXi_t\bphi(u),\bXi_t\bphi(v)\}
=\sum_{i=1}^{50}i^{-2}\phi_i(u)\phi_i(v){\rm Var}(\bxi_{ti}),
$$
provided that ${\rm Cov}(\bxi_{ti},\bxi_{ti'})=\mathbf{0}_{r\times r}$ for any $i\neq i'$. Let $\bC_i={\rm Var}(\bxi_{ti})$ and $\bC_u=\bI_r$ be the covariance matrix of the innovation $\bu_{ti}$. For weakly stationary VAR(1), it holds that
$$
\begin{aligned}\bC_i=&\bC_u+\bA\bC_u\bA^{\T}+\bA^2\bC_u(\bA^{\T})^2+\cdots=\sum_{s=0}^{\infty}\bA^s\bC_u(\bA^{\T})^s\\
=&\sum_{s=0}^{\infty}(\bA\bA^{\T})^s=\sum_{s=0}^{\infty}\bA^{2s}=(\bI_r-\bA^2)^{-1}.
\end{aligned}
$$
Similarly, $\bSigma_{\varepsilon}(u,v)= \sum_{l=1}^{25}2^{-l}\phi_l(u)\phi_l(v){\rm Var}(\bpsi_{tl})=\sum_{l=1}^{25}3^{-1}2^{2-l}\phi_l(u)\phi_l(v)\bC_{\zeta}$, provided that ${\rm Var}(\bpsi_{tl})=\sum_{s=0}^{\infty}0.5^{2s}\bC_{\zeta}=4\bC_{\zeta}/3$. Hence we have 
$$\bSigma_{y}(u,v)=\bB\bSigma_{f}(u,v)\bB^{\T}+\bSigma_{\varepsilon}(u,v)=\bB\Big\{\sum_{i=1}^{50}i^{-2}\phi_i(u)\phi_i(v)(\bI_r-\bA^2)^{-1}\Big\}\bB^{\T}+\sum_{l=1}^{25}3^{-1}2^{2-l}\phi_l(u)\phi_l(v)\bC_{\zeta}.$$ 
	
\noindent (ii) The desired result follows immediately from the proof of part~(i). 
	
\noindent (iii) To see the functional sparsity condition on $\bSigma_{\varepsilon}$, notice that $$\sigma_i(u)=\bSigma_{\varepsilon,ii}(u,u)=\sum_{l=1}^{25}3^{-1}2^{2-l}\phi_l(u)^2D_i^2(1+\tilde{\delta})\quad{\rm and\quad}\Vert\sigma_i\Vert_{\cN}=\int\sigma_i(u)\du=\tilde{c}D_i^2,$$
where $\tilde{c}=3^{-1}(4-2^{-23})(1+\tilde{\delta})$ is a constant. Then, for $q=0$ in \eqref{eq.sparsity}, we have
$$
\begin{aligned}
s_p=&\max_{i\in[p]}\sum_{j=1}^{p}\Vert\sigma_i\Vert_{\cN}^{(1-q)/2}\Vert\sigma_j\Vert_{\cN}^{(1-q)/2}\Vert\Sigma_{\varepsilon,ij}\Vert_{\cS}^q=\max_{i\in[p]}\sum_{j=1}^{p}\tilde{c}D_iD_jI\{\Vert\bSigma_{\varepsilon,ij}\Vert_{\cS}\neq0\}\\
\le& \big(\tilde{c}\max_{i\in[p]}D_i^2\big)\max_{i\in[p]}\sum_{j=1}^{p}I(\bC_{0,ij} \neq 0) \lesssim \max_{i\in[p]}\sum_{j=1}^{p}I(\breve{\bC}^{\cT}_{ij} \neq 0) \lesssim p^{1-\alpha}.
\end{aligned}
$$
\end{proof}

\section{Further derivations and definitions}
\label{supsec.E}
This section contains further derivations and definitions supporting the main context of the paper.

\subsection{Estimating FFM~(1) from a least squares perspective}
\label{digit.ls}

Similar to Section~\ref{subsec.fun_load}, we develop a least squares method to fit model \eqref{eq.model_1} with functional factors. Let $\bY(\cdot)=\{\by_1(\cdot),\dots,\by_n(\cdot)\}\in\mathbb{R}^{p\times n},$ and $\bF(u)^{\T}=\{\bbf_1(\cdot),\dots,\bbf_n(\cdot)\}\in\mathbb{R}^{r\times n}$. Consider solving the least-squares minimization problem
\begin{equation}
\label{eq.problem.DIGIT}
\arg\min_{\bB,\bF(\cdot)}\int\Vert\bY(u)-\bB\bF(u)^{\T}\Vert_{\F}^2\du=\arg\min_{\bB,\bF(\cdot)}\sum_{t=1}^{n}\Vert\by_t-\bB\bbf_t\Vert^2,
\end{equation}
subject to the normalization $p^{-1}\bB^{\T}\bB=\bI_r$. Following the similar procedure in Section~\ref{subsec.fun_load}, we obtain that, 
for each given $\bB$, the constrained least squares estimator $\widetilde \bF(\cdot)=p^{-1}\bB^{\T}\bY(\cdot).$ Plugging this into \eqref{eq.problem.DIGIT}, objective function becomes $\int{\rm tr}[(\bI_p-p^{-1}\bB\bB^{\T})\bY(u)\bY(u)^{\T}]\du$, whose minimizer is equivalent to the maximizer of ${\rm tr}\{\bB^{\T}[\int\bY(u)\bY(u)^{\T}\du]\bB\}$. Apparently, $\widehat{\bB}/\sqrt{p}$ are the eigenvectors corresponding to the $r$ largest eigenvalues of the $p\times p$ matrix $\int\bY(u)\bY(u)^{\T}\du=n\int\widehat{\bSigma}_y^{\sS}(u,u)\du$. 

For the DIGIT method, the loading matrix $\bB$ is estimated by the eigenanalysis of $\int\int\widehat{\bSigma}_y^{\sS}(u,v)\widehat{\bSigma}_y^{\sS}(u,v)^{\T}\du\dv$, while the above shows that minimizing the least squares criterion \eqref{eq.problem.DIGIT} is equivalent to performing eigenanalysis of $\int\widehat{\bSigma}_y^{\sS}(u,u)\du.$ 
By comparison, the DIGIT method contains more covariance information by taking into account not only the diagonal entries $\widehat{\bSigma}_y^{\sS}(u,u)$ but also the off-diagonal entries $\widehat{\bSigma}_y^{\sS}(u,v)$ for $u \neq v.$ Although such increased information may not alter the convergence rate of the proposed estimator, it will reduce the variance to improve the estimation efficiency.

\subsection{Relationship between two FFMs}
\label{subsec.relationship}
Notice that $\bchi_t(\cdot)=\bB\bbf_t(\cdot)$ and $\bkappa_t(\cdot)=\bQ(\cdot)\bgamma_t$ are the common components of the two FFMs~\eqref{eq.model_1} and \eqref{eq.model_2}, respectively.
The covariance matrix function of $\bchi_t(\cdot)$ is
\begin{equation}
    \label{eq.chi}
    \bSigma_{\chi}(u,v)=\bB\bSigma_{f}(u,v)\bB^{\T}=\bB\Big\{\sum_{i=1}^{\infty}\omega_i\bphi_i(u)\bphi_i(v)^{\T}\Big\}\bB^{\T}=\sum_{i=1}^{\infty}p\omega_i\bpsi_i(u)\bpsi_i(v)^{\T},
\end{equation}
where, by Mercer's theorem, $\bSigma_{f}(u,v)=\sum_{i=1}^{\infty}\omega_i\bphi_i(u)\bphi_i(v)^{\T}$ and $\bpsi_i(\cdot)=\bB\bphi_i(\cdot)/\sqrt{p}$. Suppose that Assumption~\ref{ass.ind3} is satisfied with $\bSigma_{\gamma}={\rm diag}(\widecheck{\vartheta}_1,\dots,\widecheck{\vartheta}_r).$ The covariance matrix function of $\bkappa_t(\cdot)$ is
\begin{equation}
    \label{eq.kappa}
    \bSigma_{\kappa}(u,v)=\bQ(u)\bSigma_{\gamma}\bQ(v)^{\T}=\sum_{j=1}^{r}\widecheck{\vartheta}_j\widecheck{\bq}_j(u)\widecheck{\bq}_j(v)^{\T}=\sum_{j=1}^{r}p\widecheck{\vartheta}_j\bnu_j(u)\bnu_j(v)^{\T},
\end{equation}
where $\{\widecheck{\bq}_j(\cdot)\}_{j=1}^{r}$ is the set of columns of $\bQ(\cdot)$ such that $\{\Vert\widecheck{\bq}_j\Vert\}_{j=1}^{r}$ is in a descending order, and $\bnu(\cdot)=\widecheck{\bq}_j(\cdot)/\sqrt{p}$. Note that
$$
\begin{aligned}&\int\bpsi_i(u)^{\T}\bpsi_j(u)\du=\int\bphi_i(u)^{\T}p^{-1}\bB^{\T}\bB\bphi_j(u)\du=\int\bphi_i(u)^{\T}\bphi_j(u)\du = I(i=j),\ {\rm and}\\
&\int\bnu_i(u)^{\T}\bnu_j(u)\du=p^{-1}\int\widecheck{\bq}_i(u)^{\T}\widecheck{\bq}_j(u)\du = I(i=j)\ {\rm from\ Assumption~\ref{ass.ind3}}.
\end{aligned}
$$
Consequently, $\{\bpsi_i(\cdot)\}_{i=1}^{\infty}$ are the eigenfunctions of $\bSigma_{\chi}$ with nonnegative eigenvalues $\{p\omega_i\}_{i=1}^{\infty}$, and $\{\bnu_j(\cdot)\}_{j=1}^{r}$, which can be extended to a set of orthonormal basis functions, are the eigenfunctions of $\bSigma_{\kappa}$ with nonnegative eigenvalues $\{p\widecheck{\vartheta}_j\}_{j=1}^{\infty}$ satisfying $\widecheck{\vartheta}_j=0$ when $j>r.$ 
	
In this point of view, FFM~\eqref{eq.model_1} can be converted to FFM \eqref{eq.model_2} if and only if $\omega_i=0$ when $i>r$. On the contrary, model \eqref{eq.model_2} can be regarded as a special case of model \eqref{eq.model_1} if and only if the solutions of $\{\bphi_j(\cdot)\}_{j=1}^{r}$ to the functional equations $\bB\bphi_j(\cdot)=\widecheck{\bq}_j(\cdot)$ for $j\in[r]$ exist given $\bB$ and $\{\widecheck{\bq}_j(\cdot)\}_{j=1}^{r}$. Since the rank of the space spanned by columns of matrix $\bB$ is $r$, the equivalent condition for the existence of the solutions follows that the rank of the space spanned by $\{\widecheck{\bq}_j(\cdot)\}_{j=1}^{r}$ is $r.$

\subsection{Sub-Gaussian (functional) linear process}
\label{supsybsec.subgauss}
We first define sub-Gaussian functional process.
\begin{definition}
\label{def.sub}
Let $x_t(\cdot)$ be a mean zero random variable in $\eH$ and $\Sigma_0:\eH\to\eH$ be a covariance operator. Then $x_t(\cdot)$ is a sub-Gaussian process if there exists a constant $\alpha\ge0$ such that for all $x\in\eH$, $\eE(\exp\langle x,x_t\rangle)\le\exp\{\alpha^2\langle x,\Sigma_0(x)\rangle/2\}.$
\end{definition}
To develop finite-sample theory for relevant estimators in Section \ref{sec.theory}, we focus on multivariate functional linear process with sub-Gaussian errors, namely sub-Gaussian functional linear process. Specifically, we assume $\bz_t(\cdot)=\{z_{t1}(\cdot),\dots,z_{tp}(\cdot)\}^{\T}\in \eH^p$ admits the representation
\begin{equation}
\label{flp}
\bz_t(\cdot)=\sum_{l=0}^{\infty}\bA_l(\bx_{t-l}),\ t\in\eZ,
\end{equation}
where $\bA_l=(A_{l,ij})_{p\times p}$ with each $A_{l,ij}\in\eS$ and 
$\bx_t(\cdot)=\{x_{t1}(\cdot),\dots,x_{tp}(\cdot)\}^{\T} \in \eH^p,$ whose components are independent sub-Gaussian processes satisfying Definition \ref{def.sub}, and the coefficient functions satisfy $\sum_{l=0}^{\infty}\Vert\bA_l\Vert_{\cS,\infty}=O(1)$. In Section \ref{sec.theory}, we assume that $\bbf_t(\cdot)$ in model \eqref{eq.model_1} and $\bvarepsilon_t(\cdot)$ follow sub-Gaussian functional linear processes, and $\bgamma_t$ in model \eqref{eq.model_2} follows sub-Gaussian linear process, which can be correspondingly defined from the non-functional versions of (\ref{flp}) and Definition~\ref{def.sub}. 

\subsection{Optimal functional portfolio allocation}
\label{supsec.opt_port}
In this section, we derive the optimal functional portfolio allocation $\widehat{\bw}(\cdot)$ that is required in Section~\ref{sec.real}. Specifically, we aim to solve the following constrained minimization problem:
$$
\widehat \bw = \arg\min_{\bw\in\eH^p}
\big\langle\bw,\widehat{\bSigma}_y(\bw)\big\rangle
~~\text{subject to}~~
\int_{\cU}\bw(u)^{\T}\bone_p\du=1.
$$
To solve this, we apply the method of Lagrange multipliers by defining the Lagrangian function as 
$$
L(\bw,\lambda)=\int_{\cU}\int_{\cU}\bw(u)^{\T}\widehat\bSigma_y(u,v)\bw(v)\du\dv-2\lambda\int_{\cU}\{\bw(u)^{\T}\bone_p-1\}\du,
$$
where $\lambda\in\eR$ is the Lagrange multiplier. Setting the functional derivative of $L(\bw, \lambda)$ with respect to $\bw(\cdot)$ to zero, i.e., $2\int_{\cU}\widehat\bSigma_y(u,v)\widehat{\bw}(v)\dv-2\lambda\cdot\bone_p=0$ for $u\in\cU$, we obtain that 
$$
\widehat{\bw}(u)=\lambda\cdot\widehat{\bSigma}_y^{-1}(\bone_p)(u)=\lambda\int_{\cU}\widehat{\bSigma}_y^{-1}(u,v)\bone_p\dv,~~u\in\cU.
$$
With the constraint $\int_{\cU}\widehat{\bw}(u)^{\T}\bone_p\du=1$ and the use of Moore--Penrose inverse, it yields the desired solution
$$
\widehat{\bw}(u)=\frac{\int_{\cU}\widehat{\bSigma}_y^{\dagger}(u,v)\bone_p\dv}{\int_{\cU}\int_{\cU}\bone_p^{\T}\widehat{\bSigma}_y^{\dagger}(z,v)\bone_p\dz\dv},~~u\in\cU.
$$

\subsection{Optimal portfolio allocation based on POET}
\label{supsec.opt_nonfun}
Instead of modeling the CIDR data as $p$-vector of functional time series, we can treat the data at each intraday time point $u_k$ as $p$-vector time series, i.e., $\{\by_t(u_k)\}_{t\in[n]}$ for $k\in[K],$ where $K$ denotes the number of intraday time points. For each vector time series, we assume the following factor model: 
$$
\by_t(u_k)=\bB_k\bbf_{k,t}+\bvarepsilon_{k,t},~~t\in [n], k\in[K].
$$ 
Then, the standard POET estimator (Fan et al., 2013) can be applied to obtain the estimated covariance matrix of $\by_t(u_k)$ as $\widehat\bSigma_y(u_k,u_k)$. 

To incorporate the non-functional method into our functional risk management formulation, we also need to estimate the cross-covariance matrix $\bSigma_y(u_k,u_l)=\cov\{\by_t(u_k),\by_t(u_l)\}$ for $k \neq l \in [K].$
Assuming that $\cov(\bbf_{k,t},\bvarepsilon_{l,t})=\bzero$ for $k,l\in[K],$ it follows that $\bSigma_y(u_k,u_l)=\bB_k\cov(\bbf_{k,t},\bbf_{l,t})\bB_l+\cov(\bvarepsilon_{k,t},\bvarepsilon_{l,t}).$ Thus, $\bSigma_y(u_k,u_l)$ can be estimated as 
$$
\widehat{\bSigma}_y(u_k,u_l)=\widehat{\bB}_k\big(n^{-1}\sum_{t=1}^{n}\widehat{\bbf}_{k.t}\widehat{\bbf}_{l,t}^{\T}\big)\widehat{\bB}_l^{\T}+n^{-1}\sum_{t=1}^{n}\widehat{\bvarepsilon}_{k.t}\widehat{\bvarepsilon}_{l,t}^{\T},
$$
where $\widehat{\bB}_k,\{\widehat{\bbf}_{k,t}\}_{t\in[n]}$ and $\{\widehat{\bvarepsilon}_{k,t}\}_{t\in[n]}$ are obtained by the POET method. 
With the (cross-)covariance matrix estimates $\widehat{\bSigma}_y(u_k,u_l)$ for $k,l\in[K],$ the optimal portfolio allocation can be obtained by minimizing the perceived portfolio risk:
\begin{equation}
    \label{eq.opt_non}
    \{\widehat{\bw}(u_k)\}_{k\in[K]}=\arg\min_{\{\bw(u_k)\}_{k\in[K]}}\frac{1}{K^2}\sum_{k=1}^{K}\sum_{l=1}^{K}\bw(u_k)^{\T}\widehat{\bSigma}_y(u_k,u_l)\bw(u_l)\ \ {\rm s.t.}\ \ \frac{1}{K}\sum_{k=1}^{K}\bw(u_k)^{\T}\bI_p=1.
\end{equation}
Define the matrix $\widetilde{\bSigma}_y=\big(\widehat{\bSigma}_y(u_k,u_l)\big)\in\eR^{Kp\times Kp}$, whose $(k,l)$-th block is $\widehat{\bSigma}_y(u_k,u_l).$ Following similar derivations as in Section~\ref{supsec.opt_port}, the solution of \eqref{eq.opt_non} is given by
$$
\{\widehat{\bw}(u_1)^{\T},\dots,\widehat{\bw}(u_{K})^{\T}\}^{\T}=\frac{\widetilde{\bSigma}_y^{-1}\bone_{Kp}}{\bone_{Kp}^{\T}\widetilde{\bSigma}_y^{-1}\bone_{Kp}} \in\eR^{Kp}.
$$

\section{Additional simulation results}
\label{supsec.F}
	
\begin{figure}[H]	
\centering
\includegraphics[width=0.75\textwidth]{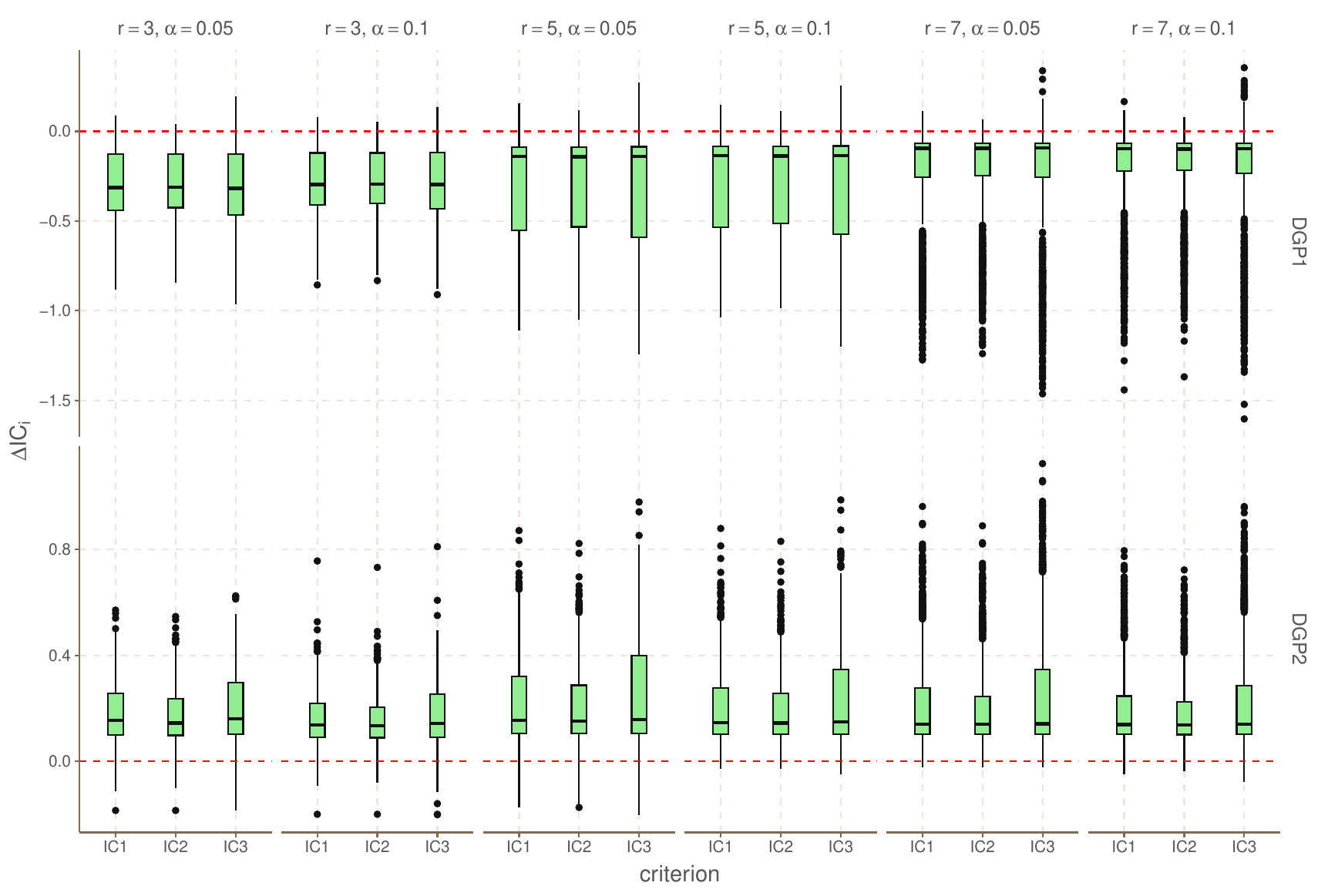}
\caption{\small The boxplots of $\Delta{\rm IC}_i$ ($i \in [3]$) for DGP1 and DGP2 with $p=100,n=50,\alpha=0.05,0.1$ and $r=3,5,7$ over 1000 simulation runs.}
\label{fig.model_select_case2}
\end{figure}

This section provides additional results supporting Section~\ref{sec.sim}. Figure~\ref{fig.model_select_case2} presents boxplots of $\Delta{\rm IC}_i$ ($i \in [3]$) for two DGPs under the setting $p=100,n=50,\alpha=0.05,0.1,$ and $r=3,5,7.$ Table~\ref{tab.S1} reports the model selection accuracies for different values of $\alpha$. The results show that the proposed criteria can select the correct model with high probability even when $\alpha$ and $n$ are relatively small.

\begin{table}[H]
\footnotesize
\centering
\caption{\small The average relative frequency estimates of $\eP\{{\rm IC}^{\cD}(\hat{r}^{\cD})<{\rm IC}^{\cF}(\hat{r}^{\cF})\}$ for DGP1, and $\eP\{{\rm IC}^{\cD}(\hat{r}^{\cD})<{\rm IC}^{\cF}(\hat{r}^{\cF})\}$ for DGP2 with $p=100,n=50,100,\alpha=0.05,0.1,0.25,0.5$ and $r=3,5,7$ over 1000 simulation runs.}
\label{tab.S1}
\vspace{-0.2cm}
\begin{tabular}{llccccccc}
\toprule
			& & & \multicolumn{2}{c}{$r=3$} & \multicolumn{2}{c}{$r=5$} & \multicolumn{2}{c}{$r=7$}\\
			$\alpha$ & $n$ & criterion & DGP1 & DGP2 & DGP1 & DGP2 &
			DGP1 & DGP2  \\
			\hline
			
			\multirow{6}{*}{$0.05$} & \multirow{3}{*}{$50$} & $\Delta{\rm IC}_1$  & 0.990  & 0.977  & 0.984  & 0.988  & 0.977  & 0.998  \\
			&       & $\Delta{\rm IC}_2$ & 0.993  & 0.980  & 0.992  & 0.989  & 0.988  & 0.998  \\
                &       & $\Delta{\rm IC}_3$ & 0.979  & 0.972  & 0.976  & 0.986  & 0.962  & 0.997  \\
			& \multirow{3}{*}{$100$} & $\Delta{\rm IC}_1$  & 0.994  & 0.974  & 0.994  & 0.992  & 0.989  & 0.997  \\
			&       & $\Delta{\rm IC}_2$ & 0.998  & 0.975  & 0.996  & 0.993  & 0.996  & 0.996  \\
                &       & $\Delta{\rm IC}_3$ & 0.993  & 0.971  & 0.991  & 0.992  & 0.985  & 0.995  \\
			\hline
			
			\multirow{6}{*}{$0.10$} & \multirow{3}{*}{$50$} & $\Delta{\rm IC}_1$  & 0.992  & 0.974  & 0.975  & 0.993  & 0.980  & 0.998  \\
			&       & $\Delta{\rm IC}_2$ & 0.998  & 0.976  & 0.984  & 0.996  & 0.984  & 0.997  \\
                &       & $\Delta{\rm IC}_3$ & 0.988  & 0.962  & 0.965  & 0.991  & 0.962  & 0.994  \\
			& \multirow{3}{*}{$100$} & $\Delta{\rm IC}_1$  & 0.996  & 0.974  & 0.995  & 0.990  & 0.995  & 0.991  \\
			&       & $\Delta{\rm IC}_2$ & 1.000  & 0.974  & 0.999  & 0.990  & 1.000  & 0.992  \\
                &       & $\Delta{\rm IC}_3$ & 0.993  & 0.971  & 0.990  & 0.989  & 0.988  & 0.991  \\
			\hline

            \multirow{6}{*}{$0.25$} & \multirow{3}{*}{$50$} & $\Delta{\rm IC}_1$  & 0.996  & 0.983  & 0.980  & 0.995  & 0.973  & 0.996  \\
			&       & $\Delta{\rm IC}_2$ & 0.996  & 0.985  & 0.985  & 0.997  & 0.984  & 0.998  \\
                &       & $\Delta{\rm IC}_3$ & 0.991  & 0.982  & 0.973  & 0.992  & 0.959  & 0.996  \\
			& \multirow{3}{*}{$100$} & $\Delta{\rm IC}_1$  & 0.998  & 0.980  & 0.993  & 0.989  & 0.993  & 0.994  \\
			&       & $\Delta{\rm IC}_2$ & 0.999  & 0.981  & 0.999  & 0.989  & 0.998  & 0.994  \\
                &       & $\Delta{\rm IC}_3$ & 0.997  & 0.978  & 0.991  & 0.989  & 0.987  & 0.992  \\
			\hline

            \multirow{6}{*}{$0.50$} & \multirow{3}{*}{$50$} & $\Delta{\rm IC}_1$  & 0.998  & 0.997  & 0.995  & 1.000  & 0.994  & 1.000  \\
			&       & $\Delta{\rm IC}_2$ & 0.999  & 0.998  & 0.996  & 1.000  & 0.995  & 1.000  \\
                &       & $\Delta{\rm IC}_3$ & 0.998  & 0.997  & 0.993  & 1.000  & 0.990  & 1.000  \\
			& \multirow{3}{*}{$100$} & $\Delta{\rm IC}_1$  & 1.000  & 0.998  & 0.999  & 1.000  & 0.998  & 1.000  \\
			&       & $\Delta{\rm IC}_2$ & 1.000  & 0.998  & 1.000  & 1.000  & 0.999  & 1.000  \\
                &       & $\Delta{\rm IC}_3$ & 1.000  & 0.998  & 0.998  & 1.000  & 0.997  & 1.000  \\
			\bottomrule
		\end{tabular}
  \vspace{-0.2cm}
	\end{table}

\begin{figure}[H]	
\centering
\includegraphics[width=0.75\textwidth]{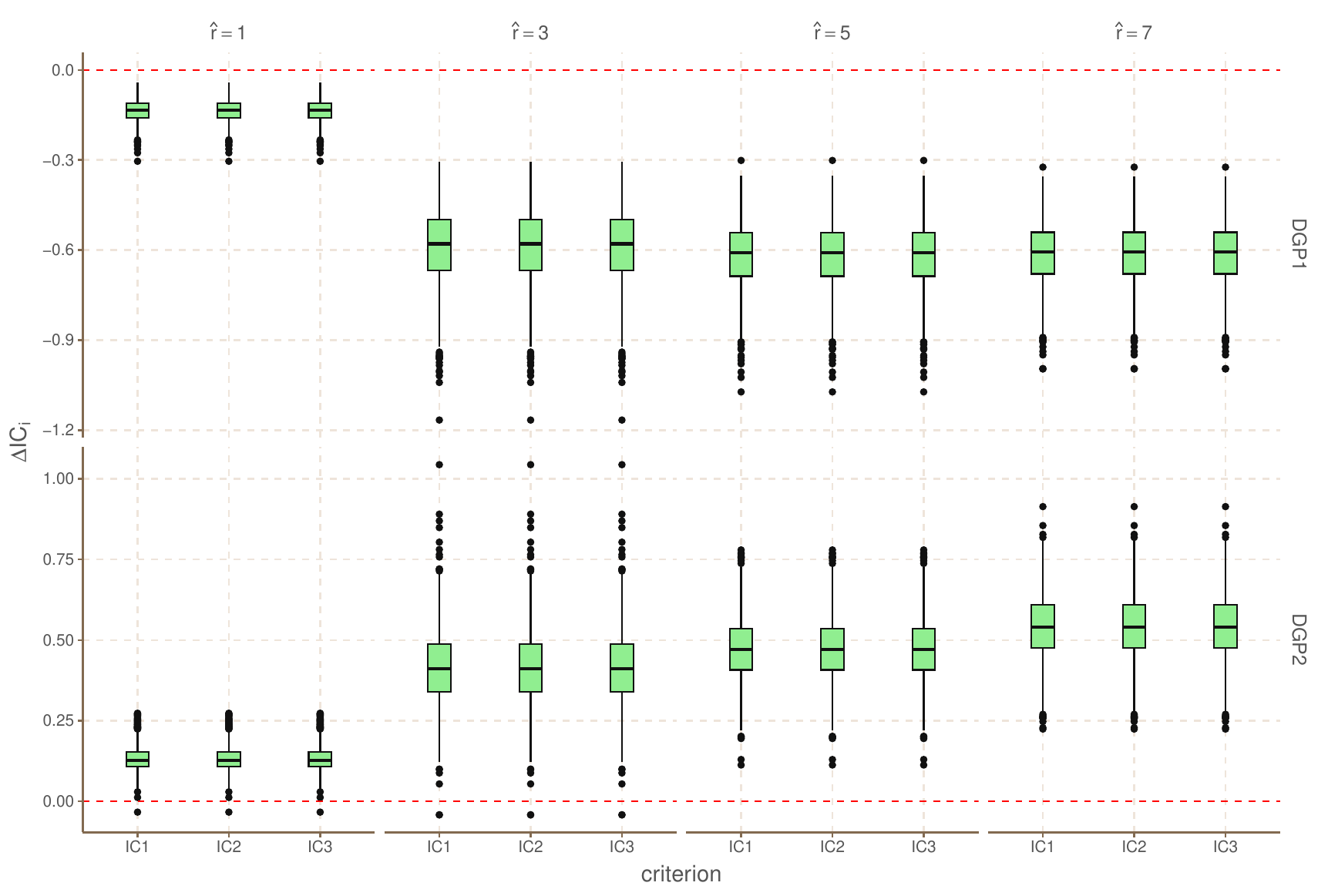}
\caption{\small The boxplots of $\Delta{\rm IC}_i$ ($i \in [3]$) for DGP1 and DGP2 with $p=100,n=50,\alpha=0.5,r=3$, and the number of factors to calculate ${\rm IC}_i^{\cD}$ and ${\rm IC}_i^{\cF}$ being fixed as $\hat{r}=1,3,5,7$ over 1000 simulation runs.}
\label{fig.model_select_case3}
\end{figure}

We conduct additional simulations with data generated using $p=100,n=50,\alpha=0.5$ and $r=3$ for DGP1 and DGP2. The identified number of factors is set to $\hat r=1,3,5,7$ when calculating ${\rm IC}_i^{\cD}$ and ${\rm IC}_i^{\cF},$ with the results reported in Figure~\ref{fig.model_select_case3}. It is observed that the proposed criteria exhibit strong robustness against the misidentification of the number of factors.

The model selection criteria assessed above use the same penalty function when calculating both ${\rm IC}^{\cD}$ and ${\rm IC}^{\cF}$. According to the discussion in Section~\ref{subsec.selection}, different penalty functions from the perspective of model complexity can be applied with $g^{\cD}(p,n)=(p+ns_n-k)/pn$ for ${\rm IC}^{\cD}(k)$ and $g^{\cF}(p,n)=(ps_n+n-k)/pn$ for ${\rm IC}^{\cF}(k)$. When $\cM_{\varepsilon}=o\{\sqrt{\log(p \wedge n)}\}$ and $p\asymp n$, it can be shown that $s_n=\log n$ guarantees that both $g^{\cD}(p,n)$ and $g^{\cF}(p,n)$ satisfy the conditions in Theorem~\ref{thm.model_sele}. Table~\ref{tab.S3} presents the numerical summaries for the model selection accuracies when different penalty functions are employed. The results show that the proposed criteria remain effective.

\begin{table}[H]
\footnotesize
\centering
\caption{\small The average relative frequency estimates of $\eP\{{\rm IC}^{\cD}(\hat{r}^{\cD})<{\rm IC}^{\cF}(\hat{r}^{\cF})\}$ for DGP1, and $\eP\{{\rm IC}^{\cD}(\hat{r}^{\cD})<{\rm IC}^{\cF}(\hat{r}^{\cF})\}$ for DGP2 with $p=100,n=50,100,\alpha=0.05,0.1,0.25,0.5,r=3,5,7$ and different penalty functions for ${\rm IC}^{\cD}$ and ${\rm IC}^{\cF}$ over 1000 simulation runs.}
\label{tab.S3}
\vspace{-0.2cm}
\begin{tabular}{llcccccc}
\toprule
			& & \multicolumn{2}{c}{$r=3$} & \multicolumn{2}{c}{$r=5$} & \multicolumn{2}{c}{$r=7$}\\
			$\alpha$ & $n$ & DGP1 & DGP2 & DGP1 & DGP2 &
			DGP1 & DGP2  \\
			\hline
			
			\multirow{2}{*}{$0.05$} & 50    & 0.994  & 0.945  & 0.971  & 0.982  & 0.961  & 0.988  \\
    & 100   & 0.998  & 0.980  & 0.991  & 0.994  & 0.985  & 0.998  \\
			\hline
			
			\multirow{2}{*}{$0.10$} & 50    & 0.991  & 0.941  & 0.979  & 0.966  & 0.973  & 0.990  \\
    & 100   & 0.991  & 0.971  & 0.993  & 0.986  & 0.987  & 0.996  \\
			\hline

            \multirow{2}{*}{$0.25$} & 50    & 0.995  & 0.956  & 0.980  & 0.987  & 0.972  & 0.992  \\
    & 100   & 0.998  & 0.979  & 0.998  & 0.992  & 0.989  & 0.997  \\
			\hline

            \multirow{2}{*}{$0.50$} & 50    & 0.999  & 0.996  & 0.991  & 0.999  & 0.993  & 1.000  \\
    & 100   & 0.999  & 1.000  & 1.000  & 1.000  & 0.997  & 1.000  \\
			\bottomrule
		\end{tabular}
  \vspace{-0.2cm}
	\end{table}

Then we examine the performance of model selection criteria when there exist moderate to even strong correlations in idiosyncratic components. Specifically, we generate new idiosyncratic components $\widetilde \bvarepsilon_t(\cdot) = c p^{\delta/2} \bvarepsilon_t(\cdot)$ for $\delta \in (0,1)$ and $c=2$ (or 0.6) for DGP1 (or DGP2) which results in $\Vert\bSigma_{\tilde \varepsilon}\Vert_{\cL}=O(p^{\delta}).$ The model selection accuracies are presented in Table~\ref{tab.S5}. The results indicate that the model selection accuracies decrease (particularly noticeable for DGP2) as $\delta$ increases.

\begin{table}[H]
\footnotesize
\centering
\caption{\small The average relative frequency estimates of $\eP\{{\rm IC}^{\cD}(\hat{r}^{\cD})<{\rm IC}^{\cF}(\hat{r}^{\cF})\}$ for DGP1, and $\eP\{{\rm IC}^{\cD}(\hat{r}^{\cD})<{\rm IC}^{\cF}(\hat{r}^{\cF})\}$ for DGP2 with $p=100,n=50,100,\delta=0.25,0.5,\alpha=0.5$ and $r=3,5,7$ over 1000 simulation runs.}
\label{tab.S5}
\vspace{-0.2cm}
\begin{tabular}{llccccccc}
\toprule
			& & & \multicolumn{2}{c}{$r=3$} & \multicolumn{2}{c}{$r=5$} & \multicolumn{2}{c}{$r=7$}\\
			$\delta$ & $n$ & criterion & DGP1 & DGP2 & DGP1 & DGP2 &
			DGP1 & DGP2  \\
			\hline
			
			\multirow{6}{*}{$0.25$} & \multirow{3}{*}{$50$} & $\Delta{\rm IC}_1$  & 0.964  & 0.997  & 0.961  & 1.000  & 0.941  & 1.000  \\
			&       & $\Delta{\rm IC}_2$ & 0.958  & 0.997  & 0.948  & 1.000  & 0.938  & 1.000  \\
                &       & $\Delta{\rm IC}_3$ & 0.809  & 0.997  & 0.832  & 1.000  & 0.827  & 1.000  \\
			& \multirow{3}{*}{$100$} & $\Delta{\rm IC}_1$  & 1.000  & 1.000  & 1.000  & 0.999  & 1.000  & 1.000  \\
			&       & $\Delta{\rm IC}_2$ & 0.998  & 1.000  & 1.000  & 0.999  & 1.000  & 1.000  \\
                &       & $\Delta{\rm IC}_3$ & 0.939  & 1.000  & 0.962  & 0.999  & 0.954  & 1.000  \\
			\hline
			
			\multirow{6}{*}{$0.50$} & \multirow{3}{*}{$50$} & $\Delta{\rm IC}_1$  & 0.956  & 0.767  & 0.925  & 0.878  & 0.892  & 0.920  \\
			&       & $\Delta{\rm IC}_2$ & 0.947  & 0.734  & 0.910  & 0.869  & 0.882  & 0.921  \\
                &       & $\Delta{\rm IC}_3$ & 0.818  & 0.779  & 0.776  & 0.873  & 0.735  & 0.916  \\
			& \multirow{3}{*}{$100$} & $\Delta{\rm IC}_1$  & 0.999  & 0.832  & 0.998  & 0.935  & 0.997  & 0.964  \\
			&       & $\Delta{\rm IC}_2$ & 0.992  & 0.819  & 0.986  & 0.925  & 0.984  & 0.958  \\
                &       & $\Delta{\rm IC}_3$ & 0.947  & 0.829  & 0.946  & 0.917  & 0.941  & 0.946  \\

			\bottomrule
		\end{tabular}
  \vspace{-0.2cm}
	\end{table}

\begin{figure}[H]	
\centering
\includegraphics[width=0.75\textwidth]{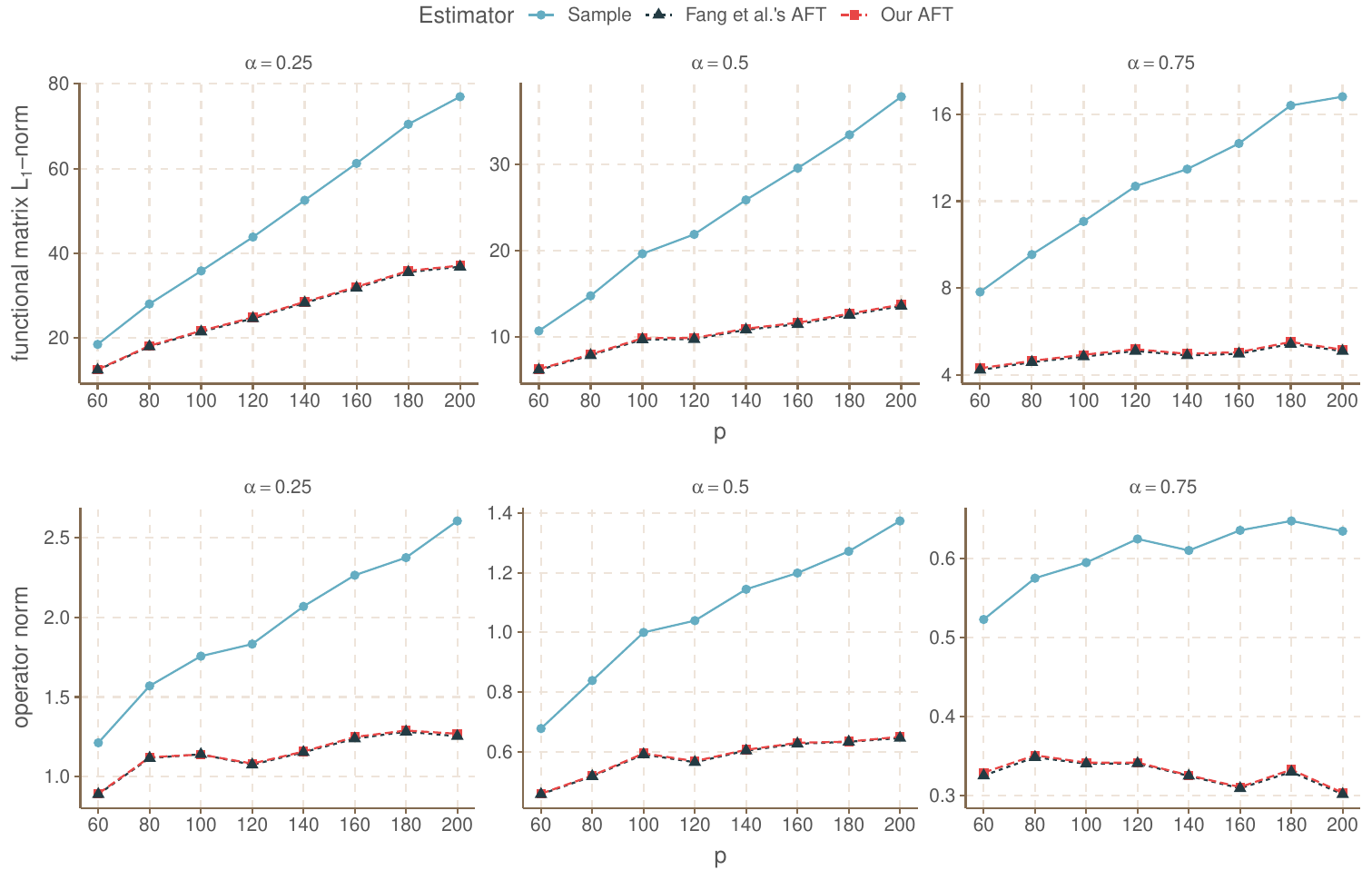}
\caption{\small The average losses of $\widehat{\bSigma}_{\varepsilon}$ in
functional matrix $\ell_1$ norm (top row) and operator norm (bottom row) for DGP1 over 1000 simulation runs with $n=p=60, 80, \dots, 200$ and $\alpha=0.25, 0.5, 0.75.$}
\label{fig.1}
\end{figure}

\begin{figure}[H]	
\centering
\includegraphics[width=0.75\textwidth]{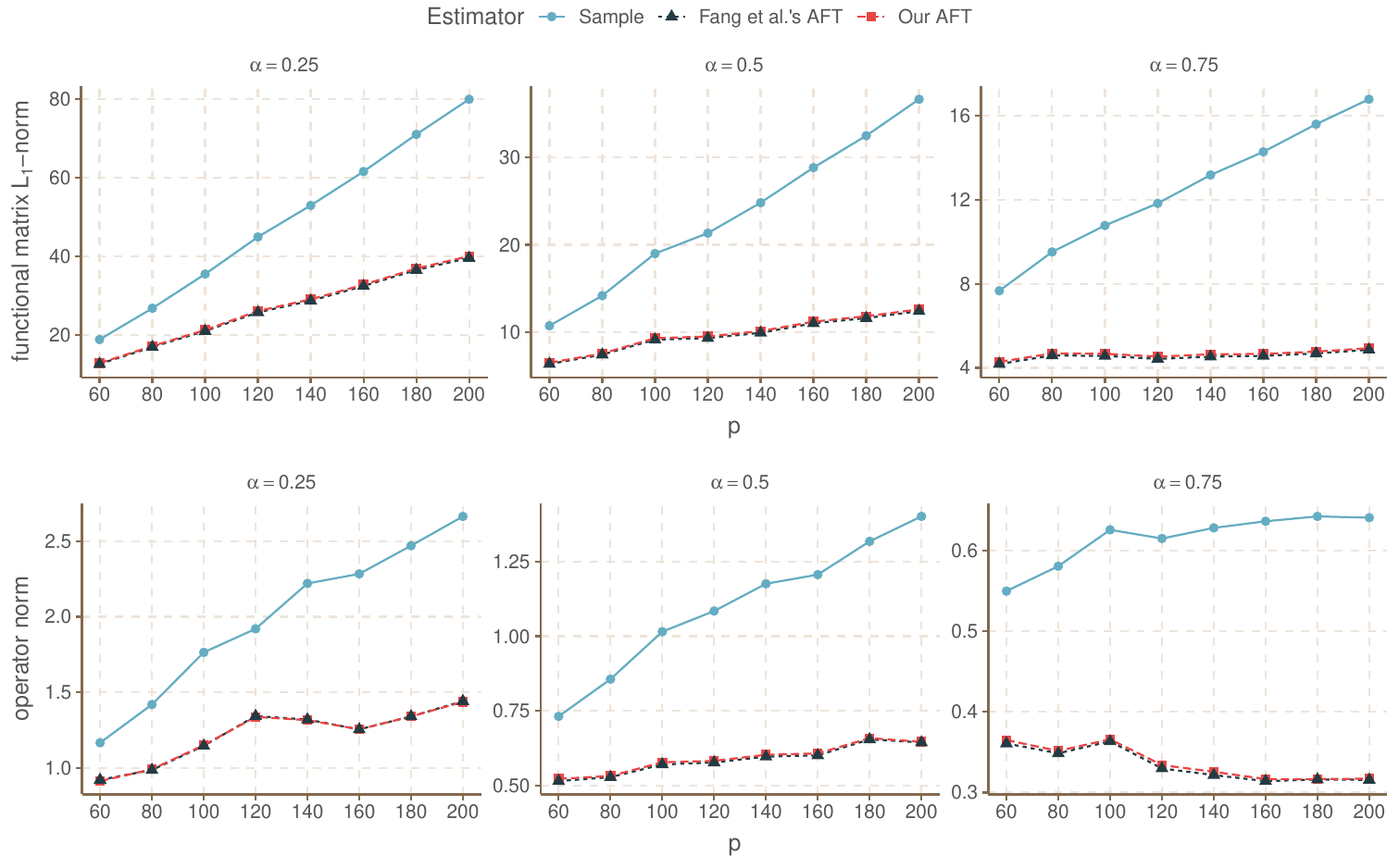}
\caption{\small The average losses of $\widehat{\bSigma}_{\varepsilon}$ in
functional matrix $\ell_1$ norm (top row) and operator norm (bottom row) for DGP2 over 1000 simulation runs with $n=p=60, 80, \dots, 200$ and $\alpha=0.25, 0.5, 0.75.$}
\label{fig.2}
\end{figure}

We compare our AFT estimator in \eqref{eq.thre} with two related methods for estimating the idiosyncratic covariance  $\bSigma_{\varepsilon}$, specifically,
the sample covariance estimator defined as $\widehat{\bSigma}_{\varepsilon}^{\sS}(u,v)=n^{-1}\sum_{t=1}^{n}\widehat{\bvarepsilon}_t(u)\widehat{\bvarepsilon}_t(v)^{\T},$ and \textcolor{blue}{Fang et al.} (\textcolor{blue}{2024})'s AFT estimator in \eqref{eq.thre_2}. 
Figures~\ref{fig.1} and \ref{fig.2} plot average losses of $\widehat\bSigma_{\varepsilon}$ measured by functional matrix $\ell_1$ norm and operator norm for DGP1 and DGP2, respectively, under the settings $n=p=60, 80, \dots, 200$ and $\alpha=0.25, 0.5, 0.75.$
We observe several evident patterns.
First, the estimation accuracy measured by both functional matrix norms substantially improves when using the AFT estimators compared to $\widehat{\bSigma}_{\varepsilon}^{\sS}.$
Second, despite our AFT proposal requiring weaker assumptions compared to \textcolor{blue}{Fang et al.} (\textcolor{blue}{2024})'s method, both AFT estimators exhibit very similar empirical performance. 
Third, for $\alpha=0.25$ and $0.5,$ the performance of
the sample and AFT estimators deteriorates
as $p$ increases. However, when $\alpha=0.75,$ both losses of two AFT estimators do not show significant upward trends.
This phenomenon can be attributed to the fact that $\{(\log p/n)^{1/2}+p^{-1/2}\} p ^{1-\alpha}=o(1)$ as $p,n \to \infty$ if $\alpha>0.5,$ which is implied by Theorems \ref{thm.idio_DIGIT} and \ref{thm.idio_fpoet}
under the setting $n=p, q=0, \cM_{\varepsilon}=O(1).$ 

\begin{figure}[H]	
		\centering
		\includegraphics[width=0.75\textwidth]{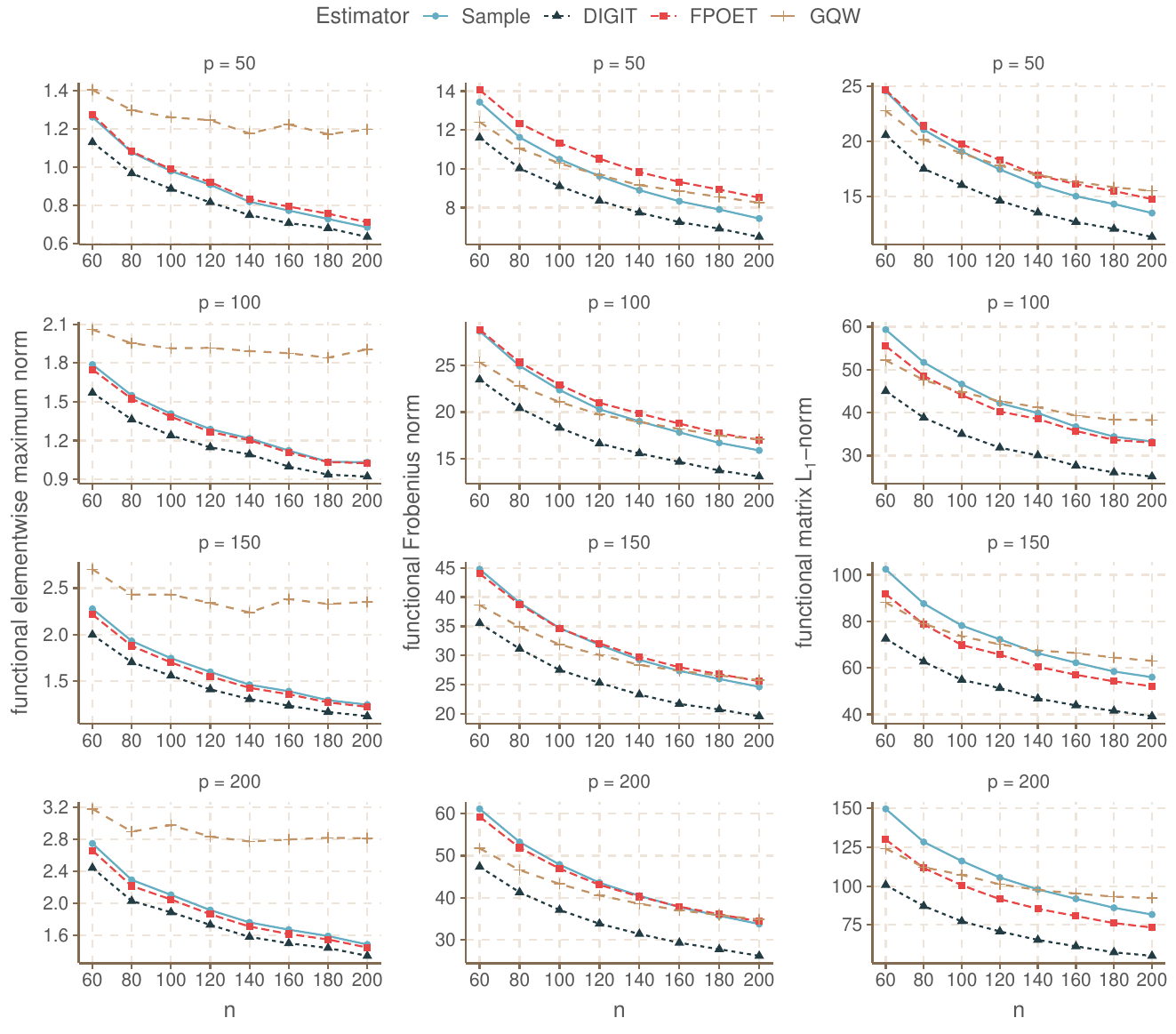}
		\caption{\small The average losses of $\widehat \bSigma_y$ in functional versions of elementwise $\ell_{\infty}$ norm (left column), Frobenius norm (middle column) and matrix $\ell_1$ norm (right column) for DGP1 with $\dot{C}=1$ over 1000 simulation runs.}
		\label{fig.7}
	\end{figure}

\begin{figure}[H]	
		\centering
		\includegraphics[width=0.75\textwidth]{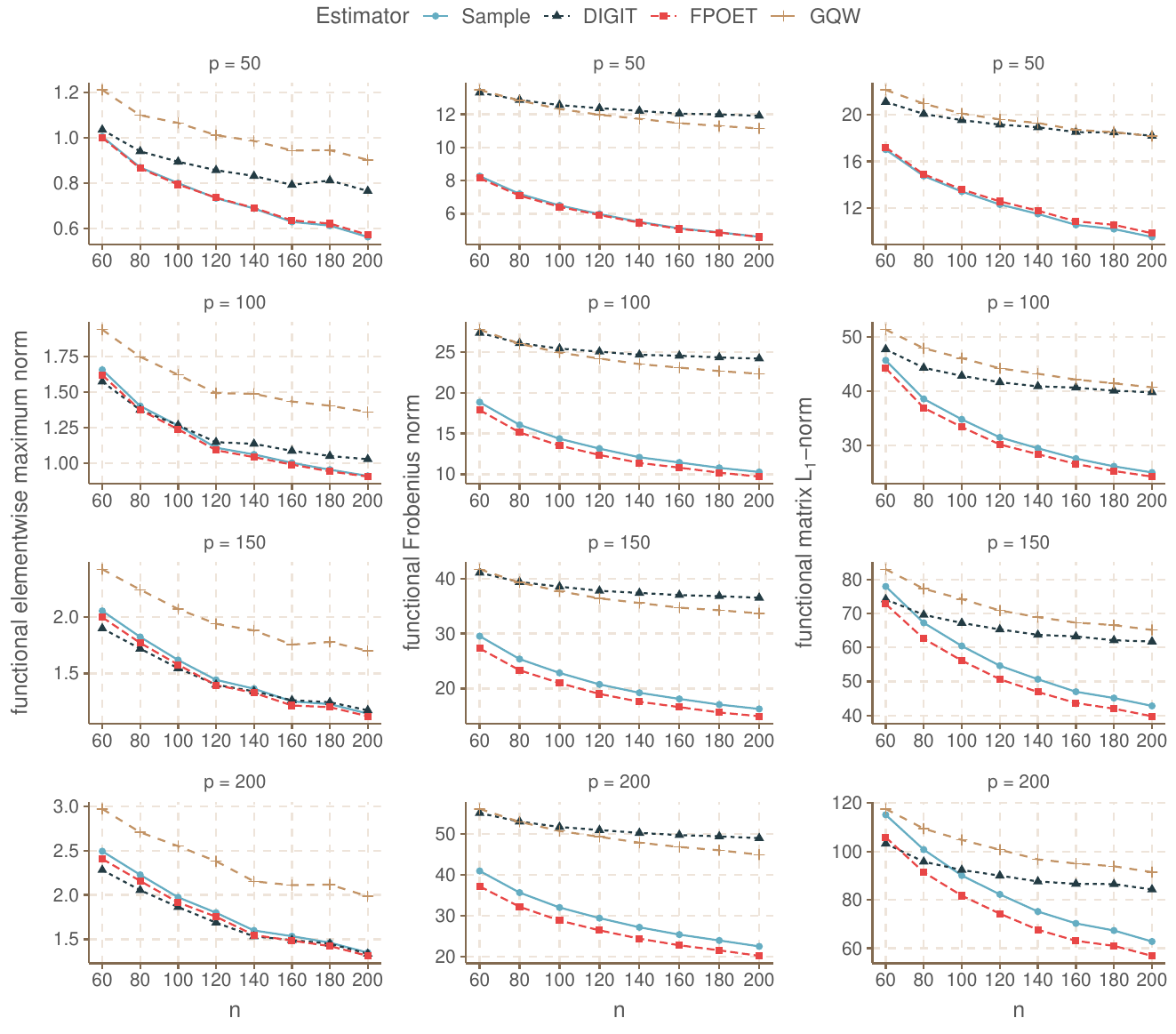}
		\caption{\small The average losses of $\widehat \bSigma_y$ in functional versions of elementwise $\ell_{\infty}$ norm (left column), Frobenius norm (middle column) and matrix $\ell_1$ norm (right column) for DGP2 with $\dot{C}=1$ over 1000 simulation runs.}
		\label{fig.8}
	\end{figure}

Figures~\ref{fig.7} and \ref{fig.8} plot average losses of $\widehat\bSigma_y$ measured by functional versions of elementwise $\ell_{\infty}$ norm, Frobenius norm and matrix $\ell_1$ norm for DGP1 and DGP2, respectively, when $\dot C=1.$ 

\section{Additional real data result}
\label{supsec.G}

Table~\ref{tab.SP100} presents the list of S\&P 100 component stocks used in Section~\ref{sec.real}.

\clearpage
\newgeometry{top=2cm,bottom=2cm, left=0.2cm, right=0.2cm}

\begin{table}[H]
\centering
\caption{\small List of S\&P~100 stocks.}
\label{tab.SP100}
\linespread{1.8}
\tiny
\begin{tabular}{lll|lll}
\toprule
Ticker & Company     &   Sector                  & Ticker & Company     & Sector                  \\ 
\hline
AAPL   & APPLE INC        & Information Technology                    & JPM    & JPMORGAN CHASE \& CO  & Financials             \\
ABBV   & ABBVIE INC       & 	Health Care                   & KHC    & KRAFT HEINZ    & 	Consumer Staples                    \\
ABT    & ABBOTT LABORATORIES    & Health Care              & KMI    & KINDER MORGAN INC     & Energy             \\
ACN    & ACCENTURE PLC CLASS A    & 	Information Technology            & KO     & COCA-COLA      & Consumer Staples                    \\
AGN    & ALLERGAN           & Health Care                  & LLY    & ELI LILLY        & Health Care                  \\
AIG    & AMERICAN INTERNATIONAL GROUP INC  & Financials     & LMT    & LOCKHEED MARTIN CORP   & Industrials            \\
ALL    & ALLSTATE CORP        & Financials                & LOW    & LOWES COMPANIES INC      & Consumer Discretionary          \\
AMGN   & AMGEN INC            & 	Health Care                & MA     & MASTERCARD INC CLASS A     & 	Information Technology        \\
AMZN   & AMAZON COM INC       & Consumer Discretionary                & MCD    & MCDONALDS CORP    & Consumer Discretionary                 \\
AXP    & AMERICAN EXPRESS     & Financials                & MDLZ   & MONDELEZ INTERNATIONAL INC CLASS A & Consumer Staples \\
BA     & BOEING               & Industrials                & MDT    & MEDTRONIC PLC           & Health Care           \\
BAC    & BANK OF AMERICA CORP   & 	Financials              & MET    & METLIFE INC          & 	Financials              \\
BIIB   & BIOGEN INC         & Health Care              & MMM    & 3M                & Industrials                 \\
BK     & BANK OF NEW YORK MELLON CORP    & Financials     & MO     & ALTRIA GROUP INC     & Consumer Staples              \\
BLK    & BLACKROCK INC       & Financials                & MON    & MONSANTO      & Materials                     \\
BMY    & BRISTOL MYERS SQUIBB   & Health Care              & MRK    & MERCK \& CO INC        & 	Health Care            \\
C      & CITIGROUP INC           & 	Financials             & MS     & MORGAN STANLEY       & 	Financials              \\
CAT    & CATERPILLAR INC        & 	Industrials              & MSFT   & MICROSOFT CORP    & 	Information Technology                 \\
CELG   & CELGENE CORP        & Health Care                & NEE    & NEXTERA ENERGY INC       & 	Utilities          \\
CHTR   & CHARTER COMMUNICATIONS INC CLASS A  & 	Communication Services  & NKE    & NIKE INC CLASS B    & 	Consumer Discretionary               \\
CL     & COLGATE-PALMOLIVE      & Consumer Staples              & ORCL   & ORACLE CORP      & Information Technology                  \\
COF    & CAPITAL ONE FINANCIAL CORP  & Financials         & OXY    & OCCIDENTAL PETROLEUM CORP   & Energy       \\
COP    & CONOCOPHILLIPS         & Energy              & PCLN   & THE PRICELINE GROUP INC  & Communication Services          \\
COST   & COSTCO WHOLESALE CORP   & 	Consumer Staples             & PEP    & PEPSICO INC          & Consumer Staples              \\
CSCO   & CISCO SYSTEMS INC     & Information Technology               & PFE    & PFIZER INC     & 	Health Care                    \\
CVS    & CVS HEALTH CORP      & Health Care                & PG     & PROCTER \& GAMBLE      & 	Consumer Staples            \\
CVX    & CHEVRON CORP        & Energy                 & PM     & PHILIP MORRIS INTERNATIONAL INC & Consumer Staples   \\
DHR    & DANAHER CORP         & 	Health Care                & PYPL   & PAYPAL HOLDINGS INC     & 	Information Technology           \\
DIS    & WALT DISNEY          & 	Communication Services                & QCOM   & QUALCOMM INC     & 	Information Technology                 \\
DUK    & DUKE ENERGY CORP     & Utilities                & RTN    & RAYTHEON      & Industrials                     \\
EMR    & EMERSON ELECTRIC    & Industrials                 & SBUX   & STARBUCKS CORP      & 	Consumer Discretionary               \\
EXC    & EXELON CORP       & Utilities                   & SLB    & SCHLUMBERGER NV    & Energy                \\
F      & F MOTOR          & 	Consumer Discretionary                    & SO     & SOUTHERN      & Utilities                     \\
FB     & FACEBOOK CLASS A INC     & Communication Services            & SPG    & SIMON PROPERTY GROUP REIT INC  & Real Estate    \\
FDX    & FEDEX CORP        & Industrials                   & T      & AT\&T INC        & 	Communication Services                  \\
FOX    & TWENTY-FIRST CENTURY FOX INC CLASS B   &  Communication Services & TGT    & TARGET CORP    & Consumer Discretionary                    \\
FOXA   & TWENTY-FIRST CENTURY FOX INC CLASS A &   Communication Services &TWX    & TIME WARNER INC      & Communication Services              \\
GD     & GENERAL DYNAMICS CORP    & Industrials            & TXN    & TEXAS INSTRUMENT INC    & Information Technology           \\
GE     & GENERAL ELECTRIC     & Industrials                & UNH    & UNITEDHEALTH GROUP INC     & Health Care        \\
GILD   & GILEAD SCIENCES INC    & Health Care              & UNP    & UNION PACIFIC CORP     & Industrials            \\
GM     & GENERAL MOTORS      & Consumer Discretionary                 & UPS    & UNITED PARCEL SERVICE INC CLASS B  & Industrials \\
GOOG   & ALPHABET INC CLASS C  & Communication Services               & USB    & US BANCORP        & Financials                 \\
GS     & GOLDMAN SACHS GROUP INC   & Financials           & UTX    & UNITED TECHNOLOGIES CORP    & Industrials       \\
HAL    & HALLIBURTON       & Energy                   & V      & VISA INC CLASS A      & 	Information Technology             \\
HD     & HOME DEPOT INC     & 	Consumer Discretionary                  & VZ     & VERIZON COMMUNICATIONS INC    & 	Communication Services     \\
HON    & HONEYWELL INTERNATIONAL INC  & Industrials        & WBA    & WALGREEN BOOTS ALLIANCE INC   & Health Care     \\
IBM    & INTERNATIONAL BUSINESS MACHINES CO & 	Information Technology  & WFC    & WELLS FARGO  & Financials                      \\
INTC   & INTEL CORPORATION CORP  & Information Technology             & WMT    & WALMART STORES INC  & Consumer Staples               \\
JNJ    & JOHNSON \& JOHNSON     & 	Health Care              & XOM    & EXXON MOBIL CORP  & Energy                 \\ 
\bottomrule
\end{tabular}
\end{table}

\clearpage
\restoregeometry
\section*{References}
\begin{description}
    \item 
    Bai, J. (2003). Inferential theory for factor models of large dimensions, {\it Econometrica} {\bf 71}: 135–171.

    \item 
    Bathia, N., Yao, Q. and Ziegelmann, F. (2010). Identifying the finite dimensionality of curve time series, {\it The Annals of Statistics} {\bf 38}: 3352 - 3386.

    \item 
    Cai, T. and Liu, W. (2011). Adaptive thresholding for sparse covariance matrix estimation, {\it Journal of the American Statistical Association} {\bf 106}: 672–684.

    \item 
    Carlsson, M. (2021). Von Neumann’s trace inequality for Hilbert–Schmidt operators, {\it Expositiones Mathematicae} {\bf 39}: 149–157.

    \item
    Fan, J., Liao, Y. and Mincheva, M. (2013). Large covariance estimation by thresholding principal orthogonal complements, {\it Journal of the Royal Statistical Society: Series B} {\bf 75}: 603–680.

    \item 
    Fang, Q., Guo, S. and Qiao, X. (2022). Finite sample theory for high-dimensional functional/scalar time series with applications, {\it Electronic Journal of Statistics} {\bf 16}: 527–591.

    \item 
    Fang, Q., Guo, S. and Qiao, X. (2024). Adaptive functional thresholding for sparse covariance function estimation in high dimensions, {\it Journal of the American Statistical Association} {\bf 119}: 1473–1485.

    \item
    Hsing, T. and Eubank, R. (2015). {\it Theoretical Foundations of Functional Data Analysis, with an Introduction to Linear Operators}, John Wiley \& Sons, Chichester.

    \item 
    Zhang, D. and Wu, W. B. (2021). Convergence of covariance and spectral density estimates for high-dimensional locally stationary processes, {\it The Annals of Statistics} {\bf 49}: 233–254.

\end{description}

\end{document}